\documentclass{aastex61}
\usepackage{subfigure}

\submitjournal{ApJS}

\shorttitle{The \ion{H}{2} Regions between $l=207.7\arcdeg$ and $l=211.7\arcdeg$}
\shortauthors{Li et al.}

\begin{document}

\title{The Molecular Clouds associated with the \ion{H}{2} Regions/Candidates between $l=207.7\arcdeg$ and $l=211.7\arcdeg$ }

\correspondingauthor{Chong Li}
\email{chongli@pmo.ac.cn}

\author{Chong Li}
\affil{Purple Mountain Observatory and Key Laboratory of Radio Astronomy, Chinese Academy of Sciences, 10 Yuanhua Road, Nanjing 210023, China}
\affil{University of Chinese Academy of Sciences, 19A Yuquan Road, Shijingshan District, Beijing 100049, China}

\author{Hongchi Wang}
\affil{Purple Mountain Observatory and Key Laboratory of Radio Astronomy, Chinese Academy of Sciences, 10 Yuanhua Road, Nanjing 210023, China}
\affil{School of Astronomy and Space Science, University of Science and Technology of China, 96 Jinzhai Road, Hefei 230026, China}

\author{Miaomiao Zhang}
\affil{Purple Mountain Observatory and Key Laboratory of Radio Astronomy, Chinese Academy of Sciences, 10 Yuanhua Road, Nanjing 210023, China}

\author{Yuehui Ma}
\affil{Purple Mountain Observatory and Key Laboratory of Radio Astronomy, Chinese Academy of Sciences, 10 Yuanhua Road, Nanjing 210023, China}
\affil{University of Chinese Academy of Sciences, 19A Yuquan Road, Shijingshan District, Beijing 100049, China}

\author{Lianghao Lin}
\affil{Purple Mountain Observatory and Key Laboratory of Radio Astronomy, Chinese Academy of Sciences, 10 Yuanhua Road, Nanjing 210023, China}
\affil{School of Astronomy and Space Science, University of Science and Technology of China, 96 Jinzhai Road, Hefei 230026, China}



\begin{abstract}

Using the PMO-13.7 m millimeter telescope at Delingha in China, we have conducted a large-scale simultaneous survey of $^{12}$CO, $^{13}$CO, and C$^{18}$O $J=1-0$ emission toward the sky region centered at $l$=$209.7\arcdeg$, $b$=$-$2.25$\arcdeg$ with a coverage of $4.0\arcdeg \times 4.5\arcdeg$. The majority of the emission in the region comes from the clouds with velocities lying in the range from $-$3 km s$^{-1}$ to 55 km s$^{-1}$, at kinematic distances from 0.5 kpc to 7.0 kpc. The molecular clouds in the region are concentrated into three velocity ranges. The molecular clouds associated with the ten \ion{H}{2} regions/candidates are identified and their physical properties are presented. Massive stars are found within Sh2-280, Sh2-282, Sh2-283, and BFS54, and we suggest them to be the candidate excitation sources of the \ion{H}{2} regions. The distributions of excitation temperature and line width with the projected distance from the center of \ion{H}{2} region/candidate suggest that the majority of the ten \ion{H}{2} regions/candidates and their associated molecular gas are three-dimensional structures, rather than two-dimensional structures.

\end{abstract}




\section{Introduction} \label{sec:intro}

Large-scale surveys of molecular clouds in the Milky Way revealed that the major part of molecular clouds is accumulated into cloud complexes called giant molecular clouds (GMCs) \citep{1985ApJ...295..422C,1987ApJ...322..706D,1998ApJS..115..241H,2001ApJ...547..792D,2006ApJS..163..145J,2013PASA...30...44B}. GMCs have a hierarchical and complex structure which can be divided into substructures as clouds, clumps, and cores \citep{1999ASIC..540....3B}. Previous studies found that GMCs are gravitationally bound while the constituent clumps of GMCs and isolated molecular clouds with M $<$ $10^3$ M$_\odot$ are not in self-gravitational equilibrium \citep{1987ApJ...319..730S, 2001ApJ...551..852H}. The mass function of molecular clouds follows a power law, $dN/dM \propto M^{\gamma}$, with an index of around $-$1.7 \citep{1997ApJ...476..166W, 2005PASP..117.1403R} and there exists a scaling relation between the line-width and the size of molecular clouds \citep{1981MNRAS.194..809L}.

Massive stars are influential in Galactic evolution by ionizing and dynamically affecting the interstellar medium and also by chemically enriching heavy elements. Stellar feedback, in particular that from young OB clusters, has strong influence on the evolution of molecular clouds through the expansion of \ion{H}{2} regions \citep{2008ApJ...681.1341W,2010ApJ...716.1478W}, stellar winds \citep{2006ApJ...649..759C}, and the supernova events \citep{2015Sci...347..526M}. The stellar feedback on the surrounding interstellar medium may trigger the formation of the new generation of stars. Recent studies show that the formation of 14$\%$ to 22$\%$ massive young stellar objects (YSOs) in the Milky Way may be triggered by the expanding \ion{H}{2} regions \citep{2012ApJ...755...71K}. Through searching for the characteristic mid-infrared (MIR) ring-like morphology,\citet{2014ApJS..212....1A} identified 8399 Galactic \ion{H}{2} regions and \ion{H}{2} region candidates, which is the most complete catalog of massive star forming regions in the Milky Way. However, the physical properties of molecular clouds in \ion{H}{2} regions are still unclear. For example, \cite{2010ApJ...709..791B} found that the molecular gas around the infrared bubbles created by young massive stars lies in a ring, rather than a sphere, whereas \cite{2011ApJS..194...32A} showed that the majority of the bubbles in their sample are three-dimensional structures. Meanwhile, the dynamic effect of \ion{H}{2} regions on the surrounding molecular gas is an important factor to understand the origins of turbulence in molecular clouds and the role of triggered star formation \citep{2011ApJ...742..105A,2017ApJ...849..140X}. Observations of molecular line emission are essential to address these important questions. To investigate the spatial distribution of molecular gas in \ion{H}{2} regions and the dynamical interaction between \ion{H}{2} regions and molecular clouds, we have conducted a large-scale survey of $^{12}$CO, $^{13}$CO, and C$^{18}$O $J=1-0$ emission toward the region of Galactic longitude of 207.7$^{\circ}$ $< l <$ 211.7$^{\circ}$ and Galactic latitude of $-$4.5$^{\circ}$ $< b <$ 0$^{\circ}$ (4.0\arcdeg $\times$ 4.5\arcdeg).

The region surveyed in this work did not obtain much attention in previous surveys of molecular clouds. \citet{1953ApJ...118..362S,1959ApJS} and \citet{1982ApJS...49..183B} identified four \ion{H}{2} regions (Sh2-280, Sh2-282, Sh2-283, and BFS54) and \citet{2014ApJS..212....1A} identified six \ion{H}{2} region candidates in this field. Five of the ten \ion{H}{2} regions/candidates mentioned above are spatially coincident with radio continuum emission. Sh2-282, also called LBN 978, is located near the OB-association Mon OB 2 and is a curved nebula. The O9.7Ib star HD 47432 \citep{2001KFNT...17..409K,2011ApJS..193...24S} has been proposed as the ionizing source of Sh2-282 \citep{1959ApJS,1981A&A...100...28F}. Several brightened-rims faced to the exciting star are identified in this region \citep{2006A&A...445L..43C}. No remarkable molecular clouds are found in this region in previous large-scale low sensitivity surveys \citep{2001ApJ...547..792D,2004PASJ...56..313K}. BFS54 is located in an isolated molecular cloud of a few thousand solar mass \citep{1986ApJ...303..375M}. Although BFS54 is about 3$^{\circ}$ away from Mon OB2, it is probably associated with this OB association. BFS54 is catalogued in surveys for outer Galactic \ion{H}{2} regions \citep{1982ApJS...49..183B,1984NInfo..56...59A,1993ApJS...86..475F,1995AZh....72..168K} and is also listed as a reflection nebula (NGC 2282) \citep{1966AJ.....71..990V,1968AJ.....73..233R,1980ApJ...237..734K,1984A&A...135L..14C}. BFS54 hosts a star cluster which was first studied by \citet{1997AJ....113.1788H} with near-infrared (NIR) data. Based on the optical and near-infrared color-magnitude diagrams and disc fraction ($\sim$58 percent) of stars, the age of the BFS54 star cluster was determined to be $2-5$ Myr \citep{2018MNRAS.476.2813D} and the masses of the YSOs are $0.1-2.0$ M$_{\odot}$ \citep{2015MNRAS.454.3597D}.


 This paper is organized as follows. The survey is described in Section \ref{observation} and the results are presented in Section \ref{results}. We discuss our results in Section \ref{discussion} and present a summary in Section \ref{summary}.


\section{Observations and Data Reduction}\label{observation}
\subsection{PMO-13.7 m CO Data}

The sky region has been observed as part of the Milky Way Imaging Scroll Painting (MWISP \footnote{\url{http://www.radioast.nsdc.cn/mwisp.php}}) project which aims to survey molecular gas along the northern Galactic plane. The simultaneous observations of $^{12}$CO, $^{13}$CO, and C$^{18}$O $J=1-0$ emission presented in this work were carried out from 2012 January to 2016 June using the PMO-13.7 m millimeter telescope at Delingha in China. A superconducting spectroscopic array receiver (SSAR) containing 3 $\times$ 3 beams was used as the front-end. The receiver is a two sideband Superconductor-Insulator-Superconductor (SIS) mixer. A specific local oscillator (LO) frequency was carefully selected so that the upper sideband is centered at the $^{12}$CO $J=1-0$ line while the lower sideband covers the $^{13}$CO and C$^{18}$O $J=1-0$ lines \citep{2012ITTST...2..593S}. For each sideband, a Fast Fourier Transform Spectrometer (FFTS) containing 16384 channels with a bandwidth of 1 GHz was used as the back-end. The effective spectral resolution of each FFTS is 61.0 KHz, corresponding to a velocity resolution of 0.16 km s$^{-1}$ at the 115 GHz frequency of the $^{12}$CO $J=1-0$ line. The observations were conducted in the position-switched on-the-fly (OTF) mode with a scanning rate of 50$\arcsec$ per second and a dump time of 0.3 seconds. The survey area is split into cells of the size of 30$\arcmin$ $\times$ 30$\arcmin$. Each cell was scanned at least twice, once along the Galactic longitude and once along the Galactic latitude, to reduce scanning effects. The pointing of the telescope has an RMS accuracy of about 5$\arcsec$ and the beam widths are about 55\arcsec and 52\arcsec at 110 GHz and 115 GHz, respectively.

The Data are processed using the CLASS package of the GILDAS \footnote{\url{http://www.iram.fr/IRAMFR/GILDAS}} software. The raw data are re-grided and converted to FITS files. All FITS files related to the same survey cells are then combined to produce the final FITS data cubes. The spatial pixel of the FITS data cube has a size of 30$\arcsec$ $\times$ 30$\arcsec$. The antenna temperature ($T_A$) is converted to the main-beam temperature with the relation $T_{mb}$ = $T_A$/$B_{eff}$, where the beam efficiency $B_{eff}$ is 0.51 at 115 GHz and 0.56 at 110 GHz according to the status report of the PMO 13.7 m telescope. The calibration accuracy is estimated to be within 10$\%$. The typical system temperature during the observation is about 350 K for the upper sideband and 250 K for the lower sideband. The sensitivity of our observation is estimated to be around 0.5 K for the $^{12}$CO $J=1-0$ emission and around 0.3 K for the $^{13}$CO and C$^{18}$O $J=1-0$ emission. Throughout this paper, all velocities are given with respect to the local standard of rest (LSR).

\subsection{Archival Data}

The complementary infrared data used in this work were obtained from the Wide-field Infrared Survey Explorer ($WISE$) \citep{2010AJ....140.1868W}. $WISE$ covers the entire sky in four photometric bands: W1 (3.4 $\mu$m), W2 (4.6 $\mu$m), W3 (12 $\mu$m), and W4 (22 $\mu$m), at angular resolutions of 6.1$\arcsec$, 6.4$\arcsec$, 6.5$\arcsec$, and 12$\arcsec$, respectively. The 5$\sigma$ point-source sensitivities at the four bands are 0.08, 0.11, 1, and 6 mJy, respectively. The $WISE$ data were retrieved from the NASA/IPAC Infrared Science Archive (IRSA)\footnote{\url{http://irsa.ipac.caltech.edu/frontpage/}}. Data for the ionized emission used in this work are taken from the NVSS 1.4 GHz map \citep{1998AJ....115.1693C}. We also make use of data products from the Southern H$\alpha$ Sky Survey Atlas (SHASSA) \citep{2001PASP..113.1326G}, which consists of 2168 images covering 542 fields south of declination 15$^{\circ}$.

\section{Results}\label{results}

\subsection{Overall Distribution of Molecular Clouds in the Region}

\begin{figure}[h] 
  \centering
  \includegraphics[width=0.35\textwidth,angle=270]{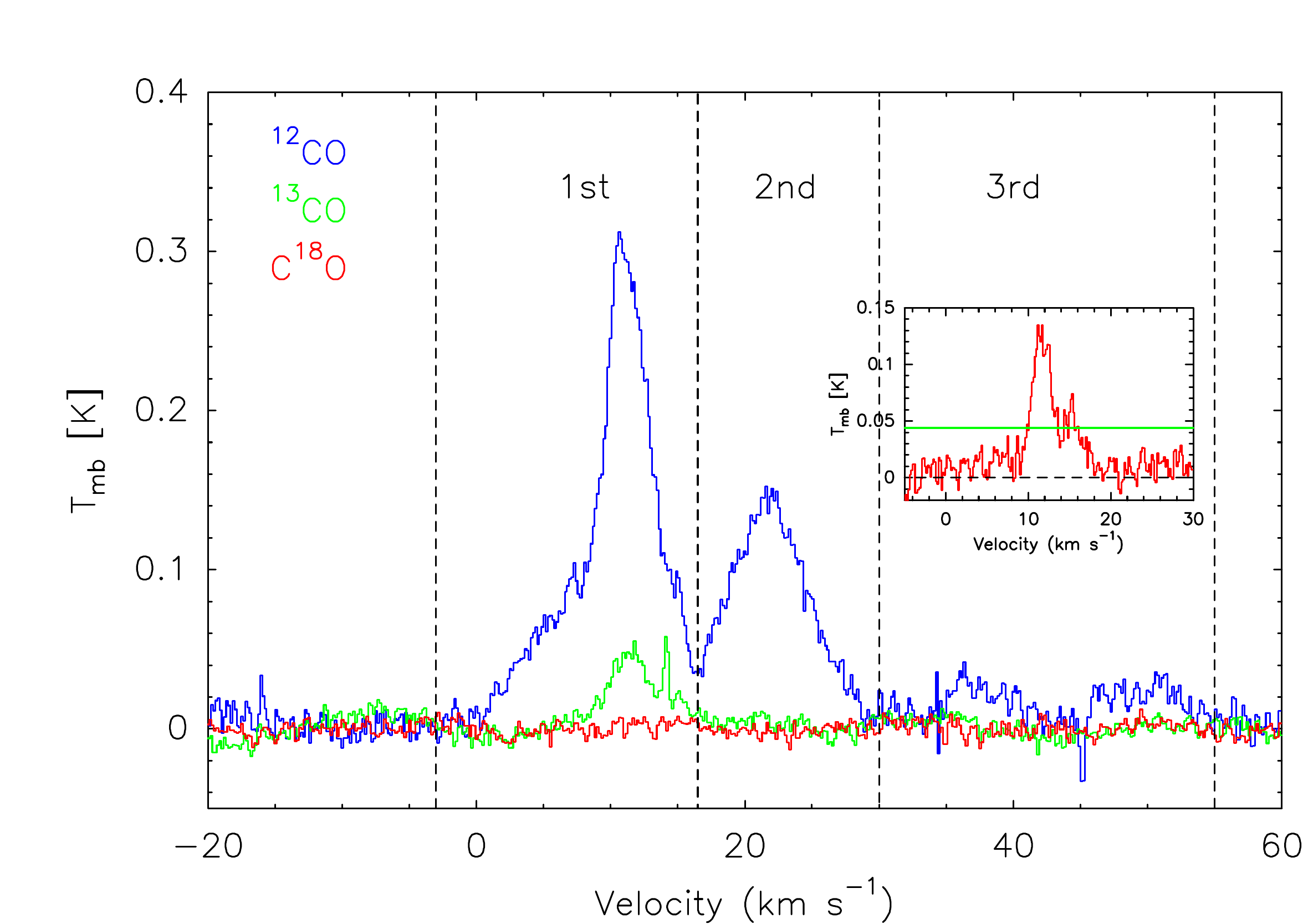}
  \caption{Average spectra of the 4.0$^{\circ}$ $\times$ 4.5$^{\circ}$ region with the blue, green, and red indicating the $^{12}$CO, $^{13}$CO and C$^{18}$O emission, respectively. The velocity range of the $^{12}$CO emission is roughly divided into three components ($-$3 to 16.5, 16.5 to 30, and 30 to 55 km s$^{-1}$), which are indicated with vertical dashed lines. The inset shows the average of the C$^{18}$O spectra in the 4.0$^{\circ}$ $\times$ 4.5$^{\circ}$ region that have at least three contiguous channels with C$^{18}$O emission above 3$\sigma$. The green line in the inset shows the 3$\sigma$ noise level of the average spectrum. From this average spectrum, it can be seen that the C$^{18}$O emission is detected in the velocity range from 10 to 16 km s$^{-1}$.}
  \label{fig:avespec}
\end{figure}

Figure \ref{fig:avespec} shows the average spectra of the CO, $^{13}$CO, and C$^{18}$O $J=1-0$ emission toward the sky region of Galactic longitude of 207.7$^{\circ}$ $< l <$ 211.7$^{\circ}$ and Galactic latitude of $-$4.5$^{\circ}$ $< b <$ 0$^{\circ}$. Among the spectra, the $^{12}$CO emission shows the highest brightness temperature while C$^{18}$O shows the lowest. As shown in Figure \ref{fig:avespec}, the $^{12}$CO average spectrum can be divided into three velocity components, i.e., $-$3 km s$^{-1}$ to 16.5 km s$^{-1}$ (first velocity component), 16.5 km s$^{-1}$ to 30 km s$^{-1}$ (second velocity component), and 30 km s$^{-1}$ to 55 km s$^{-1}$ (third velocity component). Most of the molecular clouds have velocities ranging from $-$3 km s$^{-1}$ to 30 km s$^{-1}$ (first and second velocity components). The major emission of $^{13}$CO has velocities ranging from 5 km s$^{-1}$ to 16.5 km s$^{-1}$. Comparatively, the average C$^{18}$O spectrum of this region is weak. The inset in Figure \ref{fig:avespec} shows the average of the C$^{18}$O spectra. The C$^{18}$O emission is detected mainly in the eastern part of the Rosette Molecular Cloud in the velocity range from 10 to 16 km s$^{-1}$, but rarely in other part of the surveyed area.

\begin{figure}[h] 
  \centering
  \includegraphics[width=0.45\textwidth]{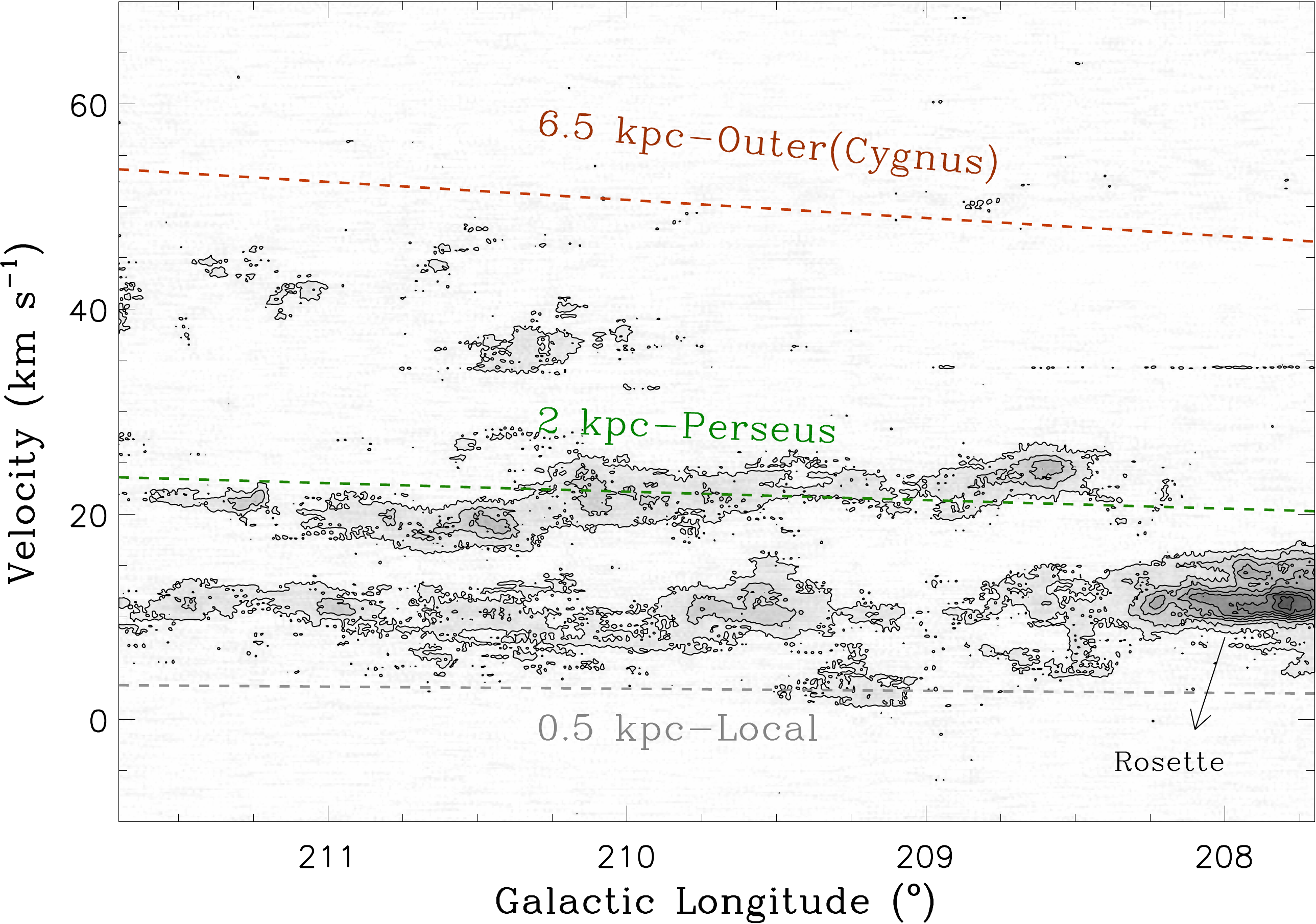}
  \caption{The longitude-velocity map of $^{12}$CO ($J=1-0$) emission from 207.7$^{\circ}$ to 211.7$^{\circ}$. The dashed lines indicate the spiral arms derived from the rotation model A5 of \citet{2014ApJ...783..130R}. The overlaid contours are $^{12}$CO emission with the minimal level and interval of the overlaid contours both being 0.18 K.}
  \label{fig:lv_arm}
\end{figure}

The position-velocity map of $^{12}$CO emission along the Galactic longitude is displayed in Figure \ref{fig:lv_arm}. According to the Galactic rotation model A5 of \citet{2014ApJ...783..130R}, we calculated the relationship between distance and velocity for the spiral arms between $l=207.7\arcdeg$ and $l=211.7\arcdeg$. The corresponding spiral arms are displayed with dashed lines. As shown in Figure \ref{fig:lv_arm}, the molecular clouds in this region exhibit multiple velocity components, with corresponding kinematic distance ranging from 0.5 kpc to $\sim$7.0 kpc. Most of the molecular clouds with velocities from $-$3 km s$^{-1}$ to 16.5 km s$^{-1}$ (first velocity component) are located between the Local and the Perseus Arms. We can see that the molecular clouds of the first velocity component are distributed to the east of the Rosette molecular cloud (RMC) \citep{2018ApJS..238...10L}. The distance to the RMC has been estimated to be 1.4$-$1.7 kpc with stellar photometry \citep{1981PASJ...33..149O,2002AJ....123..892P} and 1.39 $\pm$ 0.1 kpc with optical spectroscopy \citep{2000A&A...358..553H}. As in \cite{2018ApJS..238...10L}, we adopt a distance of 1.4 kpc for the first velocity component in this work.

\begin{figure}[h] 
  \centering
 \includegraphics[width=0.32\textwidth]{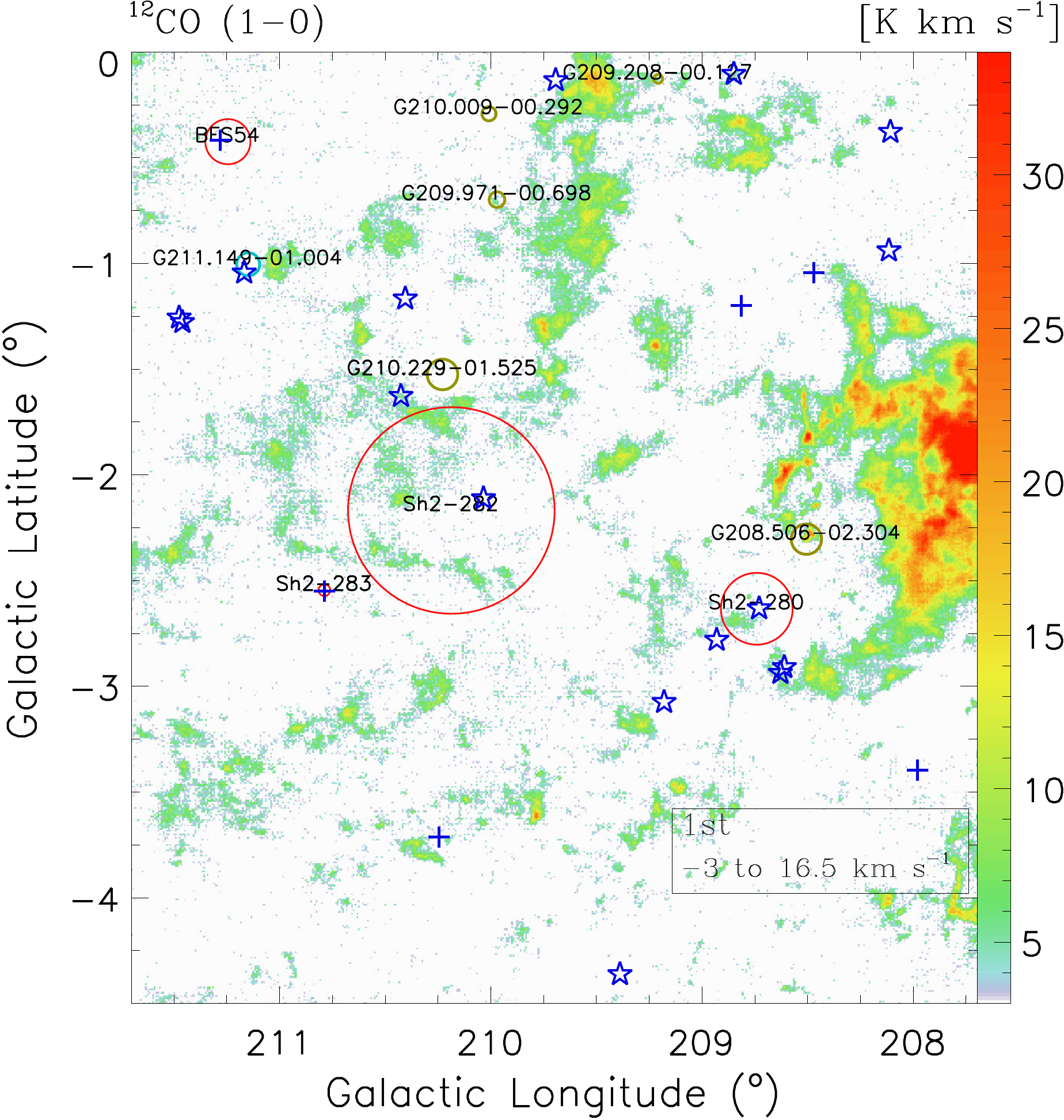}
 \includegraphics[width=0.32\textwidth]{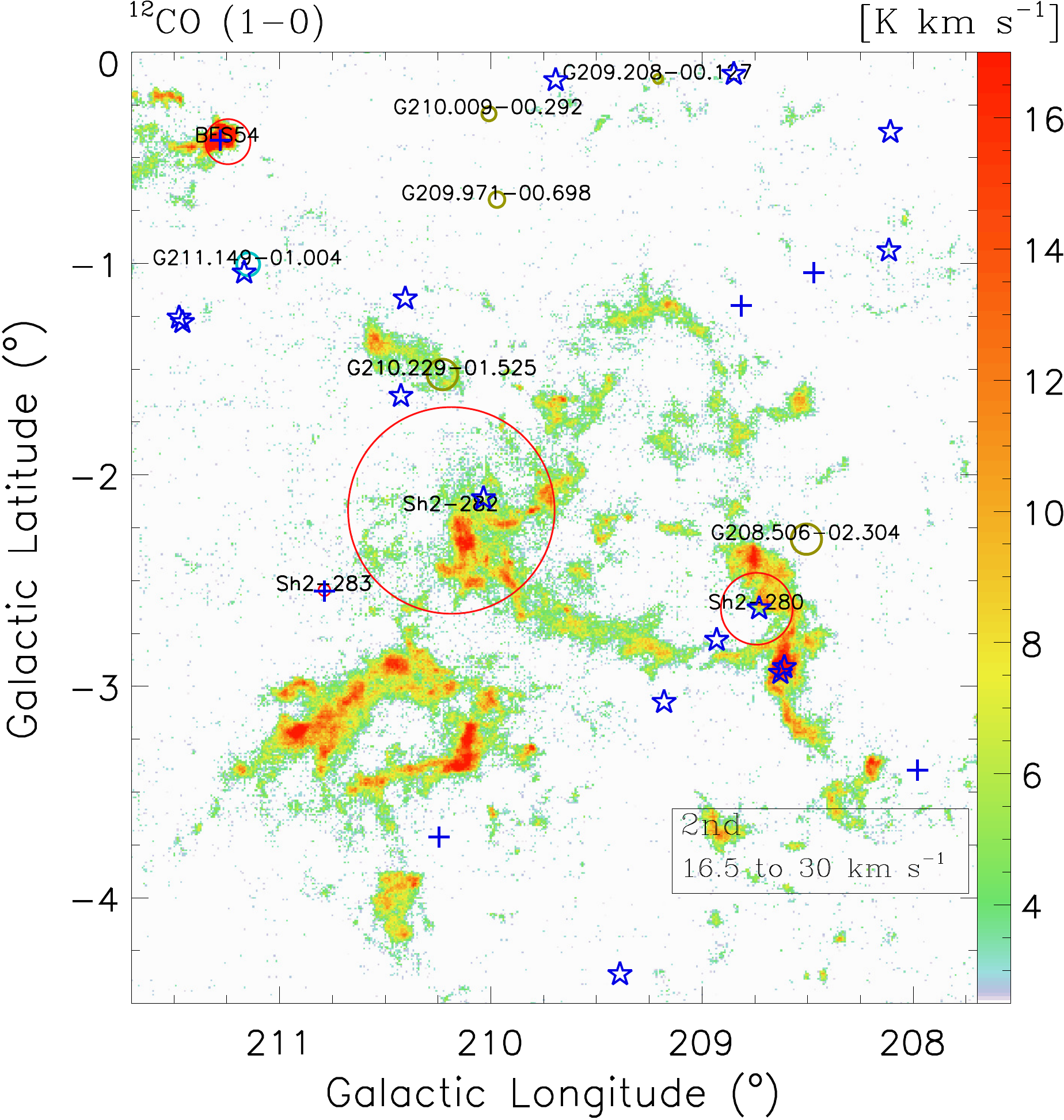}
 \includegraphics[width=0.32\textwidth]{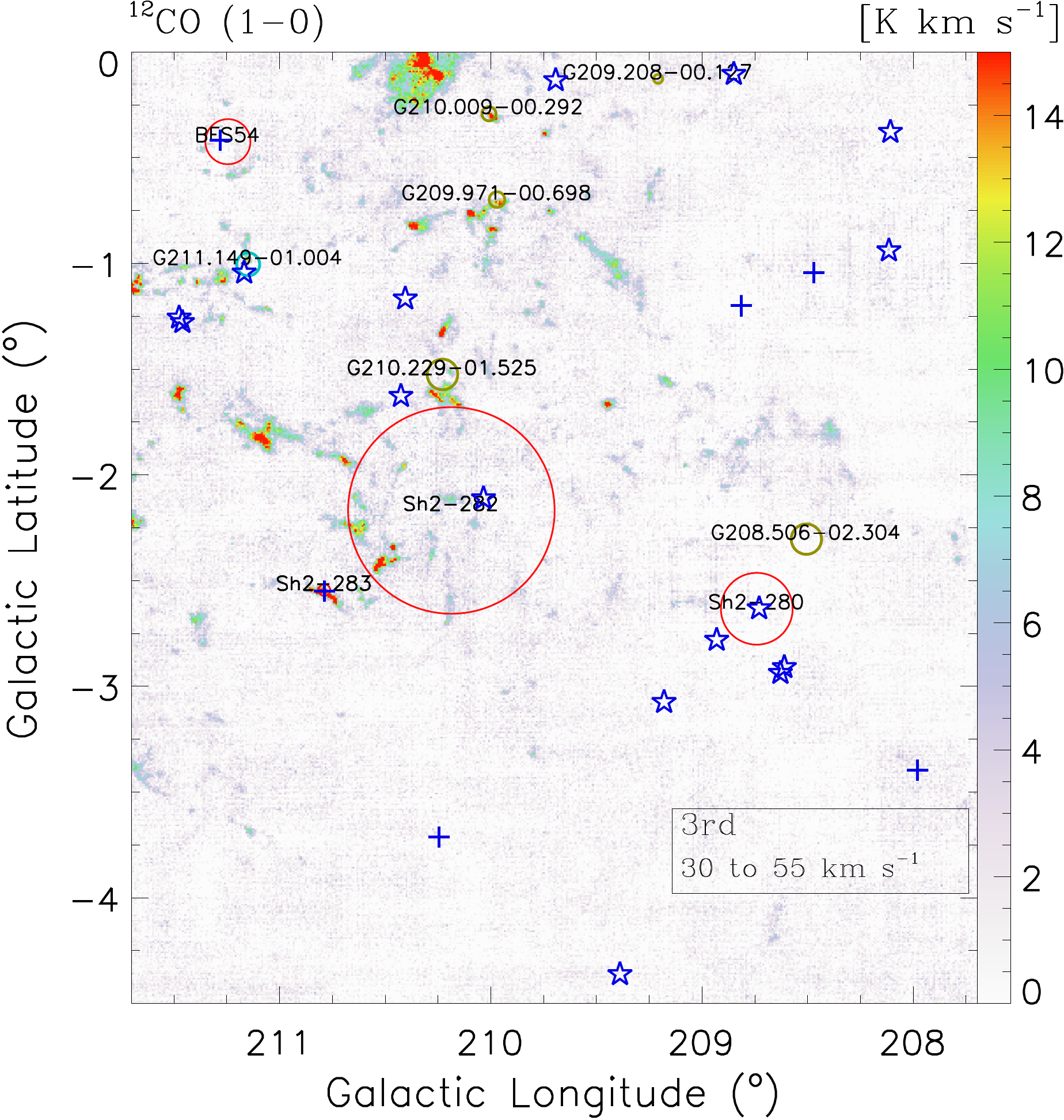}
 \includegraphics[width=0.32\textwidth]{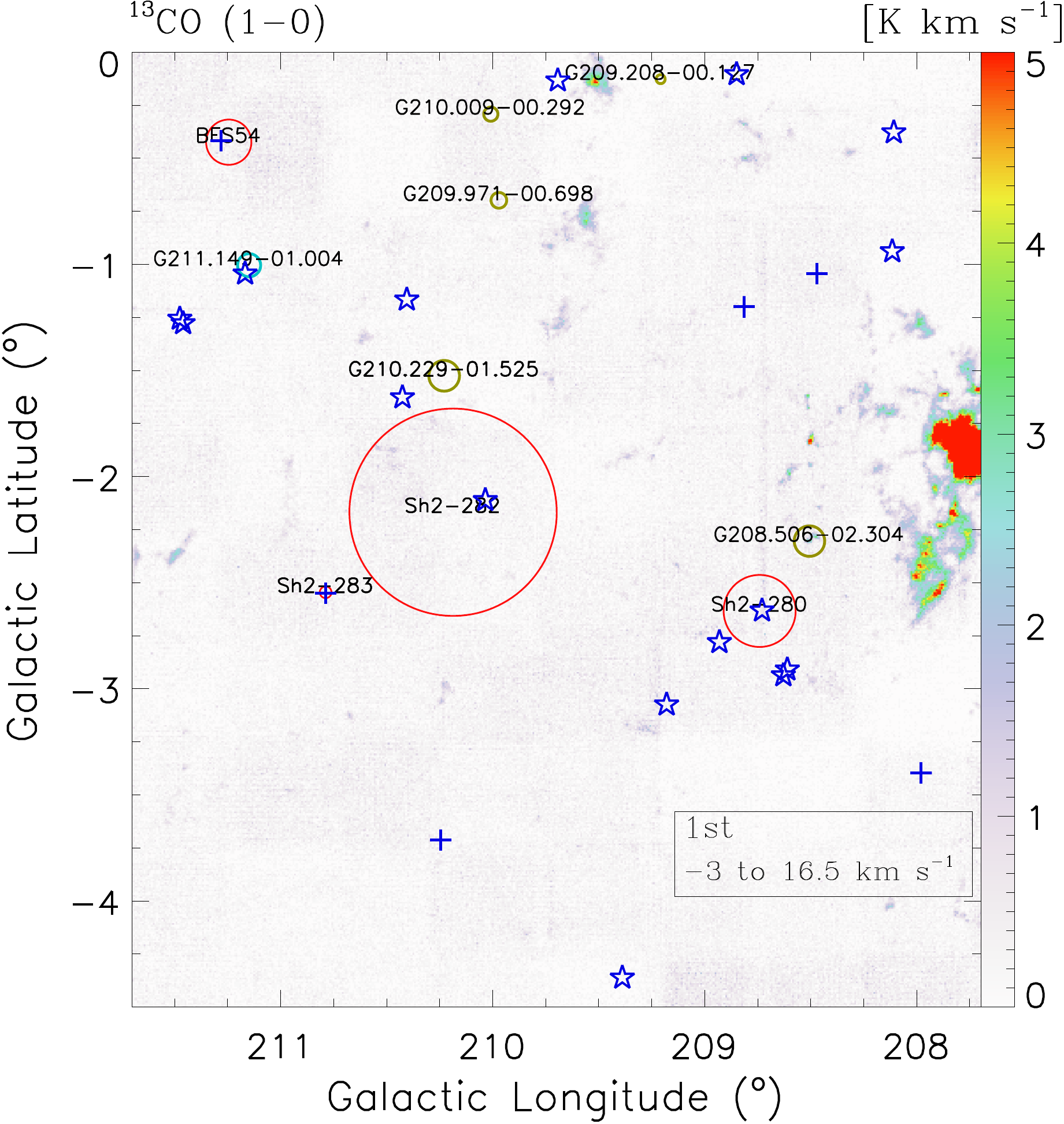}
 \includegraphics[width=0.32\textwidth]{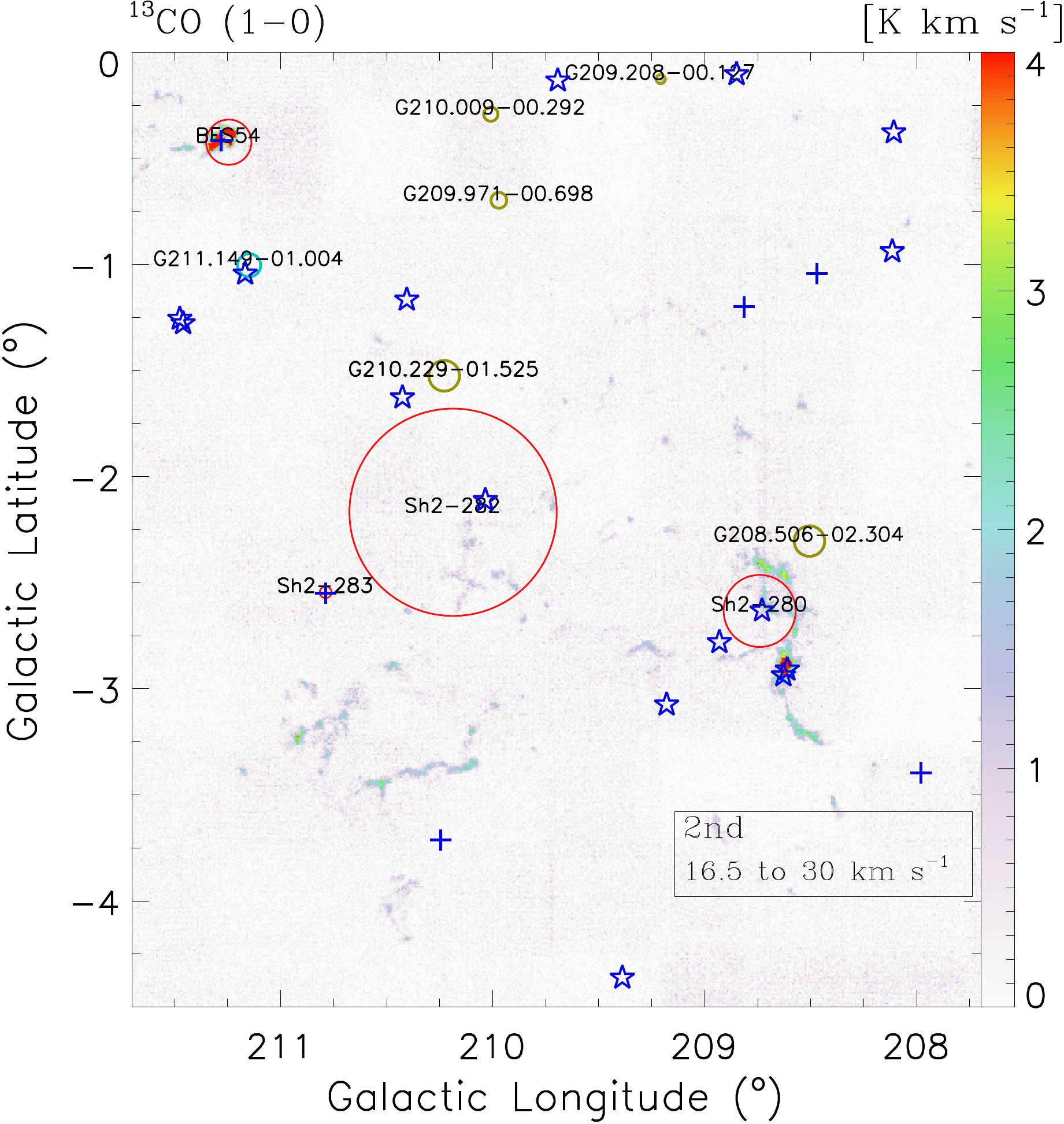}
 \includegraphics[width=0.32\textwidth]{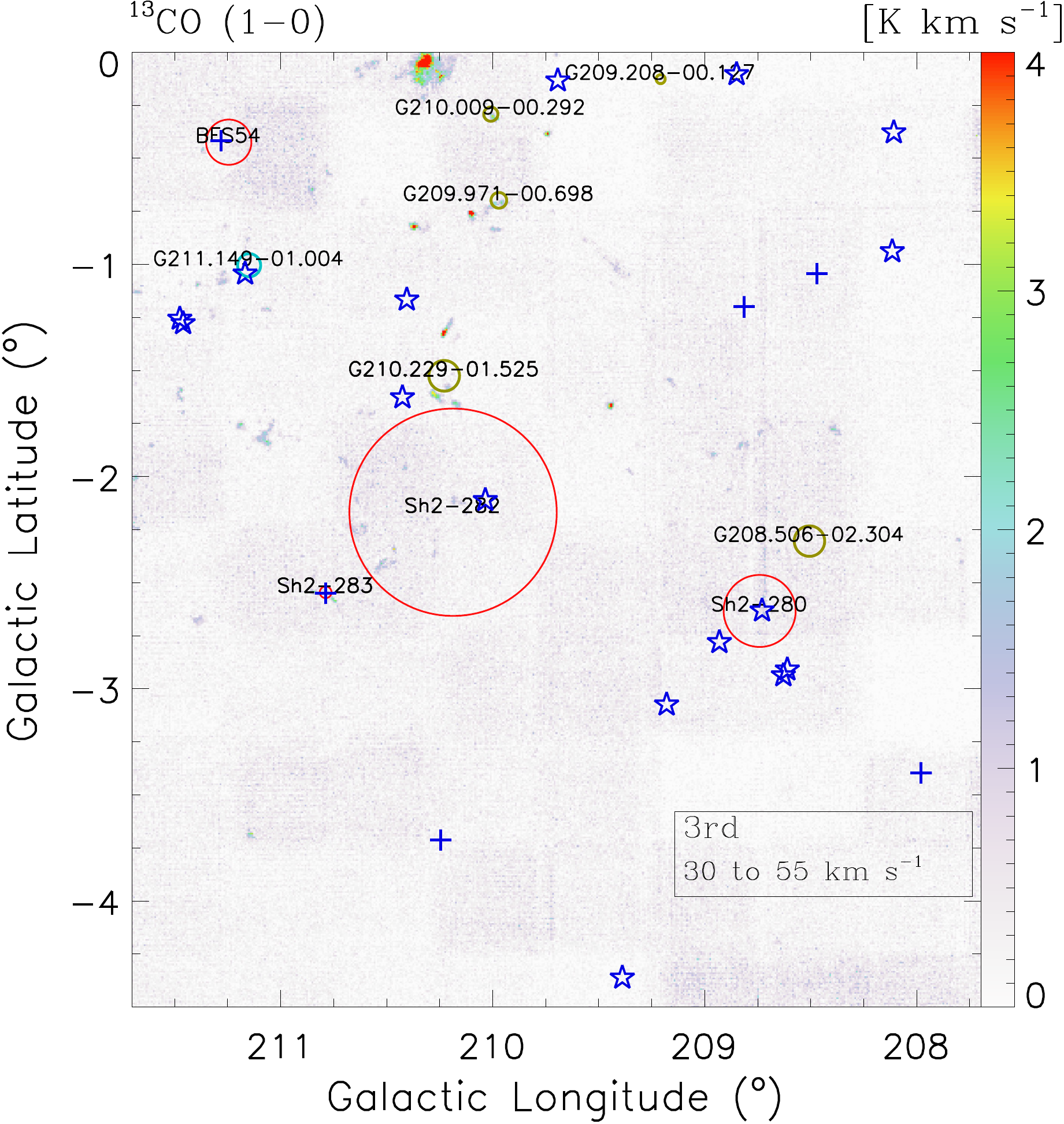}
  \caption{$^{12}$CO and $^{13}$CO $J = 1-0$ integrated intensity maps of molecular clouds in the region. The panels in the upper row show the $^{12}$CO emission and those in the lower row show the $^{13}$CO emission. Left: molecular clouds of the first velocity component. Middle: molecular clouds of the second velocity component. Right: molecular clouds of the third velocity component. The red, yellow, and cyan circles indicate the locations of the known, candidate, and radio quiet \ion{H}{2} regions. The circle sizes approximate the radius of the \ion{H}{2} regions from \citet{2014ApJS..212....1A}. The blue pentagrams and crosses indicate the O and B0 stars from the SIMBAD database, respectively.}
  \label{fig:inter}
\end{figure}

Molecular clouds of the second velocity component are located on the Perseus Arm. Three known \ion{H}{2} regions (Sh2-280, Sh2-282, and BFS54) are associated with molecular clouds of the second velocity component (Figure \ref{fig:inter}). From spectrophotometric observations, the distance of the cluster in BFS54 is 1.65 kpc \citep{2018MNRAS.476.2813D}. \citet{2006A&A...445L..43C} studied the Sh2-282 \ion{H}{2} region and suggested that this \ion{H}{2} region is photo-ionized by the O type star HD 47432. From its $V$ magnitude, $(B-V)$ color, and spectral type, the star HD 47432 is estimated to be at a distance of $\sim$1.25 kpc \citep{2006A&A...445L..43C}, which is similar to the results of \citet{1981A&A...100...28F} and \citet{1982ApJS...49..183B}. However, the parallax of HD 47432 from Gaia satellite data release 2 (DR2) \citep{2018AJ....156...58B,2018A&A...616A...1G} is 0.38 milliarcseconds, which corresponds to 2.6 kpc. The \ion{H}{2} region Sh2-280 is associated with the O type star HD 46573 (see Figures \ref{fig:S280}, \ref{fig:Kinematics_S280} and Section \ref{Kinematics}) which has a parallax of 0.64 milliarcseconds (1.6 kpc). Taking all these available distance measurements together, we adopt in this work the parallax distance of HD 46573 for the distance of molecular clouds of the second velocity component. For molecular clouds with velocities higher than 30 km s$^{-1}$ (third velocity component), we adopt kinematic distances from the rotation model A5 of \citet{2014ApJ...783..130R}. These kinematic distances are in the range from 3.0 to 7.0 kpc, therefore indicating that the clouds of the third velocity component are located between the Perseus and the Outer (Cygnus) Arms.

The $^{12}$CO and $^{13}$CO $J = 1-0$ integrated intensity maps of the molecular clouds of the three velocity components are displayed in Figures \ref{fig:inter} and the velocity channel map of $^{12}$CO emission of the region is displayed in Figure \ref{channel} in the Appendix. The majority of the molecular clouds of the first velocity component belong to the eastern part of the RMC. Other molecular clouds of the first velocity component exhibit diffuse morphology. Weak $^{12}$CO emission is associated with the known \ion{H}{2} regions Sh2-280 and Sh2-282 and the candidate \ion{H}{2} regions G208.506-02.304, G209.208-00.127, and G209.971-00.698. The molecular clouds in the first velocity component exhibit rare $^{13}$CO emission except for the eastern part of RMC.

The molecular clouds of the second velocity component clearly exhibit filamentary structures. Most of the clouds are located in the Sh2-280, Sh2-282, BFS54, and the southeastern region. The $^{12}$CO emission shows the highest brightness in the BFS54 \ion{H}{2} region. Elephant trunk and cometary structures are seen in the molecular clouds within the Sh2-282 \ion{H}{2} region. These morphologies are frequently found in \ion{H}{2} regions \citep{2006A&A...454..201G,2017A&A...605A..82M}, and are predicted by the radiation driven implosion (RDI) simulations \citep{1989ApJ...346..735B,2011ApJ...736..142B}. As shown in the $^{12}$CO and $^{13}$CO emission maps (Figure \ref{fig:inter}), the stellar winds and radiation from the massive stars in the BFS54 and Sh2-280 \ion{H}{2} regions have apparently destroyed molecular clouds within the central parts of these \ion{H}{2} regions and have excavated a circle-like (BFS54) or semicircle-like (Sh2-280) cavity, similar to the S287 \citep{2016A&A...588A.104G} and the N4 \citep{2017ApJ...838...80C} regions. As shown in Figure \ref{fig:inter}, the morphology of the molecular clouds of the third velocity component is more fragmentary and clumpy as compared to those of the first and second velocity components, which may be caused by the farther distances to the clouds of the third velocity component (3.0$-$7.0 kpc).

\begin{figure}[h] 
  \centering
 \includegraphics[width=0.32\textwidth]{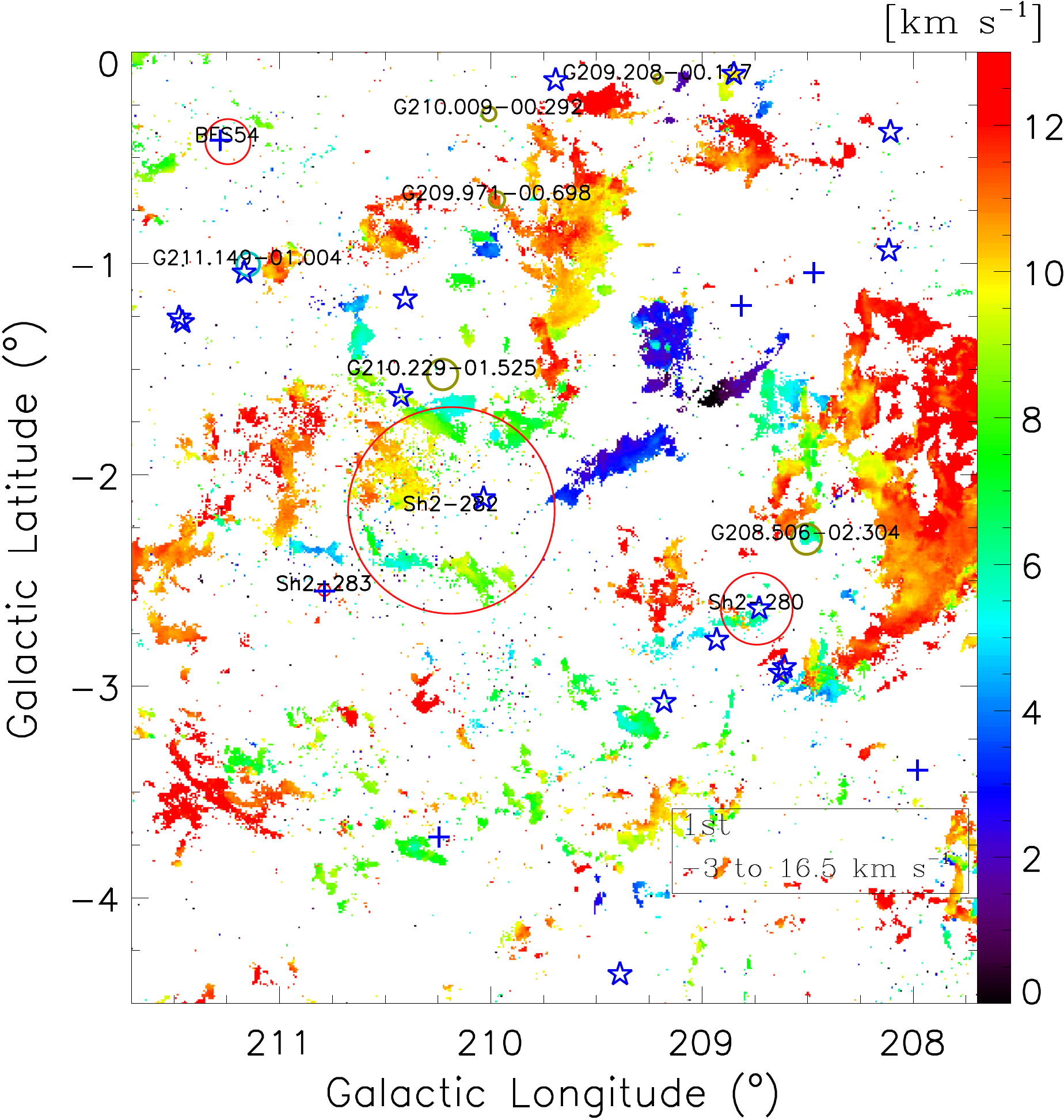}
 \includegraphics[width=0.32\textwidth]{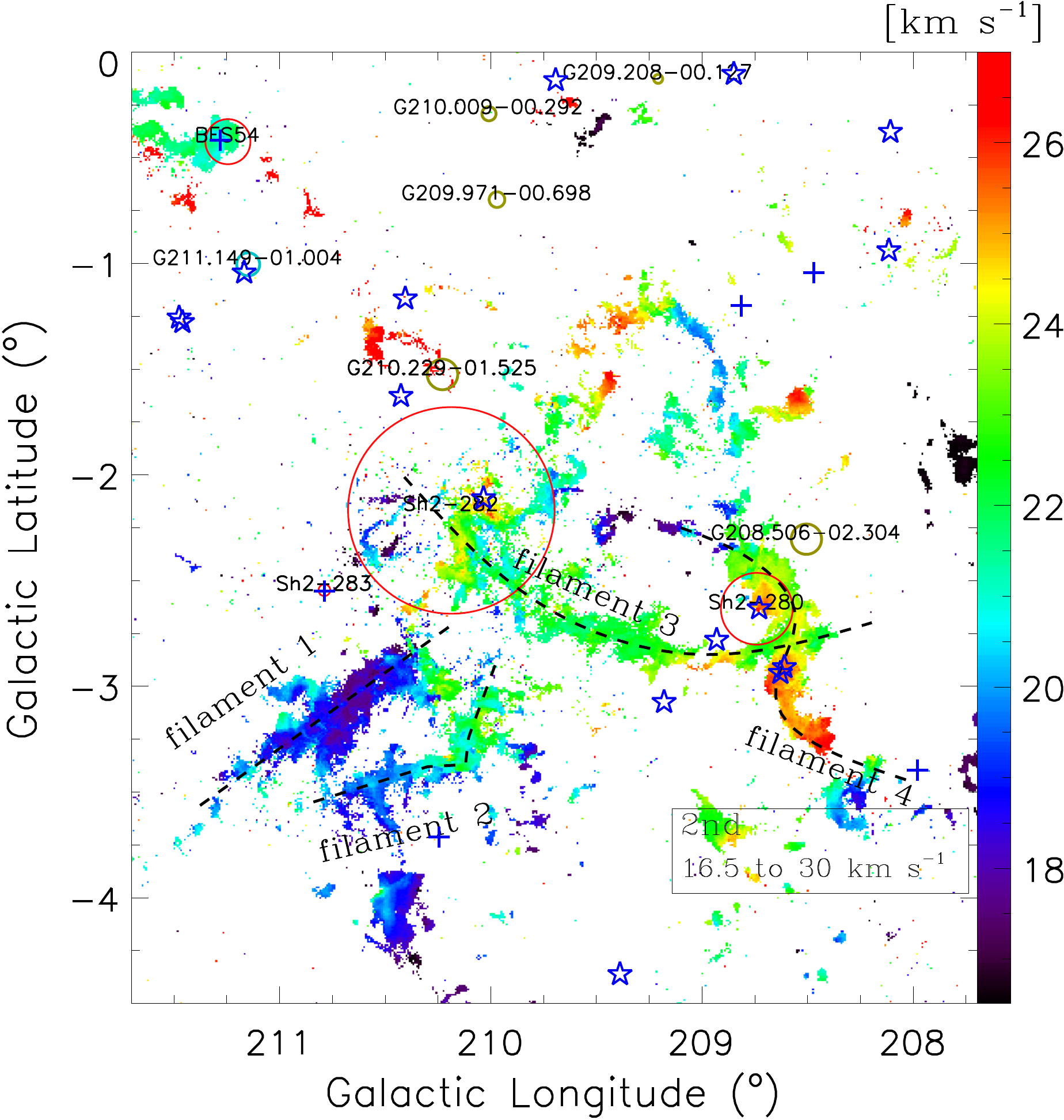}
 \includegraphics[width=0.32\textwidth]{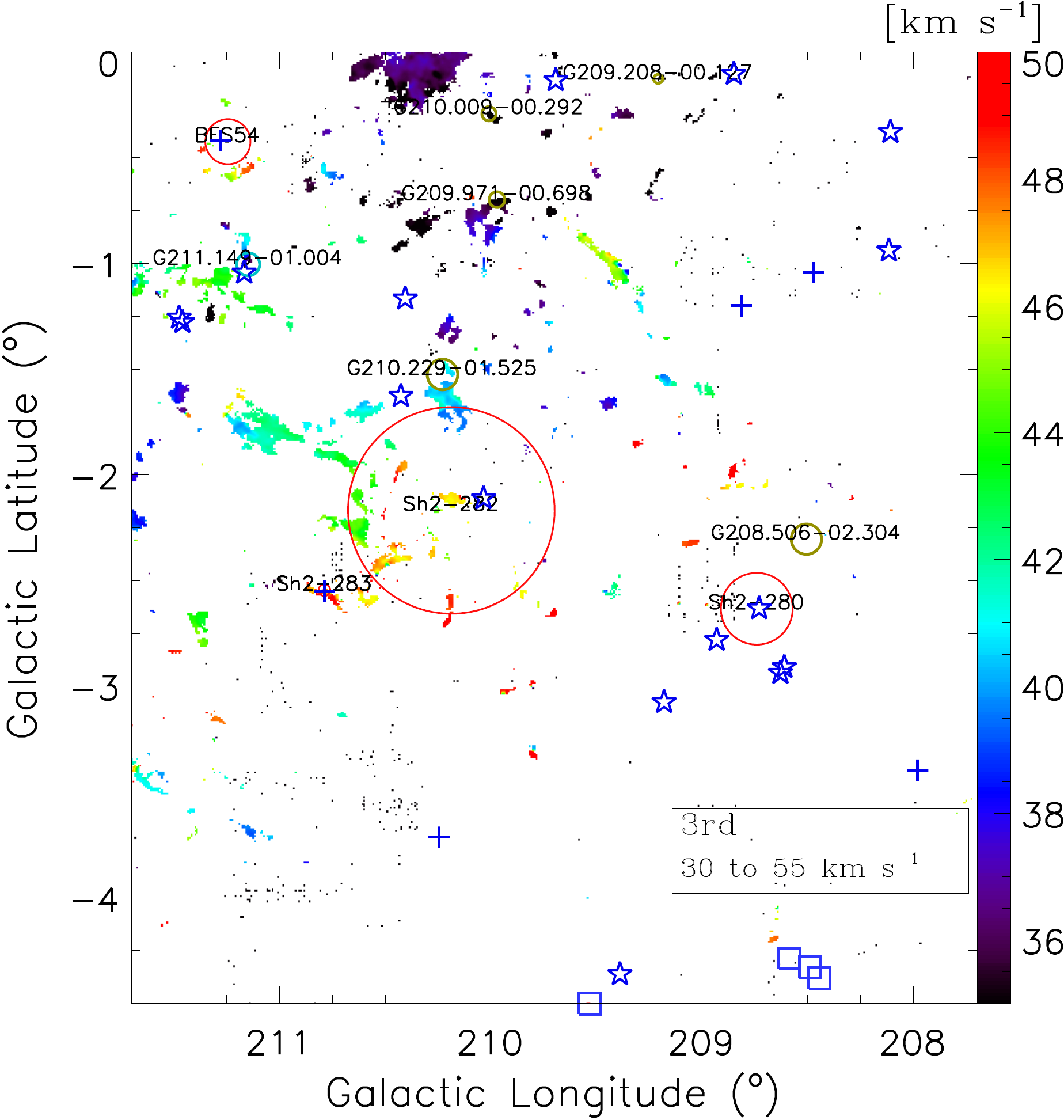}
  \caption{First moment maps of $^{12}$CO emission for molecular clouds of each velocity component. The dashed lines in the middle pannel indicate the four main filaments in the second velocity component. The blue squares near the bottom in the right pannel indicate the locations of the four clumps (shown in Figure \ref{fig:clump}) that are far below the Galactic mid-plane. The red, yellow, and cyan circles indicate the locations of the known, candidate, and radio quiet \ion{H}{2} regions. The blue pentagrams and crosses mark the O and B0 stars in this region from the SIMBAD database, respectively.}
  \label{fig:m1}
\end{figure}

\begin{figure}[h]
  \centering%
\includegraphics[width=0.23\textwidth]{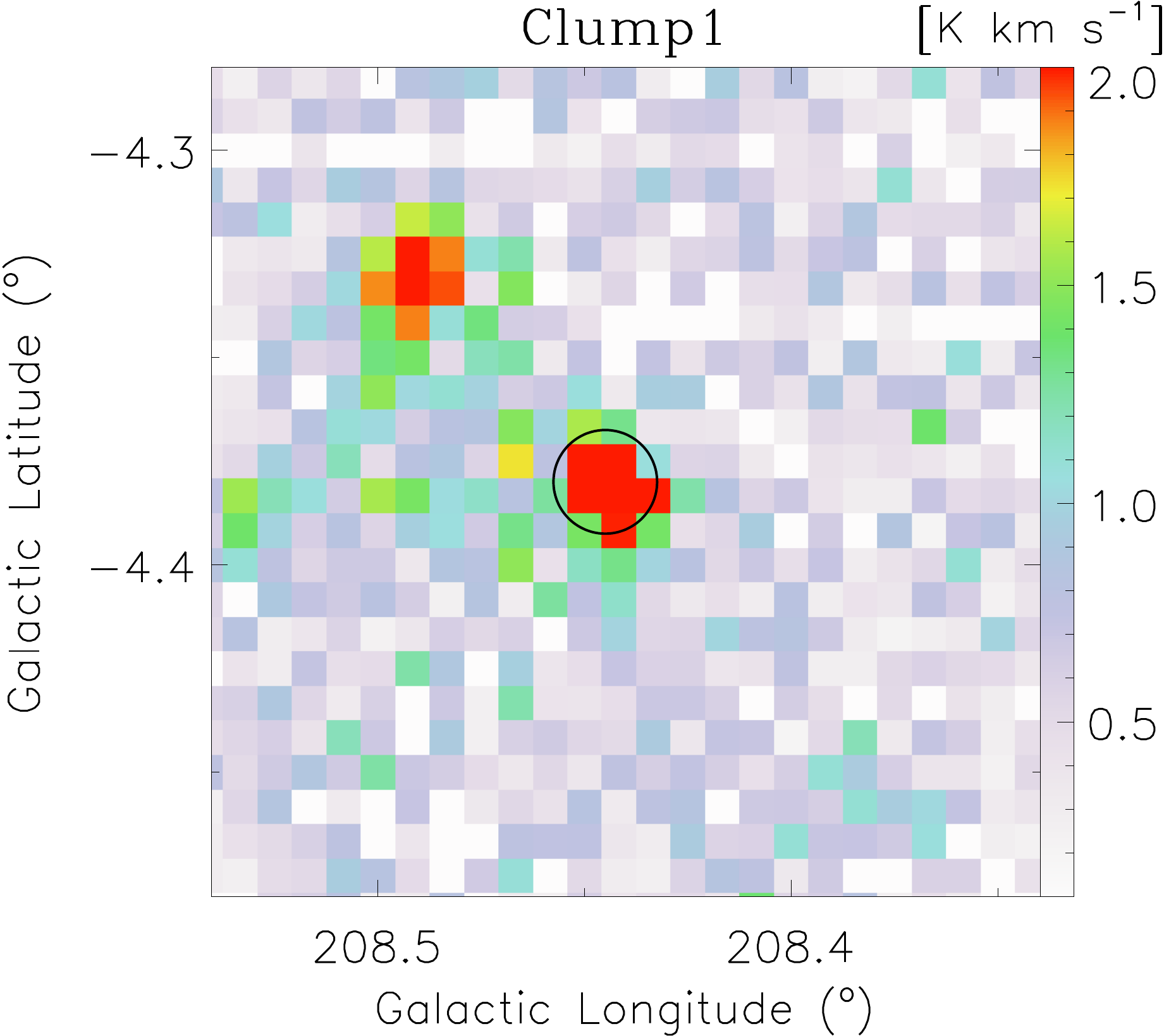}
\includegraphics[width=0.23\textwidth]{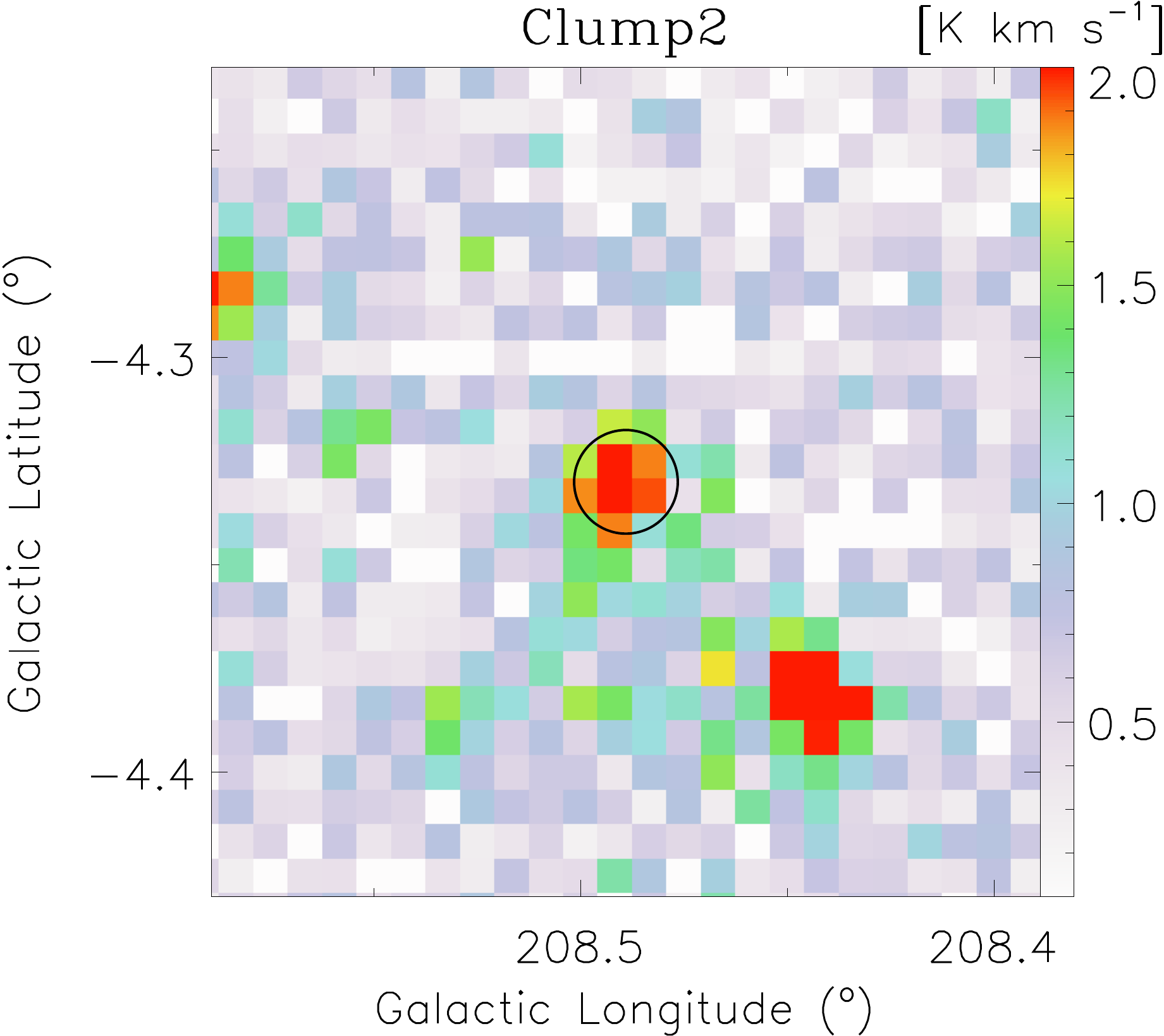}
\includegraphics[width=0.23\textwidth]{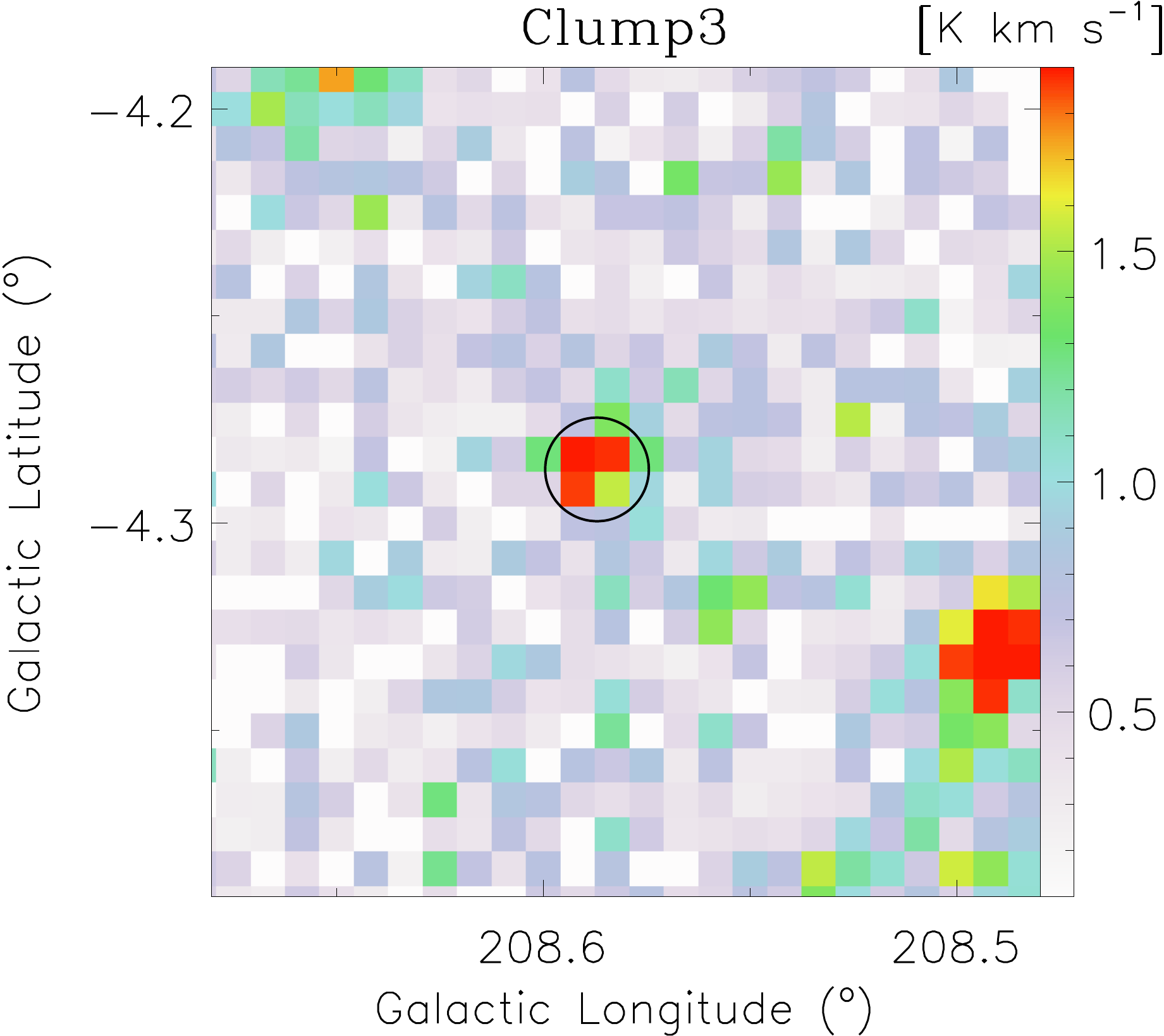}
\includegraphics[width=0.23\textwidth]{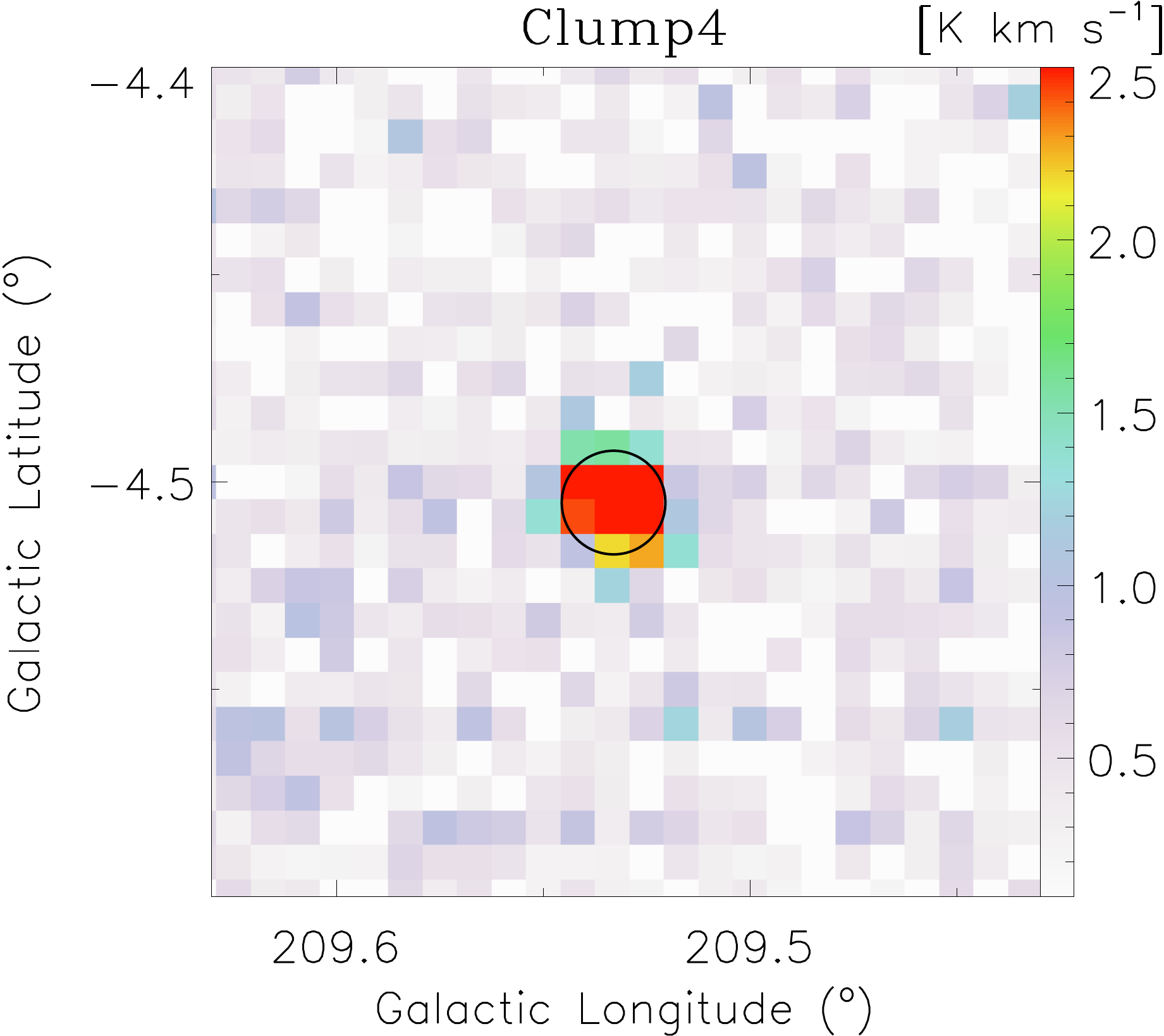}\\
\includegraphics[width=0.21\textwidth]{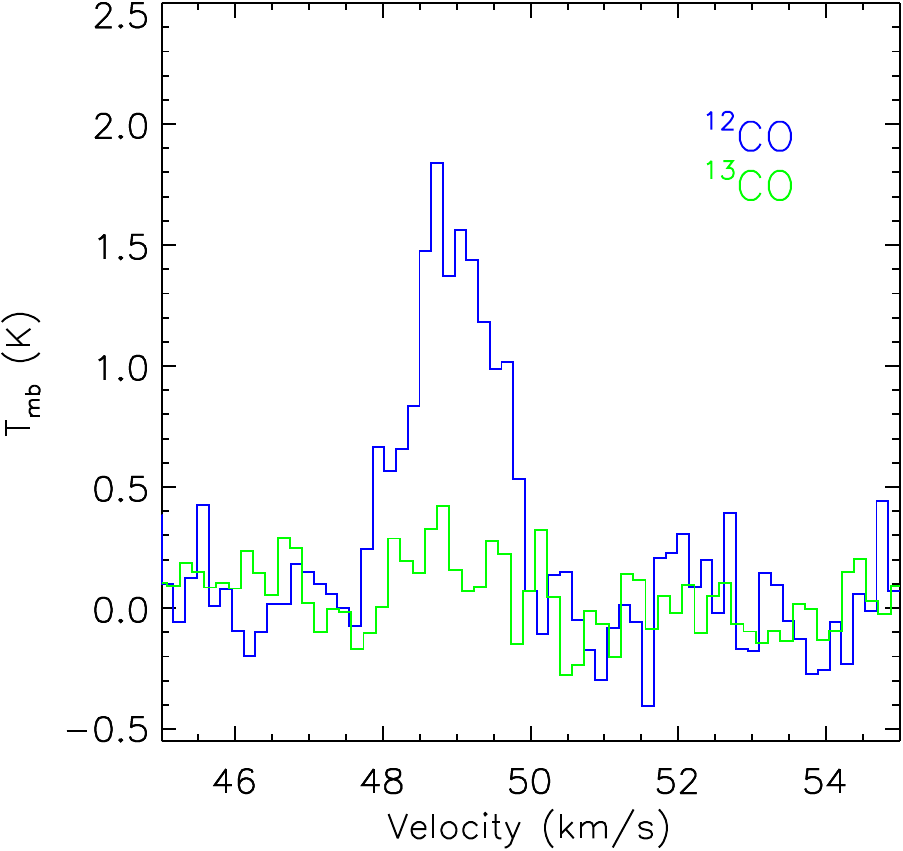}\hspace{1.3em}%
\includegraphics[width=0.21\textwidth]{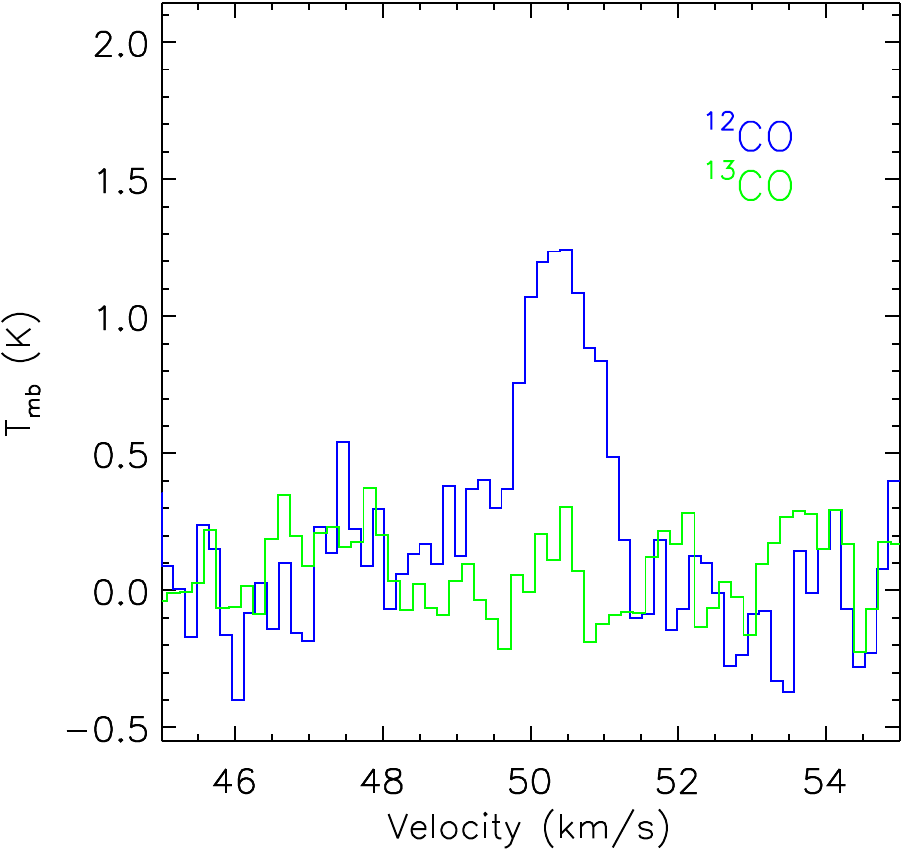}\hspace{1.3em}%
\includegraphics[width=0.21\textwidth]{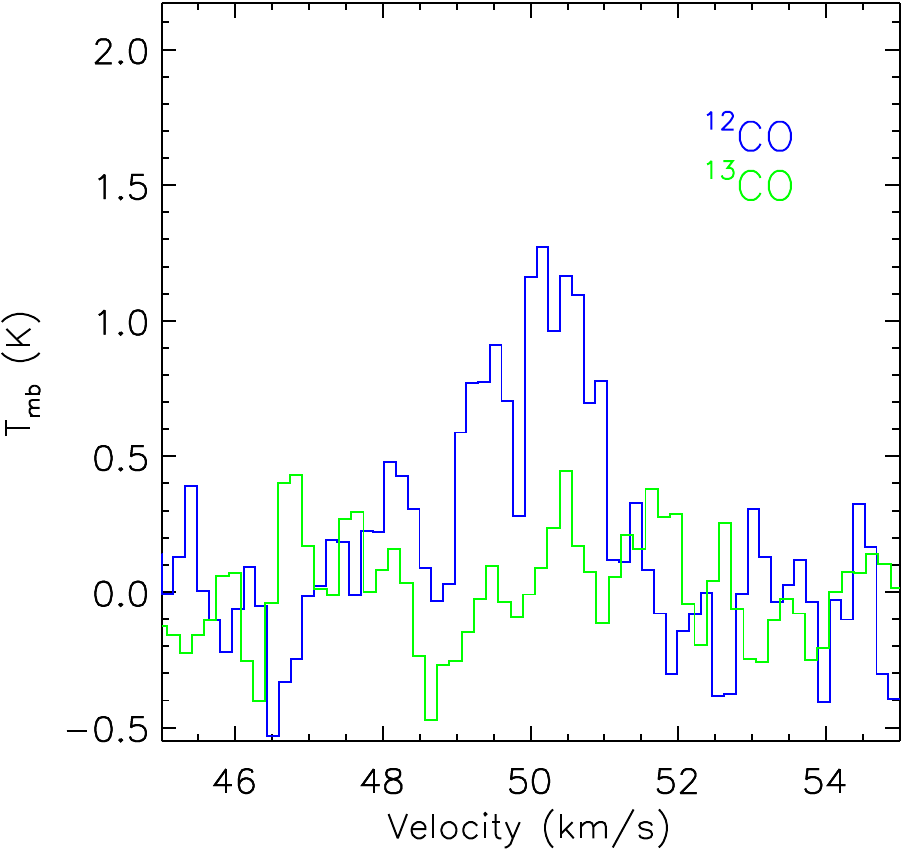}\hspace{1.3em}%
\includegraphics[width=0.21\textwidth]{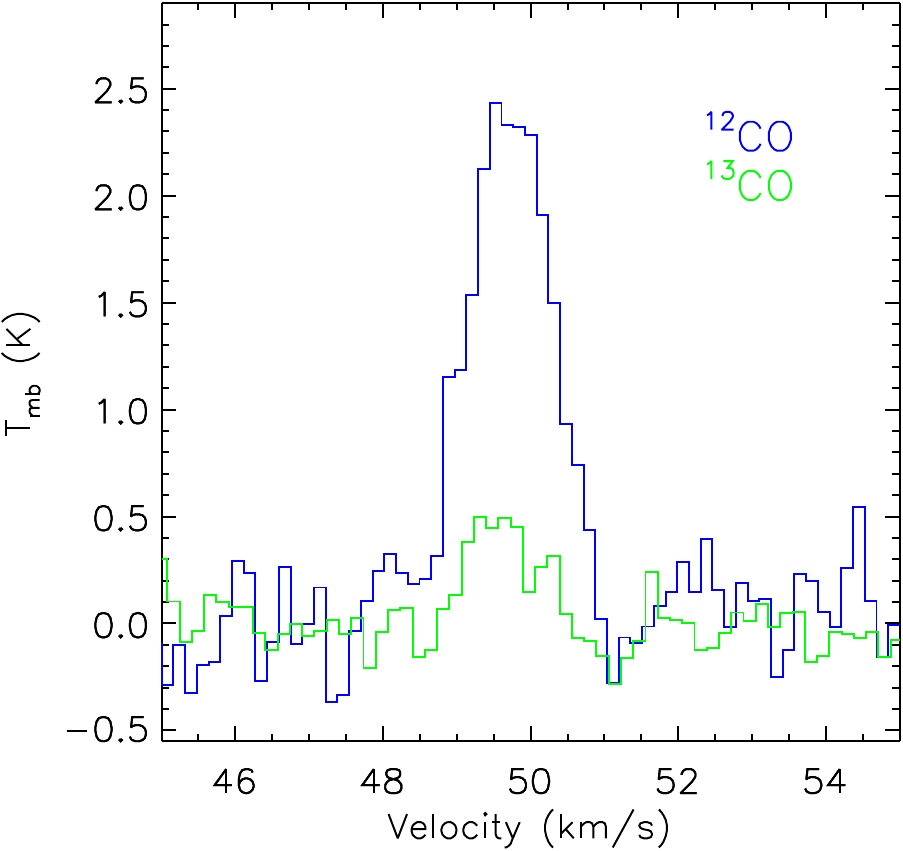}\hspace{1.3em}%
 \caption{Top: $^{12}$CO $J = 1-0$ integrated intensity maps of the four clumps located far below the Galactic mid-plane. The black circles indicate the positions of the four clumps. Bottom: average spectra of the 1.5$\arcmin$ $\times$ 1.5$\arcmin$ region of the four clumps with blue and green indicating the $^{12}$CO and $^{13}$CO emission, respectively.}
  \label{fig:clump}
\end{figure}

Figure \ref{fig:m1} shows the $^{12}$CO emission intensity weighted centroid velocity distributions of the molecular clouds. As shown in Figure \ref{fig:m1}, most of the molecular clouds of the first velocity component have velocities in the range from 5 km s$^{-1}$ to 13 km s$^{-1}$. Molecular clouds in the region of Galactic longitude of $208.5^{\circ} < l < 209.7^{\circ}$ and Galactic latitude of $-2.0^{\circ} < b < -1.0^{\circ}$ have significantly lower velocities compared to other molecular clouds (see the left panel of Figure \ref{fig:m1}). The molecular cloud within the G208.506-02.304 \ion{H}{2} region is located near the eastern part of the Rosette molecular cloud, but with a velocity of about 6 km s$^{-1}$, which is 5 km s$^{-1}$ lower than the velocity of the RMC. For molecular clouds of the second velocity component, a velocity gradient can be seen in the direction along the Galactic longitude, with the western molecular clouds possessing a larger ($20-27$ km s$^{-1}$) velocity than the eastern clouds ($16.5-20$ km s$^{-1}$). Moreover, many filament exhibits coherent and narrow velocity distribution, so these filaments are coherent structures both in the spatial and in the velocity dimensions, rather than just chance projection of individual clumps of different velocities. Four main filaments (filament 1$-$4) are marked with dashed lines in the middle panel of Figure \ref{fig:m1}. The molecular clouds of the third velocity component have velocities higher than 30 km s$^{-1}$ and their kinematic distances are larger than 3.0 kpc. As shown in Figure \ref{fig:m1}, these molecular clouds as a whole exhibit a velocity gradient in the direction from the north (35$-$40 km s$^{-1}$) to the south (40$-$50 km s$^{-1}$). According to the rotation model A5 of \citet{2014ApJ...783..130R}, the distances of the northern molecular clouds are around 4 kpc, while the southern molecular clouds are located farther away ($\sim$5.5 kpc). In the datacube, we also identified four molecular clumps with both low Galactic latitudes and high velocities (Figure \ref{fig:clump}), which implies that these clumps are located far away from the Galactic mid-plane. The locations of the four clumps are displayed with blue squares in the right panel of Figure \ref{fig:m1}. According to the rotation model A5 of \citet{2014ApJ...783..130R}, the kinematic distances of the four clumps are calculated to be from 6.8 to 7.3 kpc. Combining the Galactic longitude/latitude of the clumps and the distance of the sun from the Galactic center (8.34 kpc), the distances of the four clumps from the Galactic center are calculated to be $\sim$14 kpc. Taking the offset angle and the distance of the Sun above the Galactic physical mid-plane to be 0.072$^{\circ}$ and 17.1 pc \citep{Su_2016}, respectively, the distances of these clumps from the Galactic mid-plane are calculated to be -525 to -508 pc. The FWHM thickness of the Galactic molecular gas disk at Galactic radius of $\sim$14 kpc is about 180$-$440 pc \citep{1990A&A...230...21W,Digel_1991,2015ARA&A..53..583H}, i.e., the $\sigma$ thickness of the Galactic molecular gas disk at the Galactic centric distance of 14 kpc is about 80$-$190 pc. Therefore, the four clumps are significantly far away from the Galactic mid-plane. With a high abundance and a low excitation energy and critical density, $^{12}$CO is frequently employed to measure molecular gas masses according to the relationship between the observed CO integrated intensity and the column density of molecular hydrogen. Taking the ratio between the column density of molecular hydrogen and the CO integrated intensity ($X$ factor) to be 2.0 $\times$ 10$^{20}$ H$_{2}$ cm$^{-2}$ (K km s$^{-1}$)$^{-1}$ \citep{doi:10.1146/annurev-astro-082812-140944} and applying the fluxes of the clumps (Table \ref{tab_clump}), the masses of the clumps are calculated to be $37-120$ M$_{\odot}$. 

\begin{deluxetable}{rccccccccccc}
\decimals
\tabletypesize{\footnotesize}
\tablewidth{0pt}
\tablenum{1}
\tablecaption{Clumps far away from the Galactic physical mid-plane\label{tab_clump}}
\tablehead{
\colhead{ID} & \colhead{Name} & \colhead{l} & \colhead{b} & \colhead{$\rm{V_{lsr}}$} &\colhead{D}   & \colhead{T$_{ex}$} & \colhead{Flux}& \colhead{Mass}  & \colhead{D$_{GC}$}& \colhead{D$_z$}\\[1pt]
\colhead{ }  & \colhead{ }   & \colhead{($\arcdeg$)}  & \colhead{($\arcdeg$)}  & \colhead{(km/s)}  & \colhead{
  (kpc)} & \colhead{(K)}& \colhead{(K km s$^{-1}$)} & \colhead{(M$_{\odot}$)} &\colhead{(kpc)}&\colhead{(pc)} }
\startdata
Clump1 & MWISP G208.445-4.380    &    208.445 & -4.380 & 48.8  & 7.0 & 4.5 & 17.0  & 78  & 14.5& -509 \\
Clump2 & MWISP G208.489-4.330    &    208.489 & -4.330 & 50.3  & 7.3 & 4.4 & 13.6  & 68   & 14.8& -525 \\
Clump3 & MWISP G208.587-4.287    &    208.587 & -4.287 & 50.5  & 7.3 & 4.2 & 7.4    & 37   & 14.7& -519 \\
Clump4 & MWISP G209.533-4.505    &    209.533 & -4.505 & 49.8  & 6.8 & 4.9 & 27.6  & 120 & 14.3& -508 \\
\enddata
\tablecomments{Columns 3-5 give the position centroids of the clumps in the PPV space. Column 6 gives the kinematic distance derived from model A5 in \cite{2014ApJ...783..130R}. The excitation temperature, integrated intensity, and mass derived from $^{12}$CO are list in columns 7-9. Column 10 gives the distances of the clumps from the Galactic center. Column 11 gives the vertical distances of the clumps from the Galactic physical mid-plane.}
\end{deluxetable}

\subsection{Physical Properties of Molecular Clouds in the Region}\label{Physical properties}

Assuming that the molecular clouds are under the local thermodynamic equilibrium (LTE) conditions, the mass of the molecular cloud can be calculated with the measured brightness of $^{12}$CO, $^{13}$CO, and C$^{18}$O $J=1-0$ emission. Taking the filling factor of $^{12}$CO $J=1-0$ emission to be unity and assuming that the emission is optically thick, the excitation temperature in units of Kelvin can be calculated according to the following formula \citep{1998AJ....116..336N}
\begin{equation}
T_{ex} = \frac{5.53}{ln (1+\frac{5.53}{T_{mb} (^{12}CO)+0.819})},
\end{equation}
where $T_{mb}$ is the main-beam brightness temperature of $^{12}$CO emission. The excitation temperatures derived from above equation for the four clumps far away from the Galactic physical mid-plane are listed in Table \ref{tab_clump}.

\begin{figure}[h]
  \centering%
\includegraphics[width=0.32\textwidth]{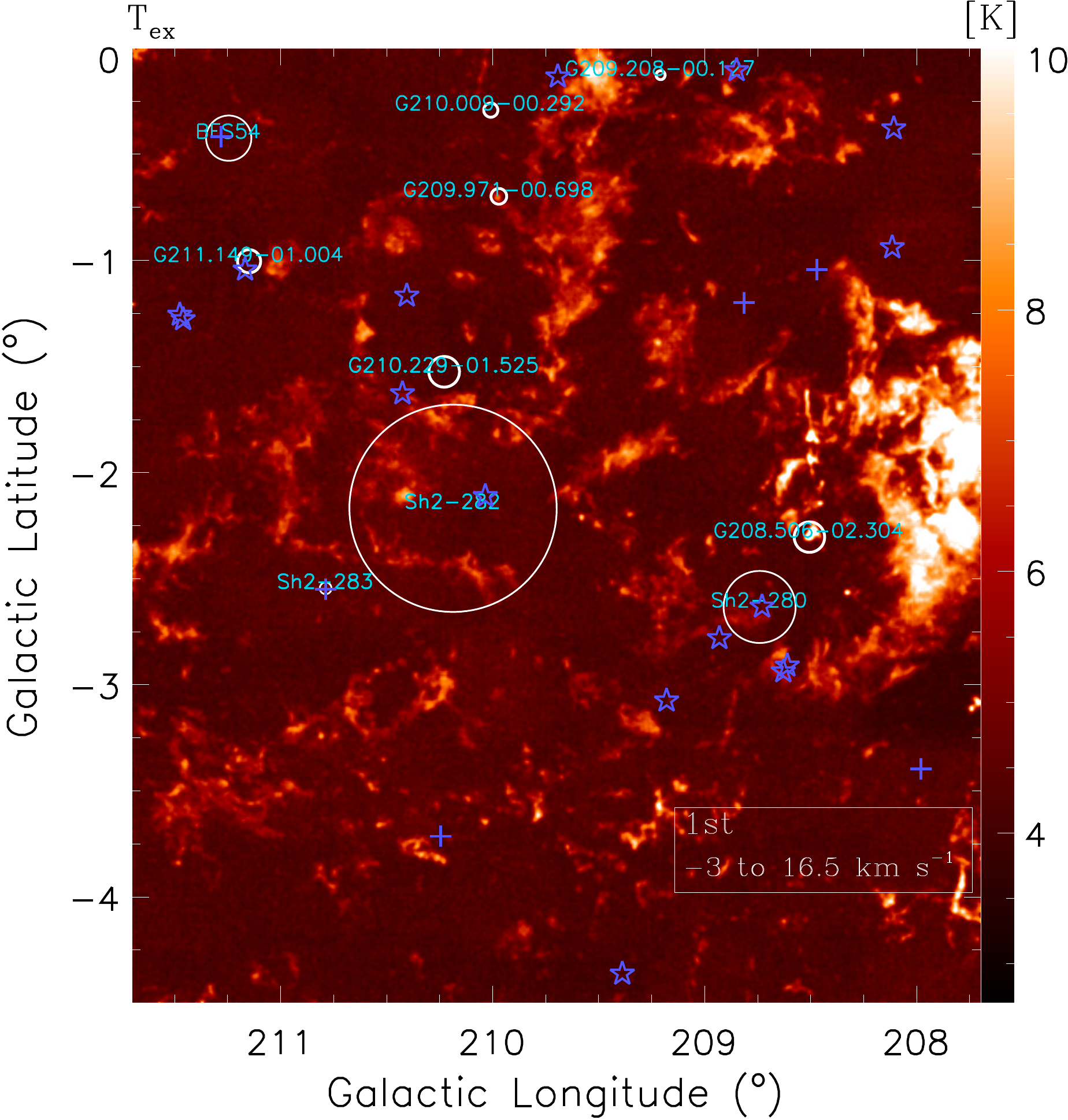}
\includegraphics[width=0.32\textwidth]{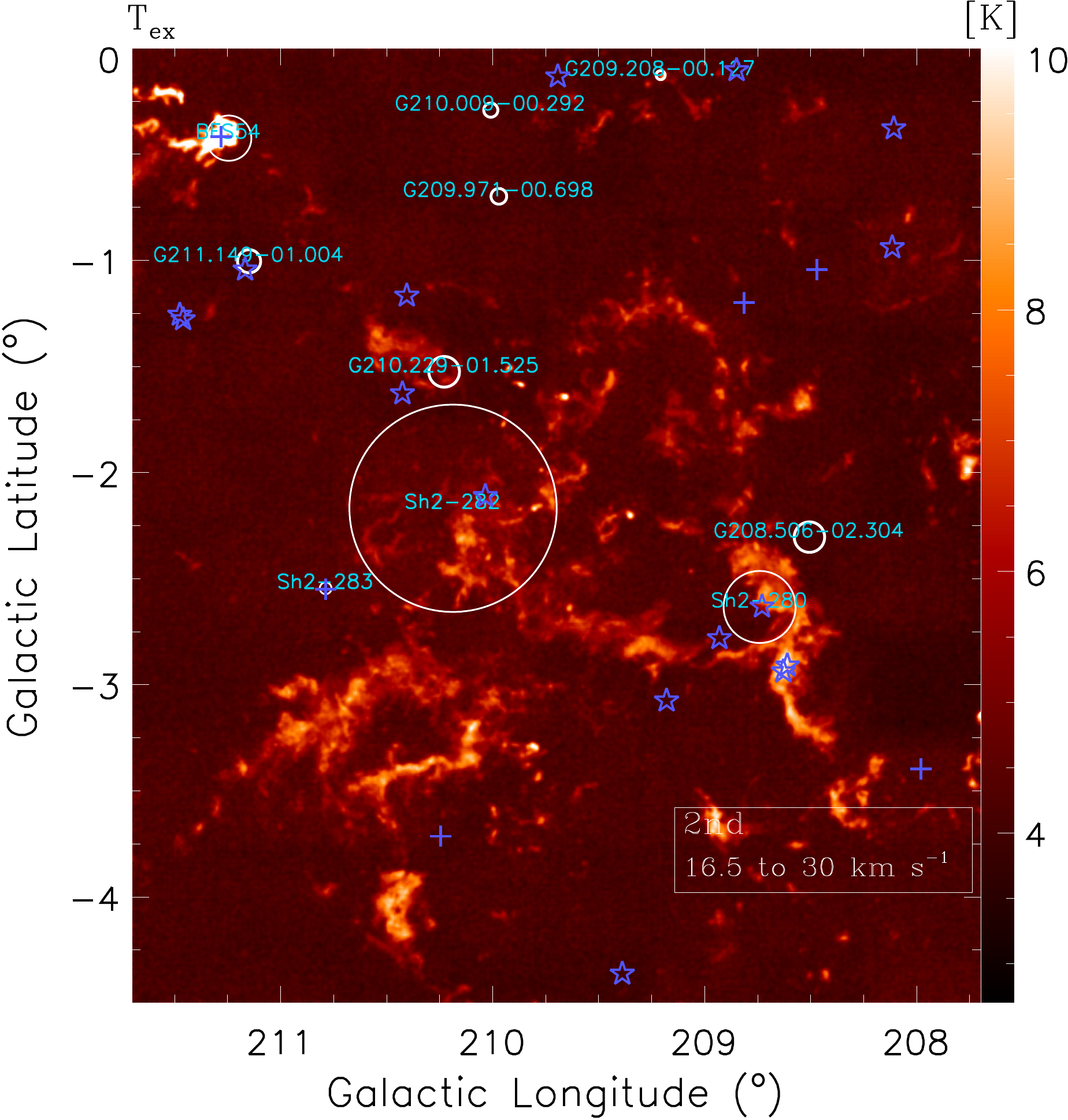}
\includegraphics[width=0.32\textwidth]{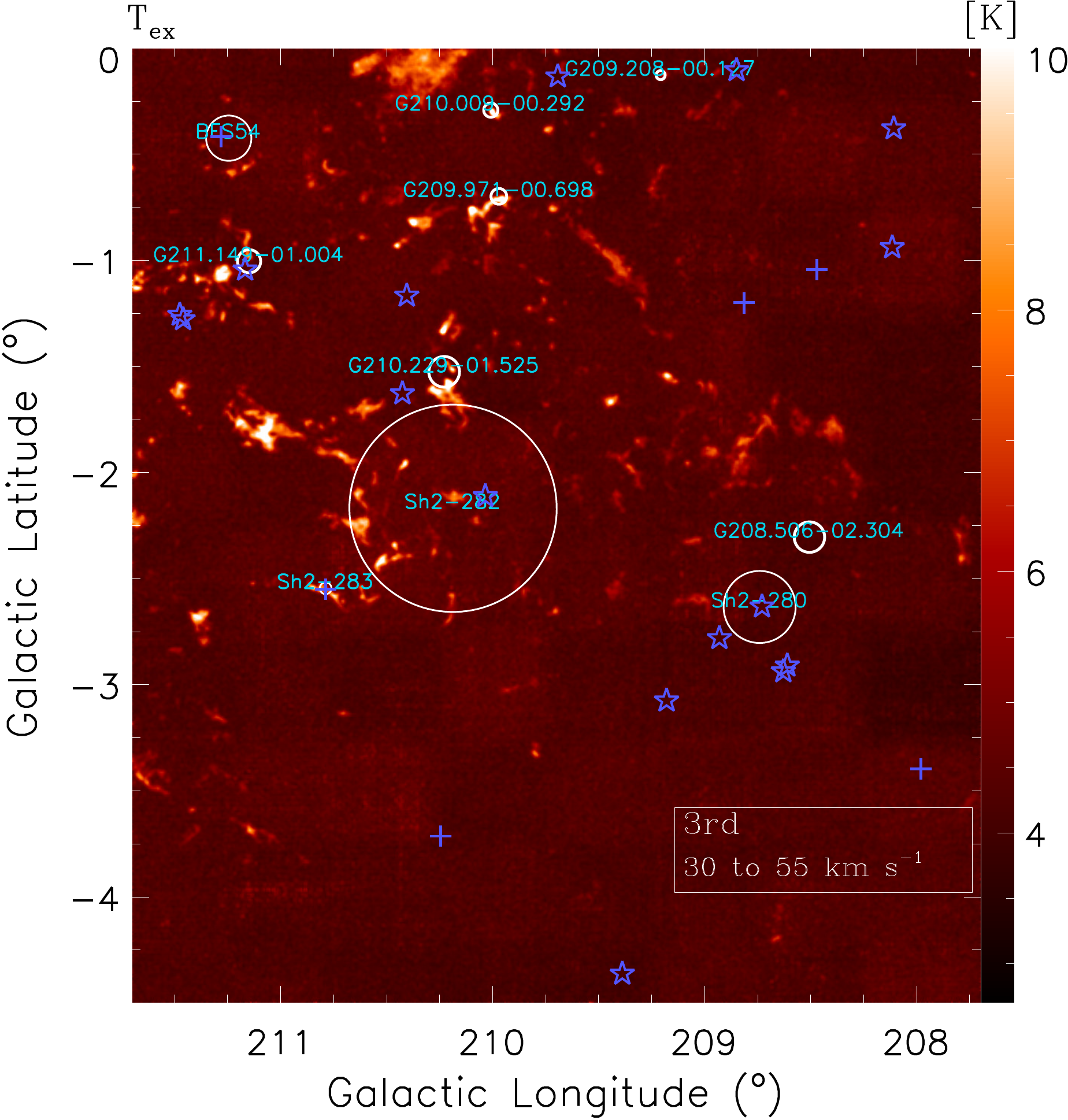}
 \caption{Distributions of excitation temperature of molecular clouds of the three velocity components in the region. The red, yellow, and cyan circles indicate the locations of the known, candidate, and radio quiet \ion{H}{2} regions. The blue pentagrams and crosses indicate the O and B0 stars in this region from the SIMBAD database, respectively.}
  \label{fig:tex_s282}
\end{figure}

\begin{figure}[h]
  \centering%
\includegraphics[width=0.6\textwidth,angle=270]{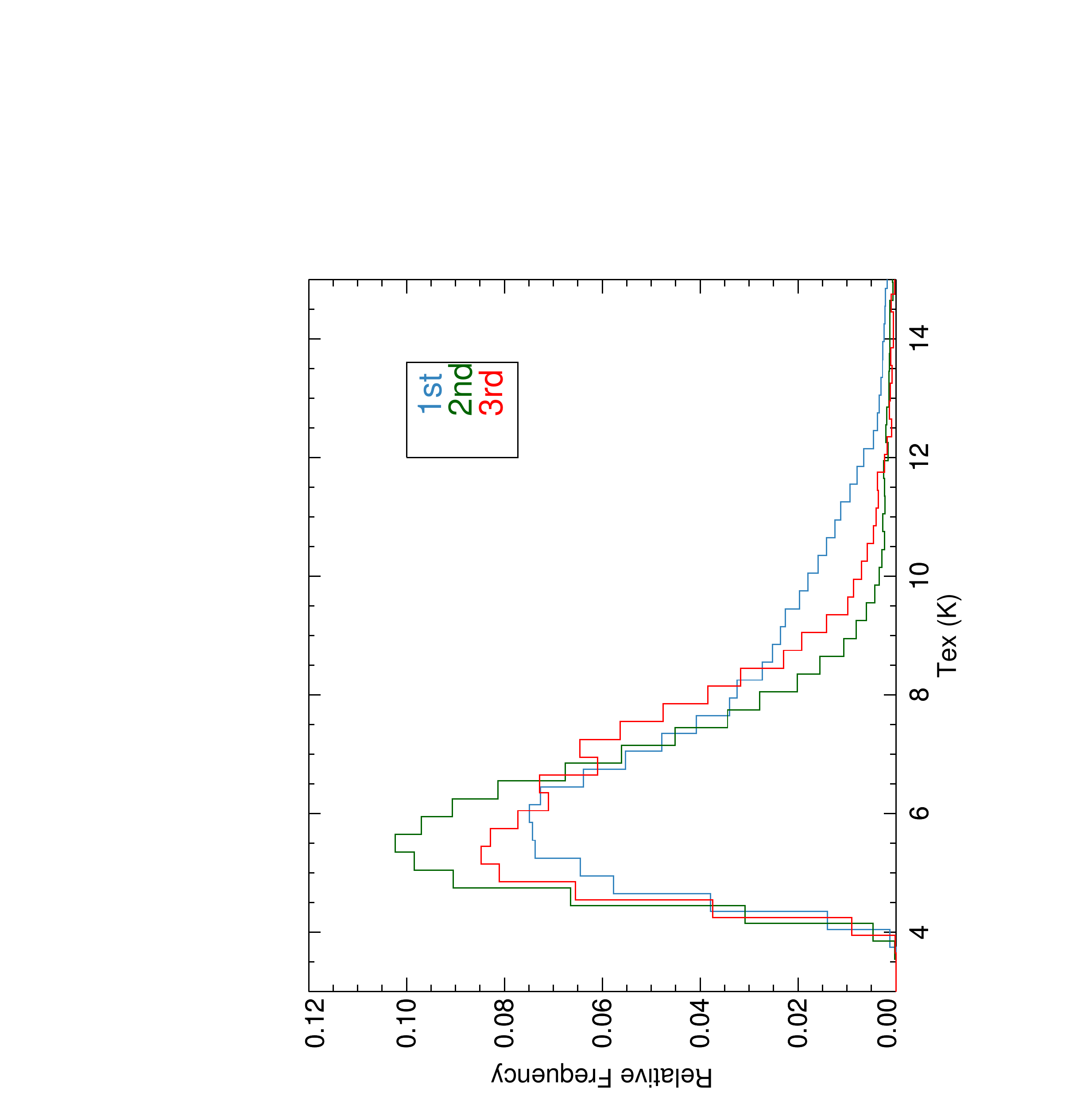}
 \caption{Probability distributions of the excitation temperatures of molecular clouds of the three velocity components. Only the spatial pixels with the integrated intensity of $^{12}$CO emission above 5$\sigma$ are used in this statistics.}
  \label{fig:tex_123}
\end{figure}

The spatial distributions of the excitation temperature of the molecular clouds for each velocity component are displayed in Figure \ref{fig:tex_s282} and the histograms are presented in Figure \ref{fig:tex_123}. For molecular clouds of the first velocity component, the eastern part of the Rosette molecular cloud possesses relatively high temperature of around 12 K. According to the map of the 21 cm radio continuum emission from \citet{1997A&AS..126..413R}, the ionization front of the Rosette \ion{H}{2} region is far away from the eastern part of the RMC. The circle-like cavity of the Rosette molecular cloud shows that the shock front just reaches the Extended Ridge and Monoceros Ridge of RMC \citep{2018ApJS..238...10L}. The relatively high temperature of the eastern part of RMC may be due to the local evolution of the molecular clouds rather than the feedback of the OB cluster of the RMC \ion{H}{2} region. The molecular cloud within G208.506-02.304 possesses the highest temperature (15 K), indicating that this cloud has been influenced by the G208.506-02.304 \ion{H}{2} region. Except for the eastern part of the RMC and the molecular cloud associated with the G208.506-02.304 \ion{H}{2} region, most of the other molecular clouds of the first velocity component have temperatures around 6 K.

Most of the molecular clouds of the second velocity component possess excitation temperatures ranging from 8 to 15 K. The highest temperature (33 K) occurs in the BFS54 region, which indicating that the molecular clouds within the BFS54 \ion{H}{2} region have been heated by the \ion{H}{2} region BFS54.

Most of the molecular clouds of the third velocity component possess temperatures ranging from 7 to 15 K, with a typical value of about 9K. Considering that the distances of these molecular clouds are from 3.0 kpc to 7.0 kpc, the derived temperatures should be heavily influenced by the effect of the filling factor. Therefore, the excitation temperatures of the molecular clouds of the third velocity component should be strongly underestimated.

As shown in Figure \ref{fig:tex_123}, the excitation temperatures of molecular clouds in the region are low as a whole. The peak probability excitation temperatures are at $5-6$ K. Above the excitation temperature of 9 K, the first velocity component has the most number of molecular clouds and most of these clouds belong to the eastern part of RMC. The second velocity component has the least number of clouds with high temperature. However, the molecular cloud within the BFS54 \ion{H}{2} region, which is a second velocity component cloud, processes the highest temperature (33 K) among the three velocity components. 

Using the method in \citet{2018ApJS..238...10L} and assuming that the molecular clouds of the first velocity component are located at the same distance as the RMC (1.4 kpc), the total mass of molecular clouds of the first velocity component is calculated to be 3.8 $\times$ 10$^{4}$ M$_{\odot}$ with the $X$ factor method, whereas 8.5 $\times$ 10$^{3}$ M$_{\odot}$ from $^{13}$CO, and 4.2 $\times$ 10$^{2}$ M$_{\odot}$ from C$^{18}$O with the LTE approach. Most of the mass of the first velocity is contained in the eastern part of the RMC, which is 3.1 $\times$ 10$^{4}$, 7.7 $\times$ 10$^{3}$, and 2.1 $\times$ 10$^{2}$ M$_{\odot}$ from $^{12}$CO, $^{13}$CO, and C$^{18}$O, respectively. Taking the distance of molecular clouds of the second velocity component to be 1.6 kpc, the total mass of the molecular clouds is calculated to be 2.1 $\times$ 10$^{4}$ M$_{\odot}$ ($^{12}$CO), 2.2 $\times$ 10$^{3}$ M$_{\odot}$ ($^{13}$CO), and 2.5 $\times$ 10$^{2}$ M$_{\odot}$ (C$^{18}$O), respectively. Considering that the distances of the molecular clouds of the third velocity component vary greatly (3.0$-$7.0 kpc), the total mass of clouds of the third velocity component is not calculated. Due to its higher abundance, $^{12}$CO could trace the diffuse clouds whereas the $^{13}$CO and C$^{18}$O emission is detectable only in relatively dense clouds. Therefore, the mass from $^{12}$CO is larger than those from $^{13}$CO and C$^{18}$O. We note that although there is less molecular mass in the surveyed region than its neighboring massive star forming regions such as Monoceros OB1 (1.3 $\times$ 10$^{5}$ M$_{\odot}$ from $^{12}$CO \citep{1996A&A...315..578O}) and RMC (2.0 $\times$ 10$^{5}$ M$_{\odot}$ from $^{12}$CO \citep{2018ApJS..238...10L}), the number of \ion{H}{2} regions/candidates (10) is similar to those of Monoceros OB1 (11) and RMC (10) \citep{2014ApJS..212....1A}.

\subsection{Molecular Clouds Associated with the \ion{H}{2} Regions/Candidates}\label{sec:HII}

\begin{figure}[h]
  \centering
  {\includegraphics[width=0.43\textwidth]{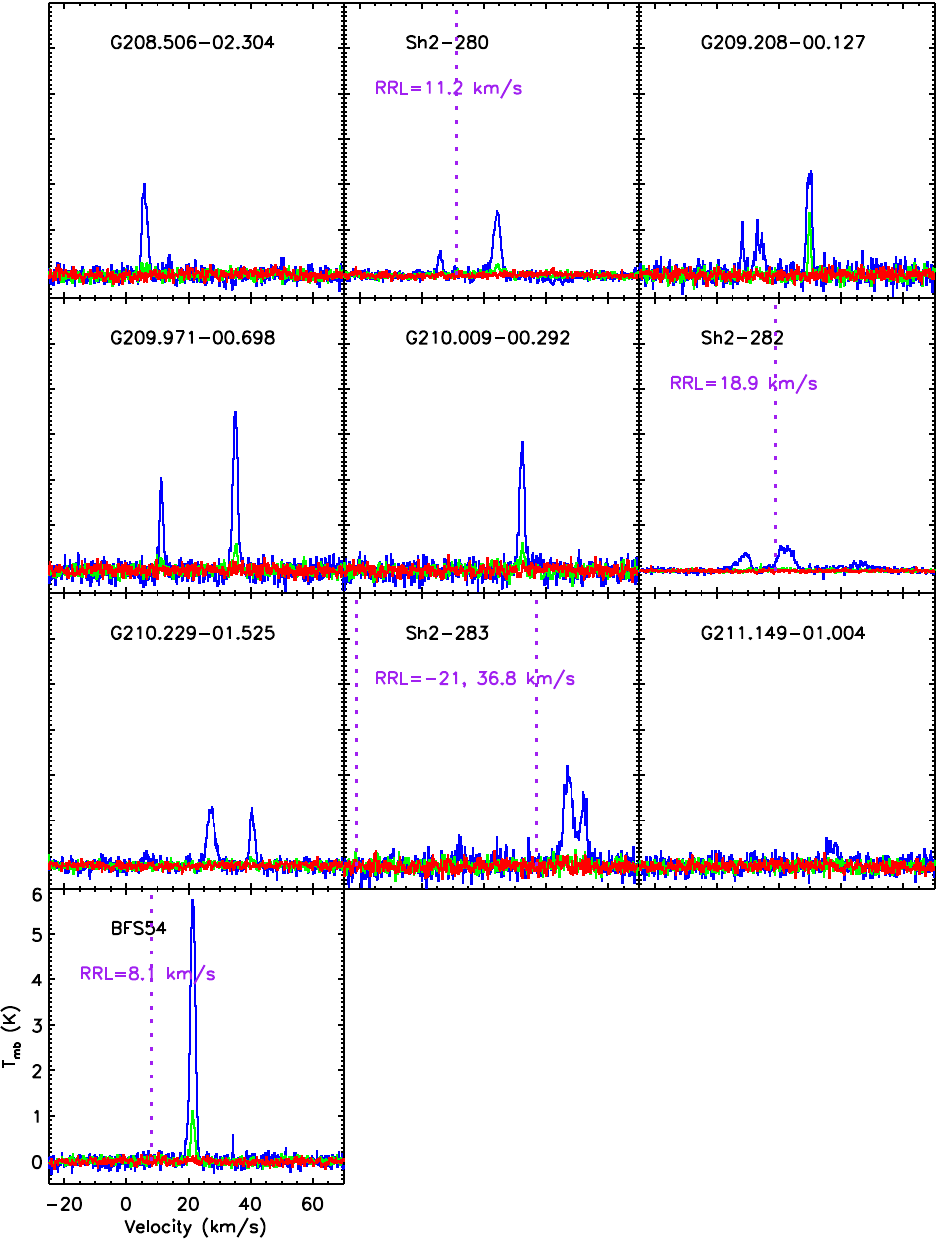}}
  \caption{Average spectra of the \ion{H}{2} regions/candidates with the blue, green, and red indicating the $^{12}$CO, $^{13}$CO and C$^{18}$O emission, respectively. The purple dashed line represents the velocity of the radio recombination line (RRL).}
  \label{fig:spec_HII}
\end{figure}

Ten \ion{H}{2} regions/candidates are located in the region, of which four \ion{H}{2} regions (Sh2-280, Sh2-282, Sh2-283, and BFS54) are identified by \citet{1953ApJ...118..362S,1959ApJS} and \citet{1982ApJS...49..183B}. By visual and automatic search of bubble morphology in the WISE 12 $\micron$ and 22 $\micron$ images, \citet{2014ApJS..212....1A} identified six \ion{H}{2} regions or \ion{H}{2} region candidates (G208.506-02.304, G209.208-00.127, G209.971-00.698, G210.009-00.292, G210.229-01.525, and G211.149-01.004). \citet{2014ApJS..212....1A} also reported the angular radii of the ten \ion{H}{2} regions/candidates on the basis of their visual check on the morphologies of the \ion{H}{2} regions/candidates in the MIR images. In this work we investigate whether there are any molecular clouds associated with these known or candidate \ion{H}{2} regions. For this purpose, we extract the mean spectra of $^{12}$CO, $^{13}$CO, and C$^{18}$O within the infrared bubble radii of the ten \ion{H}{2} regions/candidates. These spectra, ordered by Galactic longitude of \ion{H}{2} regions/candidates, are presented in Figure \ref{fig:spec_HII}. Single velocity component is present in the spectra of the \ion{H}{2} regions/candidates BFS54, G208.506-02.304, G210.009-00.292, and G211.149-01.004, while multiple velocity components exist in the other six known or candidate \ion{H}{2} regions. Radio recombination line (RRL) measurements are available for the four \ion{H}{2} regions (Sh2-280, Sh2-282, Sh2-283, and BFS54). The velocities of ionized gas from RRL \citep{2014ApJS..212....1A} are different from the velocities of the molecular clouds. The velocity offsets are 13 km s$^{-1}$  (Sh2-280), 3 km s$^{-1}$ (Sh2-282), 13 or 70 km s$^{-1}$ (Sh2-283), and 14 km s$^{-1}$ (BFS54), respectively. \citet{2009ApJS..181..255A} calculated the velocity offsets between the RRL and the $^{13}$CO emission line for 301 \ion{H}{2} regions. They found that the velocity offsets are smaller than 5 km s$^{-1}$ for most of the \ion{H}{2} regions. In our case, the velocity offsets are significantly larger than this value in three \ion{H}{2} regions (Sh2-280, Sh2-283, and BFS54).

The distributions of the molecular clouds in the \ion{H}{2} regions/candidates are presented in Figures \ref{fig:G2085-023}$-$\ref{fig:BFS54} in the Appendix. It can be seen that the 12 $\micron$ emission in G208.506-02.304 exhibits a typical bubble structure (Figure \ref{fig:G2085-023}). The molecular cloud associated with G208.506-02.304 has a similar morphology to the dust emission at 12 and 22 $\micron$. From the center of the \ion{H}{2} region to the outside, the velocity of the molecular cloud gradually increases, which may imply the influence of the \ion{H}{2} region on the molecular cloud. The molecular cloud associated with Sh2-280 consists of a giant filamentary structure surrounding the \ion{H}{2} region (Figure \ref{fig:S280}), which is similar to case of N131 studied by \citet{2016A&A...585A.117Z}. These authors suggest that the massive star in the N131 bubble was born from the disruption of the gas filament. Two compact sources of 1.4 GHz continuum emission are detected by the NRAO VLA Sky Survey (NVSS) within the infrared bubble. However, the 1.4 GHz continuum emission does not cover the central O-type star HD 46573. This may be caused by the poor (u, v) coverage in the snapshots of the NVSS observations which results in the NVSS images being insensitive to smooth radio emission structures larger than several arcminutes \citep{1998AJ....115.1693C}. For G209.208-00.127, two molecular clouds with distinct velocities are located within the radius of the \ion{H}{2} region/candidate (Figure \ref{fig:G2092-001}). We designate the lower velocity component as the 'near' cloud and the higher velocity component as the 'far' cloud. The 'near' and 'far' clouds have a velocity difference of about 20 km s$^{-1}$ . The G209.208-00.127\_far cloud has similar morphology to the mid-infrared emission and shows higher excitation temperature than the G209.208-00.127\_near cloud. Therefore, G209.208-00.127\_far is more likely to be associated with the \ion{H}{2} region than the 'near' cloud. However, Both the 'far' and the 'near' clouds are offset from the center of G209.208-00.127. It is also possible that both of the clouds are interacting with G209.208-00.127, with the 'far' cloud being pushed away and the 'near' cloud toward the Sun. Indeed, the 'far' cloud exhibits an increase of velocity from the center of G209.208-00.127 to the outside. Similar to G209.208-00.127, G209.971-00.698 is spatially associated with two clouds of distinct velocities (Figure \ref{fig:G2099-006}). The 'far' cloud has a velocity about 20 km s$^{-1}$ larger than that of the 'near' cloud. Both the 'near' and 'far' clouds are filamentary. The 'near' cloud is oriented in the northeast-southwest direction while the 'far' cloud roughly in the east-west direction when it is within the radius of G209.971-00.698 and then in the direction of northwest-southeast when it is outside of G209.971-00.698. Within the radius of G209.971-00.698, the 'far' cloud is similar to the MIR emission in morphology and it also shows a higher excitation temperature than the 'near' cloud. These facts suggest the 'far' cloud is more likely to be associated with G209.971-00.698 than the 'near' cloud. As both clouds have sizes much larger than the radius of G209.971-00.698 and that each cloud has its own coherent velocity, it is unlikely that the clouds are both accelerated by G209.971-00.698. The molecular cloud in the G210.009-00.292 region is mainly constrained within the radius of G210.009-00.292 and exhibits a clumpy morphology (Fig.\ref{fig:G2100-002}). The molecular cloud well matches the distribution of heated dust grains as traced by the MIR 22 $\micron$ emissions. The molecular clouds in the Sh2-282 region exhibit elephant trunk and cometary structures facing to the center (Fig.\ref{fig:S282}). \citet{2006A&A...445L..43C} found several bright-rimed structures in the Sh2-282\_far cloud that are faced to HD 47432, which supports that the Sh2-282\_far cloud is interacting with the Sh2-282 \ion{H}{2} region. For G210.229-01.525, as in the case of G209.208-00.127, both the 'far' and the 'near' clouds could be associated with the \ion{H}{2} region, considering that the 'near' cloud is located to the northeast and the 'far' cloud roughly to the southwest of G210.229-01.525 (Fig.\ref{fig:G2102-015}). The molecular cloud in the Sh2-283 region exhibits a filamentary morphology and is well coincident with the distribution of heated dusts as traced by the 12 and 22 $\micron$ emission (Figure \ref{fig:S283}). Therefore, the molecular cloud should be associated with the Sh2-283 \ion{H}{2} region. From Figure \ref{fig:G2111-010}, it can be seen that the molecular cloud in the G211.149-01.004 region exhibits the filamentary morphology. The velocity of this molecular filament increases from the center of G211.149-01.004 to the outside. Weak radio continuum and H$\alpha$ emission exist within the radius of the MIR bubble. According to \citet{2014ApJS..212....1A}, bubbles that have measured RRL or H$\alpha$ emission are classified as true \ion{H}{2} regions. Thus we adjust G211.149-01.004 from a candidate to a known \ion{H}{2} region in Table \ref{tab_HII}. In the BFS54 \ion{H}{2} region (Figure \ref{fig:BFS54}), the molecular cloud shows a central cavity which coincides not only with the MIR 12 and 22 $\micron$ emission but also with the radio continuum and H$\alpha$ emission, suggesting that the molecular cloud is being excavated by the BFS54 \ion{H}{2} region. The part of the molecular cloud that surrounds the cavity also possesses an unusually high temperature at around 30 K, which further supports that the BFS54 \ion{H}{2} region has a strong influence on its surrounding molecular cloud.

\startlongtable
\begin{deluxetable}{ccccccccccccccc}
\tabletypesize{\footnotesize}
\tablewidth{10pt}
\tablenum{2}
\tablecaption{Properties of molecular clouds in the regions of \ion{H}{2} regions/candidates}
\label{tab_HII}
\tablehead{
 \colhead{Name}  & \colhead{Category}& \colhead{Glon} & \colhead{Glat} & \colhead{V$_{lsr}$}
&\colhead{Distance}  & \colhead{Linear Radius} & \colhead{T$_{ex}$} & \colhead{Mass} & \colhead{$\Delta$v}  \\[1pt]
\colhead{ }&\colhead{ }& \colhead{(deg)} &
\colhead{(deg)} & \colhead{(km s$^{-1}$)}& \colhead{
  (kpc)}& \colhead{(pc)} & \colhead{(K)}& \colhead{(M$_{\odot}$)} & \colhead{(km/s)}}
\startdata
 \colhead{ }&\colhead{ }&\colhead{ }&\colhead{ }&\colhead{1st velocity component}   \\[1pt]
\hline
G208.506-02.304 &Q & 208.506 & -2.304 &  5.9  & 0.6 &  0.7 & 9.2   & 2.6 $\times$ 10$^1$ & 1.0  \\
\hline
 \colhead{ }&\colhead{ }&\colhead{ }&\colhead{ }&\colhead{2nd velocity component}   \\[1pt]
\hline
Sh2-280               &K  & 208.741 & -2.633 & 24.2 & 2.6 &  7.7 & 7.0   & 1.0 $\times$ 10$^4$ & 1.5  \\
G209.208-00.127 &Q & 209.208 & -0.127 & 29.7 & 3.3 &  1.2 & 6.8   & 1.3 $\times$ 10$^2$ & 1.1  \\
Sh2-282               &K  & 210.187 & -2.168 & 22.1 & 2.2 & 19.1 & 6.4  & 2.0 $\times$ 10$^4$ & 1.7  \\
BFS54                  &K  & 211.245 & -0.424 & 21.4 & 2.1 &  3.9 & 12.4 & 3.0 $\times$ 10$^3$ & 1.2  \\
\hline
 \colhead{ }&\colhead{ }&\colhead{ }&\colhead{ }&\colhead{3rd velocity component}   \\[1pt]
\hline
G209.971-00.698 &Q & 209.971 & -0.698 & 35.0 & 4.0 &  2.5 & 8.5.  & 1.1 $\times$ 10$^3$ & 1.1  \\
G210.009-00.292 &Q & 210.009 & -0.292 & 32.2 & 3.6 &  2.0 & 8.7   & 4.5 $\times$ 10$^2$ & 1.3  \\
G210.229-01.525 &Q & 210.229 & -1.525 & 40.5 & 4.8 &  6.1 & 8.3   & 3.6 $\times$ 10$^3$ & 1.4  \\
Sh2-283               &K  & 210.788 & -2.545 & 49.6 & 6.5 &  3.1 & 7.8   & 2.9 $\times$ 10$^3$ & 2.3  \\
G211.149-01.004 &K  & 211.149 & -1.004 & 37.1 & 4.2 &  3.9 & 7.6   & 4.9 $\times$ 10$^2$ & 1.1  \\\hline
\enddata
\tablecomments{Column 2 gives the category of the \ion{H}{2} regions/candidates, with K and Q indicating the known and radio quiet category, respectively. Column 6 gives the kinematic distance derived from model A5 in \cite{2014ApJ...783..130R}. Columns 9$-$10 give the mass and line width derived from $^{12}$CO. The physical properties are extracted within regions of two times the infrared bubble radii from \citet{2014ApJS..212....1A} because molecular clouds are generally larger than the infrared bubbles.}
\end{deluxetable}

By combining the data of dust, ionized gas, and molecular gas, the molecular clouds associated with the ten \ion{H}{2} regions/candidates are identified. The three velocity components contain one (G208.506-02.304), four (Sh2-280, G209.208-00.127, Sh2-282, and BFS54), and five (G209.971-00.698, G210.009-00.292, G210.229-01.525, Sh2-283, and G211.149-01.004) \ion{H}{2} regions/candidates, respectively. We derived the physical properties for each \ion{H}{2} region/candidate and list the results in Table \ref{tab_HII}. The masses of the \ion{H}{2} regions/candidates derived from the $X$ factor are from 26 to 2$\times 10^4$ M$_{\odot}$. Combining the angular radii from the MIR emission \citep{2014ApJS..212....1A} with the kinematic distances from the $^{12}$CO molecular line, we derived the linear radii of the ten \ion{H}{2} regions/candidates to be from 0.7 pc to 19.1 pc. The mean temperatures of the molecular clouds associated with the ten \ion{H}{2} regions/candidates are 6$-$12K, which are relatively higher than those of the other molecular clouds in the surveyed area.

Massive stars (B0 or earlier) are suggested to be the excitation sources of \ion{H}{2} regions \citep{2014ApJS..212....1A}. From the SIMBAD database, we searched for the O and B0 stars located within the \ion{H}{2} regions/candidates. Massive stars are found within Sh2-280 (Figure \ref{fig:S280}), Sh2-282 (Figure \ref{fig:S282}), Sh2-283 (Figure \ref{fig:S283}), G211.149-01.004 (Figure \ref{fig:G2111-010}), and BFS54 (Figure \ref{fig:BFS54}) and we list the properties of these massive stars in Table \ref{tab_HII_star}. Among the five massive stars list in Table \ref{tab_HII_star}, HD 46573, HD 47432, ALS 18674, and 2MASS J06465642+0116405, are approximately located within the emissions of the MIR, H$\alpha$, and radio continuum in regions of Sh2-280, Sh2-282, Sh2-283, and BFS54, respectively, and thus we suggest them to be the candidate excitation sources of the \ion{H}{2} regions. However, it is insufficient to identify the excitation sources of \ion{H}{2} regions based on the locations of the stars only and further investigation is needed to confirm the ionizing sources of the \ion{H}{2} regions. In Sh2-282, the radial velocity of HD 47432 is about 40 km s$^{-1}$ larger than that of the molecular clouds. However, the velocities of RRL and molecular clouds in this region are similar (Figure \ref{fig:spec_HII}). The large radial velocity of HD 47432 may come from its peculiar motion. No massive stars are found in the other \ion{H}{2} region candidates. As the OB-type star EM* RJHA 48 is located outside the well defined MIR bubble in the 12 and 22 $\micron$ images (Figure \ref{fig:G2111-010}), we do not consider it to be the excitation source of G211.149-01.004.

\clearpage
\startlongtable
\begin{deluxetable}{ccccccccccccccc}
\tabletypesize{\scriptsize}
\tablewidth{10pt}
\tablenum{3}
\tablecaption{Properties of the massive stars within the \ion{H}{2} regions/candidates}
\label{tab_HII_star}
\tablehead{
 \colhead{Region}  & \colhead{Category} & \colhead{Massive star}& \colhead{Glon} & \colhead{Glat} & \colhead{Radial velocity}
&\colhead{Parallax}  & \colhead{Spectral type}& \colhead{Association} \\[1pt] 
\colhead{ }&\colhead{ }&\colhead{ }& \colhead{(deg)} &
\colhead{(deg)} &\colhead{(km s$^{-1}$)}& \colhead{
  (mas)}& \colhead{ } & \colhead{ }  }
\startdata
 \colhead{ }&\colhead{ }&\colhead{ }&\colhead{ }&\colhead{1st velocity component}   \\[1pt]
\hline
G208.506-02.304  &Q&\nodata \\
\hline
 \colhead{ }&\colhead{ }&\colhead{ }&\colhead{ }&\colhead{2nd velocity component}   \\[1pt]
\hline
Sh2-280              &K  &HD 46573  & 208.730 & -02.631 & 29.3 & 0.6471 &  O7V((f))z &  Yes\\
G209.208-00.127 &Q&\nodata \\
Sh2-282              &K  &HD 47432  & 210.035 & -02.111 & 60.1 & 0.3873 & O9.7Ib  &  Yes\\
BFS54                 &K  &2MASS J06465642+0116405  & 211.282 & -00.418 & \nodata & 0.4240 &  B0.5Ve &  Yes\\
\hline
 \colhead{ }&\colhead{ }&\colhead{ }&\colhead{ }&\colhead{3rd velocity component}   \\[1pt]
\hline
G209.971-00.698  &Q&\nodata \\
G210.009-00.292  &Q&\nodata \\
G210.229-01.525  &Q&\nodata \\
Sh2-283                &K&ALS 18674 & 210.788 & -02.550 & 70.6 & \nodata& B0V &  Yes\\
G211.149-01.004  &K&EM* RJHA 48 & 211.168 & -01.043 & \nodata & \nodata &  OB &  No\\\hline
\enddata
\tablecomments{Column 3 gives the massive stars within the \ion{H}{2} regions/candidates. Columns $4-5$ give the Galactic longitude and latitude of the massive stars. Columns 6$-$8 give the radial velocity \citep{2006AstL...32..759G}, parallax \citep{2018yCat.1345....0G}, and spectral type \citep{2011ApJS..193...24S,2015MNRAS.454.3597D} of the massive stars, respectively.}
\end{deluxetable}


\section{Discussion}\label{discussion}

\subsection{Kinematics of the Molecular Clouds Associated with the \ion{H}{2} Regions/Candidates}\label{Kinematics}

Classical theories on \ion{H}{2} regions assume the interstellar medium around the \ion{H}{2} regions to be three-dimensional structures. However, \citet{2010ApJ...709..791B} did not find the associated foreground or background molecular clouds at the center for 43 infrared bubbles. Many recent studies argue that the molecular clouds associated with the infrared bubbles are two-dimensional ring-like structures \citep{2012ApJ...755...87S,2017ApJ...849..140X}.

The kinematic of molecular clouds provides unique information on the interaction between \ion{H}{2} regions and the surrounding medium and can be used to reveal the spatial relation between the molecular clouds and the \ion{H}{2} regions. We present the position-velocity maps of the ten \ion{H}{2} regions/candidates in Figures \ref{fig:Kinematics_G2085-023}$-$\ref{fig:Kinematics_BFS54} in the Appendix. In order to examine the possible influence of the \ion{H}{2} regions/candidates on the surrounding gas, the position paths of these p-v diagrams are set to pass through the centers of the \ion{H}{2} regions/candidates and along the major axes of the molecular clouds. All the paths have the same width of 5 pixels.

In the G208.506-02.304 region, the velocity of the molecular clouds along the position path is roughly constant at first and then increases toward the edge of the infrared bubble (Fig.\ref{fig:Kinematics_G2085-023}). The position-velocity map of Sh2-280 exhibits a partial cavity structure (Figure \ref{fig:Kinematics_S280}). Fitting the cavity structure in the position-velocity map with the model of expanding gas shell and taking the velocity offset of the most red-shifted gas at the middle of the position path as the expansion velocity, we estimate the expansion velocity of the Sh2-280 \ion{H}{2} region to be 3 km s$^{-1}$. This expansion velocity gives a dynamical time of 2.5 Myr for the Sh2-280 \ion{H}{2} region, assuming a uniform expansion. The velocity width of the molecular cloud at the position offset of HD 46573 is broadened, showing the feedback of HD 46573 on the cloud and supporting that HD 46573 is the excitation source of the Sh2-280 \ion{H}{2} region. The p-v diagram of G209.208-00.127\_far (Fig.\ref{fig:Kinematics_G2092-001}) shows two clumps with a velocity difference of about 1.5 km s$^{-1}$, supporting the idea that the molecular cloud is interacting with the \ion{H}{2} region candidate G209.208-00.127. Along the position path, the velocity of G209.971-00.698\_far is nearly constant (Fig.\ref{fig:Kinematics_G2099-006}), implying that G209.971-00.698\_far is either the background cloud or the foreground cloud to the \ion{H}{2} region candidate G209.971-00.698, but unlikely to contain the both. The velocity of the molecular cloud in G210.009-00.292 increases gradually from the center to the outside (Fig.\ref{fig:Kinematics_G2100-002}). The cloud associated with Sh2-282, Sh2-282\_far, is clumpy in the p-v diagram and the clumps show broadened velocity range (Figure \ref{fig:Kinematics_S282}), indicating the influence of the O-type star HD 47432 on the molecular cloud. As in G210.009-00.292, the the 'far' molecular cloud in G210.229-01.525 shows a velocity increase from the center of the region to the outside (Fig.\ref{fig:Kinematics_G2102-015}). In Sh2-283, the velocity at first decreases from the outside to the center of the region and then increases from the center to the outside (Figure \ref{fig:Kinematics_S283}). In G211.149-01.004 (Figure \ref{fig:Kinematics_G2111-010}) the velocity increases obviously from the center to the outside, whereas in BFS54 (Fig.\ref{fig:Kinematics_BFS54}), the velocity gradient is relatively small.

\subsection{Radial Temperature and Line Width Profiles of the \ion{H}{2} Regions/Candidates}
\begin{figure}[h]
  \centering
    \subfigure[]{
  \label{intensity_radius_tex}
\includegraphics[width=0.45\textwidth]{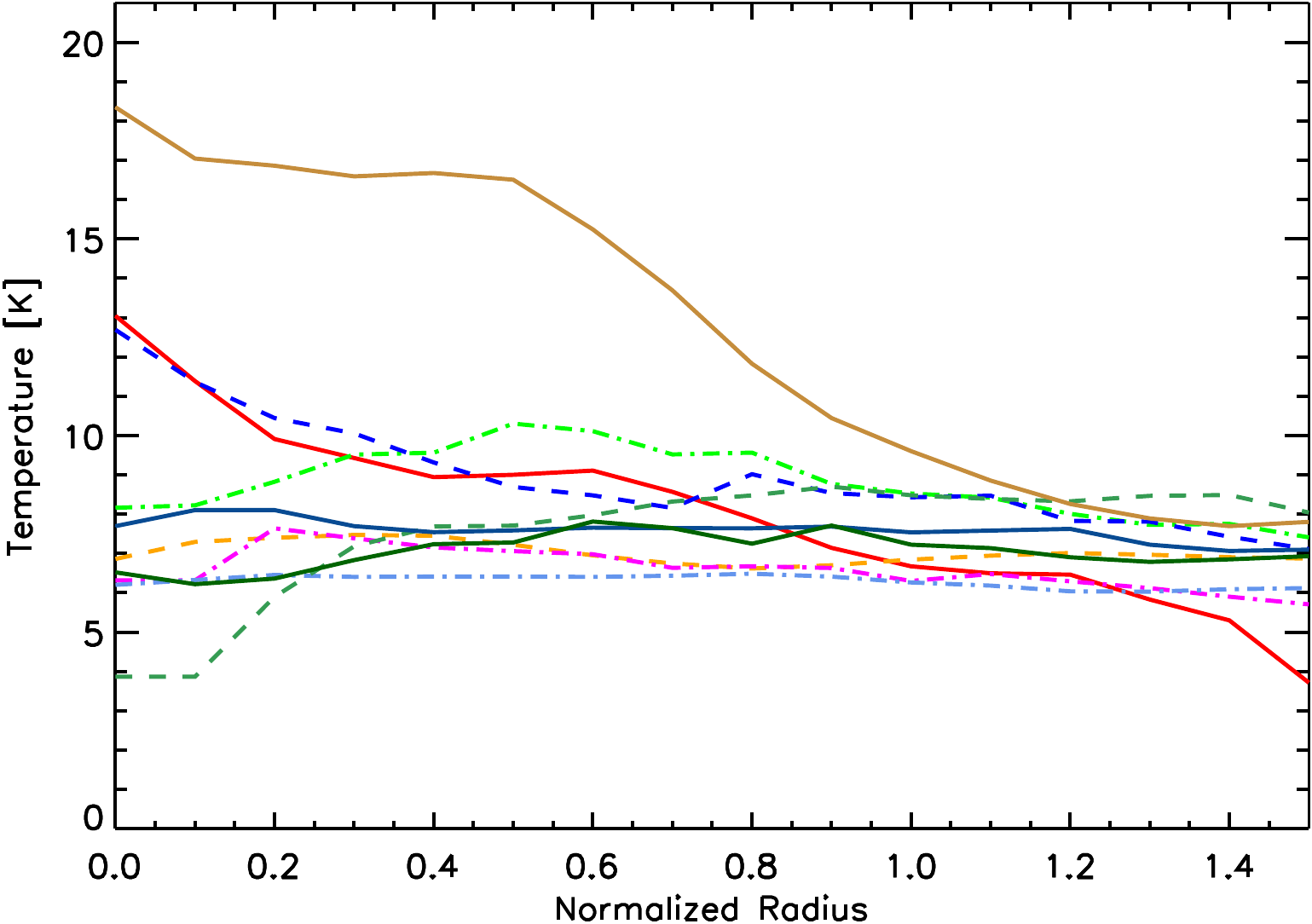}}
  \subfigure[]{
  \label{intensity_radius_m2}
\includegraphics[width=0.443\textwidth]{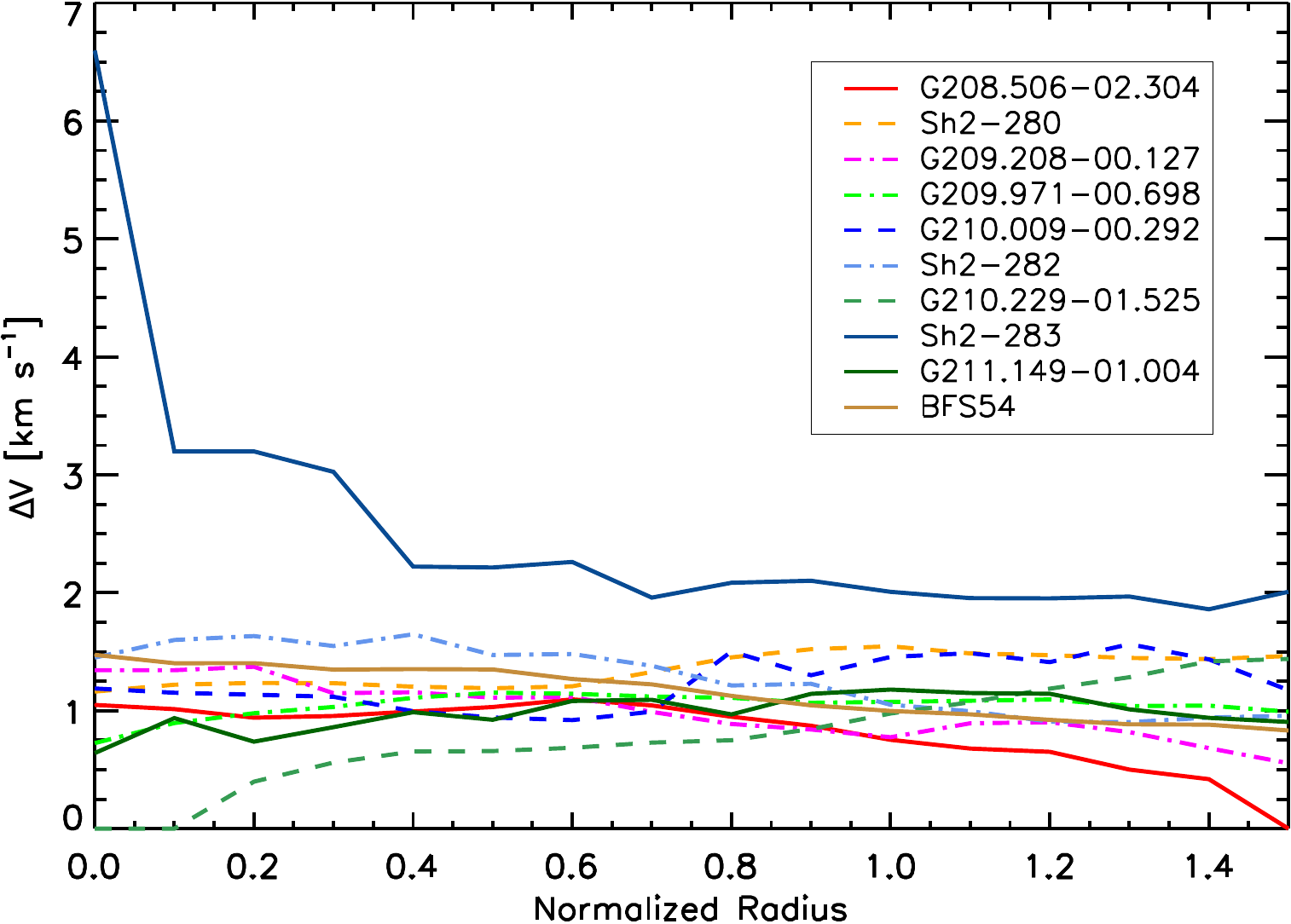}}
  \caption{Azimuthally averaged radial profiles of excitation temperature (a) and line width (b) of the molecular clouds in the ten \ion{H}{2} regions/candidates.}
  \label{fig:intensity_radius_tex_m2}
\end{figure}

Expanding \ion{H}{2} regions feed momentum and energy to the surrounding interstellar medium, which affects the temperature and turbulence of the surrounding molecular clouds. Assuming three-dimensional structure and uniform density for the surrounding gas, \citet{2006ApJ...646..240H} calculated the dynamical expansion of \ion{H}{2} regions, including the evolution of velocity, density, temperature, and pressure of the gas. Their results show that the peak of expansion velocity occurs at the position of the shock front and the temperature of gas decreases with the distance from the excitation star (see their Figs 3 and 4). If the \ion{H}{2} regions and the surrounding gas are two-dimensional structures, it is reasonable to assume that the results for the molecular gas are qualitatively similar to those in the three-dimensional case. In the two-dimensional case, the projected distance to the excitation star is closely related to the physical distance. Therefore we could expect that a line width maximum would occur around the radius of the \ion{H}{2} region and that the gas temperature would decrease with the projected distance from the center of the \ion{H}{2} region. In the three-dimensional case, however, the projected distance does no directly represent the physical distance, and therefore the trends in two-dimensional case may not appear. To examine whether these two trends appear in the observed molecular line emission images, we calculated the azimuthally averaged radial profiles of excitation temperature and line width for the 10 \ion{H}{2} regions/candidates and show the results in Figure \ref{fig:intensity_radius_tex_m2}. The temperatures in three regions, BFS54, G208.506-02.304, and G210.009-00.292 gradually decrease with the distance to the center of the \ion{H}{2} regions, whereas no such trend is found for the other seven \ion{H}{2} regions/candidates. Except for Sh2-283, the ten \ion{H}{2} regions/candidates show smooth distributions of line width with the projected distance. Sh2-283 shows decreasing line width with the projected distance and has a higher line width across the whole range of distance than the other \ion{H}{2} regions/candidates. For all the ten \ion{H}{2} regions/candidates, no obvious maximum of line width occurs around the nominal radius. Therefore, our data favor the scenario that \ion{H}{2} regions and their surrounding molecular gas are three-dimensional structures.

\section{Summary}\label{summary}

Based on the MWISP project, we performed $^{12}$CO, $^{13}$CO, and C$^{18}$O observations covering the region of Galactic longitude of 207.7$^{\circ}$ $< l <$ 211.7$^{\circ}$ and Galactic latitude of $-$4.5$^{\circ}$ $< b <$ 0$^{\circ}$ (4.0$^{\circ}$ $\times$ 4.5$^{\circ}$). The velocity resolution is 0.16 km s$^{-1}$ and the corresponding sensitivities are 0.5 K for $^{12}$CO emission and 0.3 K for $^{13}$CO and C$^{18}$O emission. For the first time, we find abundant molecular clouds in this region. We summarize our results as follows,

1. According to the velocity distribution, the molecular clouds are divided into three groups of different velocity components, i.e., $-$3 km s$^{-1}$ to 16.5 km s$^{-1}$ (first velocity component), 16.5 km s$^{-1}$ to 30 km s$^{-1}$ (second velocity component), and 30 km s$^{-1}$ to 55 km s$^{-1}$ (third velocity component), with kinematic distances from 0.5 kpc to 7.0 kpc. The molecular clouds of the second velocity component typically have filamentary morphology. Four molecular clumps with velocities of $48-51$ km s$^{-1}$ at Galactic latitudes of around -4.4$^{\circ}$ are estimated to be -525 to -508 pc from the Galactic mid-plane, which are significantly greater than the $\sigma$ thickness of the Galactic molecular gas disk (80$-$190 pc) at Galactic radius of $\sim$14 kpc.

2. By combining the data of dust, ionized gas, and molecular gas, the molecular clouds associated with the ten infrared bubbles in this region are identified and the physical properties of these molecular clouds are obtained. The linear radii of the ten \ion{H}{2} regions/candidates are from 0.7 pc to 19.1 pc and the masses derived from the $X$ factor are from 26 to 2$\times 10^4$ M$_{\odot}$. We suggest G211.149-01.004 to be a true \ion{H}{2} region based on its detection of radio continuum and H$\alpha$ emission. Massive stars are found within Sh2-280, Sh2-282, Sh2-283, and BFS54, and  we suggest them to be the candidate excitation sources of the \ion{H}{2} regions. No massive stars with spectral type earlier than B0 have been found in the other six \ion{H}{2} regions/candidates.

3. The distributions of excitation temperature and line width with the projected distance from the center of \ion{H}{2} region/candidate suggest that the majority of the ten \ion{H}{2} regions/candidates and their associated molecular gas are three-dimensional structures, rather than two-dimensional structures.

\section{Acknowledgements}

We would like to thank the PMO-13.7 m telescope staff for their supports during the observation and thank Min Fang, Yang Su, Zhiwei Chen, and Shaobo Zhang for the helpful discussions. We thank the anonymous referee for valuable comments and suggestions that helped to improve this paper. This work is supported by the National Key R\&D Program of China (NO. 2017YFA0402701) and Key Research Program of Frontier Sciences of CAS under grant QYZDJ-SSW-SLH047. C. Li acknowledges supports by NSFC grants 11973091, 11503086, and 11503087. H. Wang acknowledges supports by NSFC grant 11973091. This work makes use of data products from the Wide-field Infrared Survey Explorer, which is a joint project of the University of California, Los Angeles, and the Jet Propulsion Lab-oratory/California Institute of Technology, funded by the National Aeronautics and Space Administration. This work makes use of data products from the Southern H-Alpha Sky Survey Atlas (SHASSA), which is supported by the National Science Foundation. This work has also made use of the SIMBAD database, operated at CDS, Strasbourg, France.

\appendix
\label{appendix}

\begin{figure}[ht]
\centering
\includegraphics[width=1.0\textwidth]{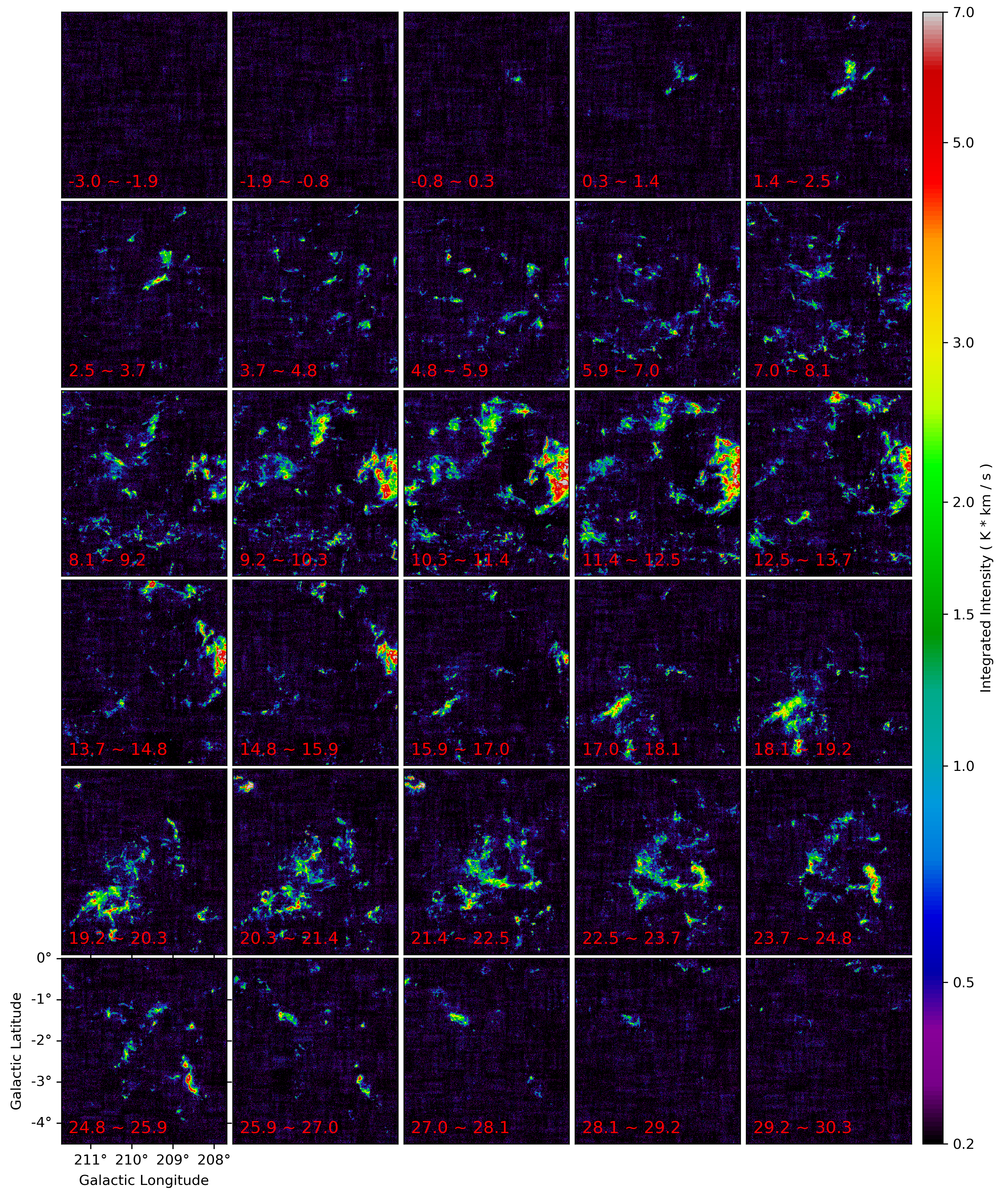}
\caption{Velocity channel map of $^{12}$CO emission of the surveyed area.}
\label{channel}
\end{figure}
\clearpage

\begin{figure}[h]
  \centering
\includegraphics[width=0.21\textwidth]{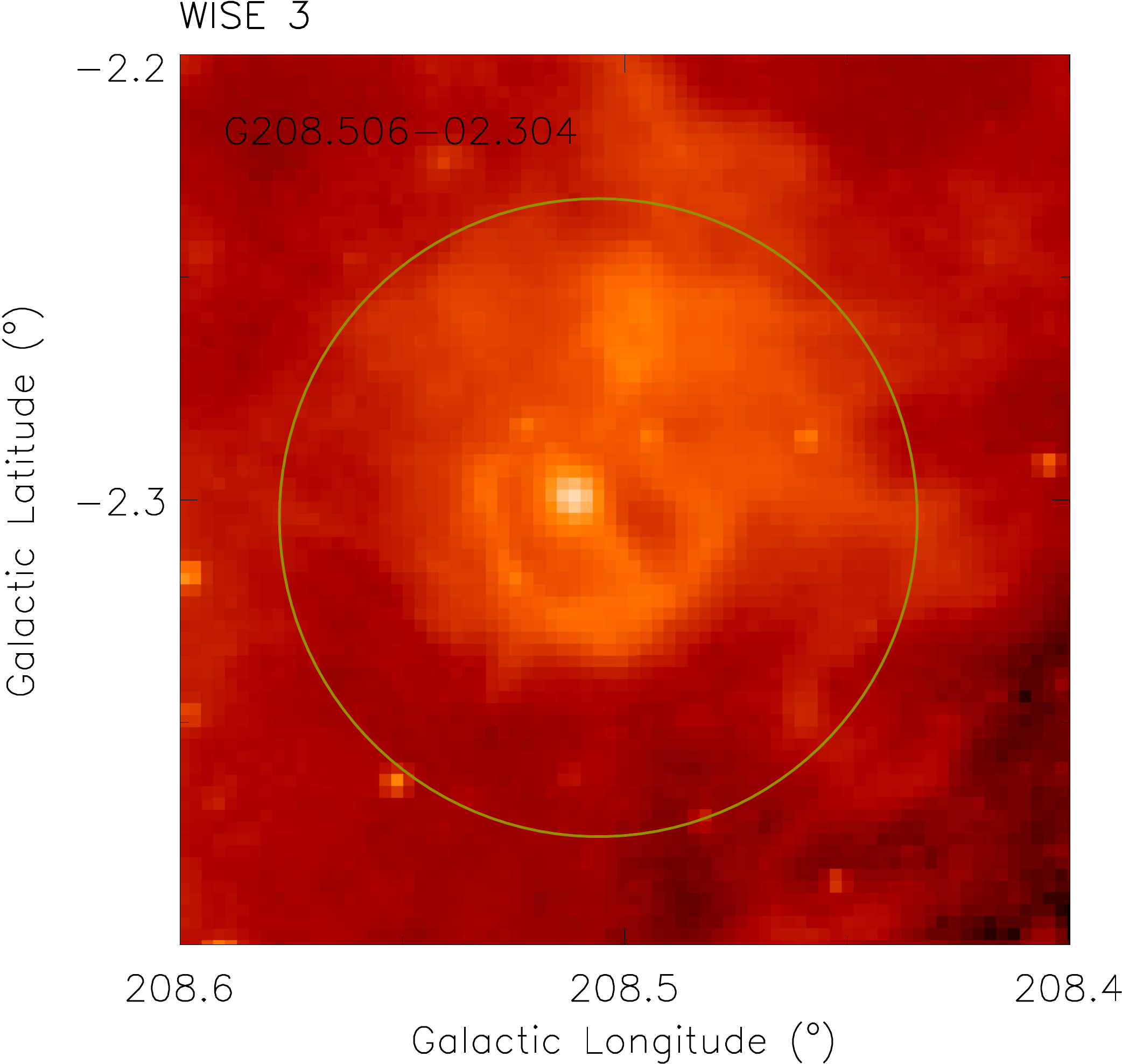}
\includegraphics[width=0.21\textwidth]{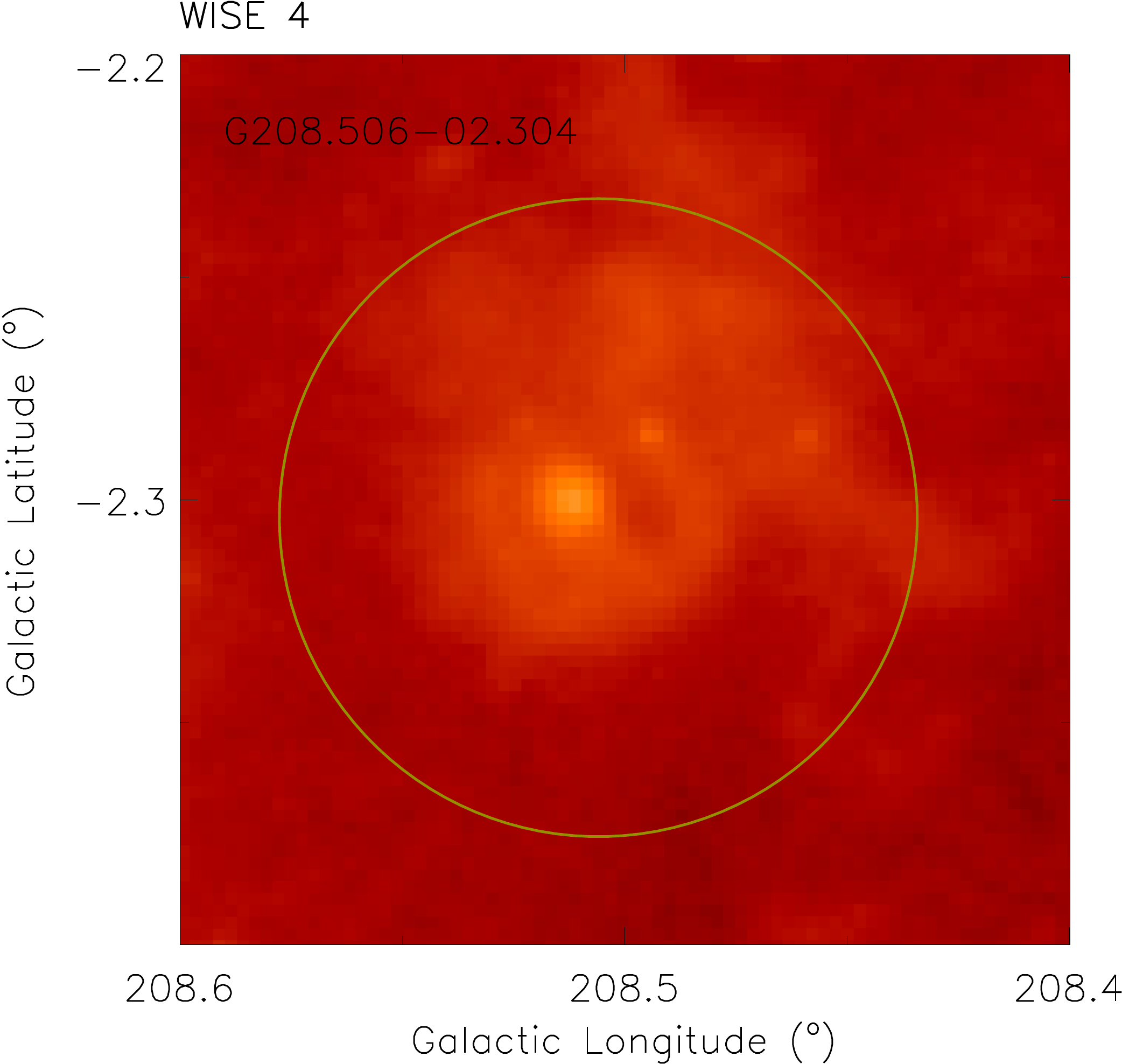}
\includegraphics[width=0.21\textwidth]{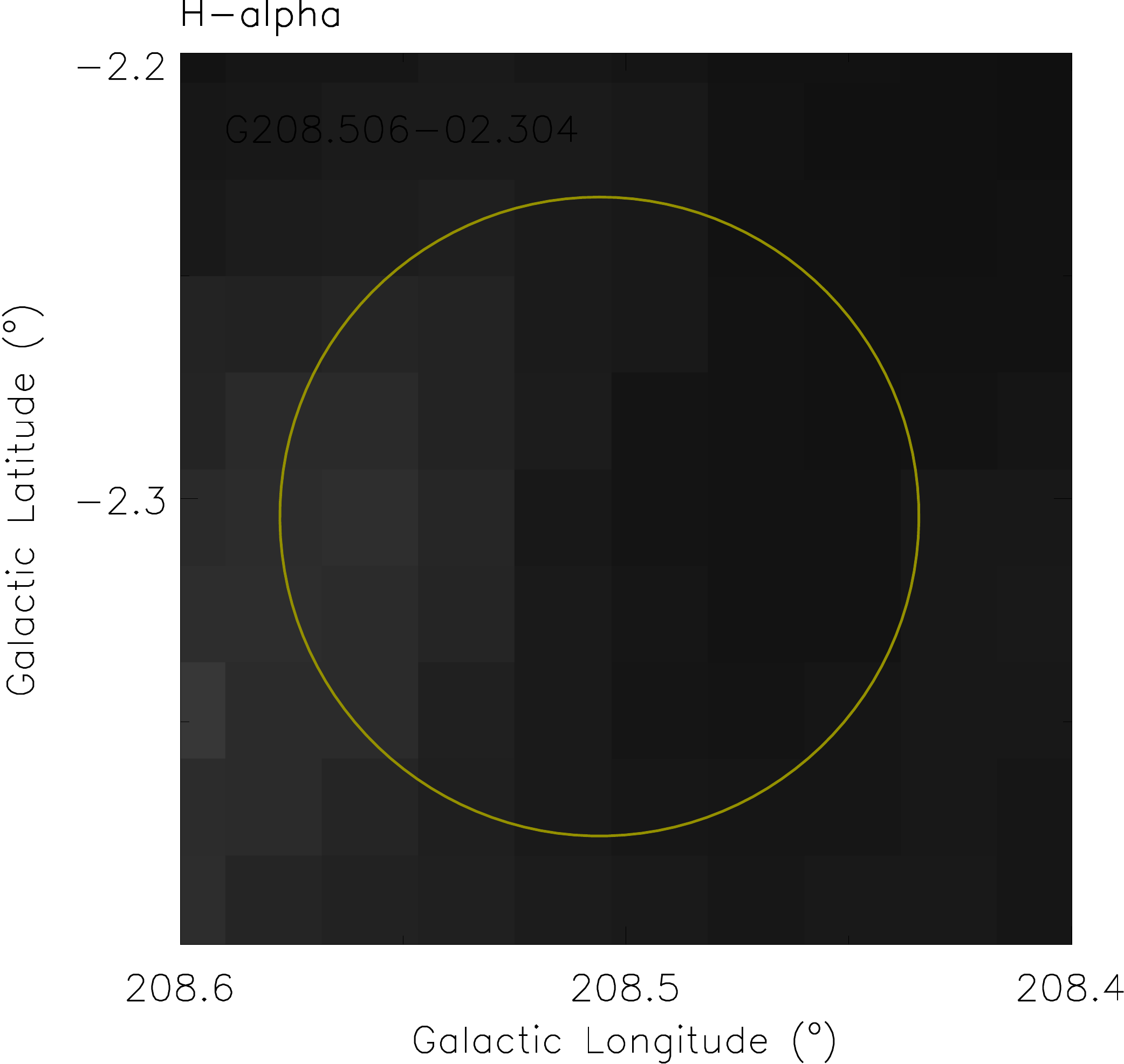}\\
\includegraphics[width=0.24\textwidth]{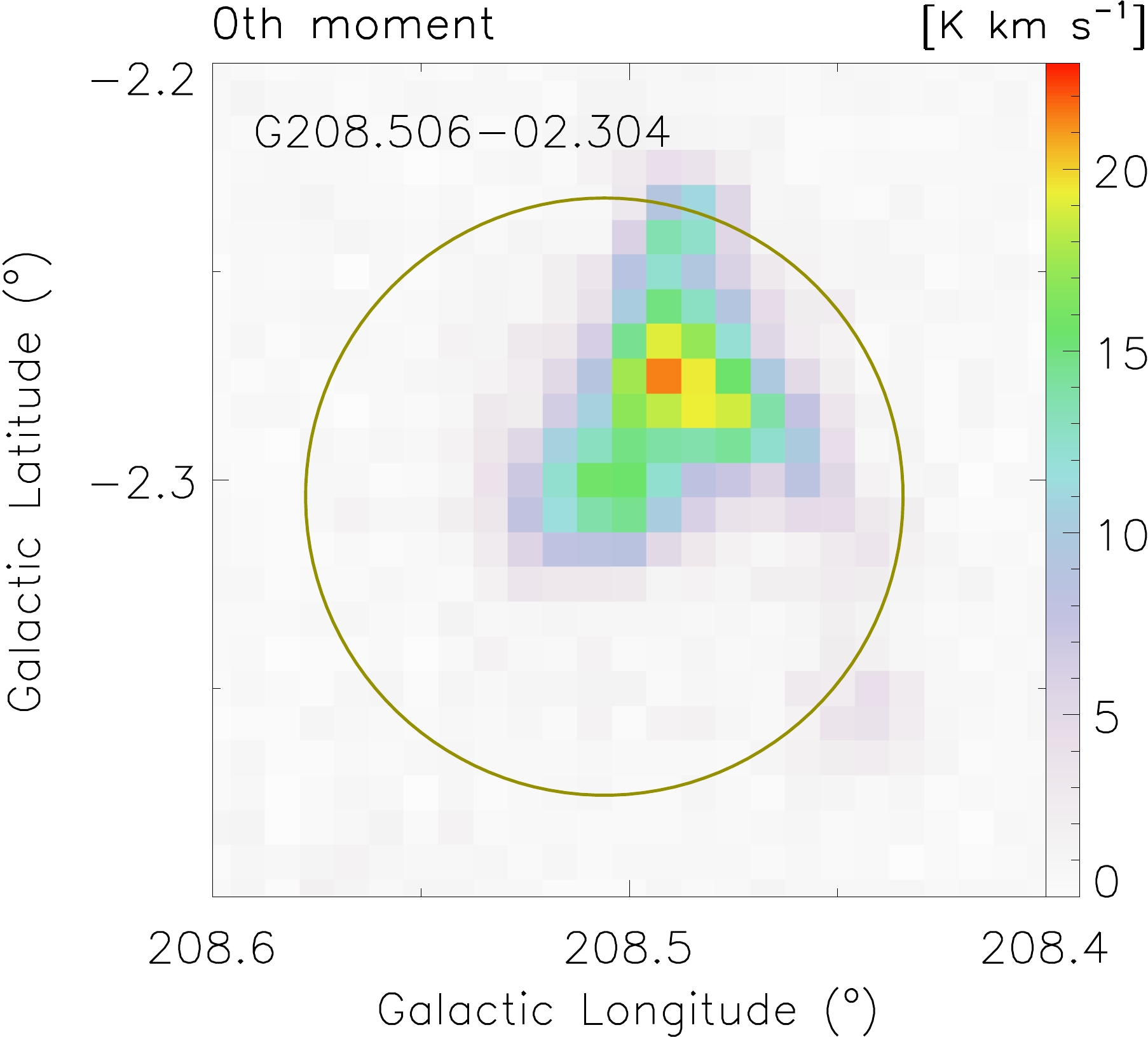}
\includegraphics[width=0.24\textwidth]{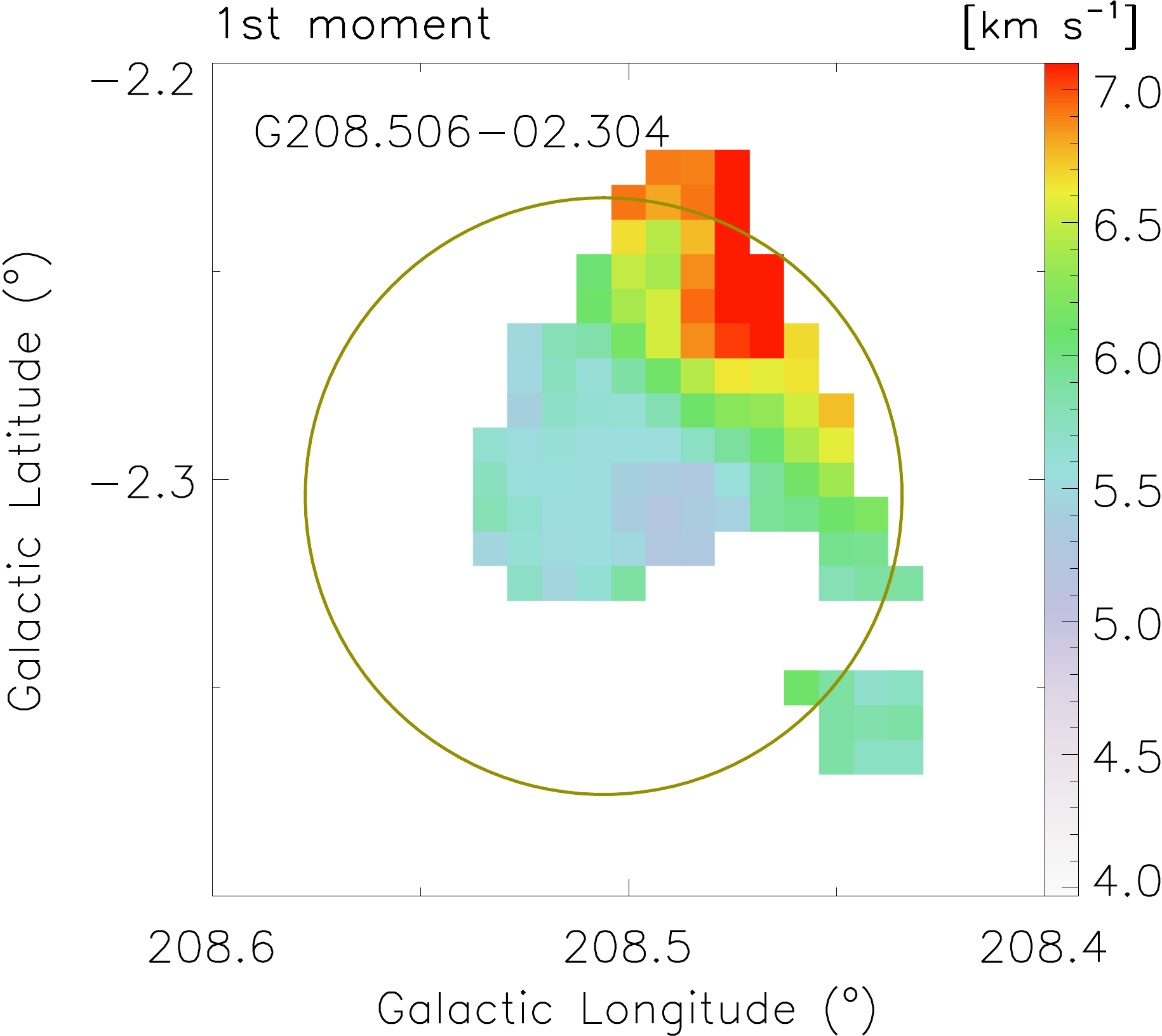}
\includegraphics[width=0.24\textwidth]{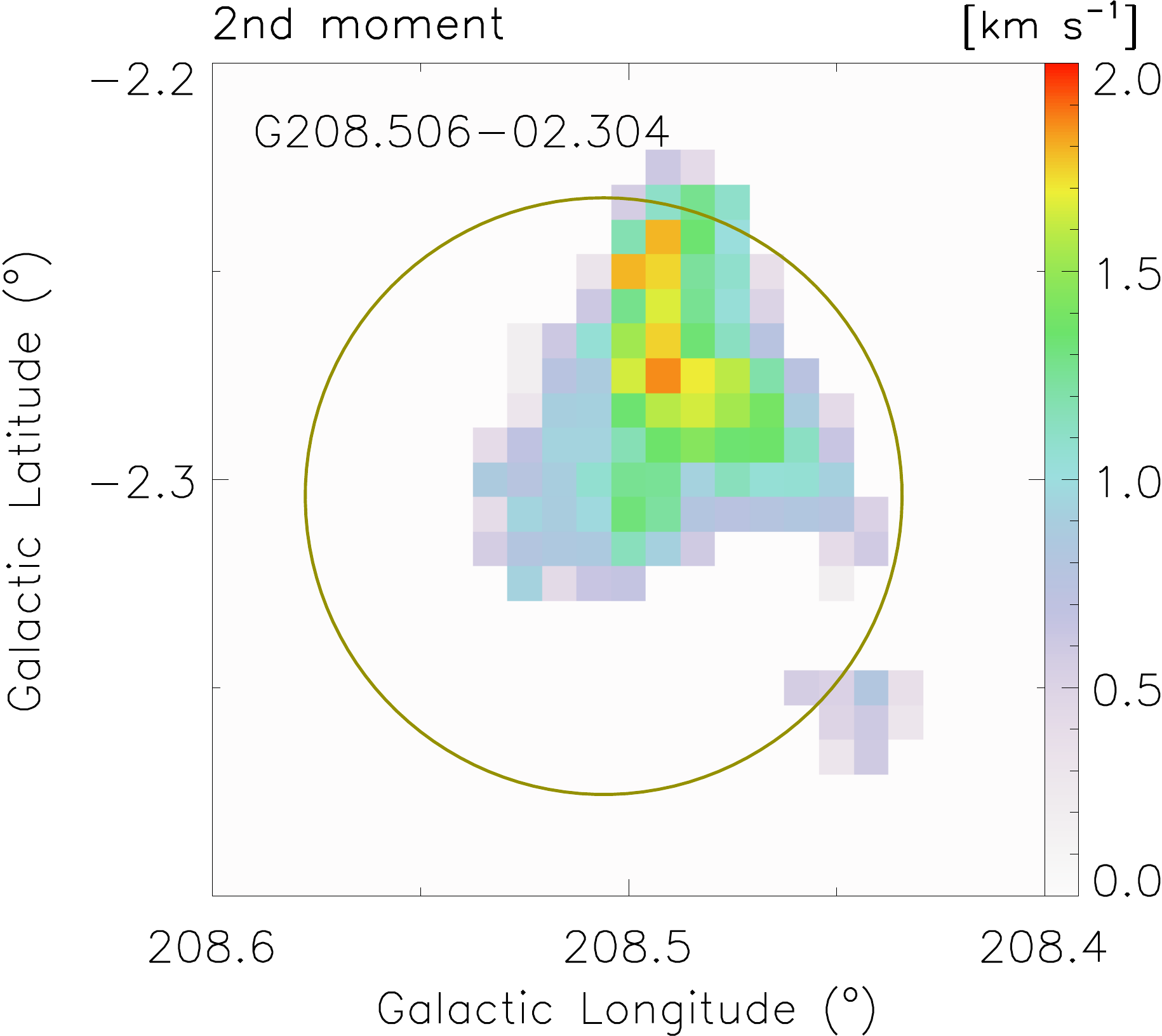}
\includegraphics[width=0.24\textwidth]{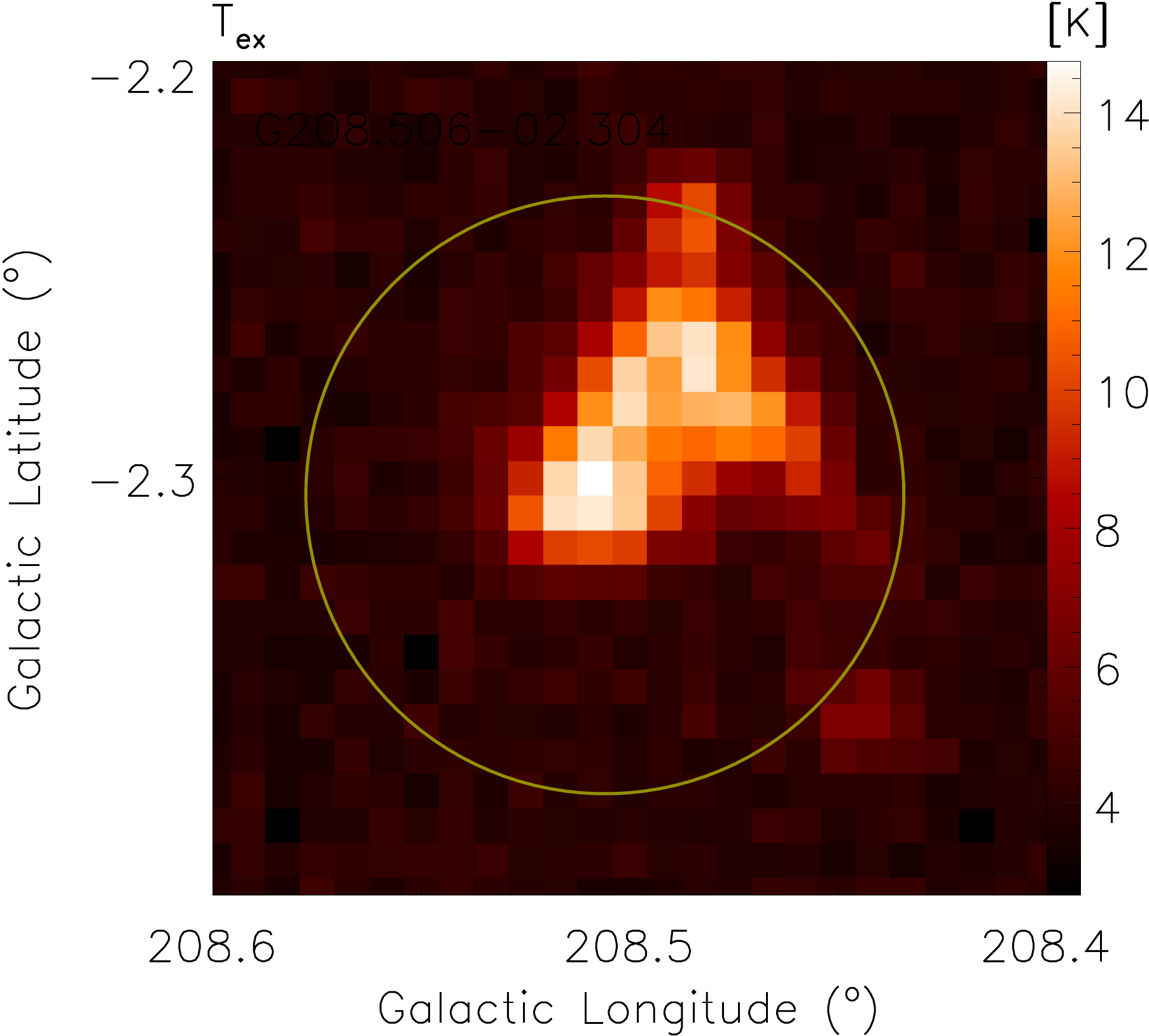}
  \caption{Morphology of G208.506-02.304 in various tracers. Top left: $WISE$ 3 (12 $\mu m$). Top middle: $WISE$ 4 (22 $\mu m$). Top right: H$\alpha$ emission from the SHASSA survey \citep{2001PASP..113.1326G}. Bottom: integrated intensity (left 1), velocity distribution (left 2), line width of velocity (left 3), and excitation temperature (left 4) images of $^{12}$CO. The circle size approximates the radius of the \ion{H}{2} region from \citet{2014ApJS..212....1A}. The velocity range for intensity integration is from 4 km s$^{-1}$ to 8 km s$^{-1}$.}
  \label{fig:G2085-023}
\end{figure}

\begin{figure}[h]
  \centering
\includegraphics[width=0.21\textwidth]{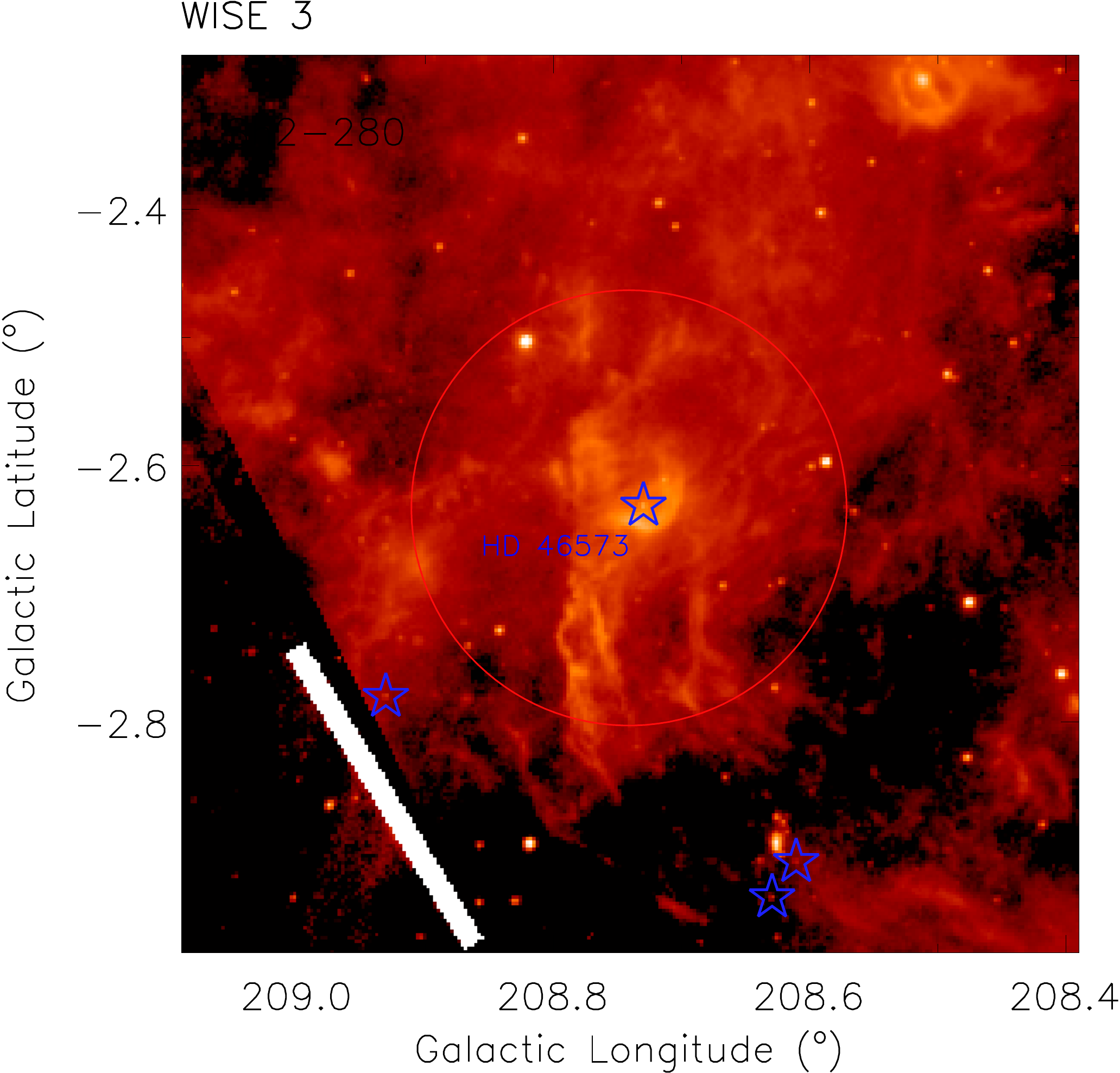}
\includegraphics[width=0.21\textwidth]{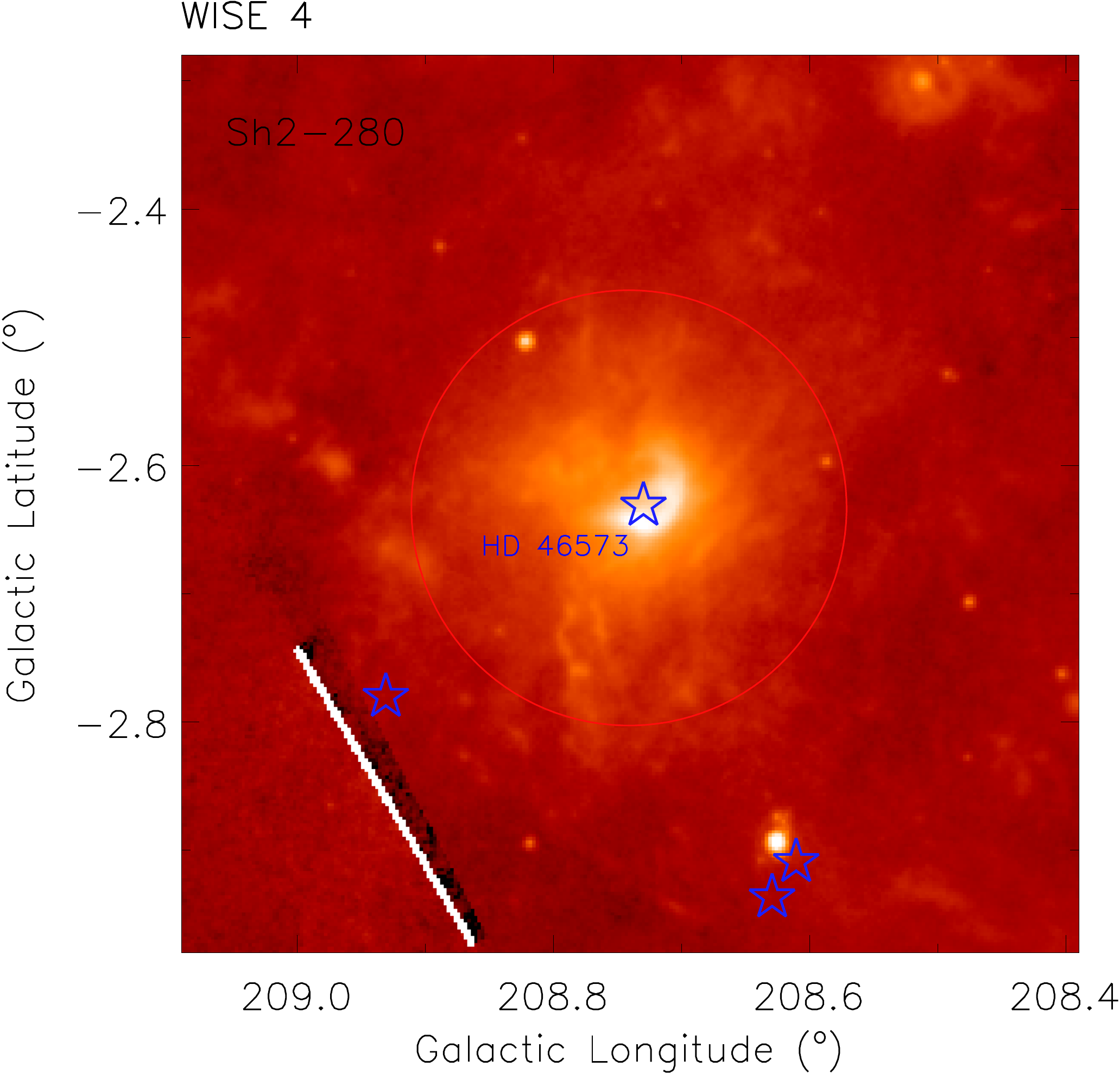}
\includegraphics[width=0.21\textwidth]{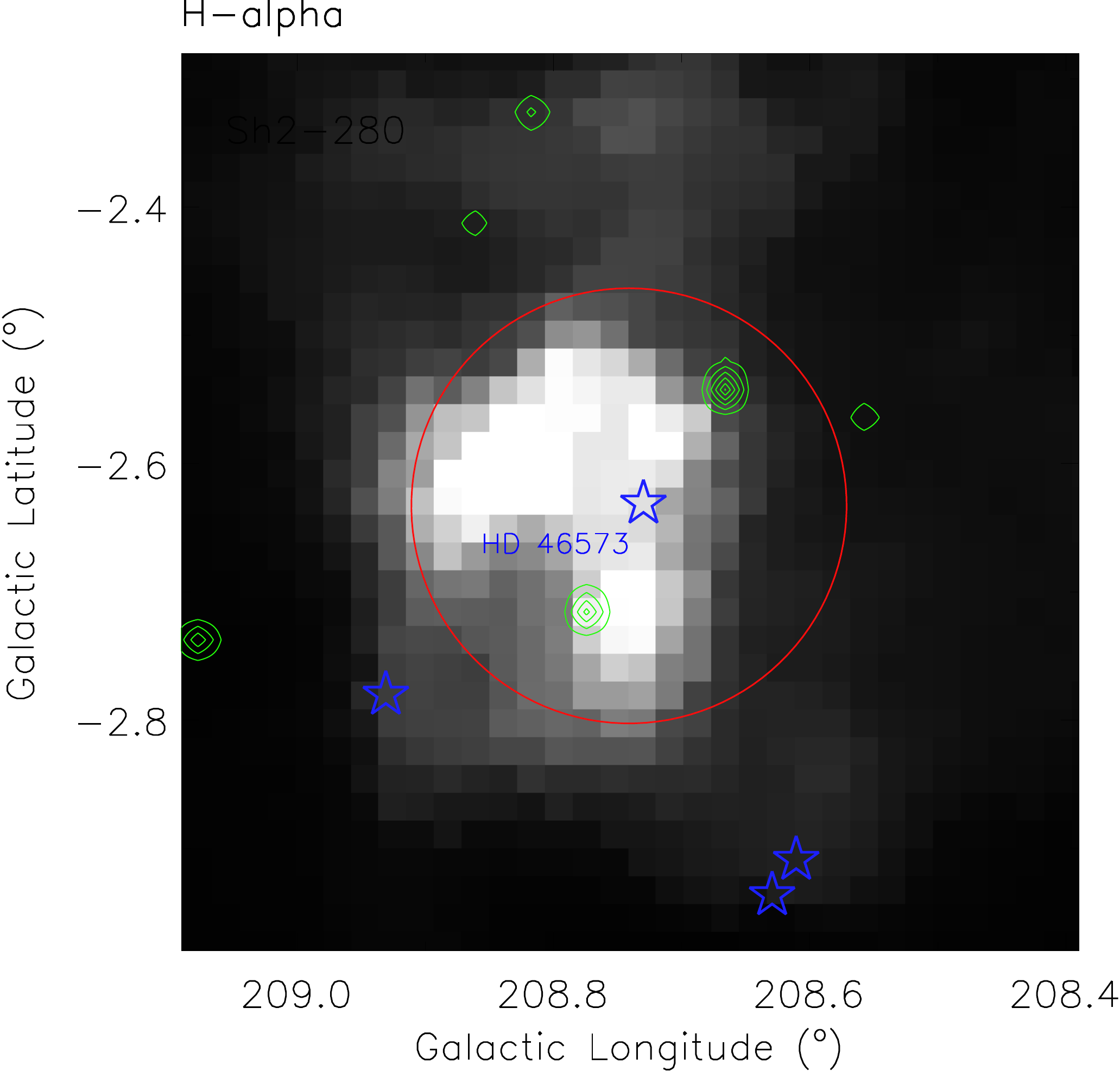}\\
\includegraphics[width=0.24\textwidth]{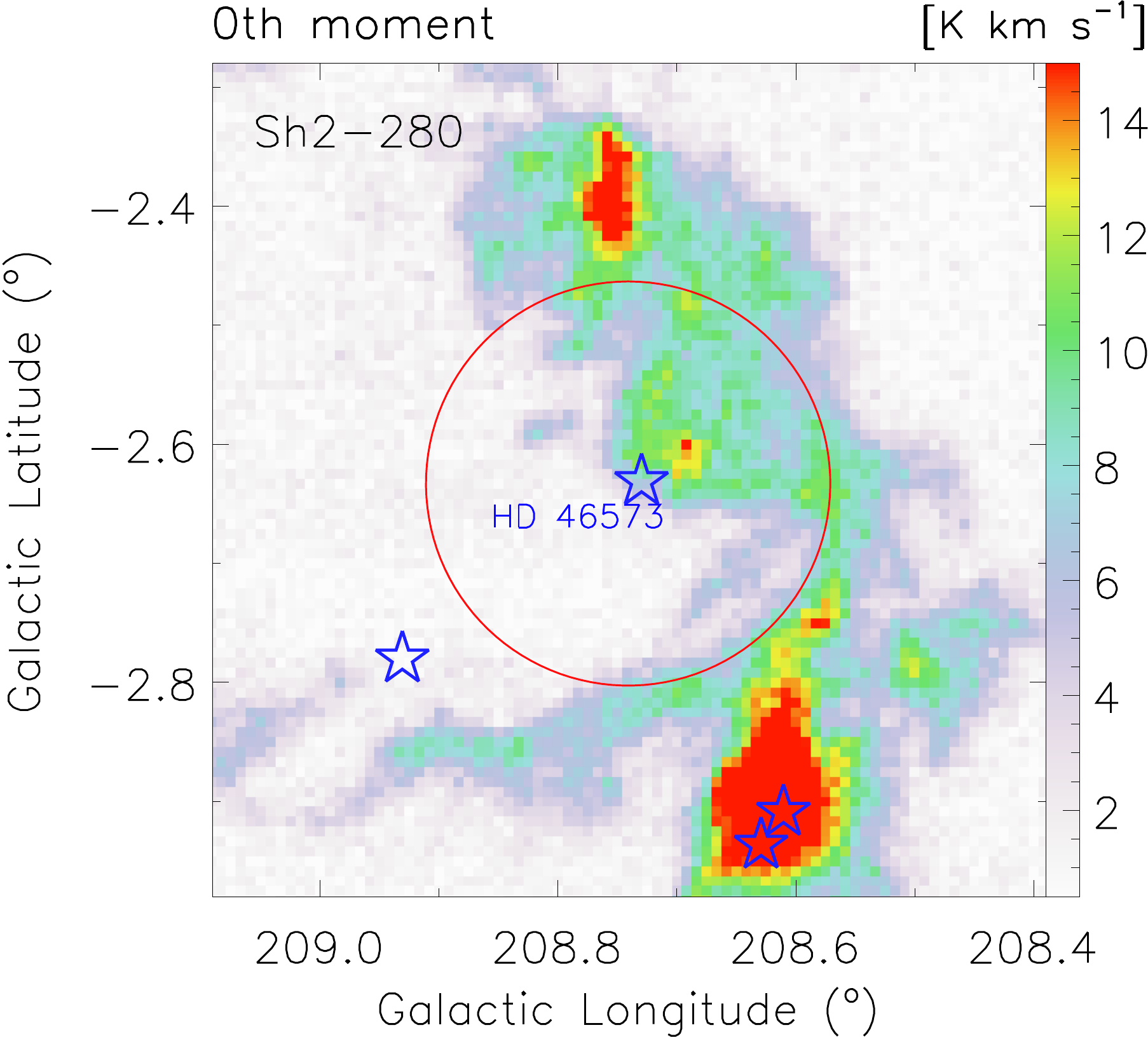}
\includegraphics[width=0.24\textwidth]{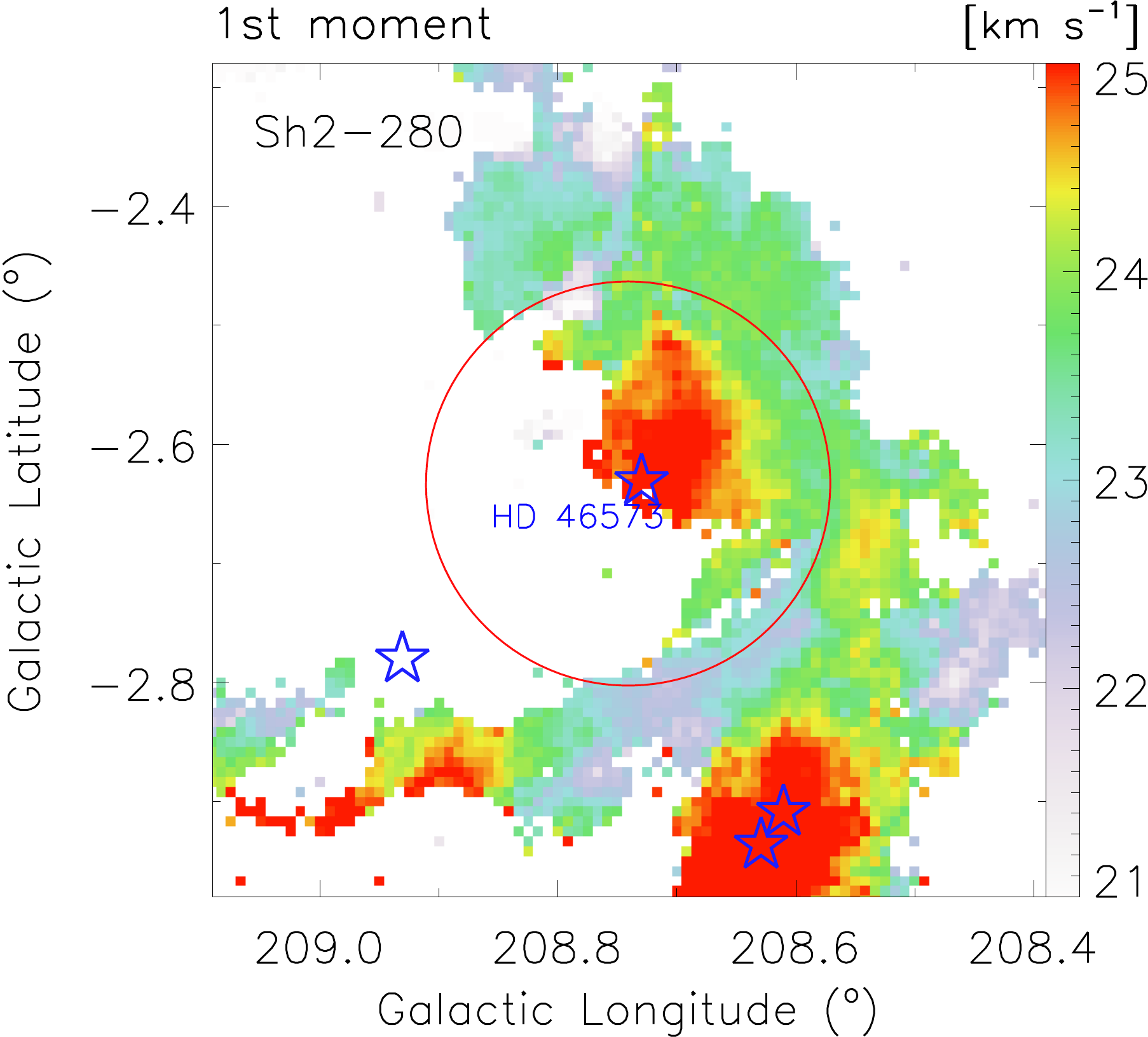}
\includegraphics[width=0.24\textwidth]{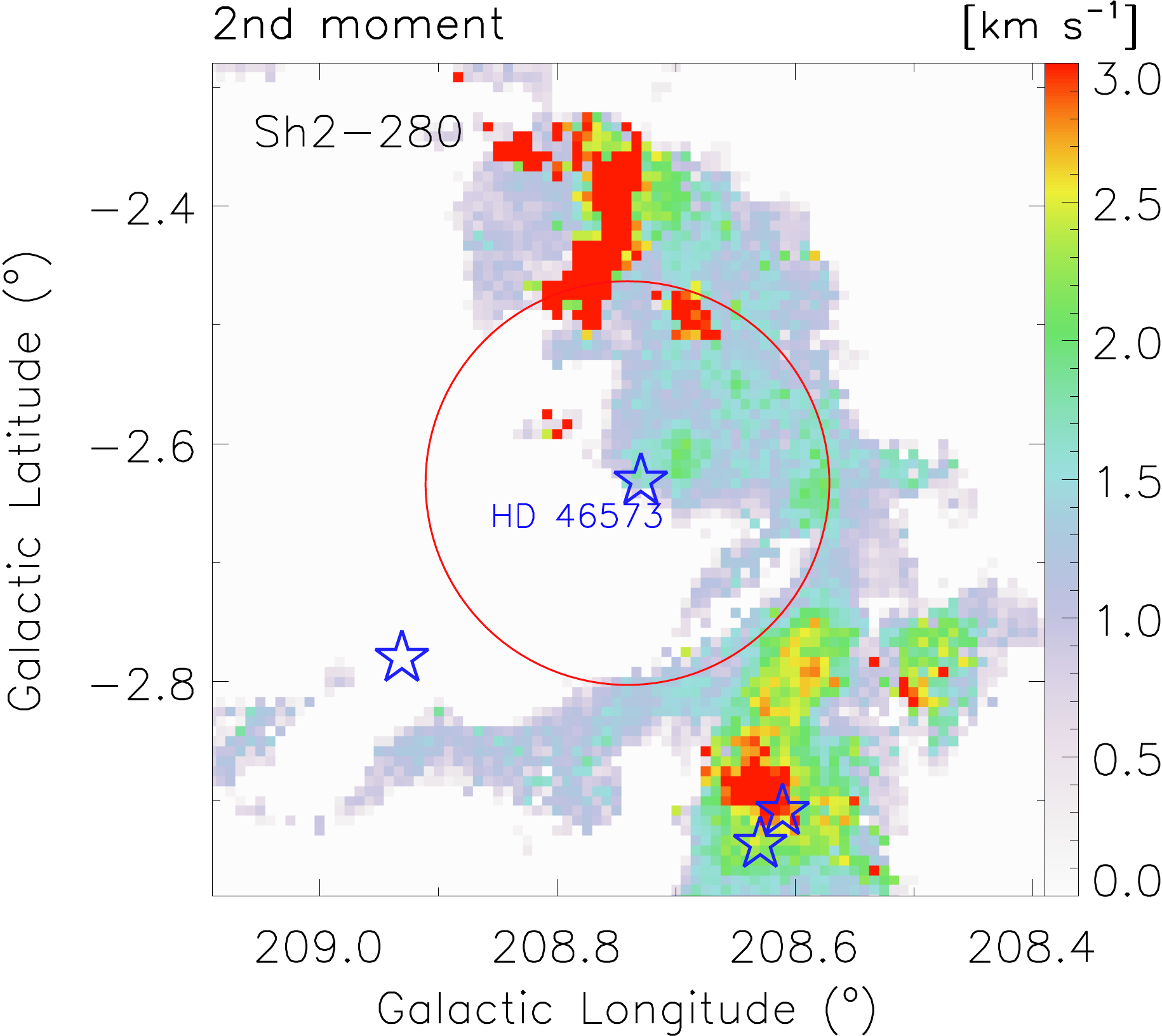}
\includegraphics[width=0.24\textwidth]{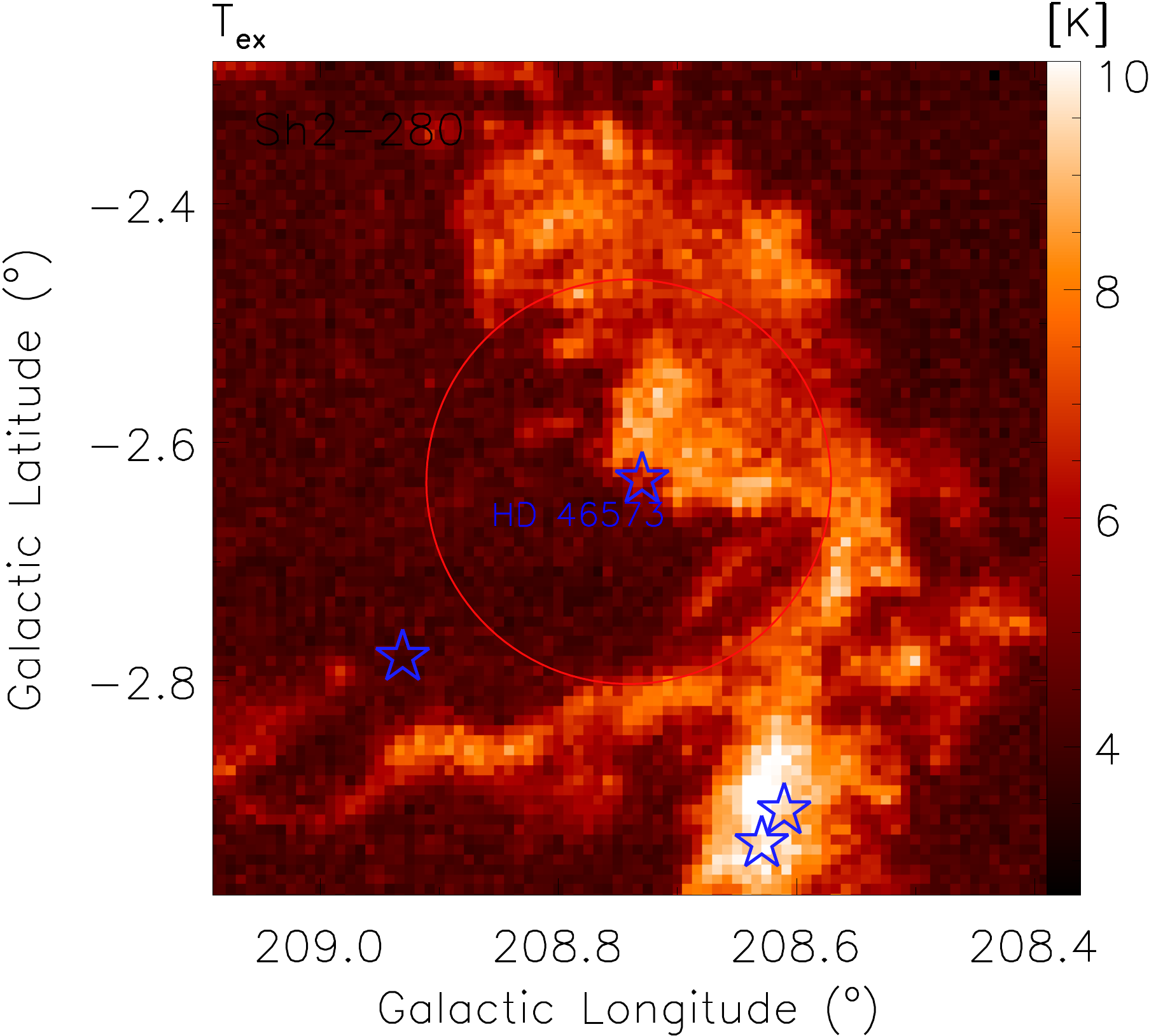}
  \caption{Morphology of Sh2-280 in various tracers. Top left: $WISE$ 3 (12 $\mu m$). Top middle: $WISE$ 4 (22 $\mu m$). Top right: NVSS 1.4 GHz radio continuum emission (green contours) overlaid on the H$\alpha$ emission from the SHASSA survey \citep{2001PASP..113.1326G}. The minimal level and the interval of the contours are 15 and 5 mJy/beam, respectively. Bottom: integrated intensity (left 1), velocity distribution (left 2), line width of velocity (left 3), and excitation temperature (left 4) images of $^{12}$CO. The circle size approximates the radius of the \ion{H}{2} region from \citet{2014ApJS..212....1A}. The velocity range for intensity integration is from 19 km s$^{-1}$ to 28 km s$^{-1}$. The blue pentagram signs indicate the O stars in this region from the SIMBAD database.}
  \label{fig:S280}
\end{figure}

\begin{figure}[h]
  \centering
\includegraphics[width=0.21\textwidth]{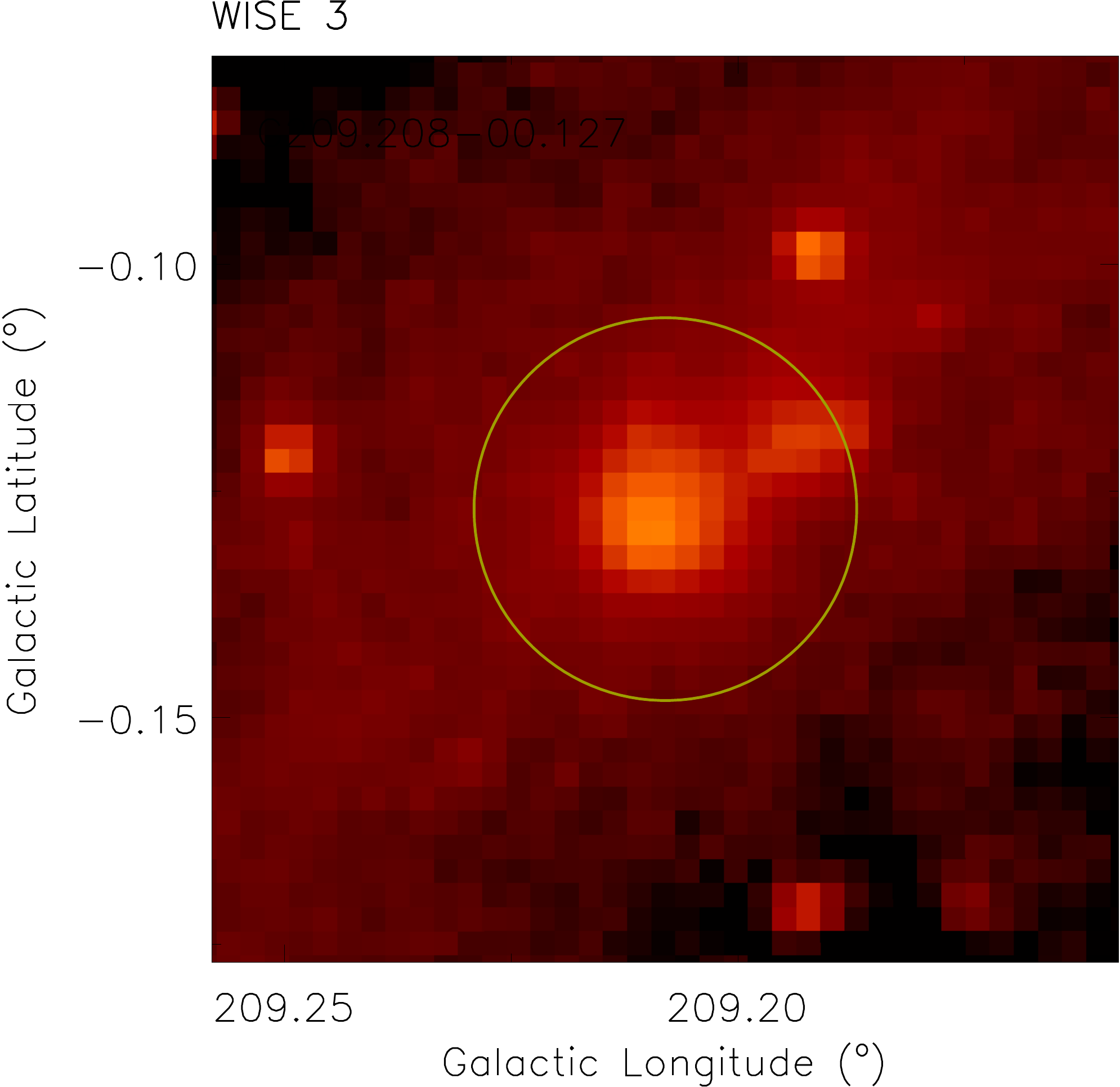}
\includegraphics[width=0.21\textwidth]{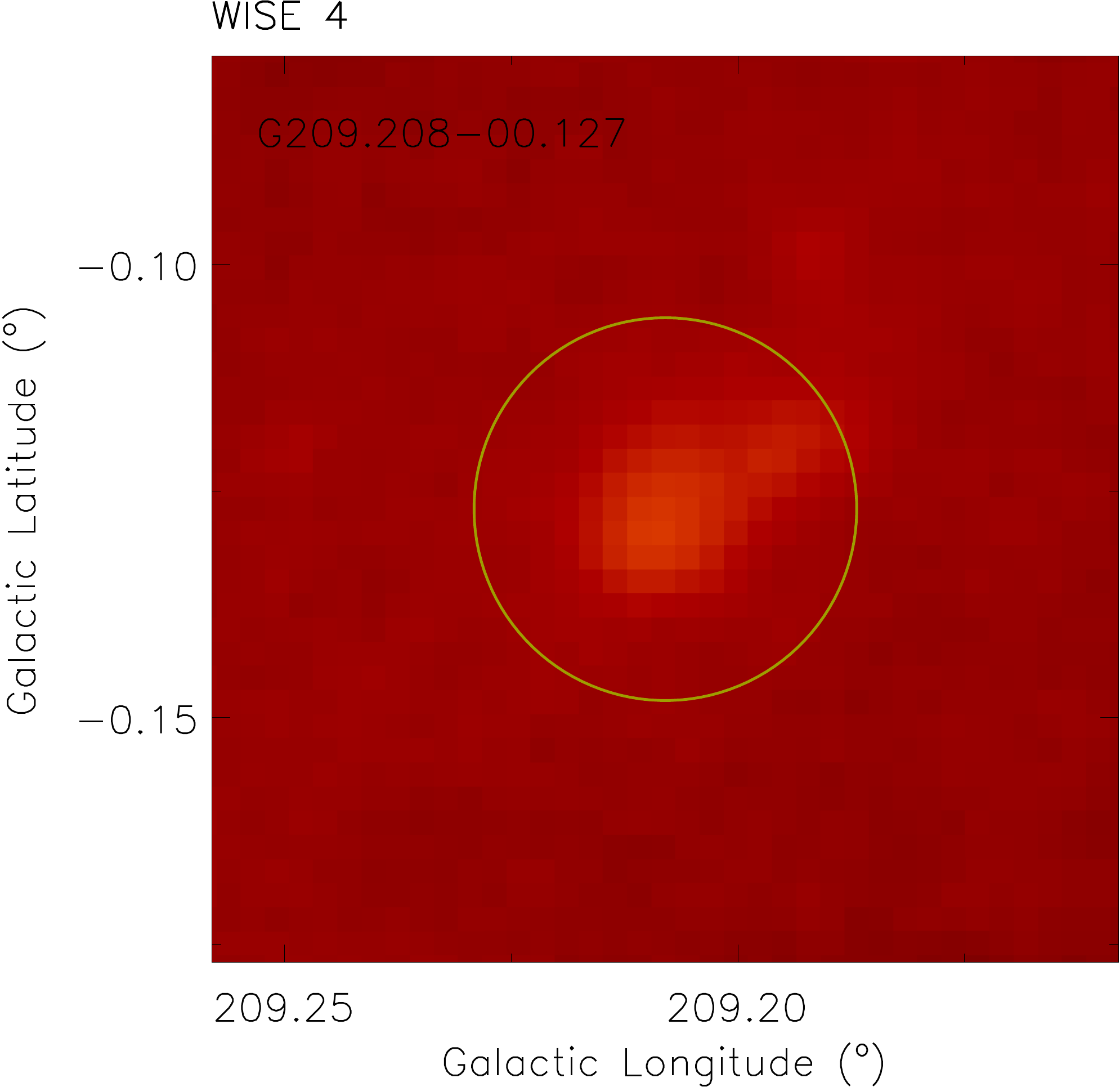}
\includegraphics[width=0.21\textwidth]{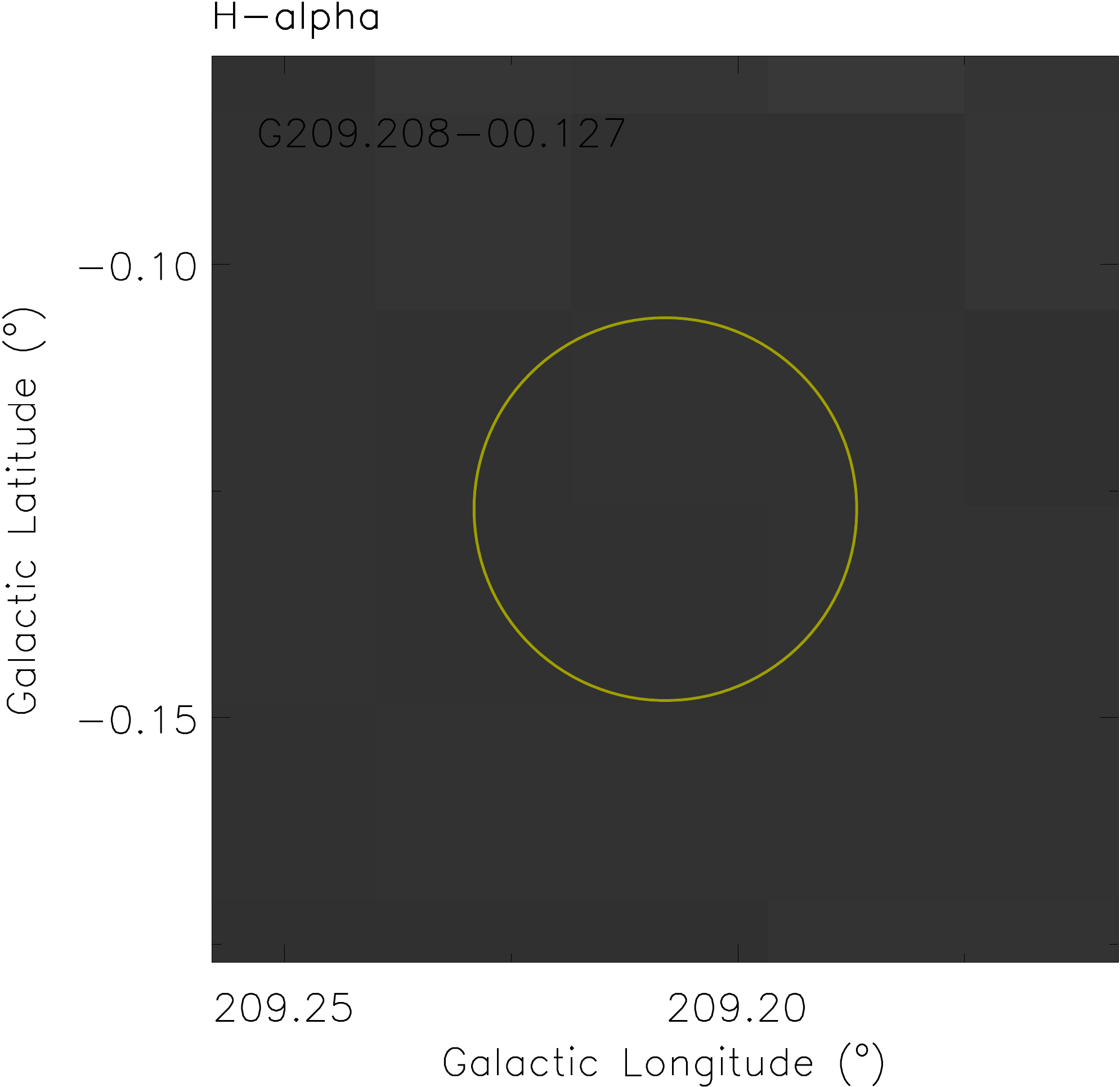}\\
\includegraphics[width=0.237\textwidth]{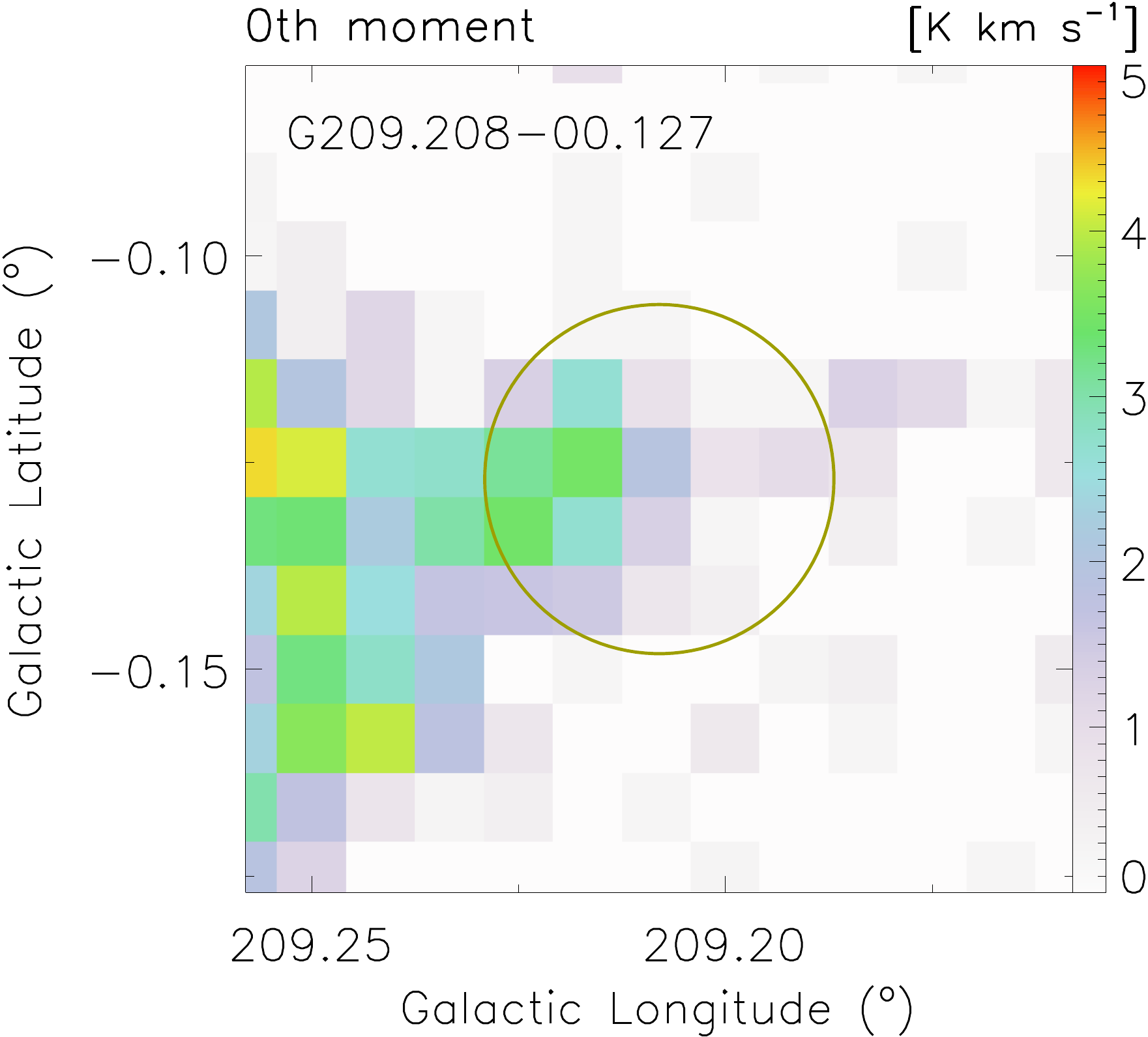}
\includegraphics[width=0.24\textwidth]{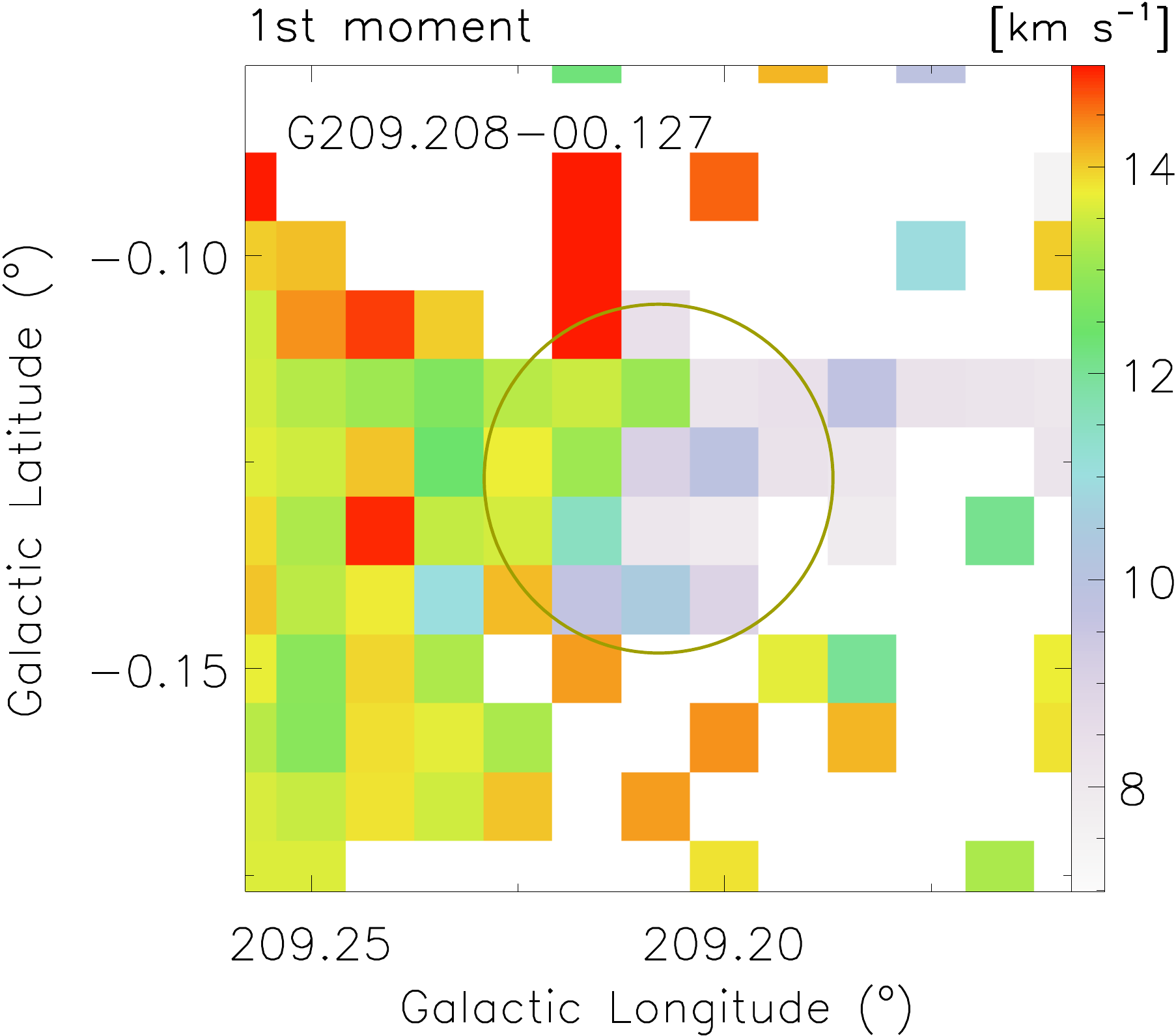}
\includegraphics[width=0.24\textwidth]{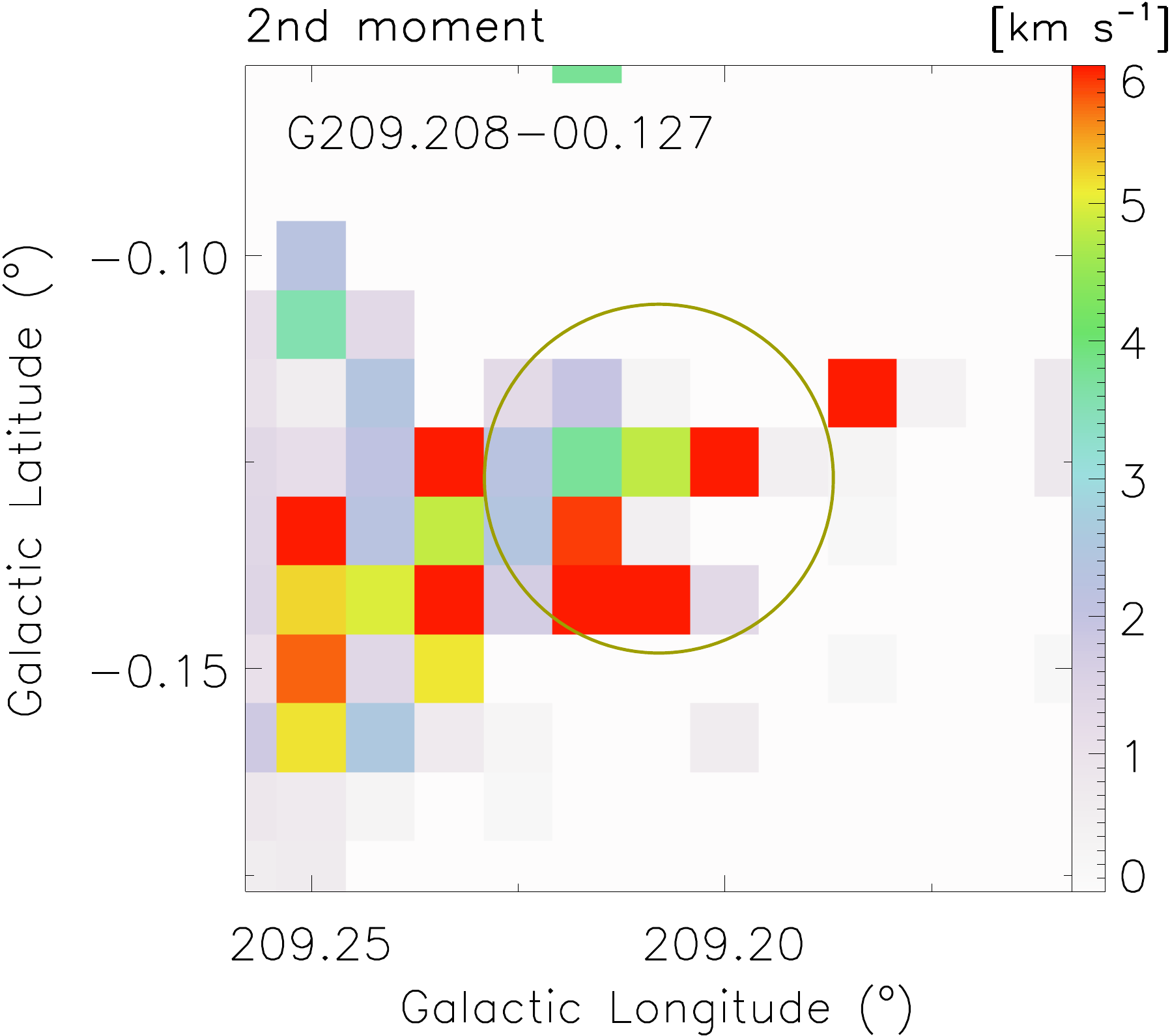}
\includegraphics[width=0.237\textwidth]{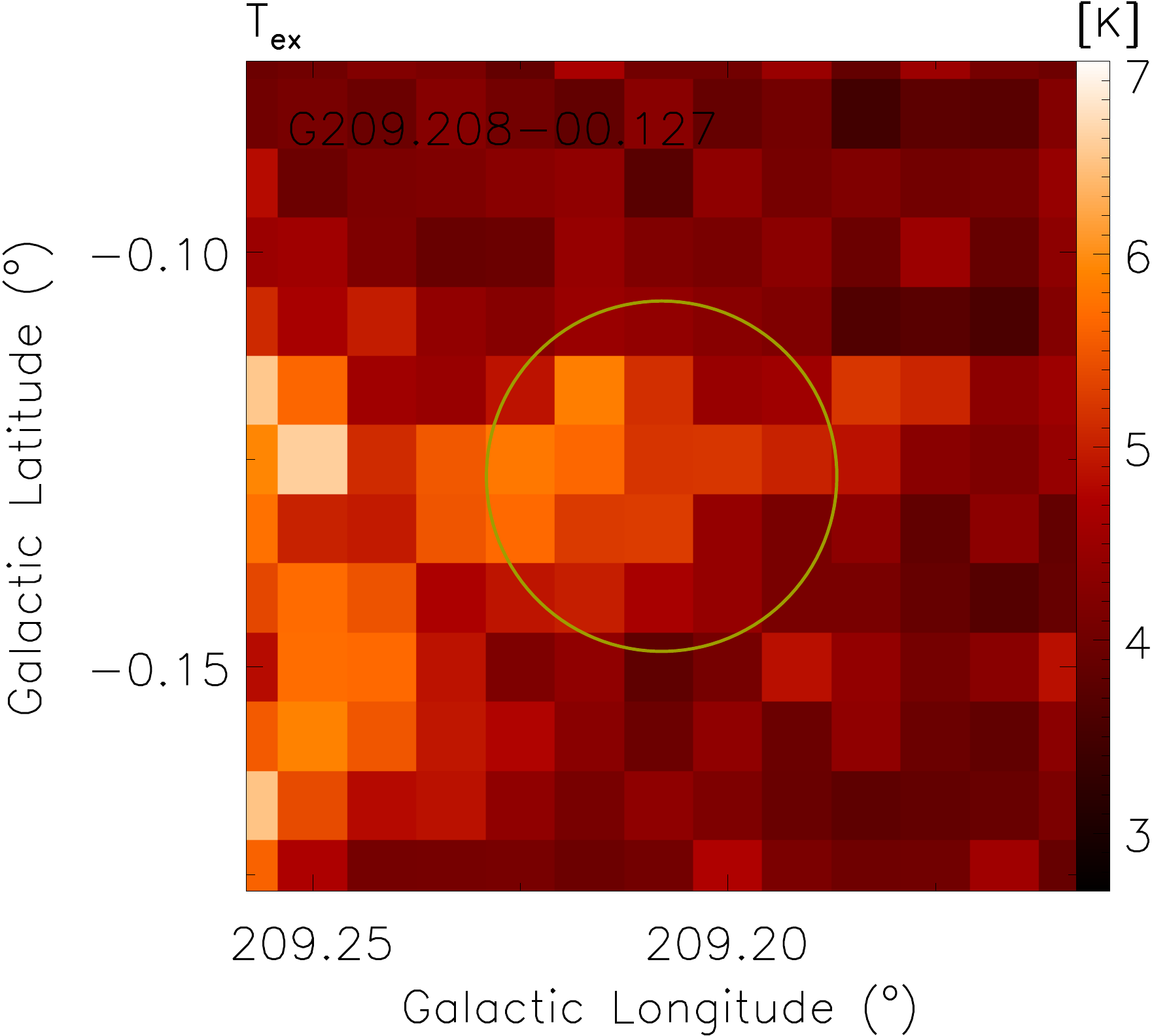}\\
\includegraphics[width=0.239\textwidth]{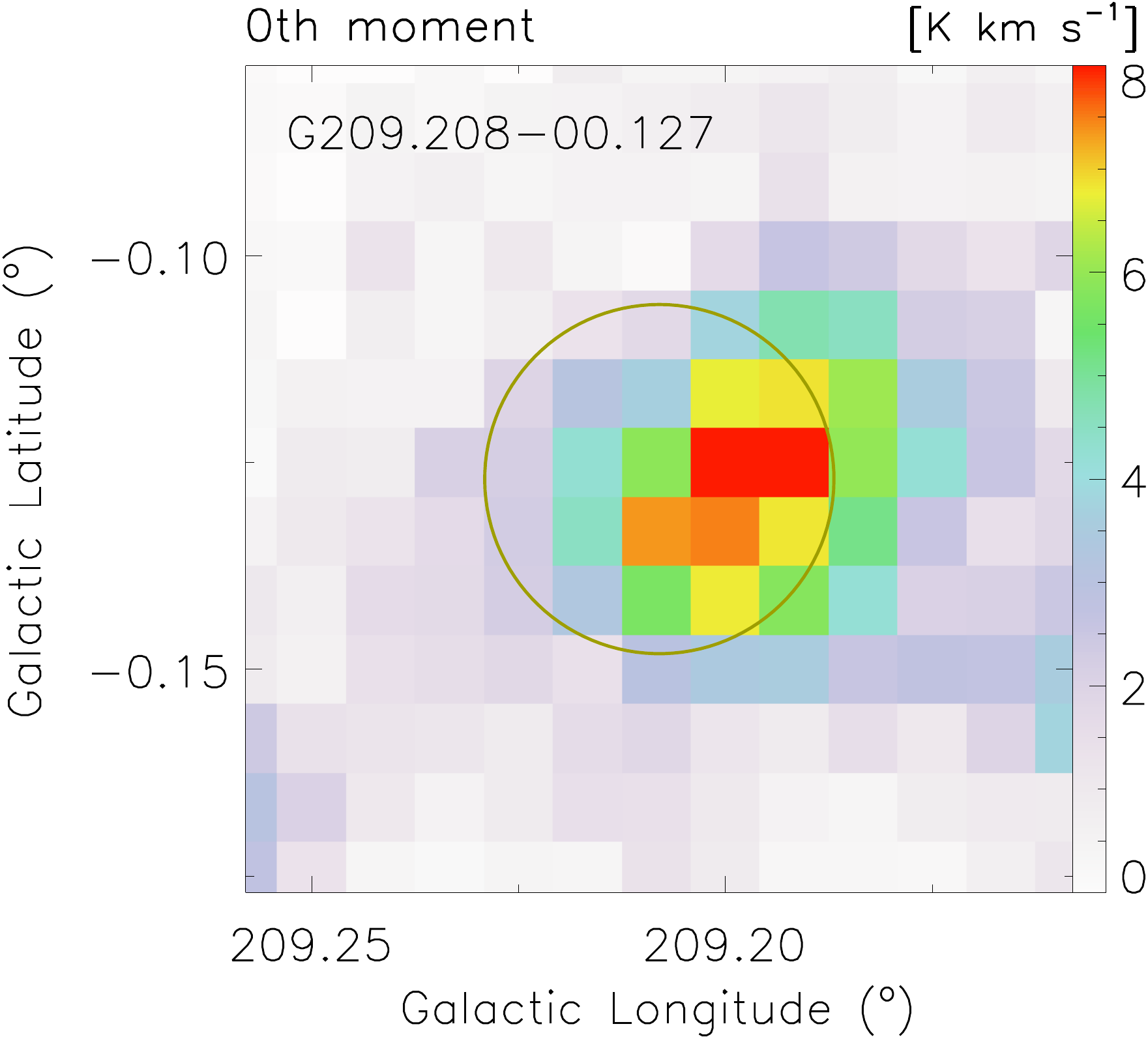}
\includegraphics[width=0.25\textwidth]{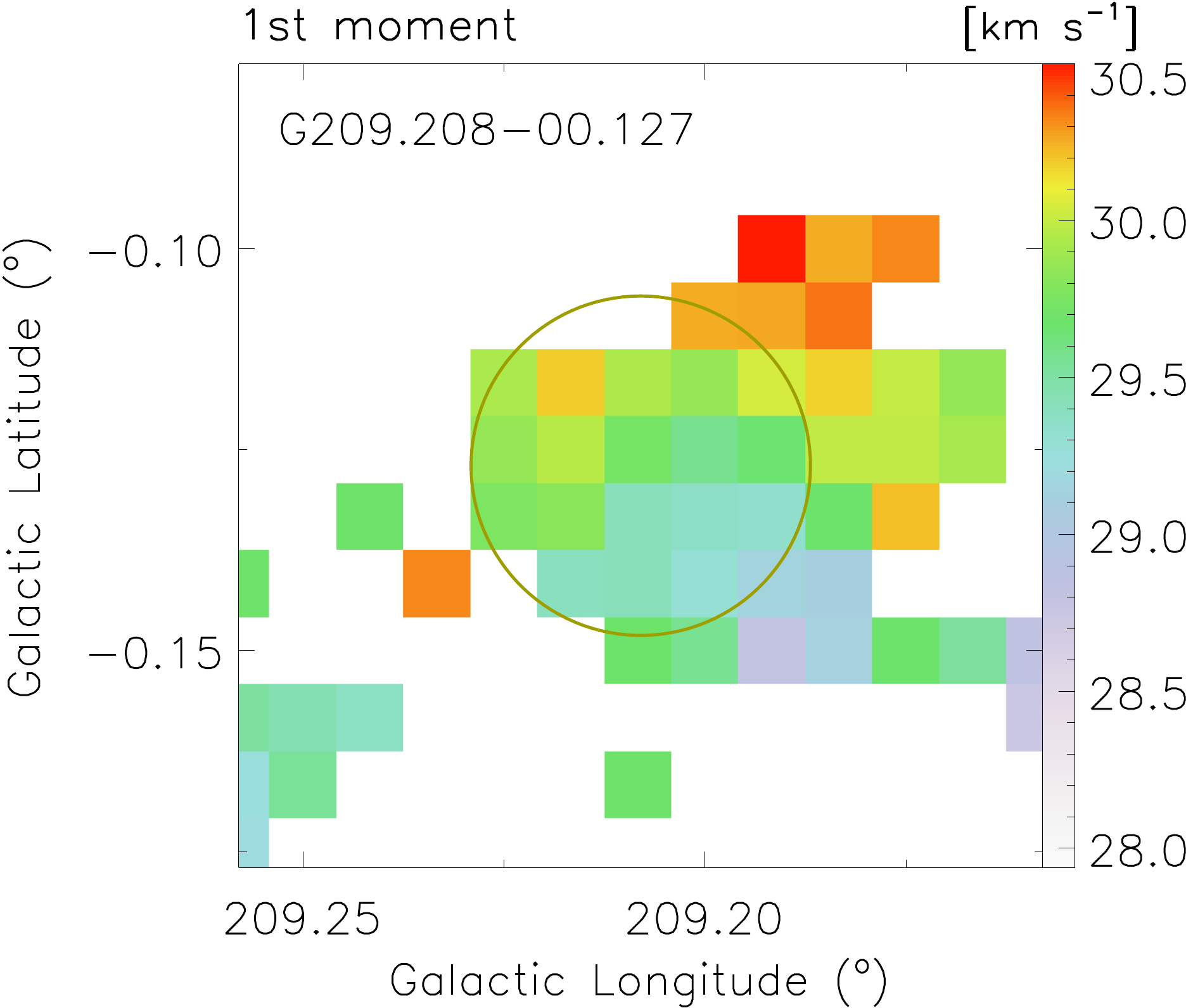}
\includegraphics[width=0.245\textwidth]{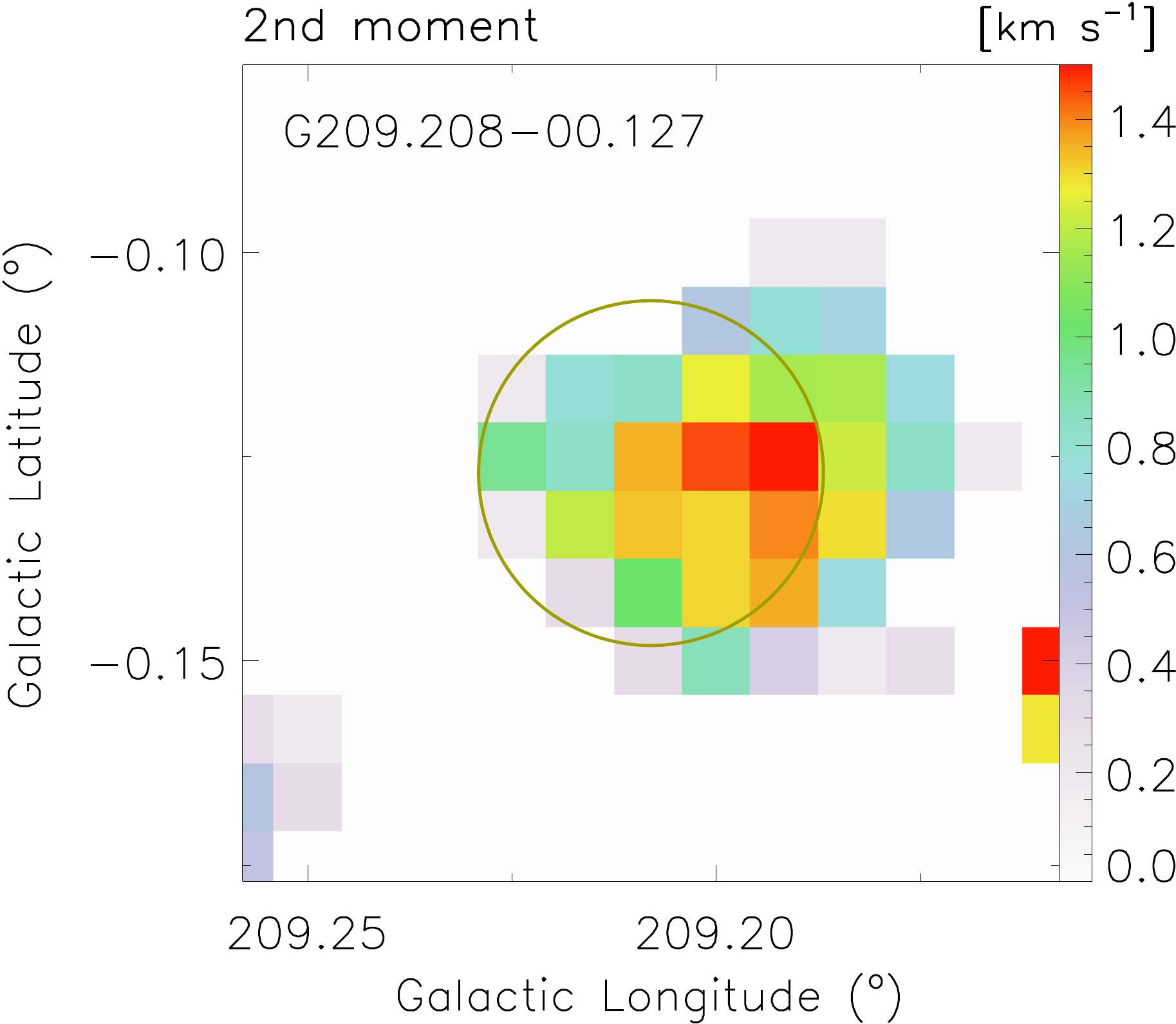}
\includegraphics[width=0.243\textwidth]{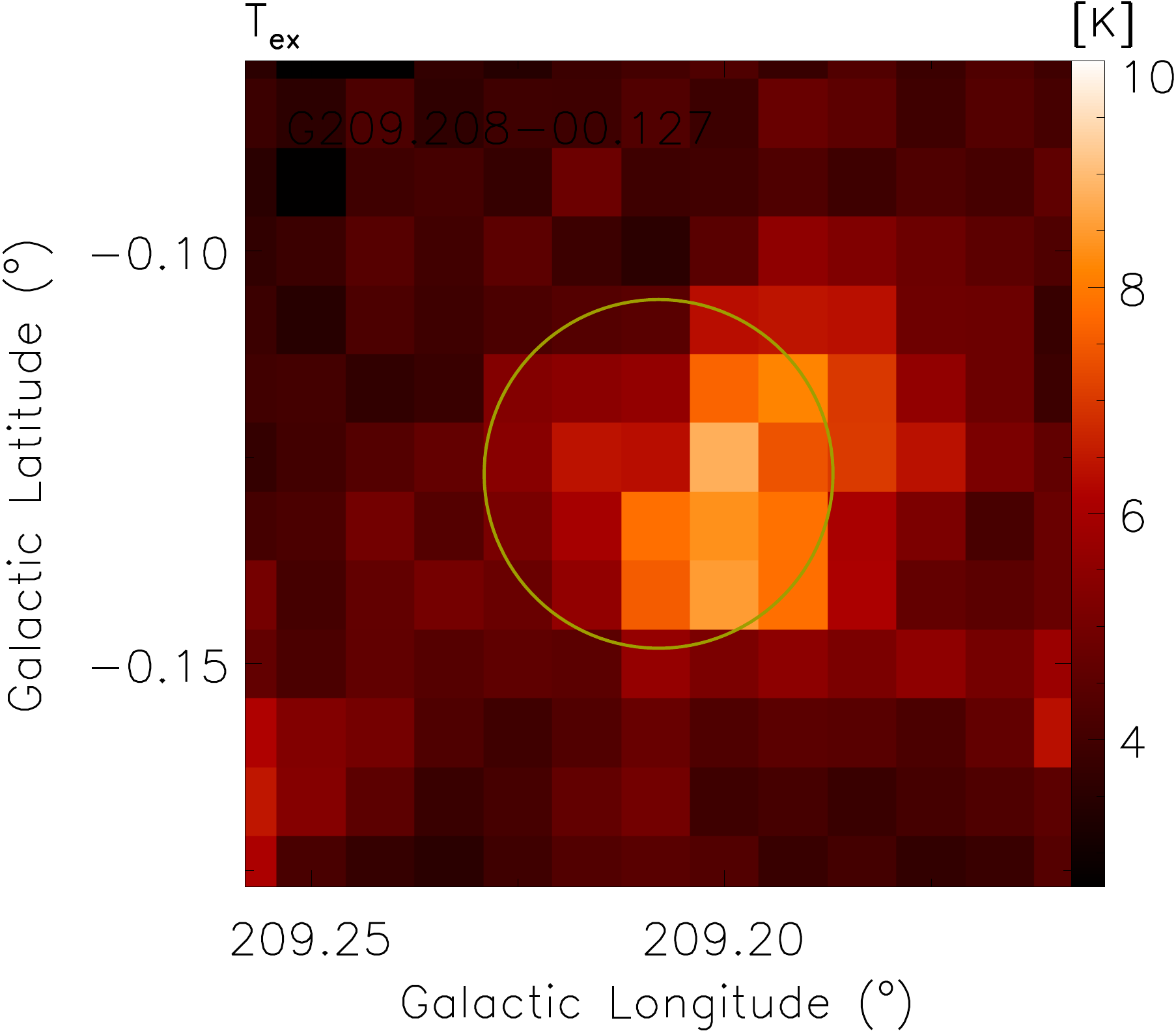}\\
  \caption{Morphology of G209.208-00.127 in various tracers. Middle: images of G209.208-00.127\_near with integrated velocity range from 7 to 17 km s$^{-1}$. Bottom: images of G209.208-00.127\_far with integrated velocity range from 28 to 32 km s$^{-1}$. All the others are the same as in Figures \ref{fig:G2085-023}.}
  \label{fig:G2092-001}
\end{figure}

\begin{figure}[h]
  \centering
\includegraphics[width=0.21\textwidth]{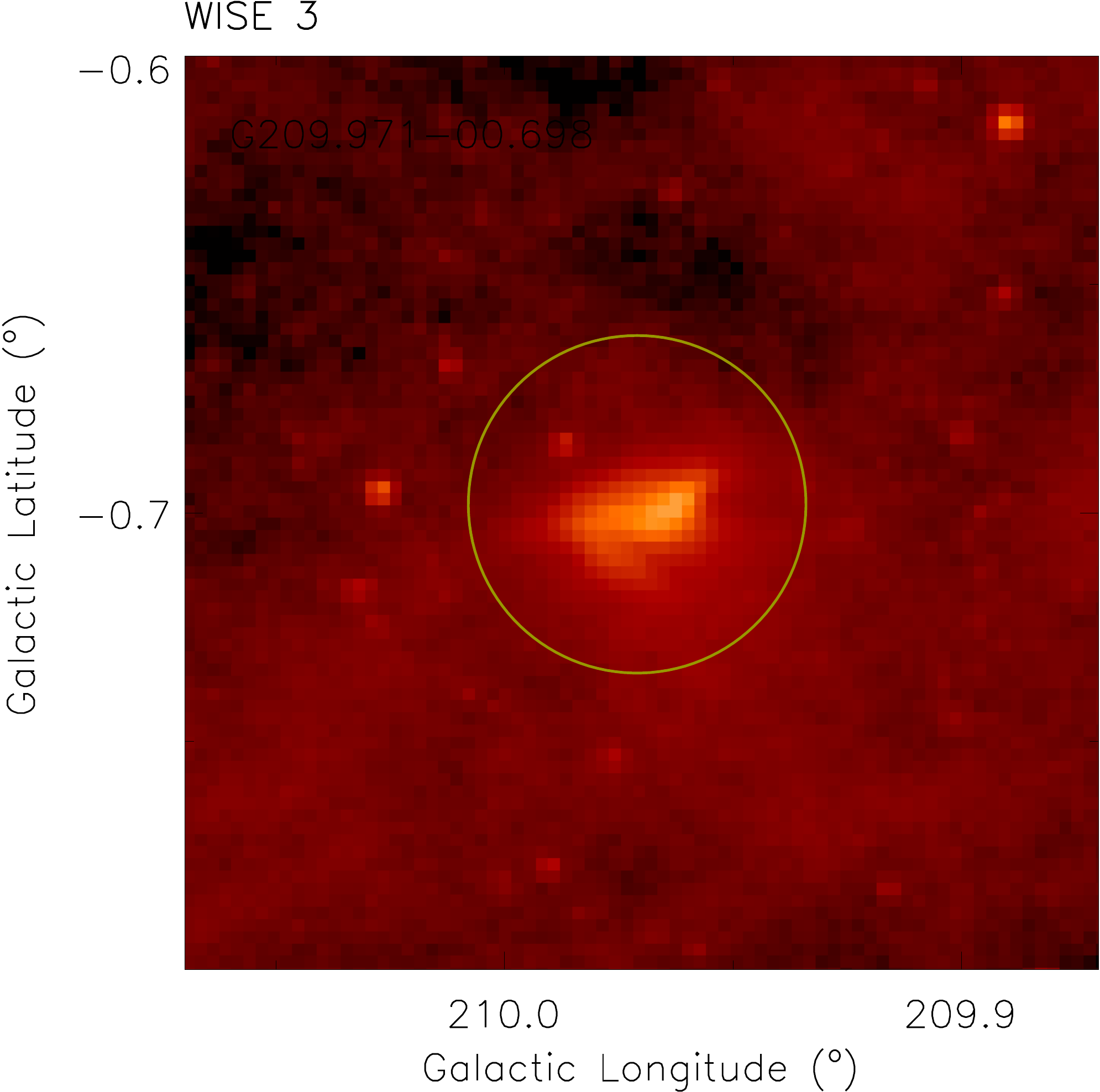}
\includegraphics[width=0.21\textwidth]{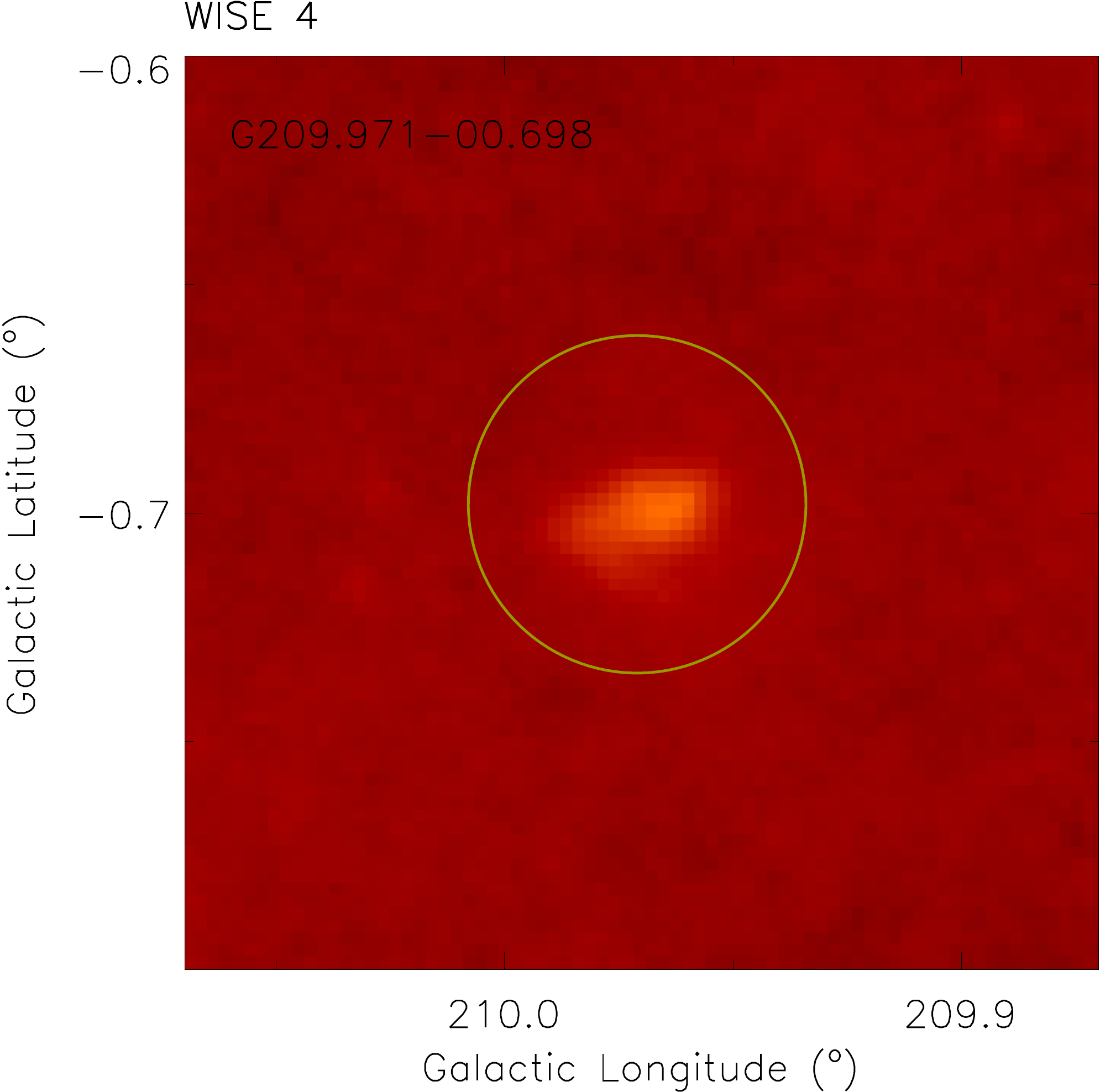}
\includegraphics[width=0.21\textwidth]{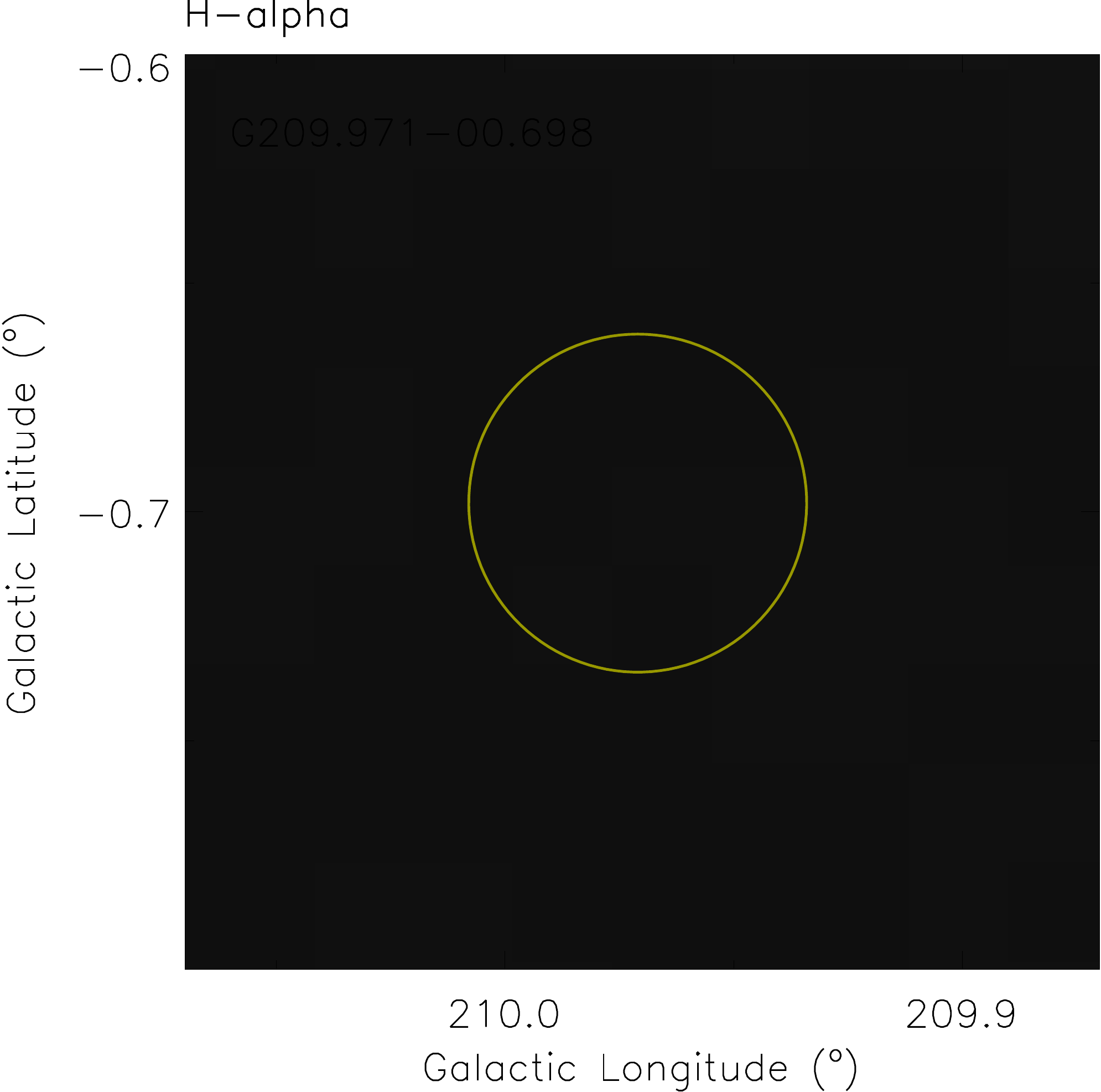}\\
\includegraphics[width=0.23\textwidth]{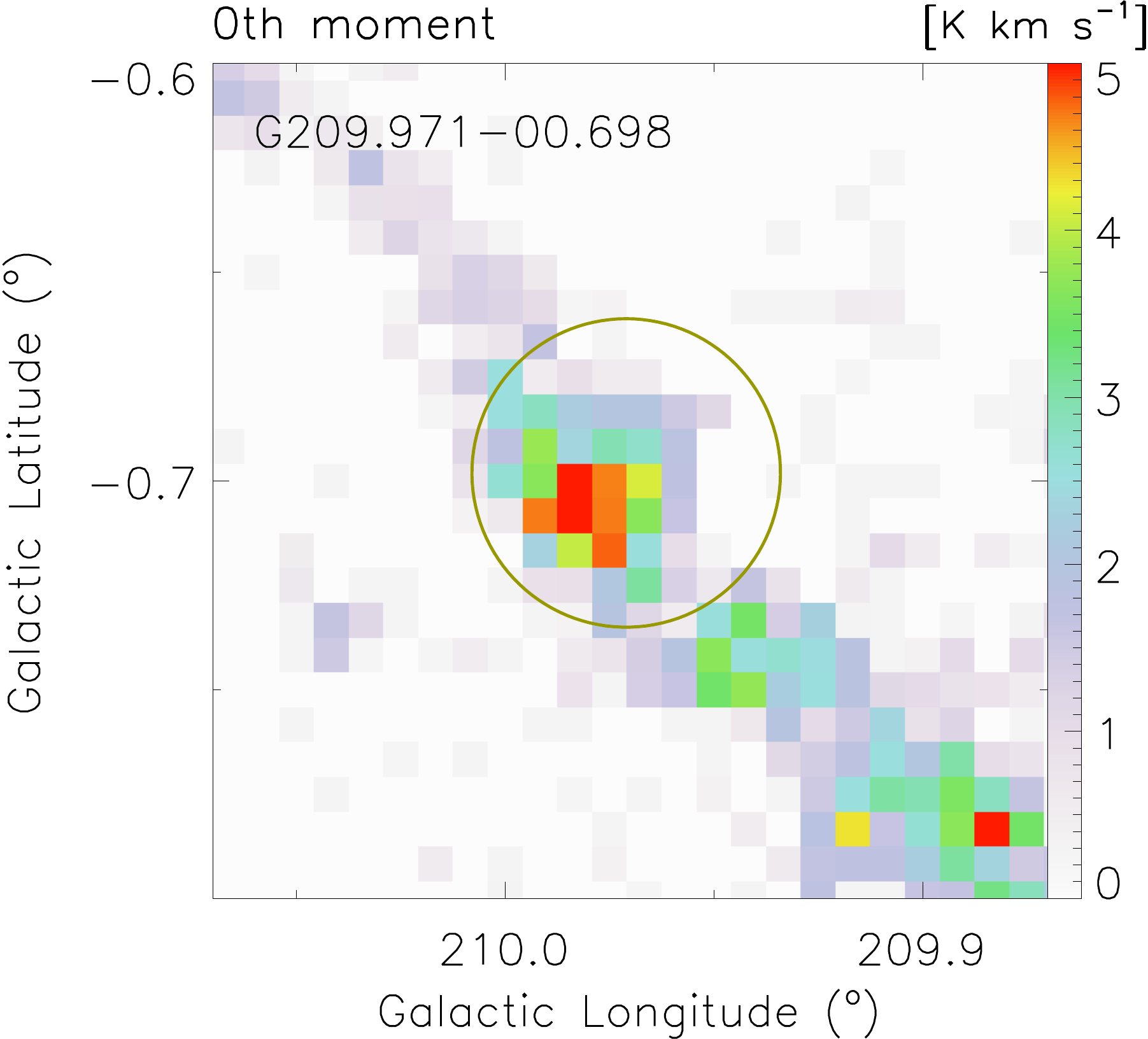}
\includegraphics[width=0.24\textwidth]{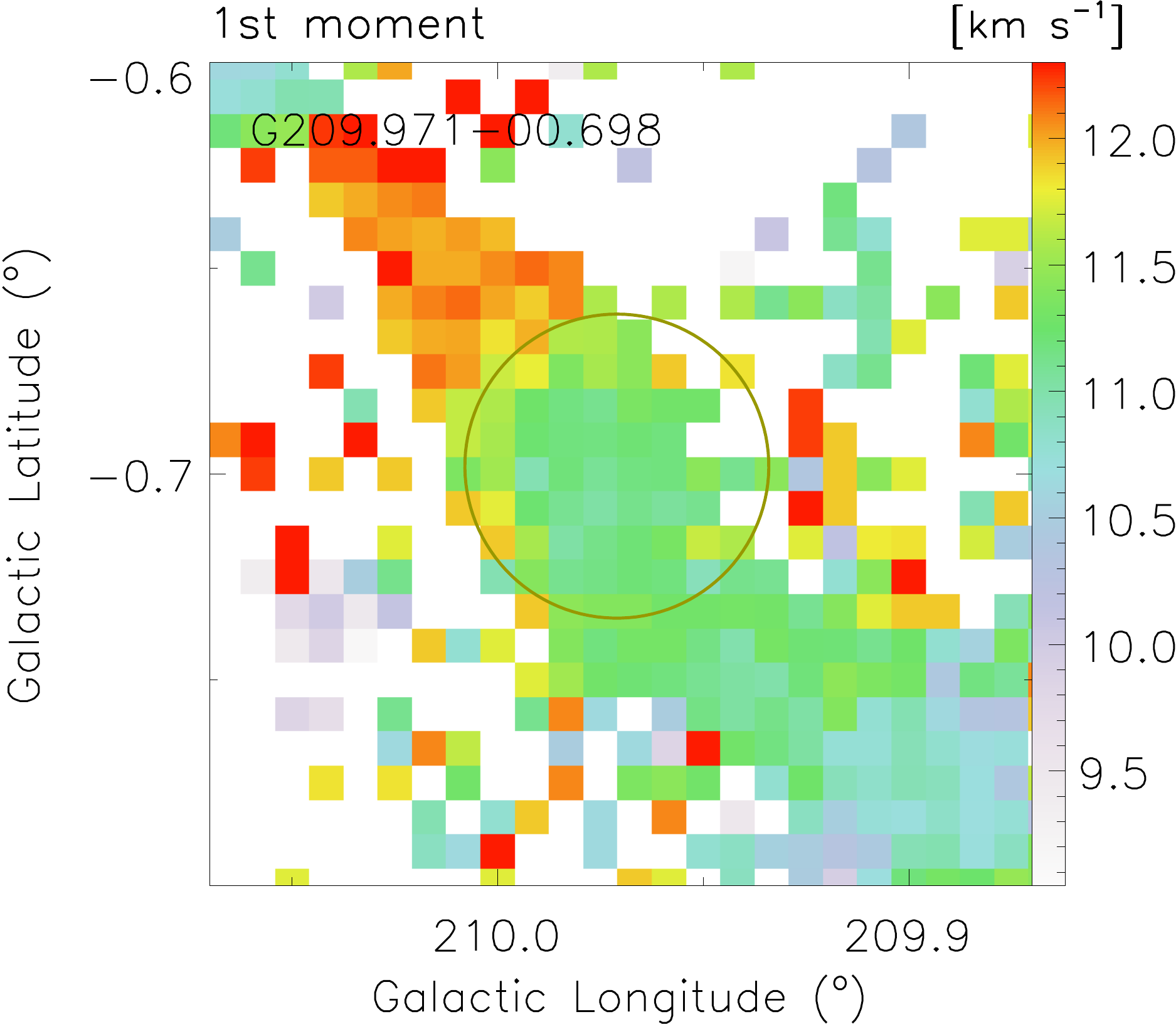}
\includegraphics[width=0.24\textwidth]{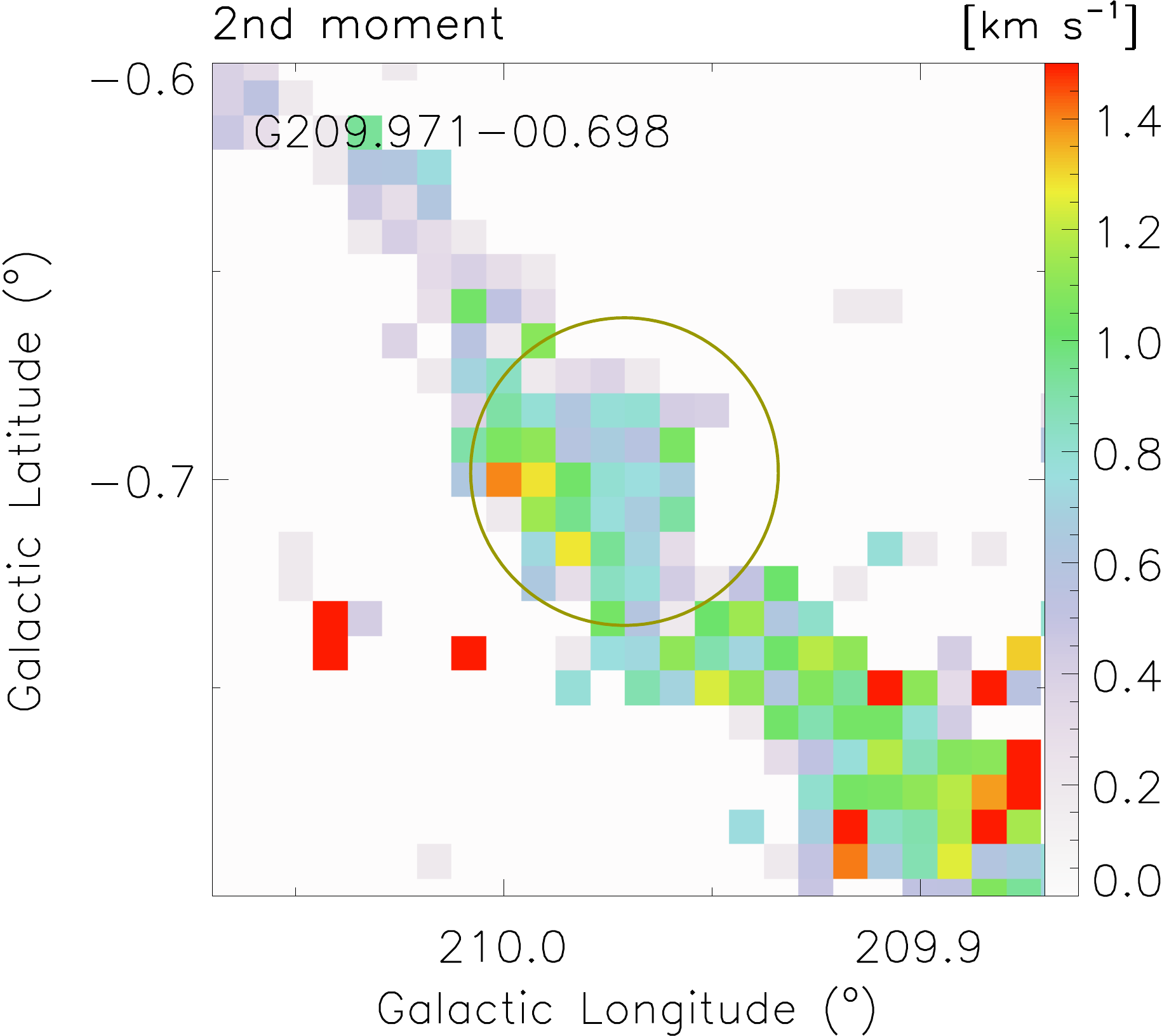}
\includegraphics[width=0.24\textwidth]{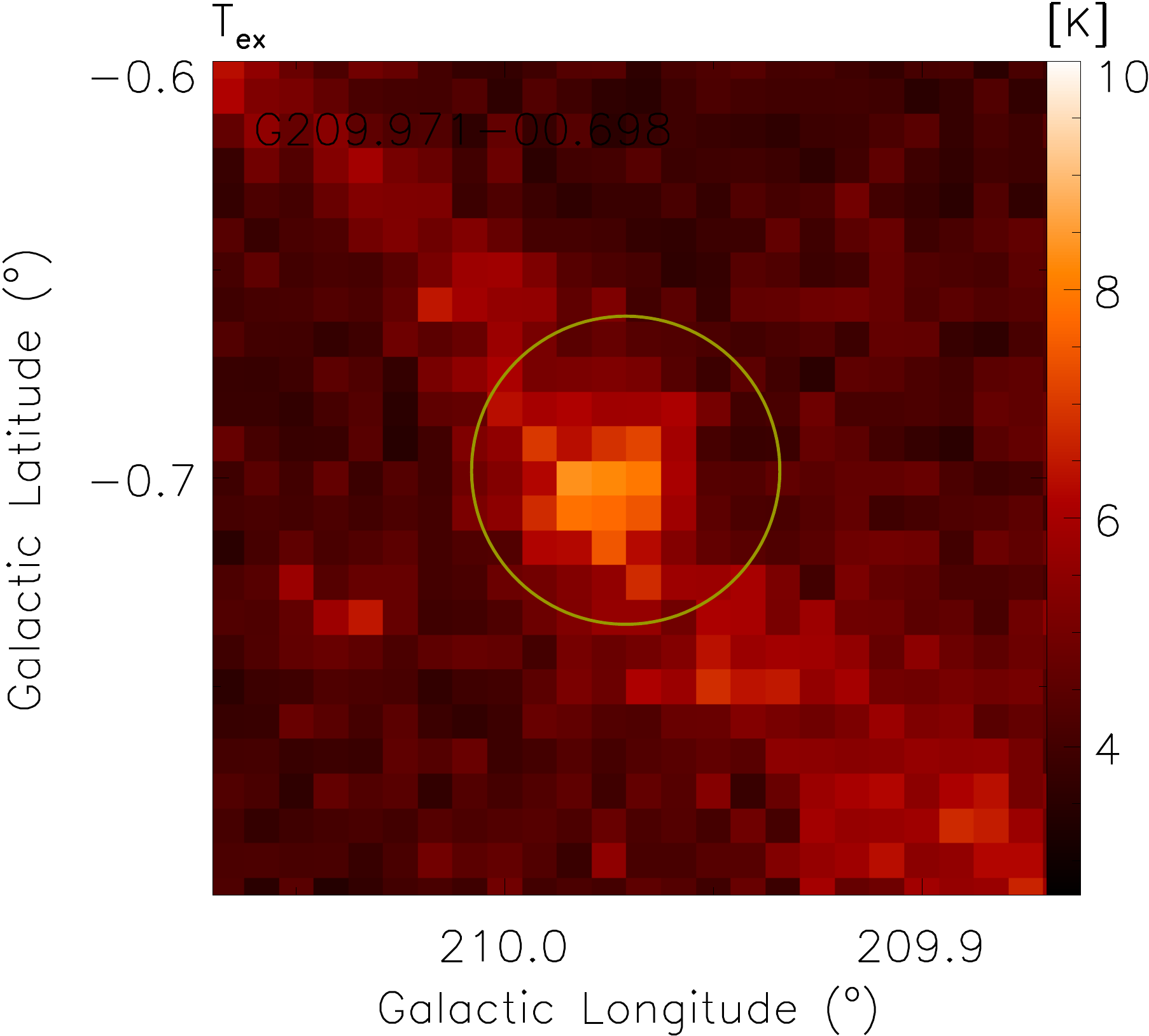}\\
\includegraphics[width=0.237\textwidth]{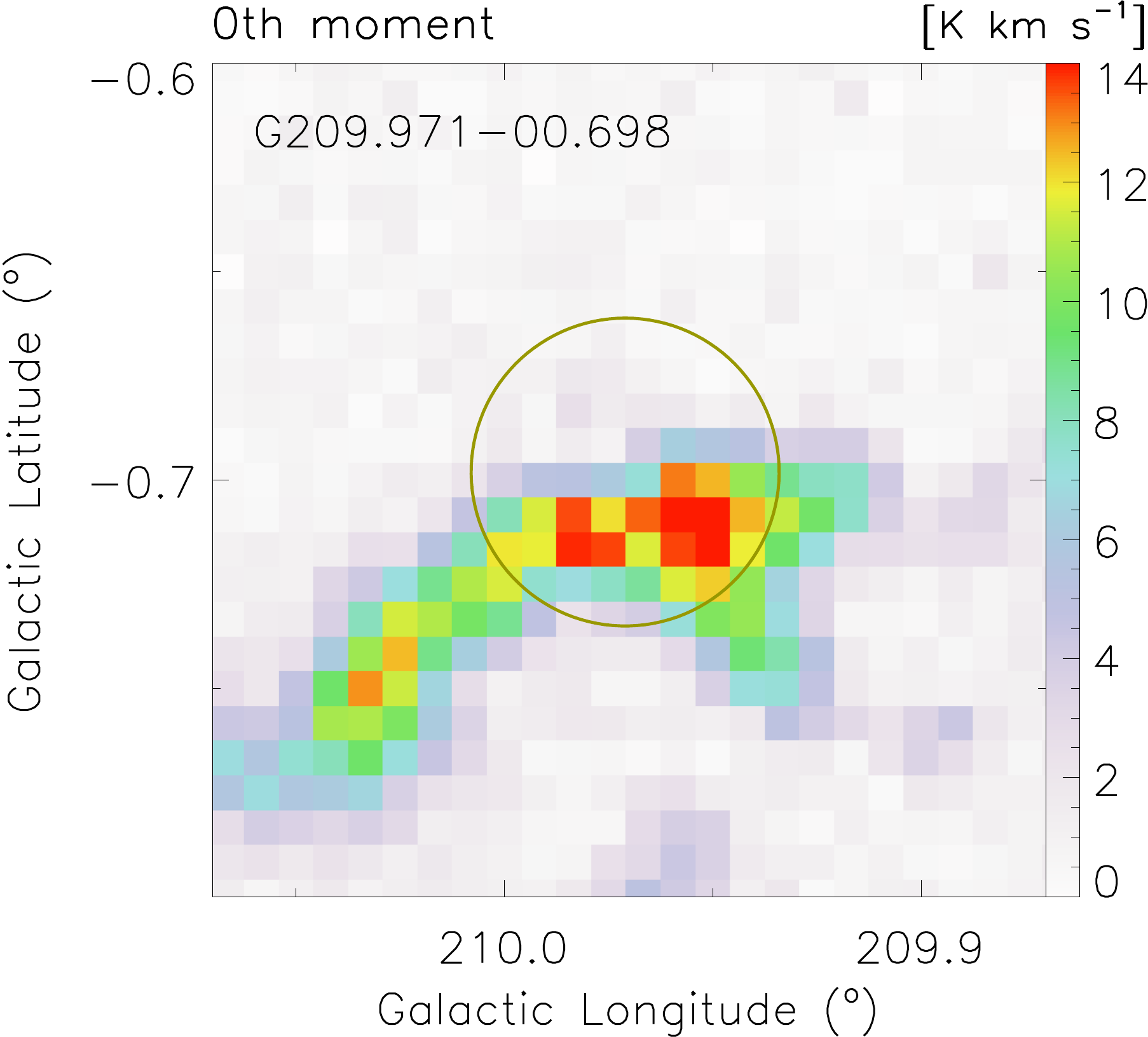}
\includegraphics[width=0.242\textwidth]{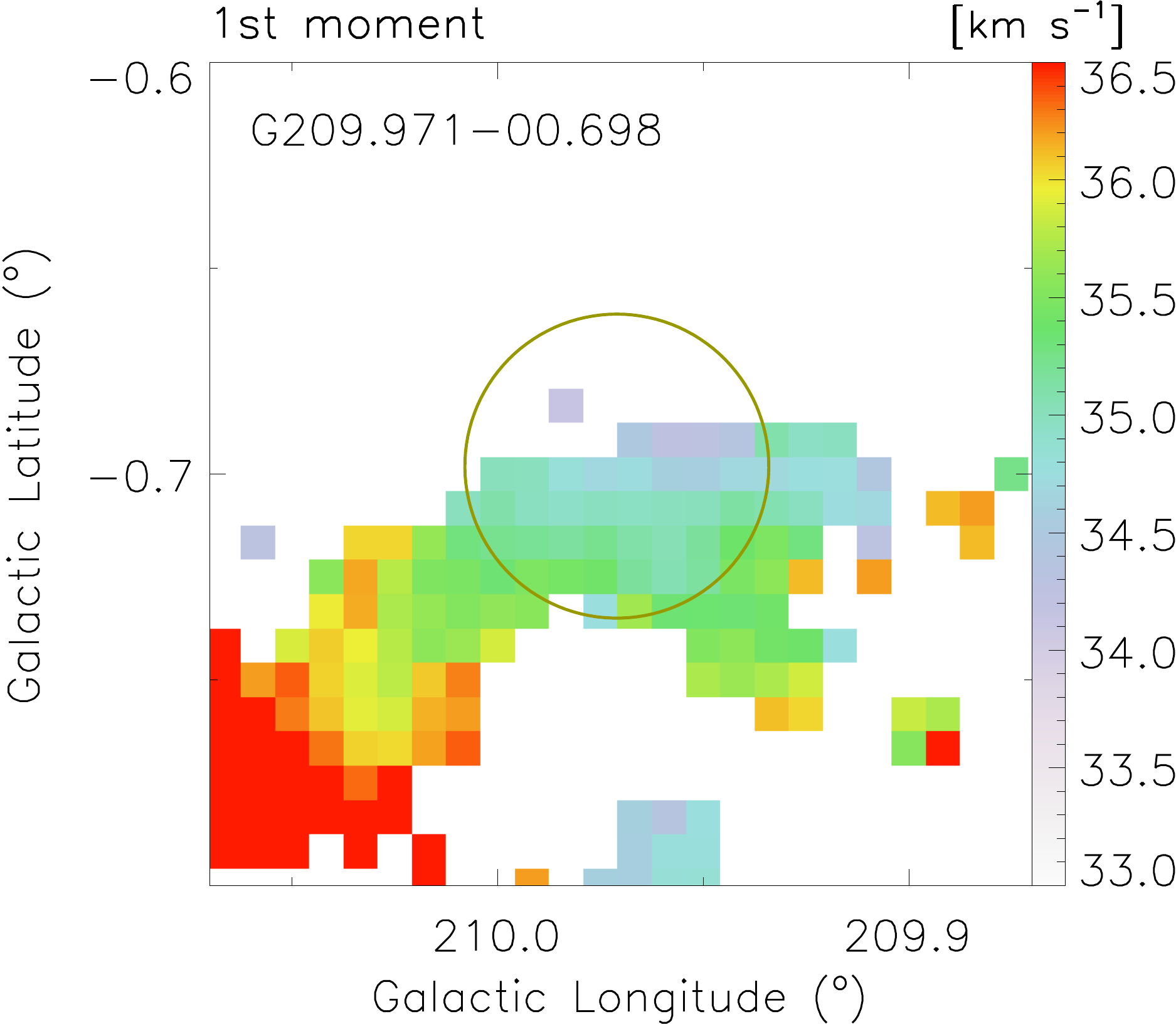}
\includegraphics[width=0.24\textwidth]{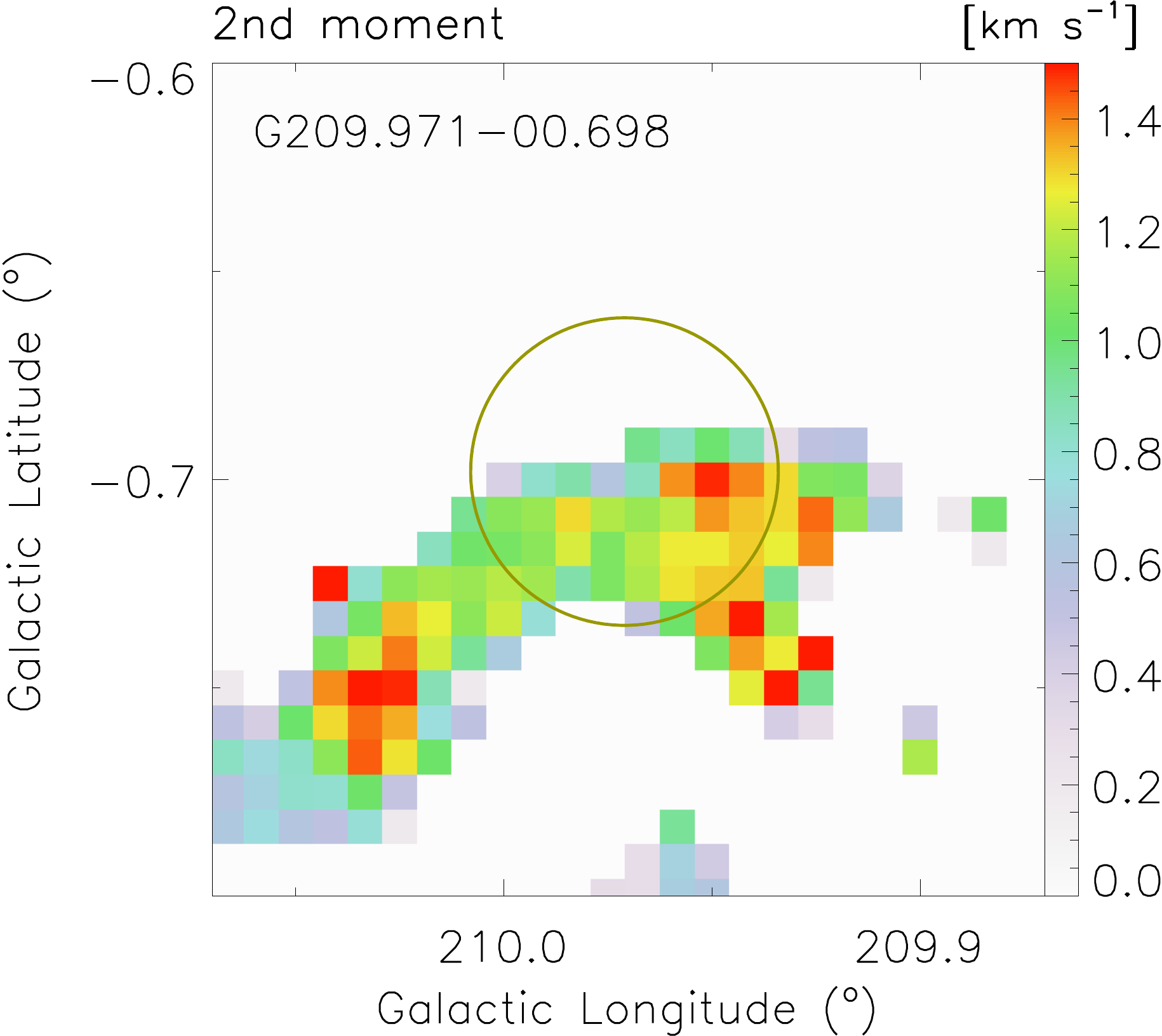}
\includegraphics[width=0.24\textwidth]{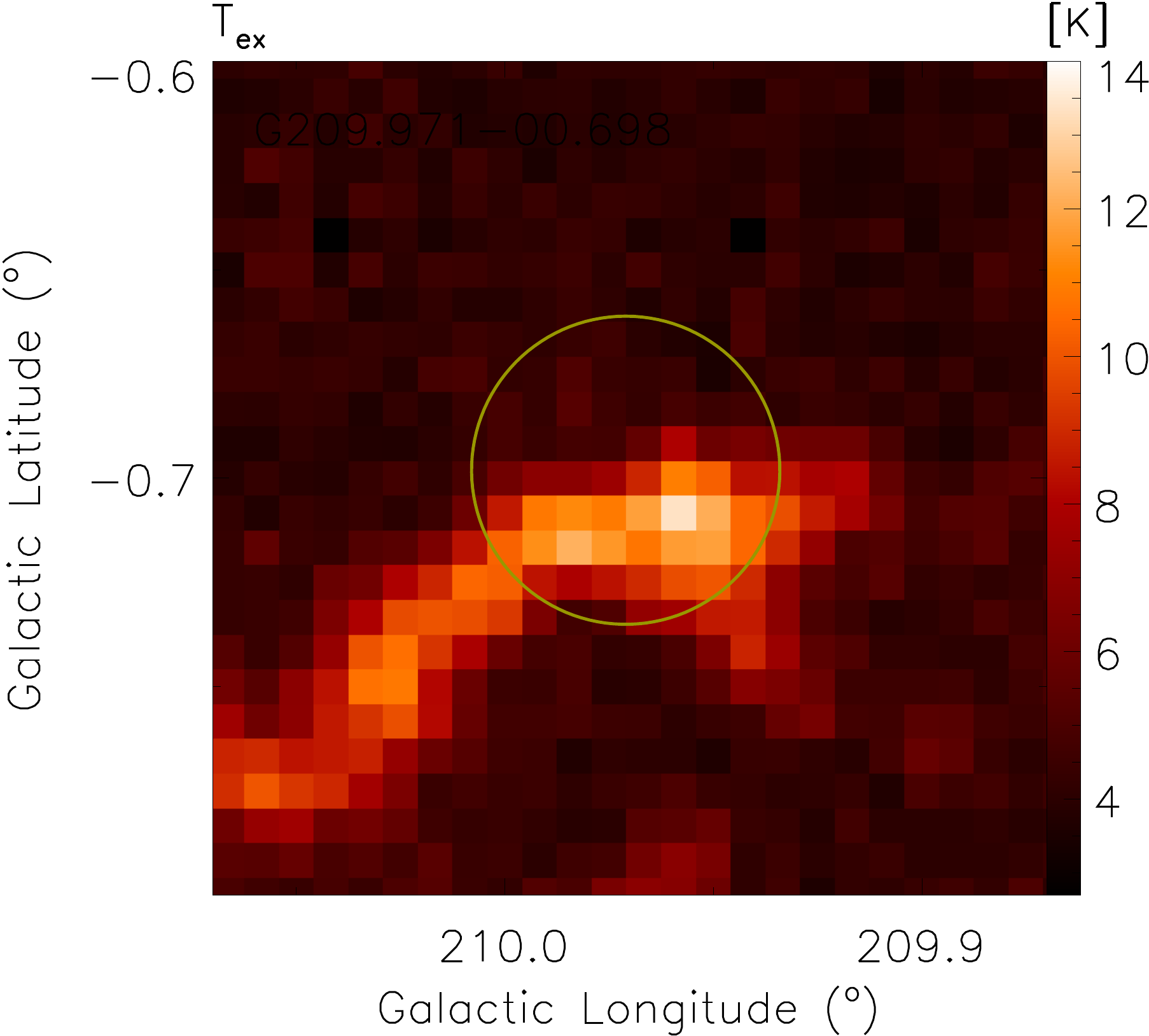}\\
  \caption{Morphology of G209.971-00.698 in various tracers. Middle: images of G209.971-00.698\_near with integrated velocity range from 9 to 13 km s$^{-1}$. Bottom: images of G209.971-00.698\_far with integrated velocity range from 32 to 37 km s$^{-1}$. All the others are the same as in Figures \ref{fig:G2085-023}.}
  \label{fig:G2099-006}
\end{figure}

\begin{figure}[h]
  \centering
\includegraphics[width=0.22\textwidth]{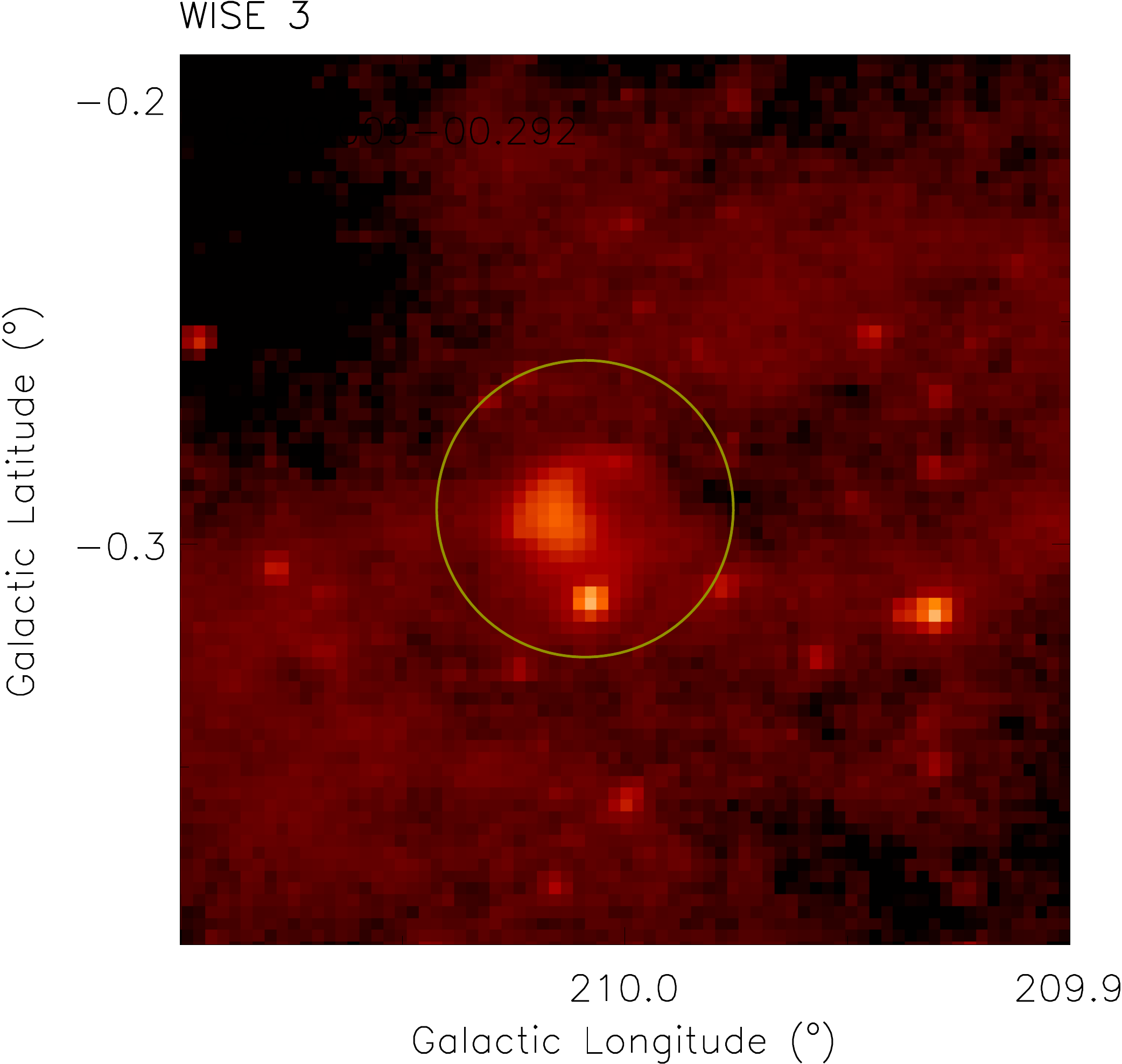}
\includegraphics[width=0.22\textwidth]{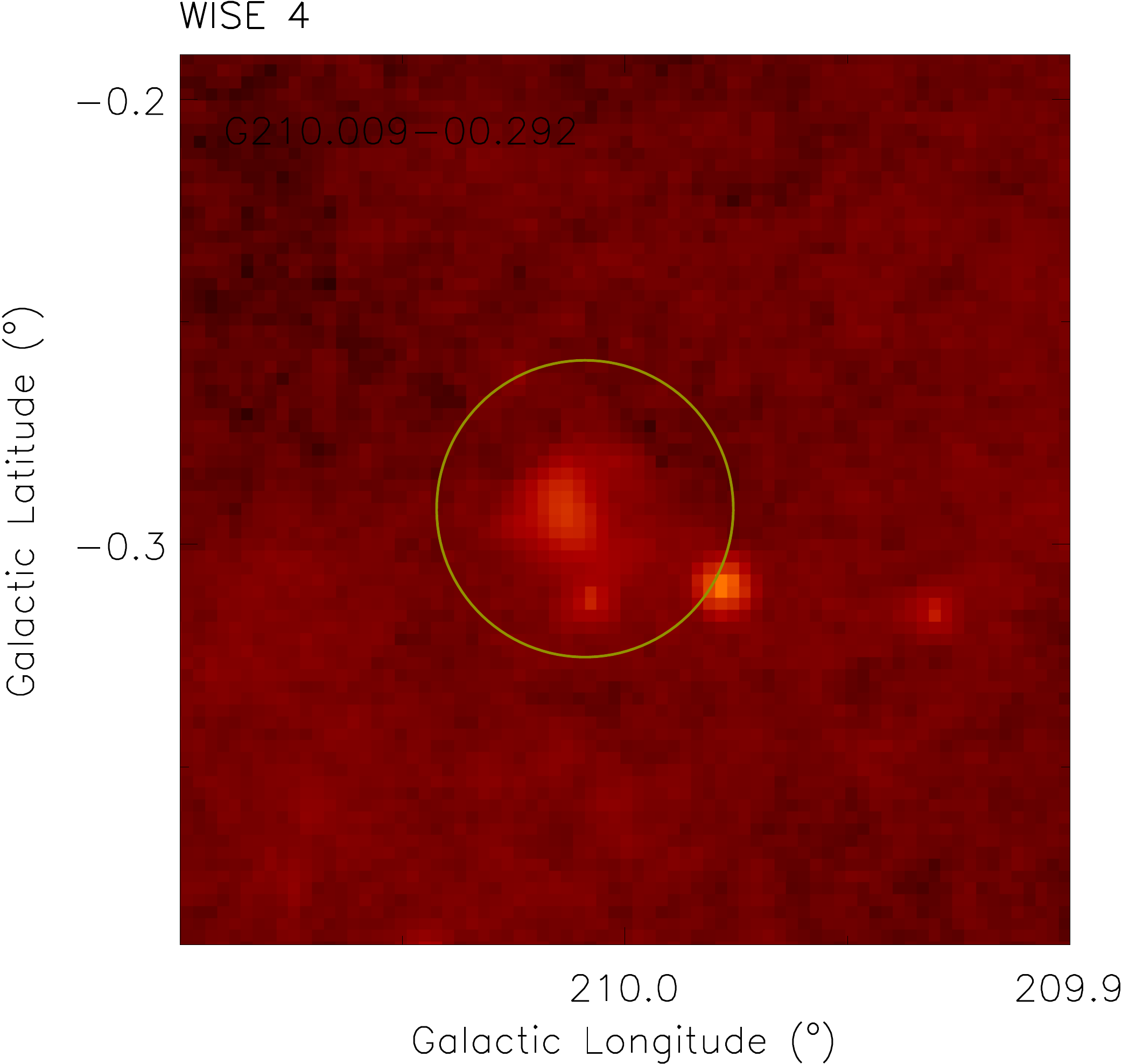}
\includegraphics[width=0.22\textwidth]{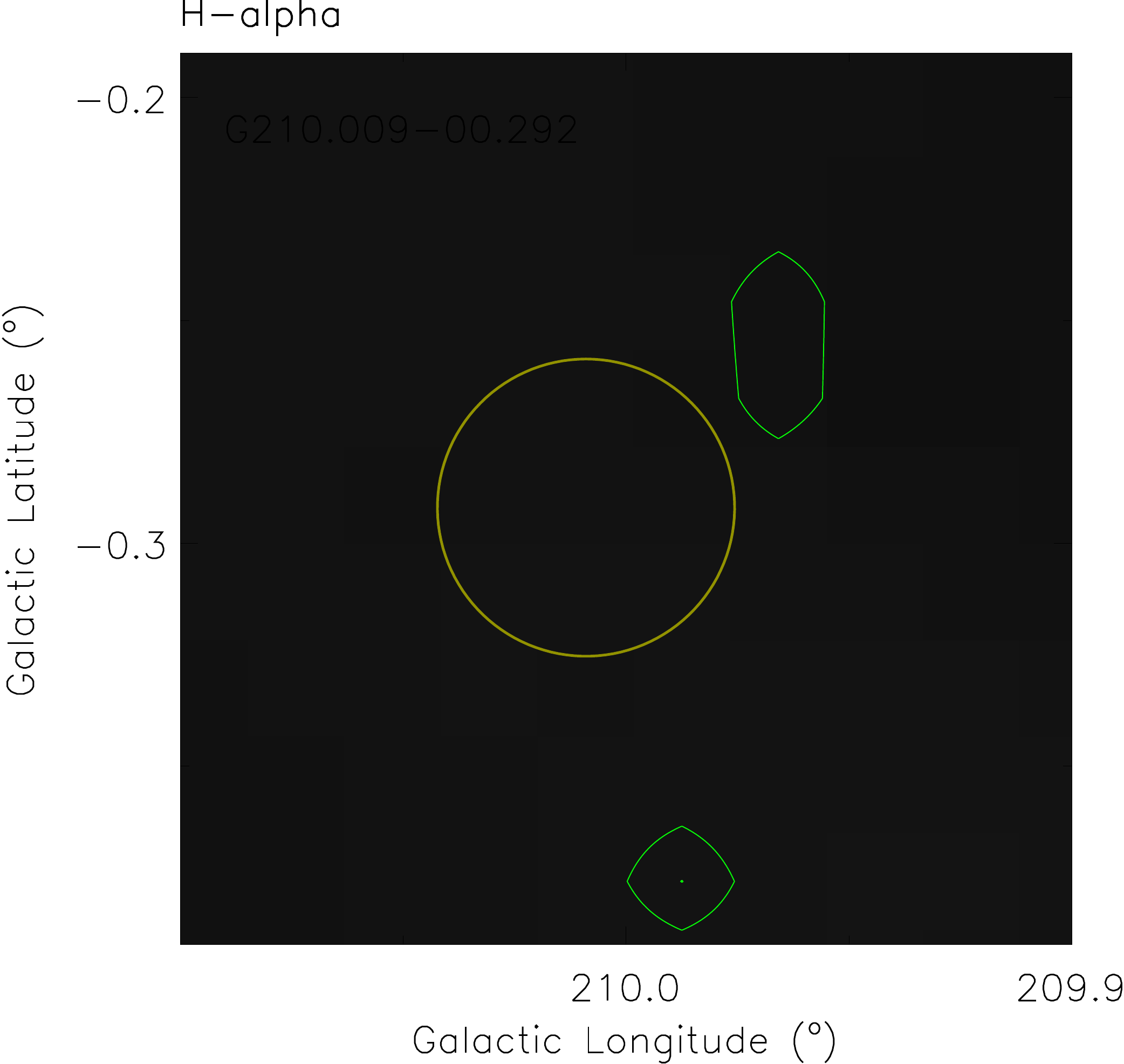}\\
\includegraphics[width=0.24\textwidth]{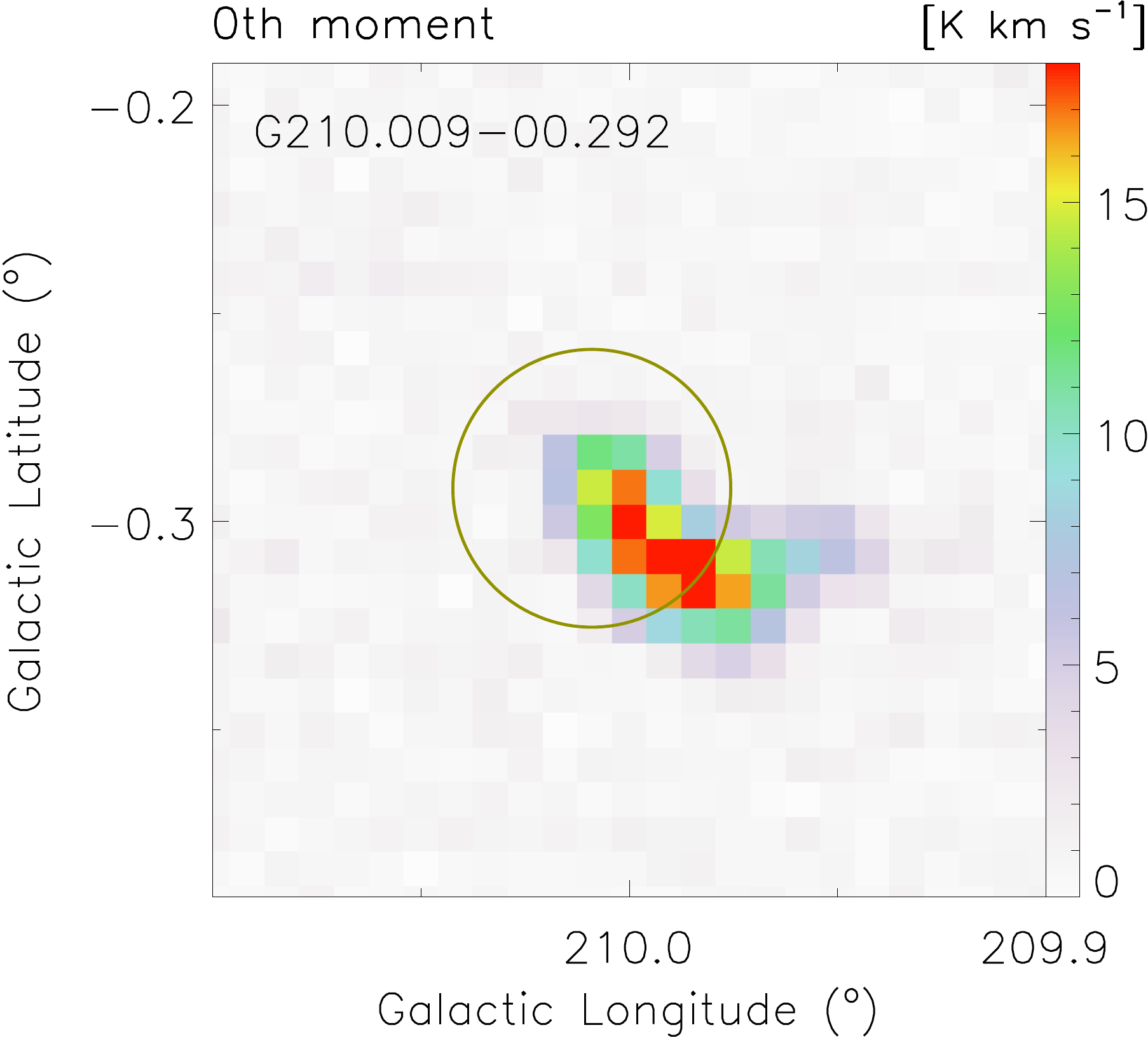}
\includegraphics[width=0.24\textwidth]{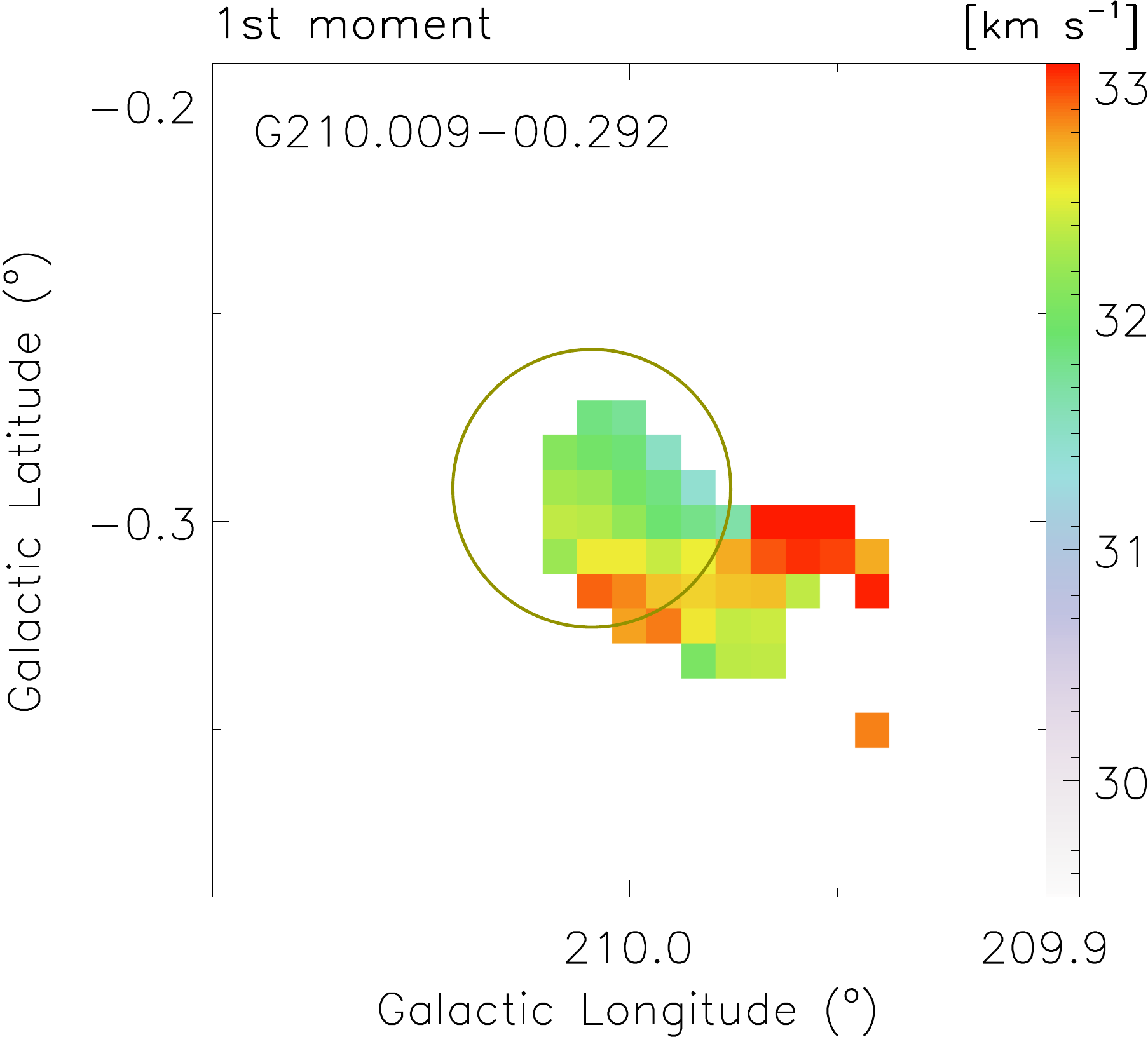}
\includegraphics[width=0.24\textwidth]{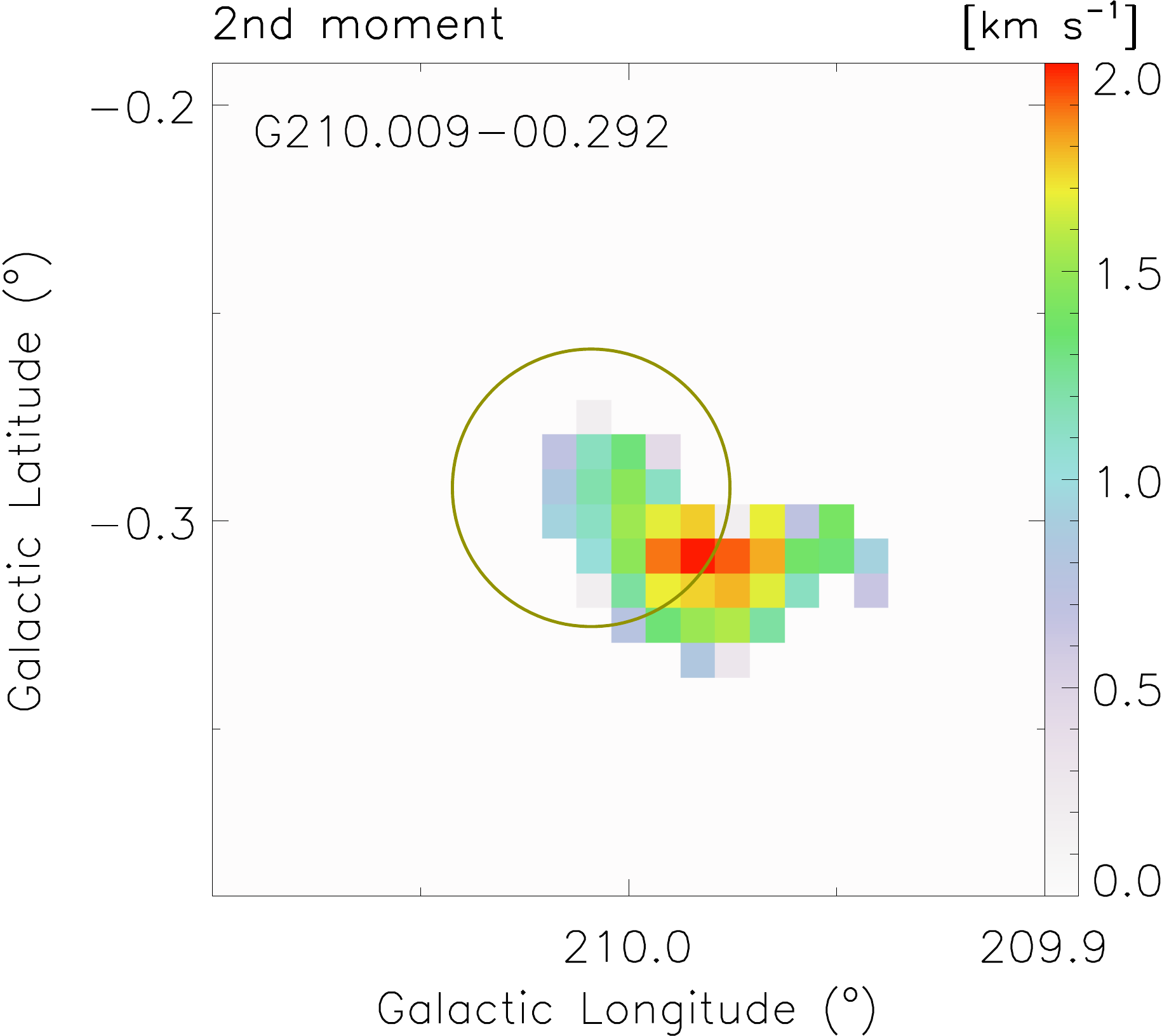}
\includegraphics[width=0.24\textwidth]{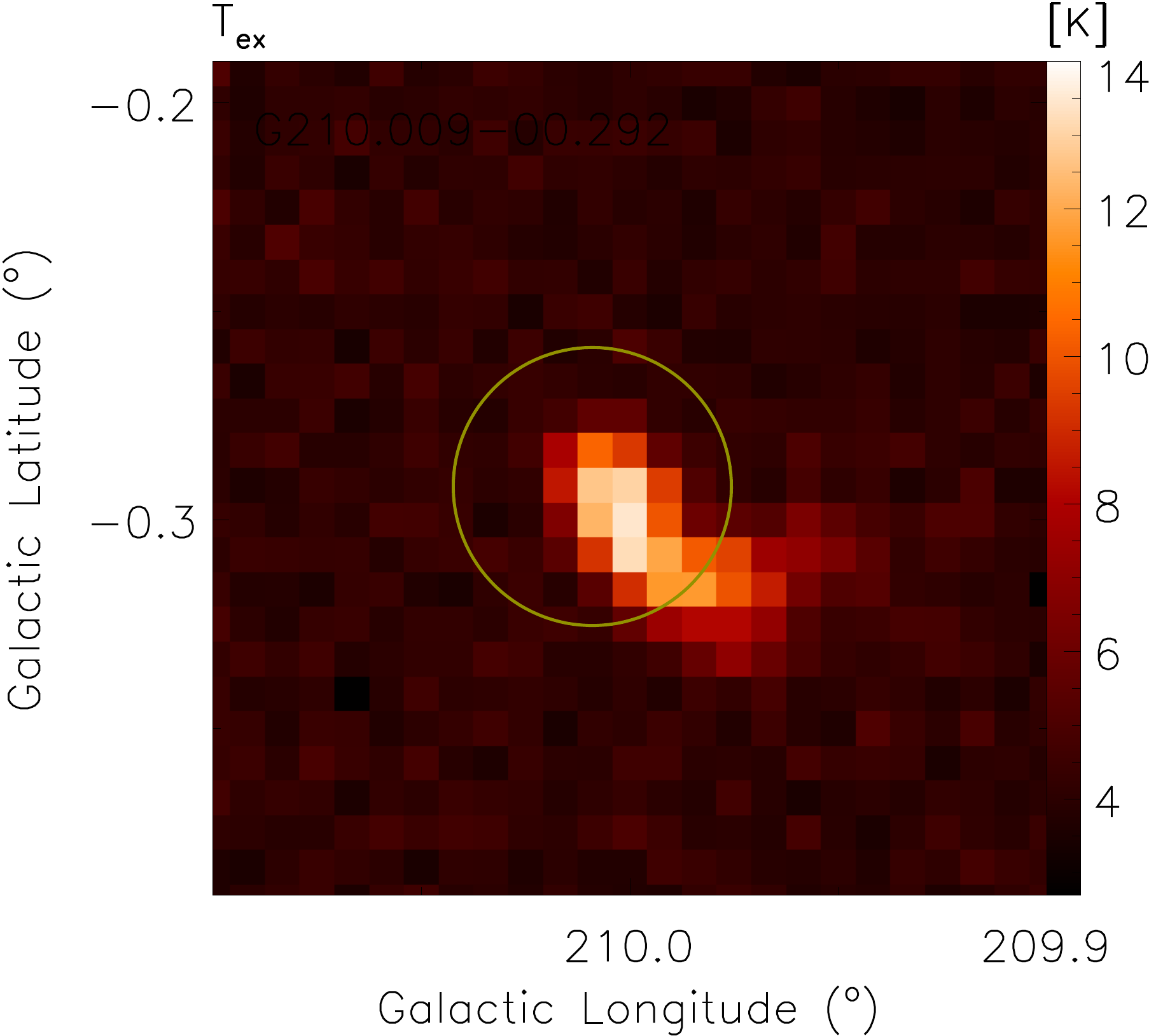}
  \caption{Morphology of G210.009-00.292 in various tracers. The velocity range for intensity integration is from 29 km s$^{-1}$ to 35 km s$^{-1}$. All the others are the same as in Figures \ref{fig:G2085-023}.}
  \label{fig:G2100-002}
\end{figure}

\begin{figure}[h]
  \centering
\includegraphics[width=0.21\textwidth]{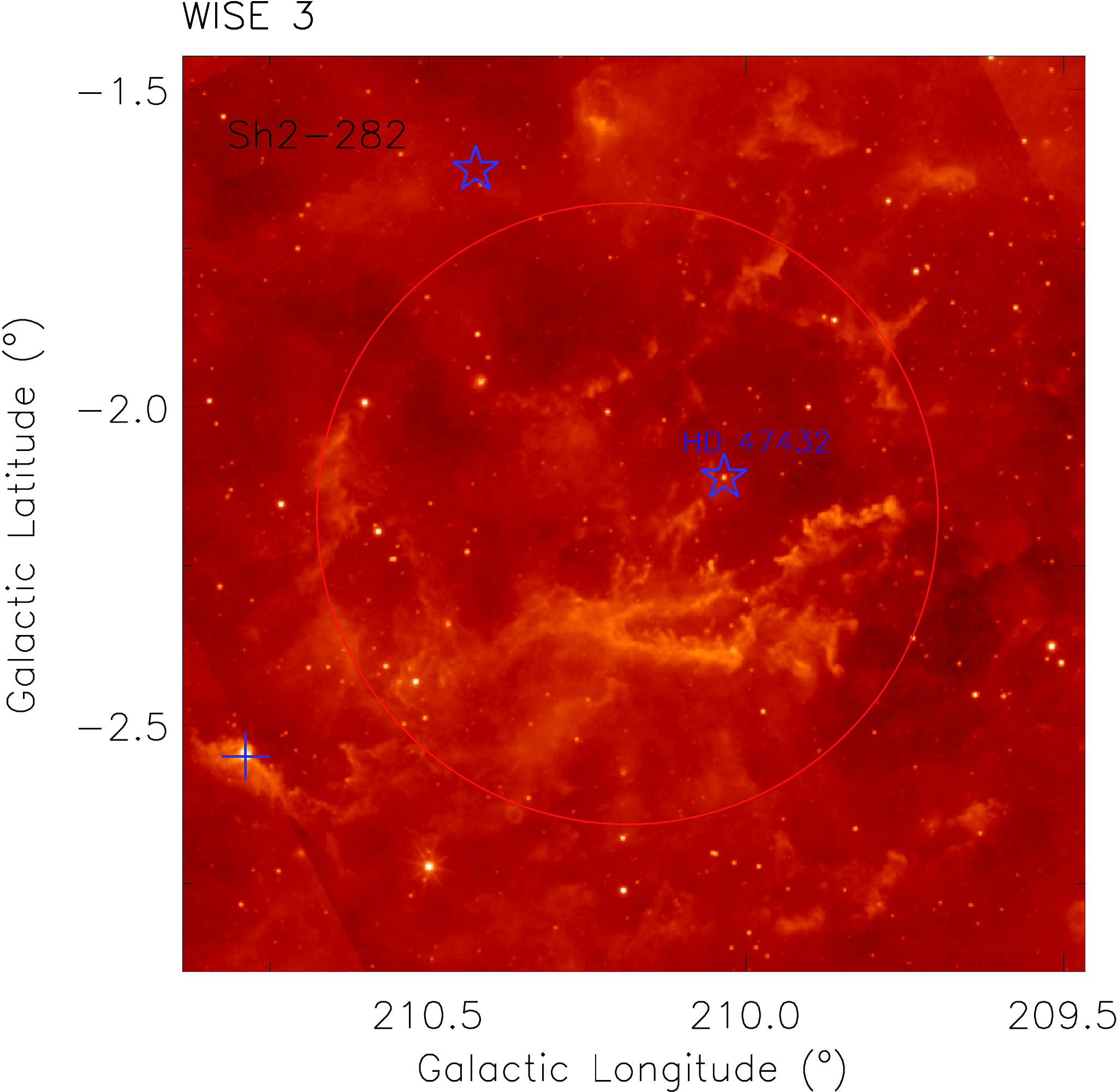}
\includegraphics[width=0.21\textwidth]{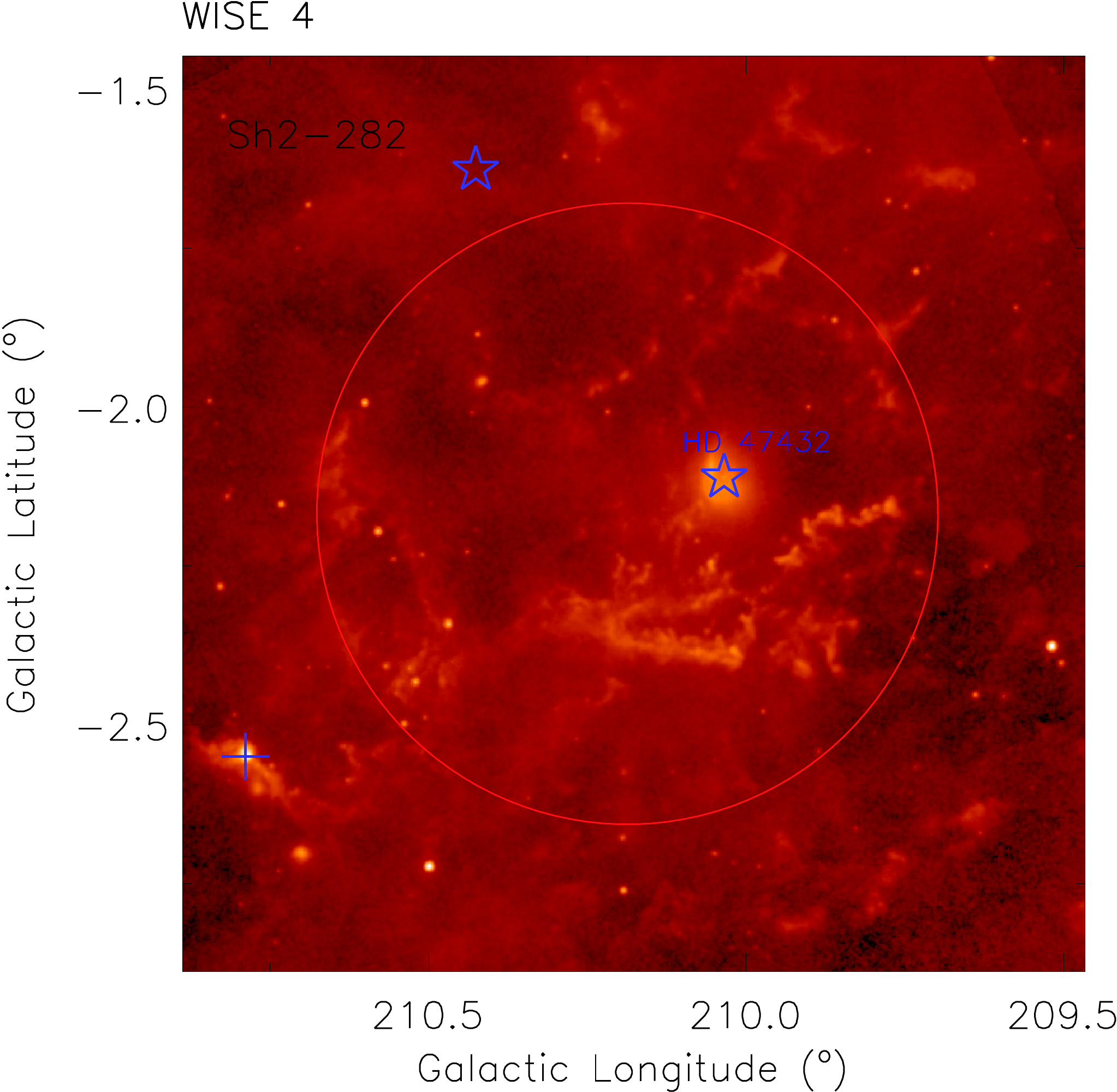}
\includegraphics[width=0.21\textwidth]{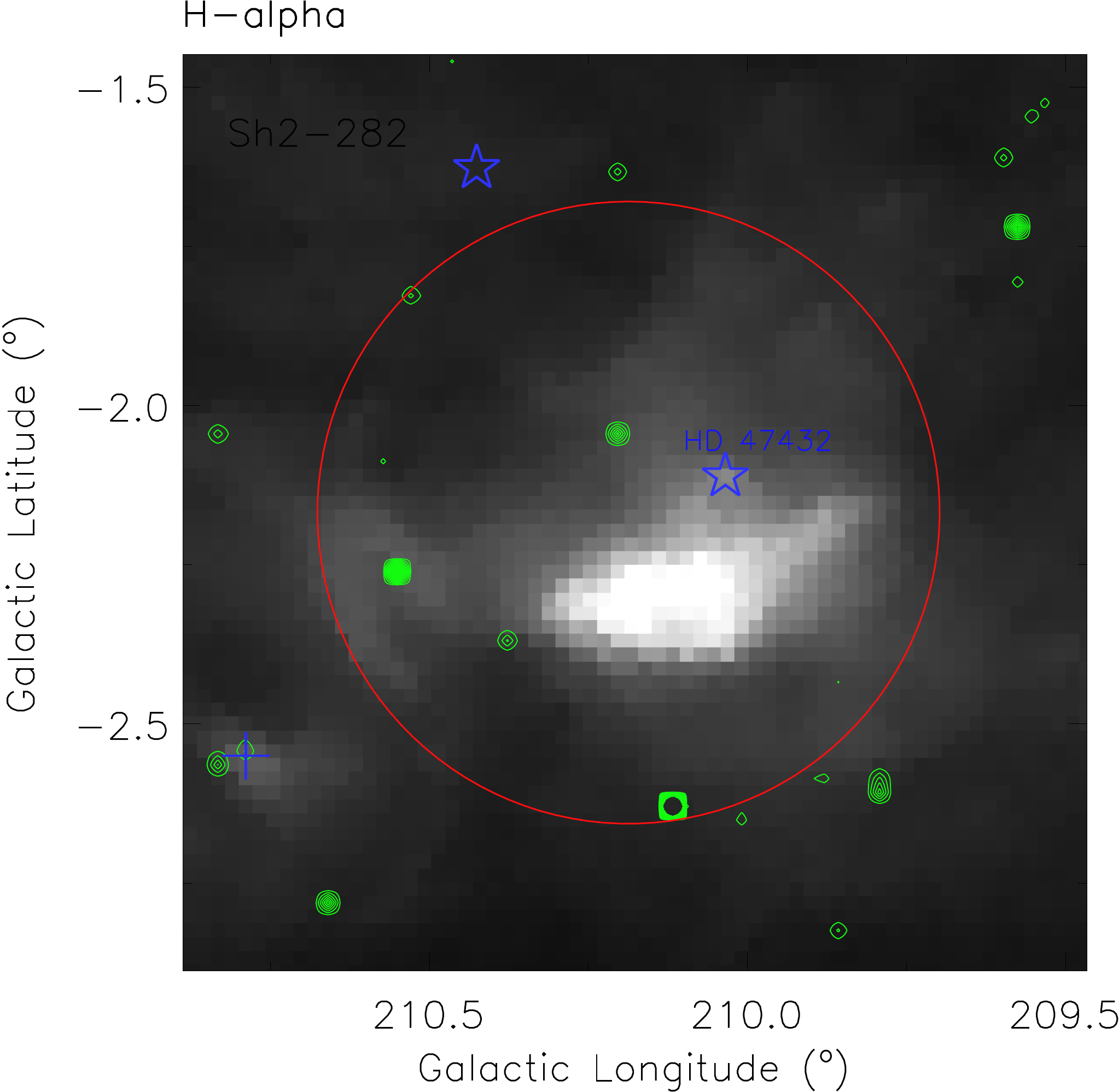}\\
\includegraphics[width=0.24\textwidth]{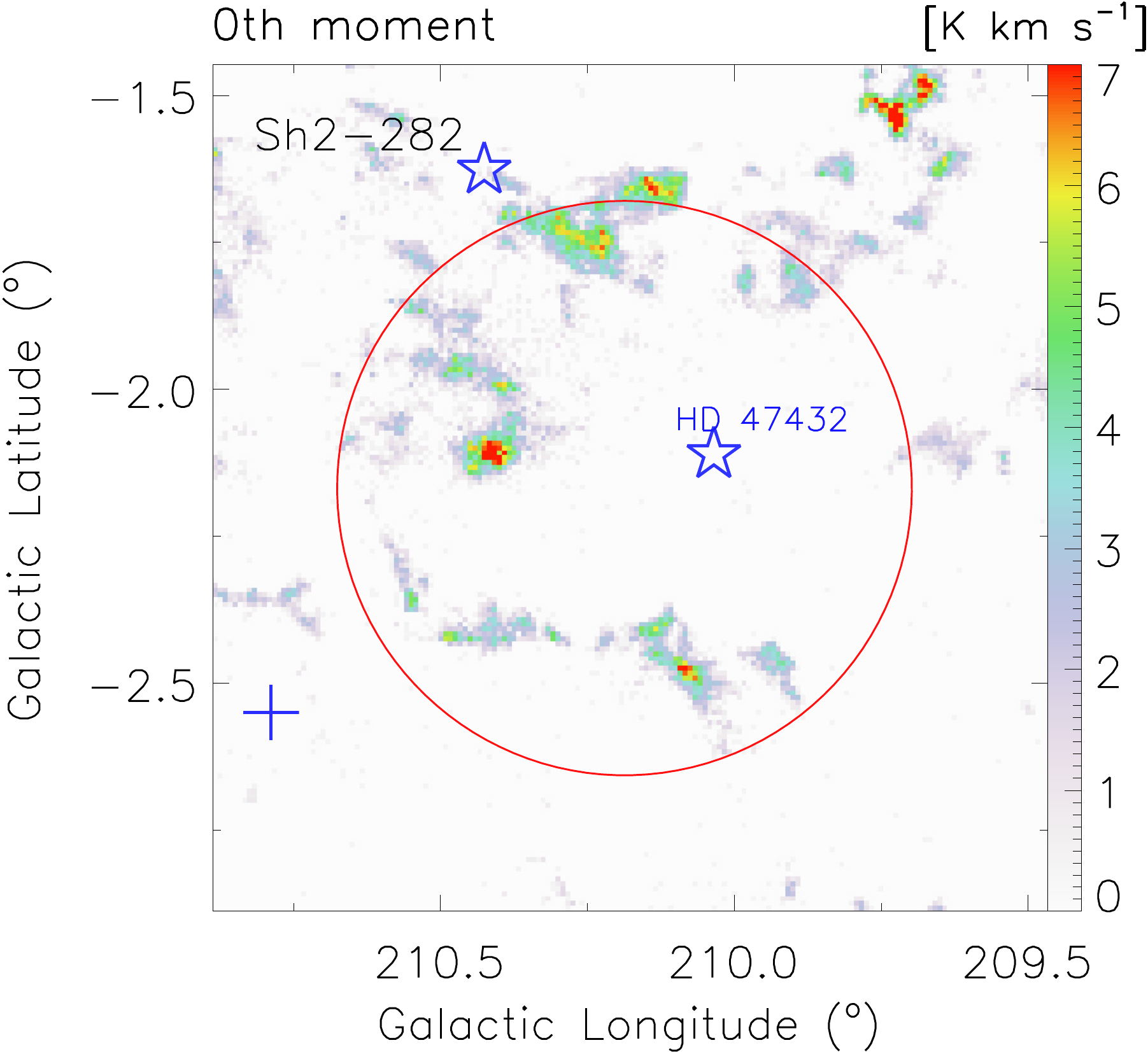}
\includegraphics[width=0.24\textwidth]{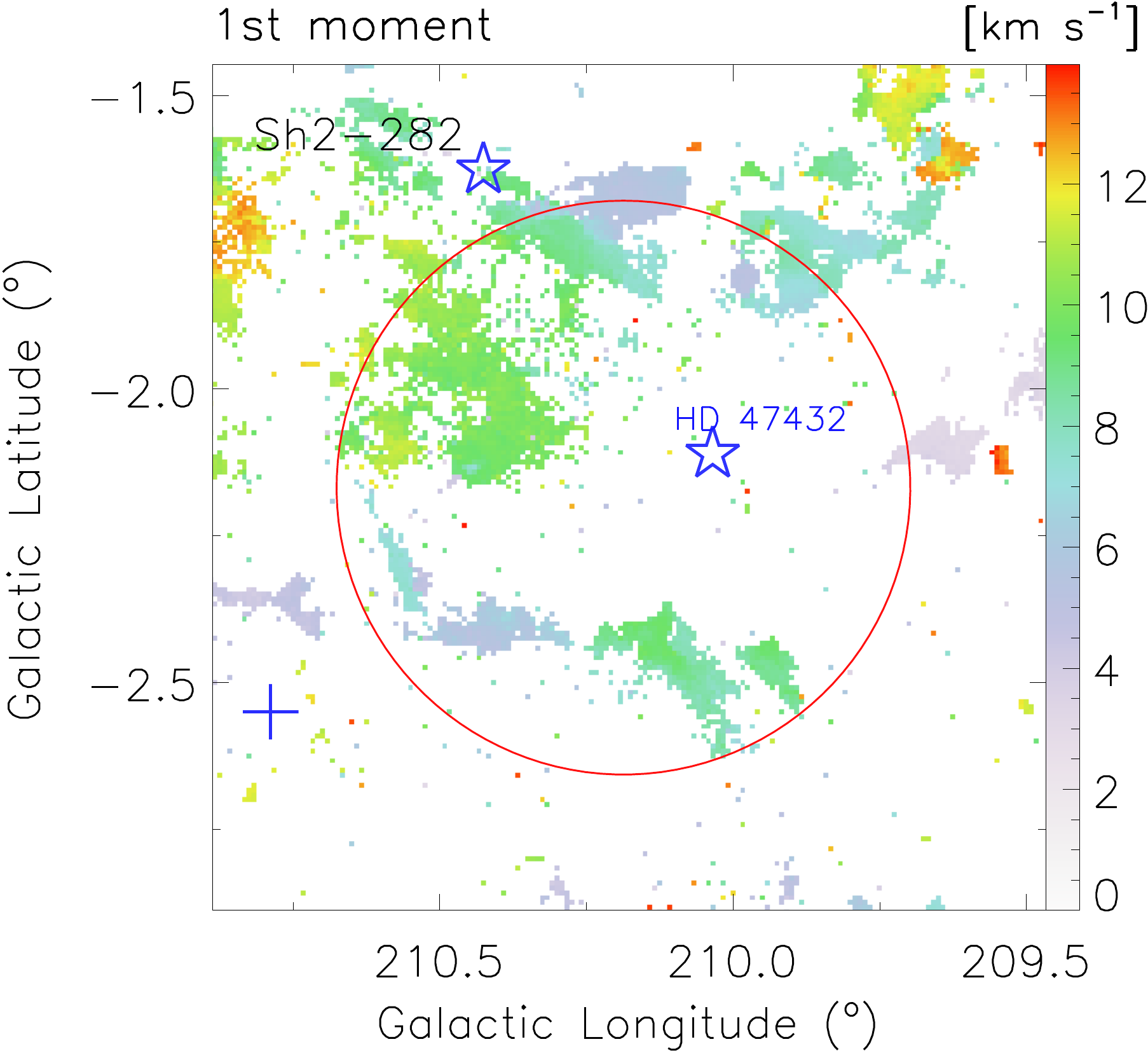}
\includegraphics[width=0.24\textwidth]{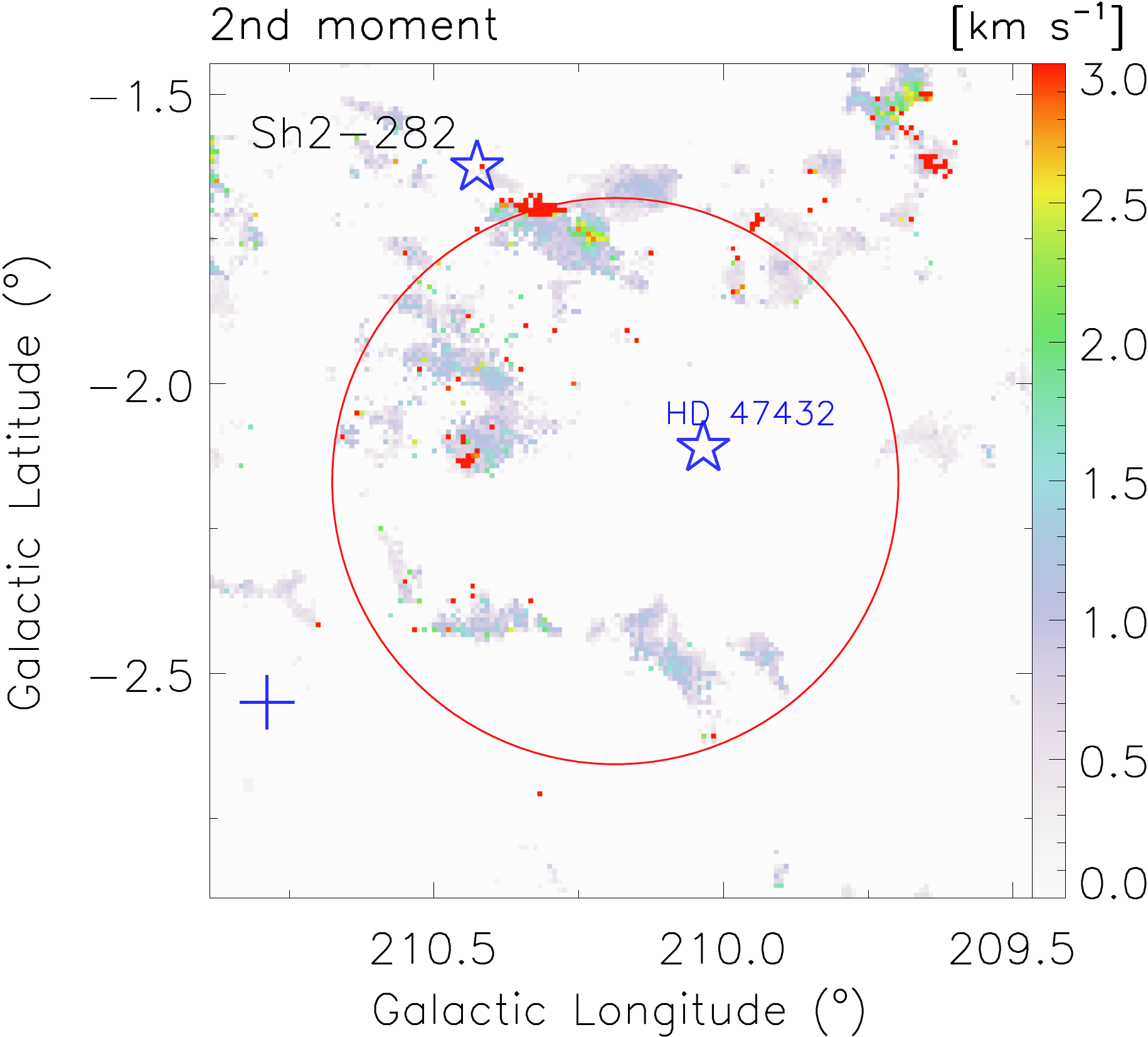}
\includegraphics[width=0.24\textwidth]{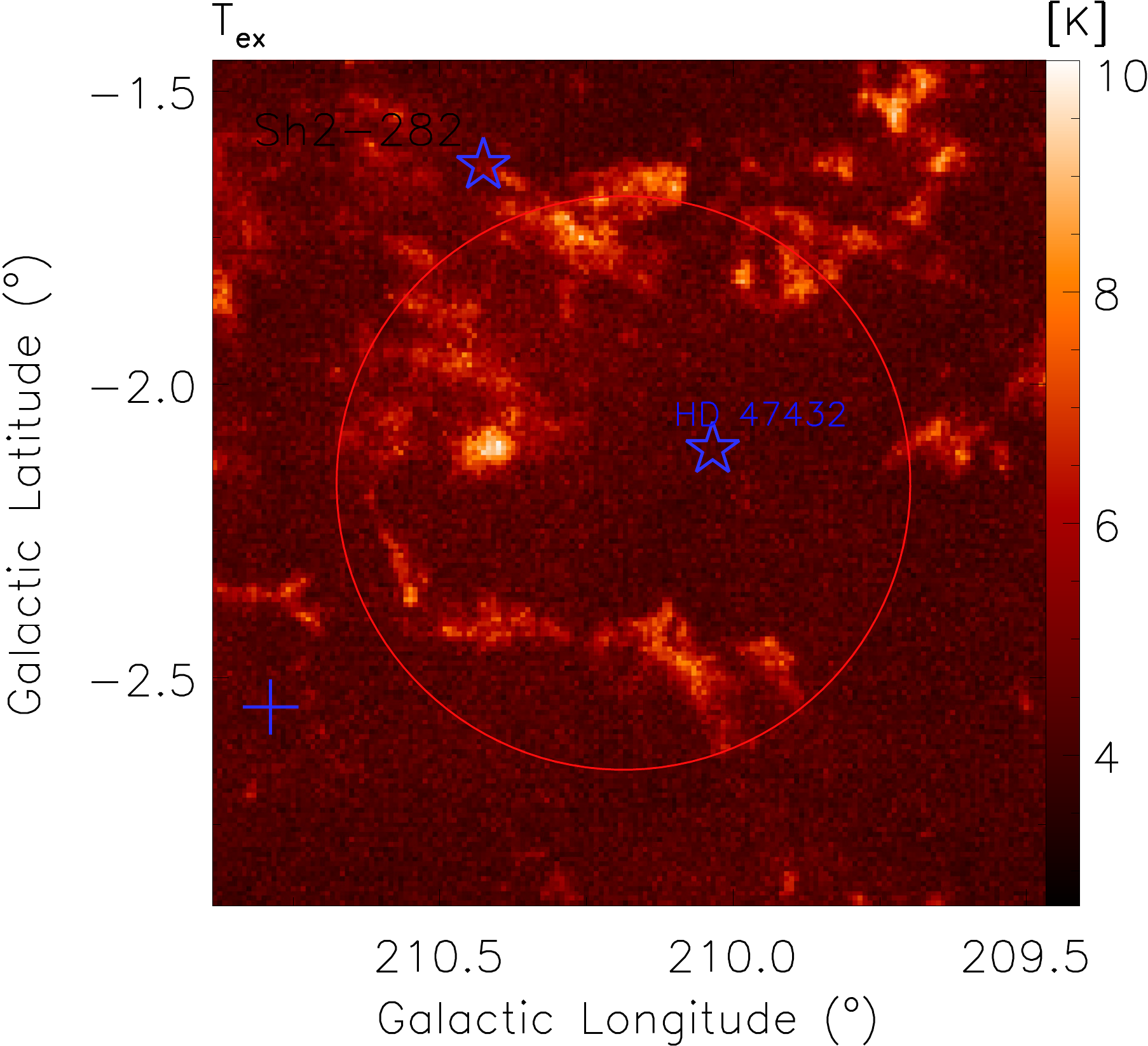}\\
\includegraphics[width=0.24\textwidth]{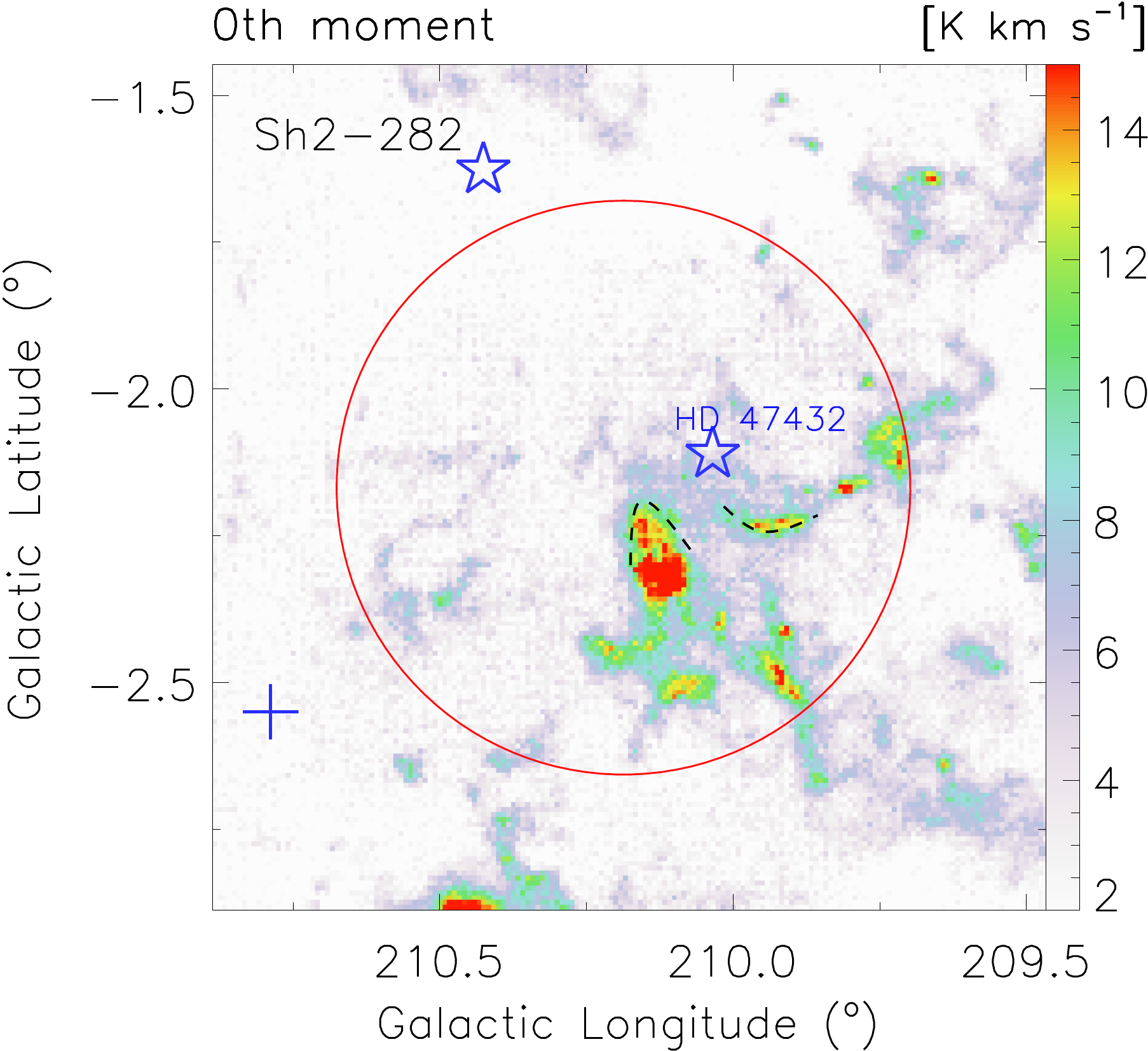}
\includegraphics[width=0.24\textwidth]{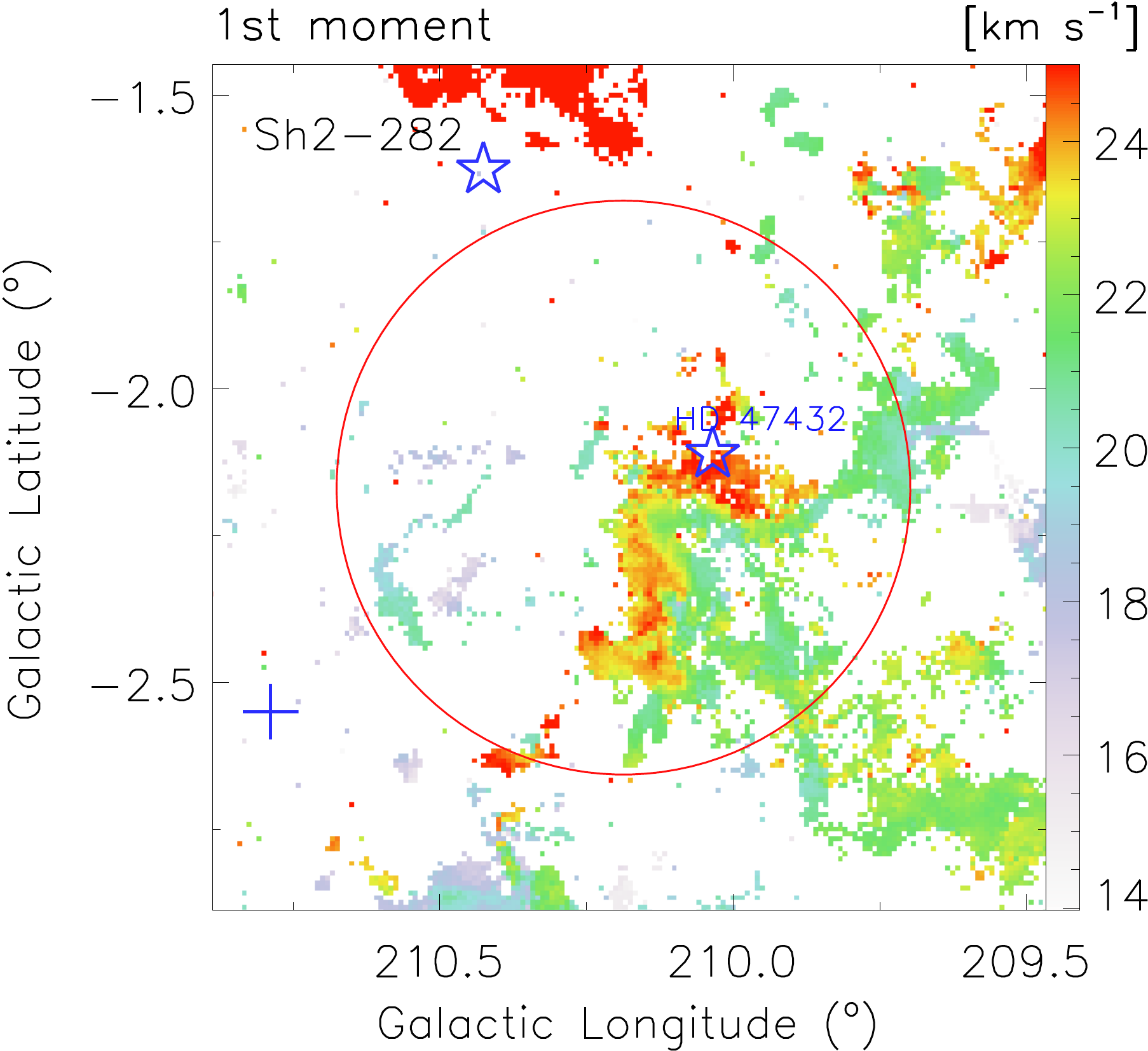}
\includegraphics[width=0.24\textwidth]{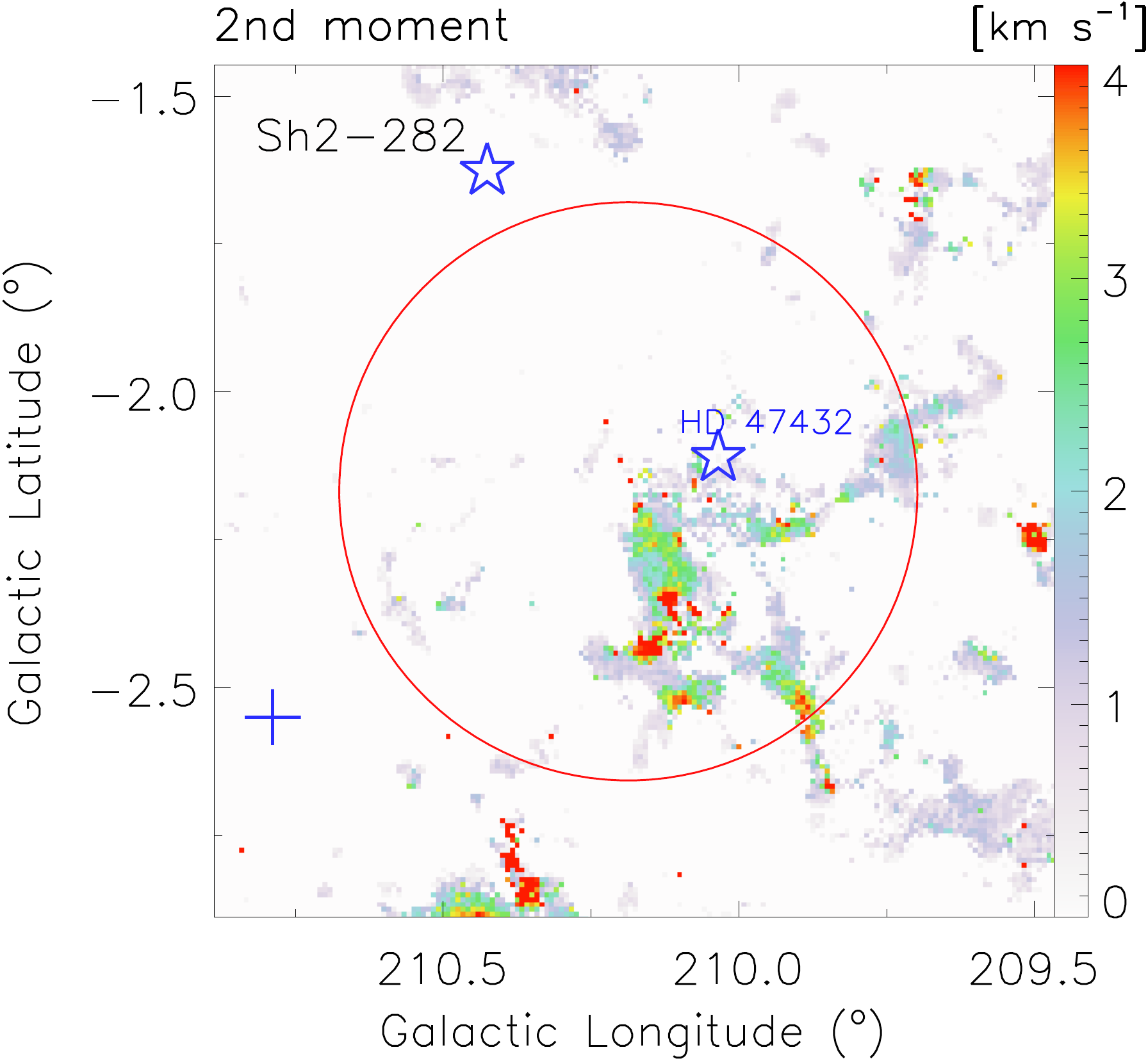}
\includegraphics[width=0.24\textwidth]{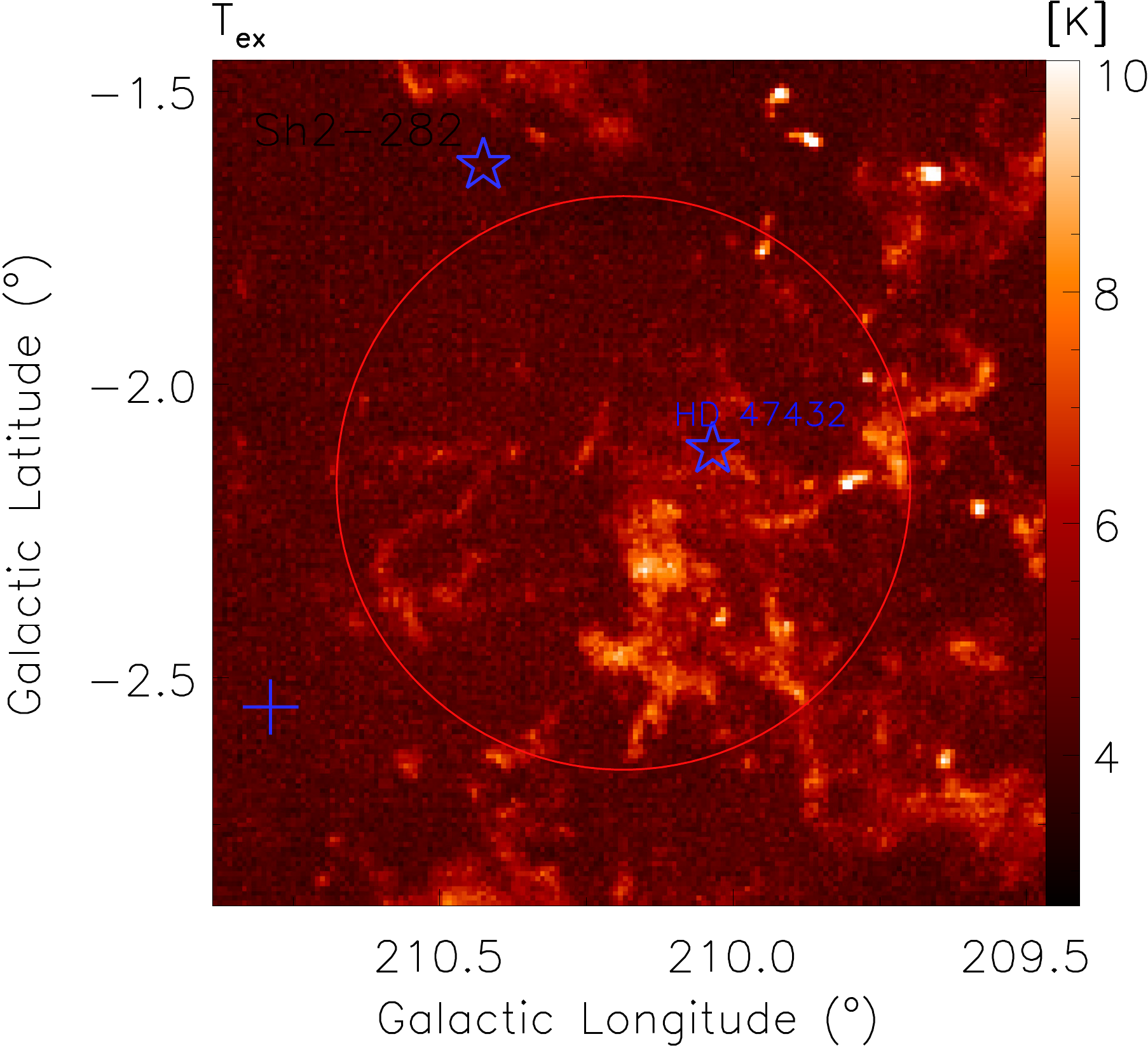}\\
  \caption{Morphology of Sh2-282 in various tracers. Middle: images of Sh2-282\_near with integrated velocity range from 3 to 14 km s$^{-1}$. Bottom: images of Sh2-282\_far with integrated velocity range from 14 to 30 km s$^{-1}$. The green contours indicate the radio continuum emission, with the minimal level and the interval of the contours are 15 and 5 mJy/beam, respectively. The dashed lines in bottom left panel indicate the elephant trunk and cometary features. The blue pentagram and cross signs indicate, respectively, the O and B0 stars in this region from the SIMBAD database.}
  \label{fig:S282}
\end{figure}

\begin{figure}[h]
  \centering
\includegraphics[width=0.21\textwidth]{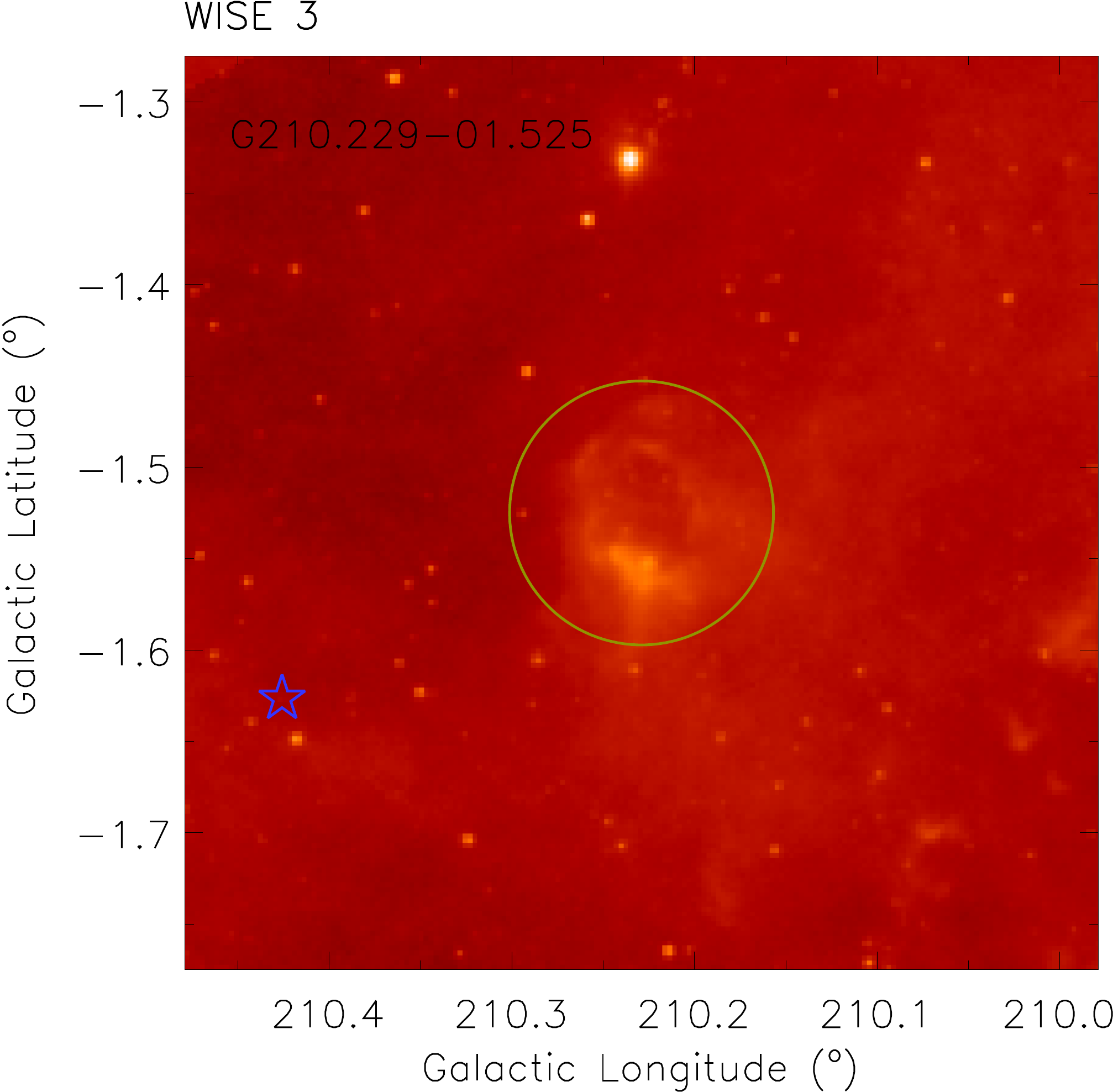}
\includegraphics[width=0.21\textwidth]{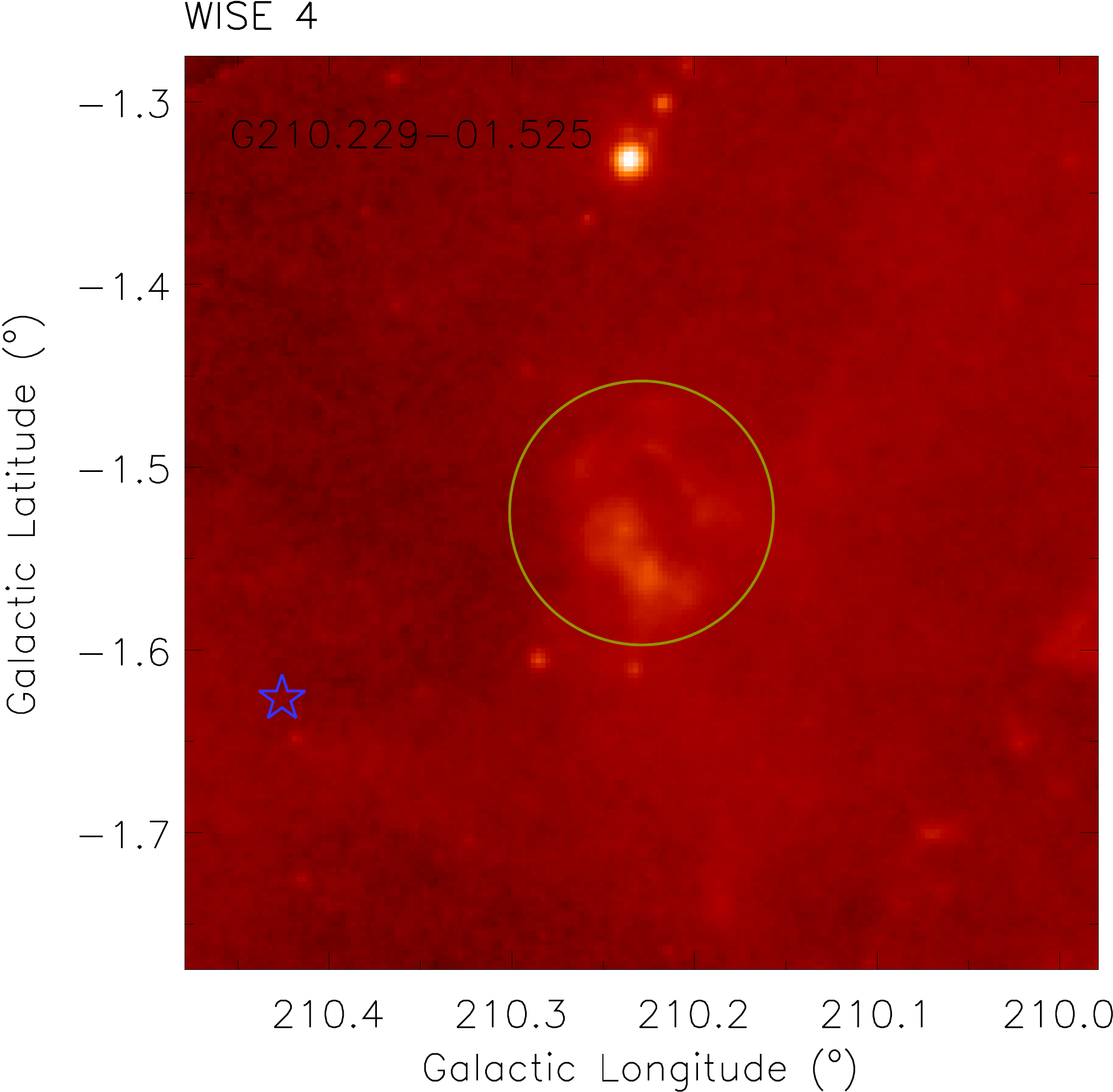}
\includegraphics[width=0.21\textwidth]{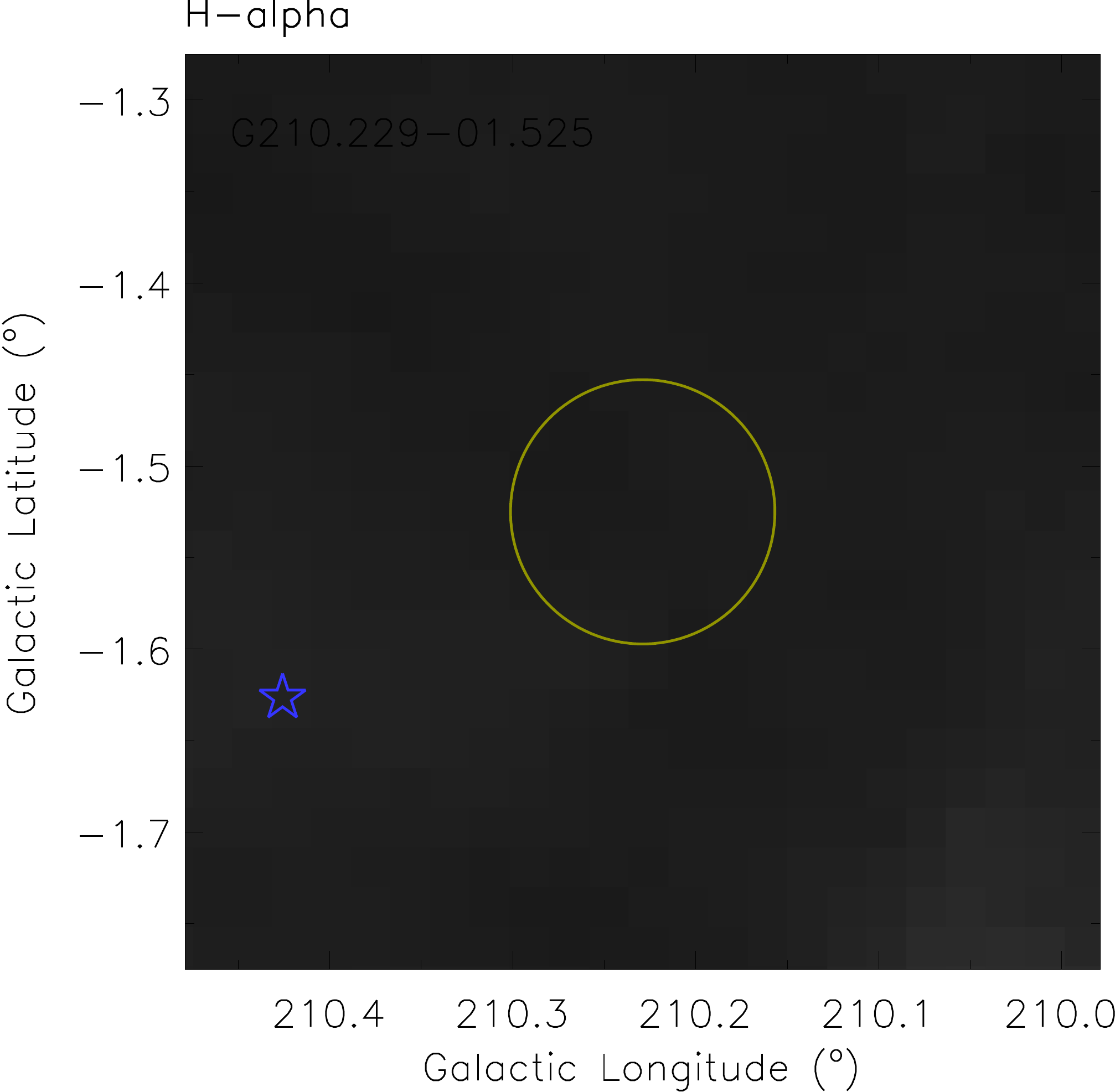}\\
\includegraphics[width=0.24\textwidth]{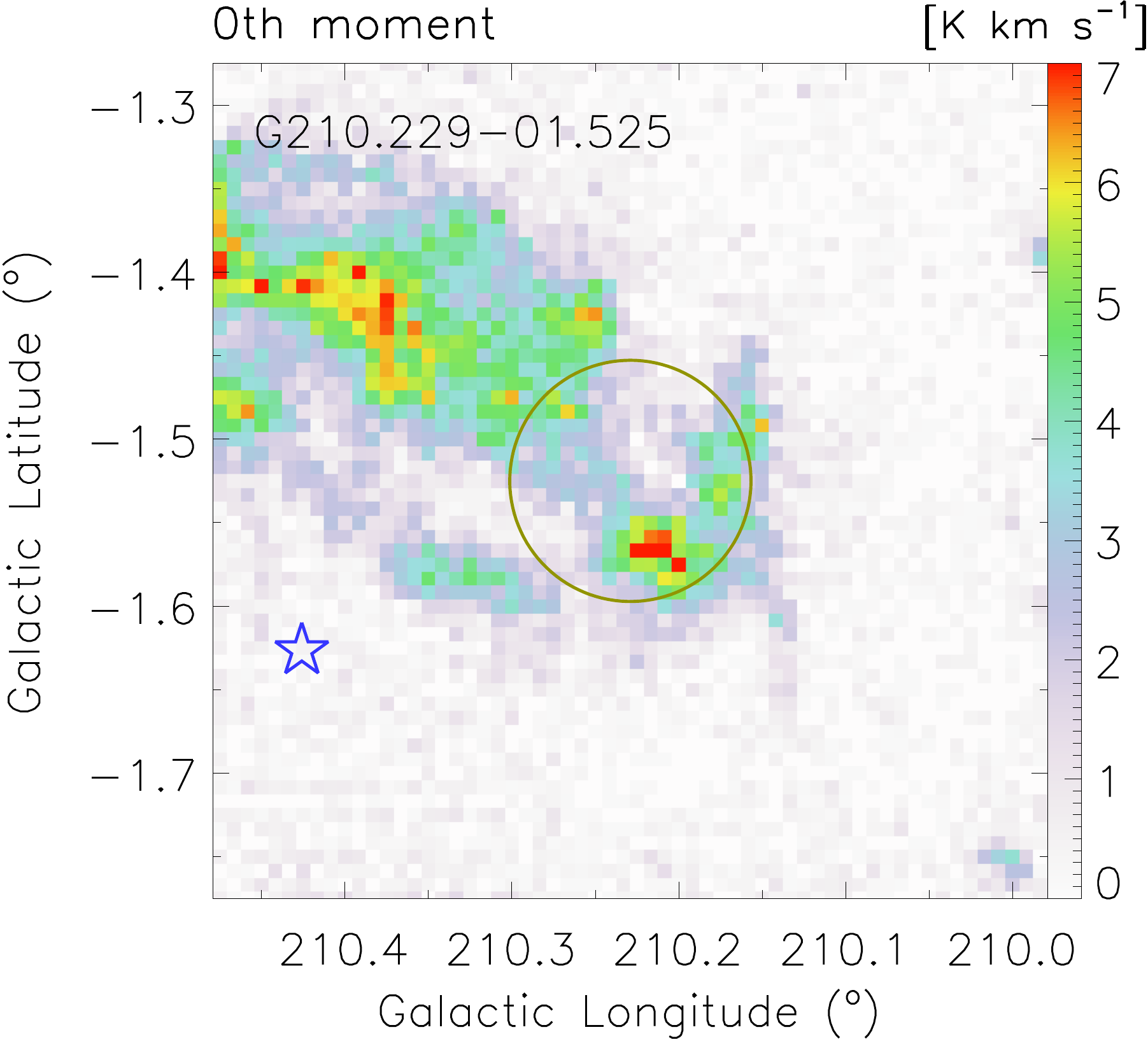}
\includegraphics[width=0.24\textwidth]{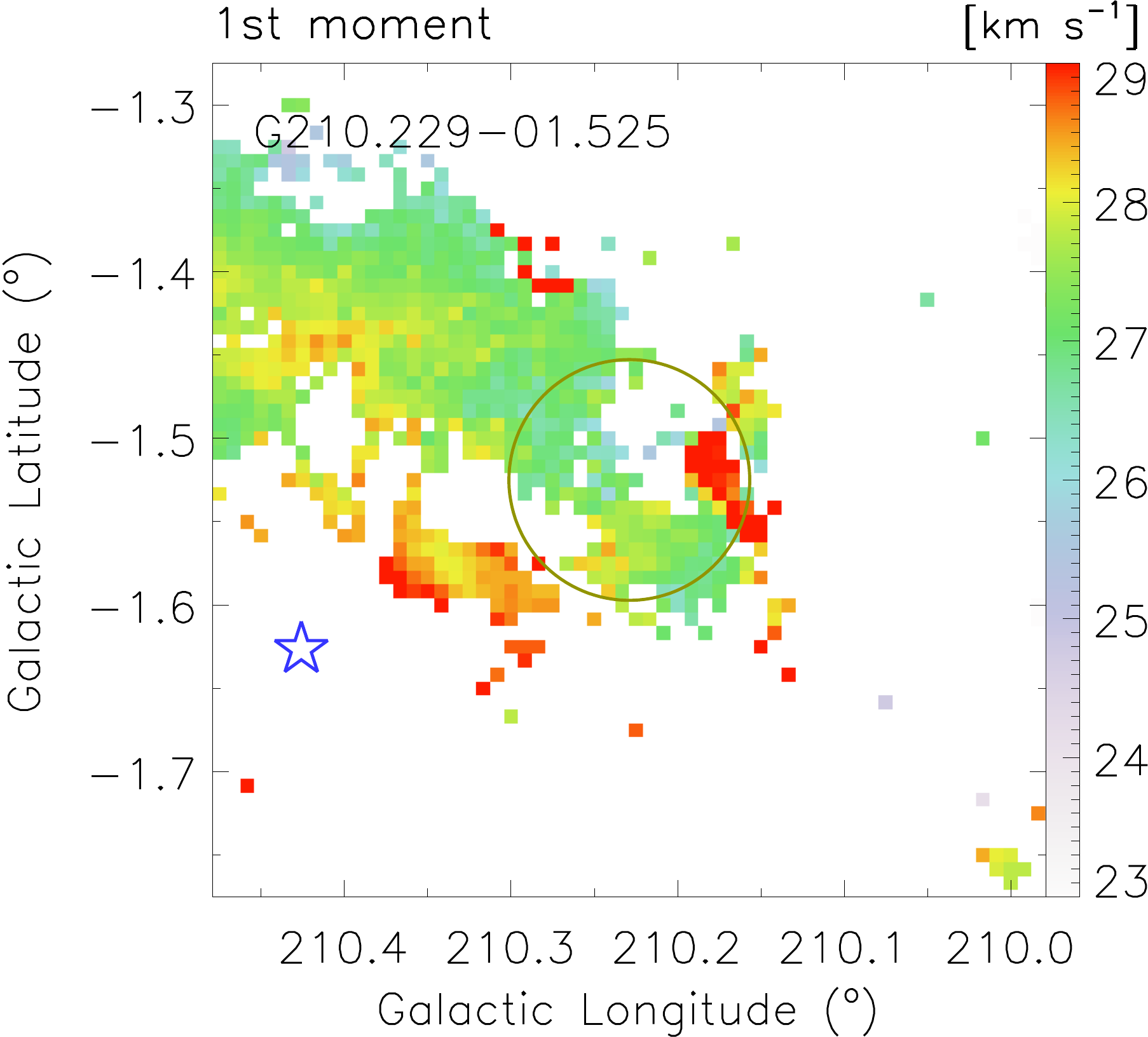}
\includegraphics[width=0.24\textwidth]{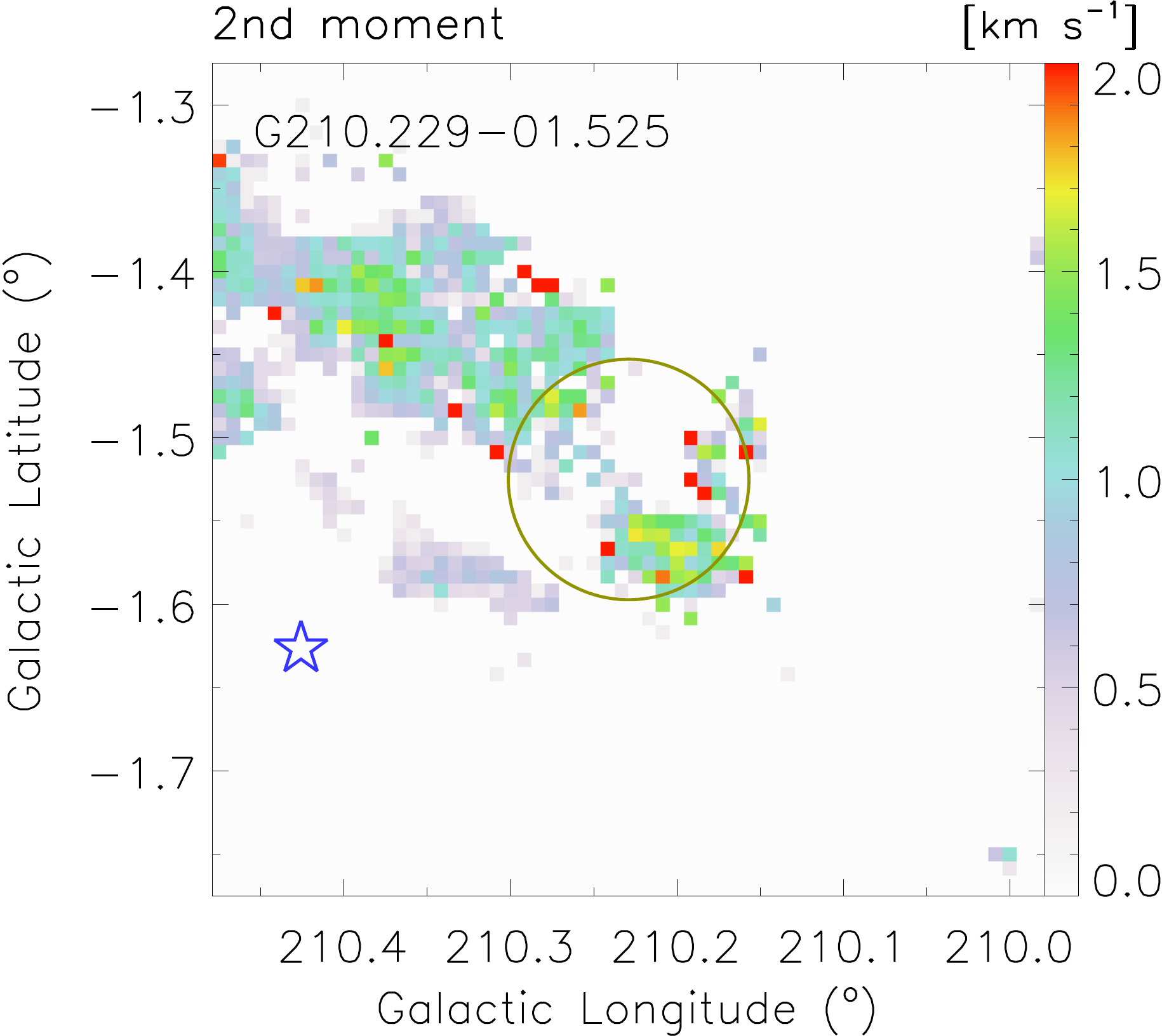}
\includegraphics[width=0.23\textwidth]{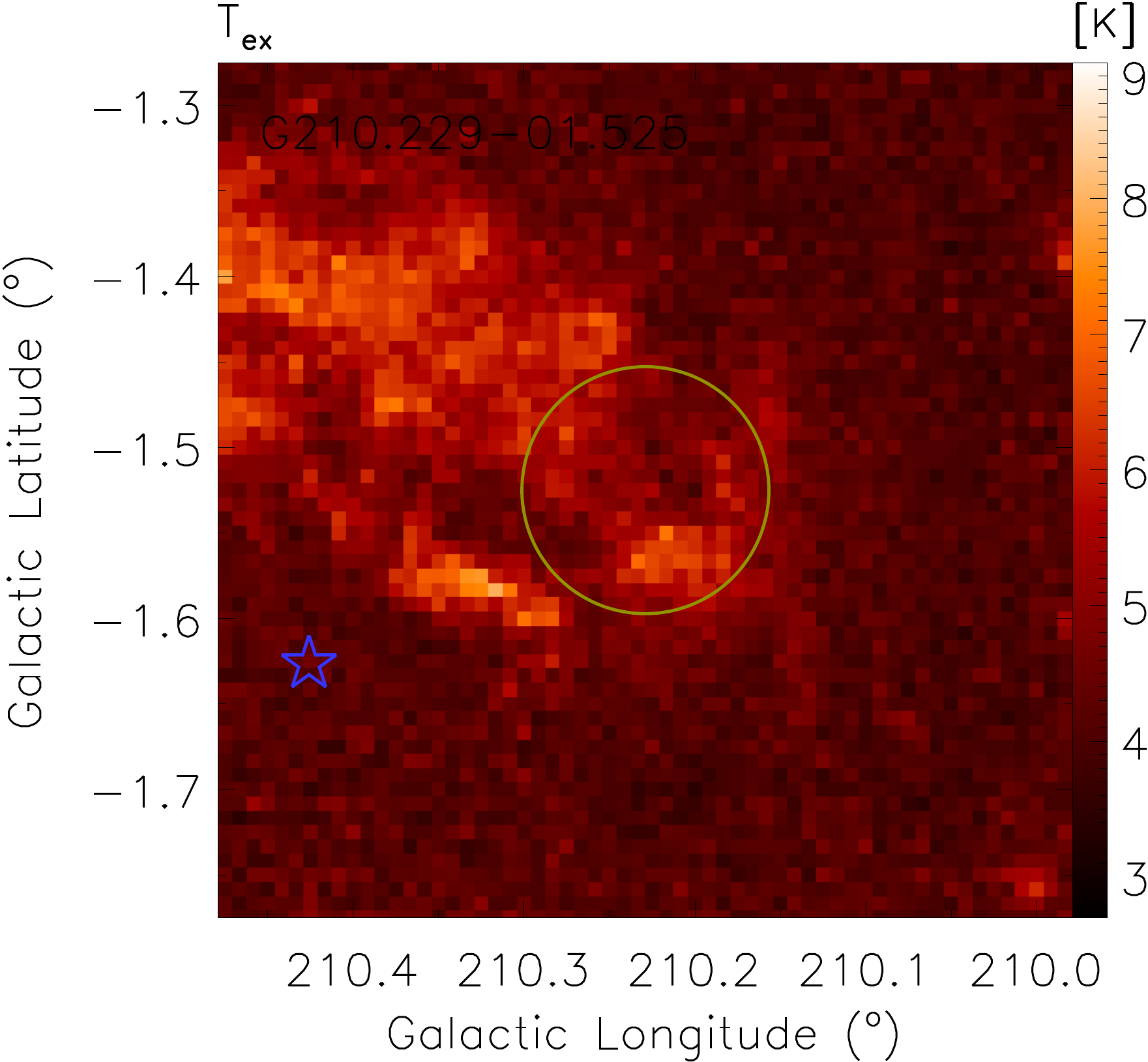}\\
\includegraphics[width=0.24\textwidth]{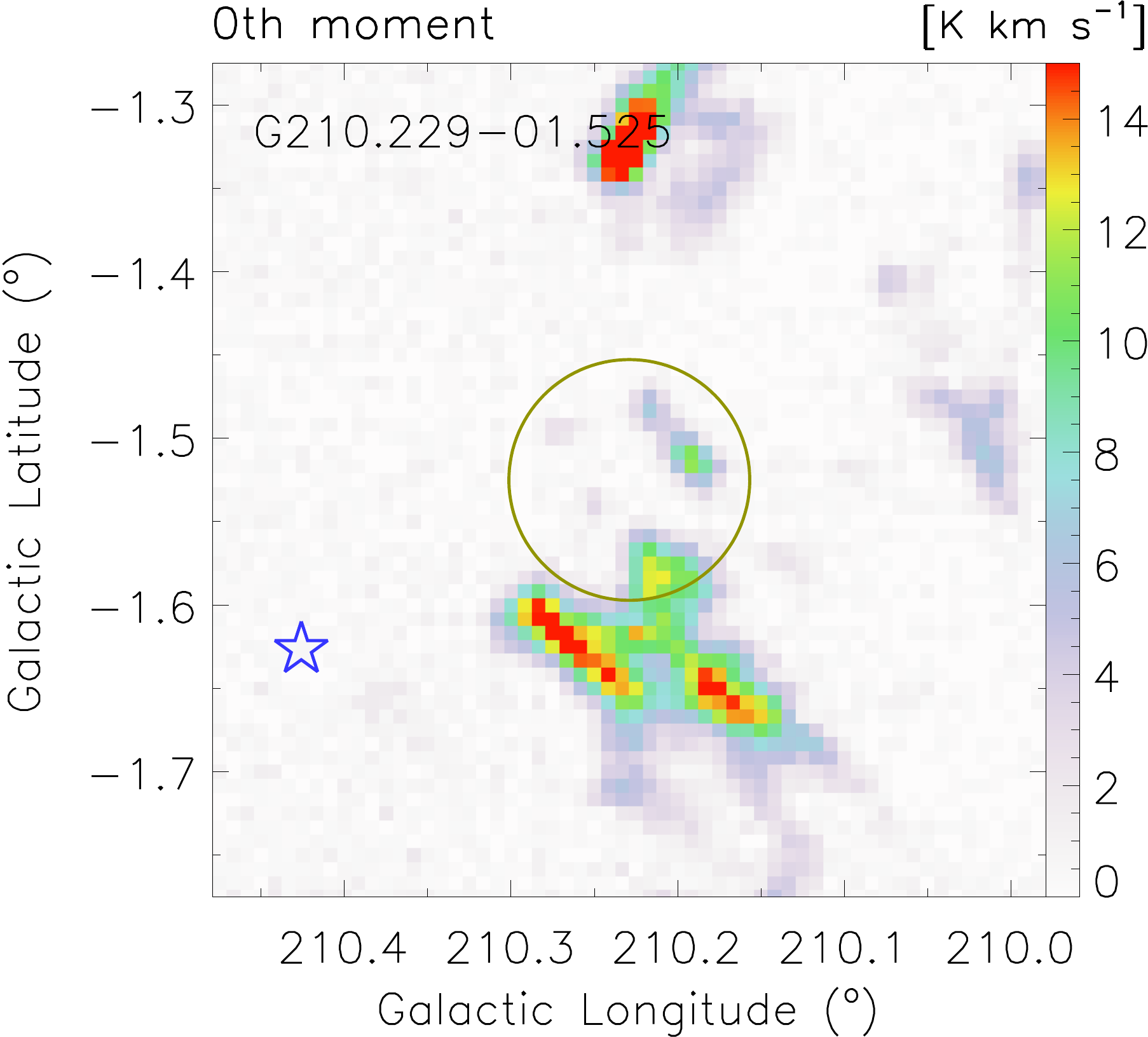}
\includegraphics[width=0.24\textwidth]{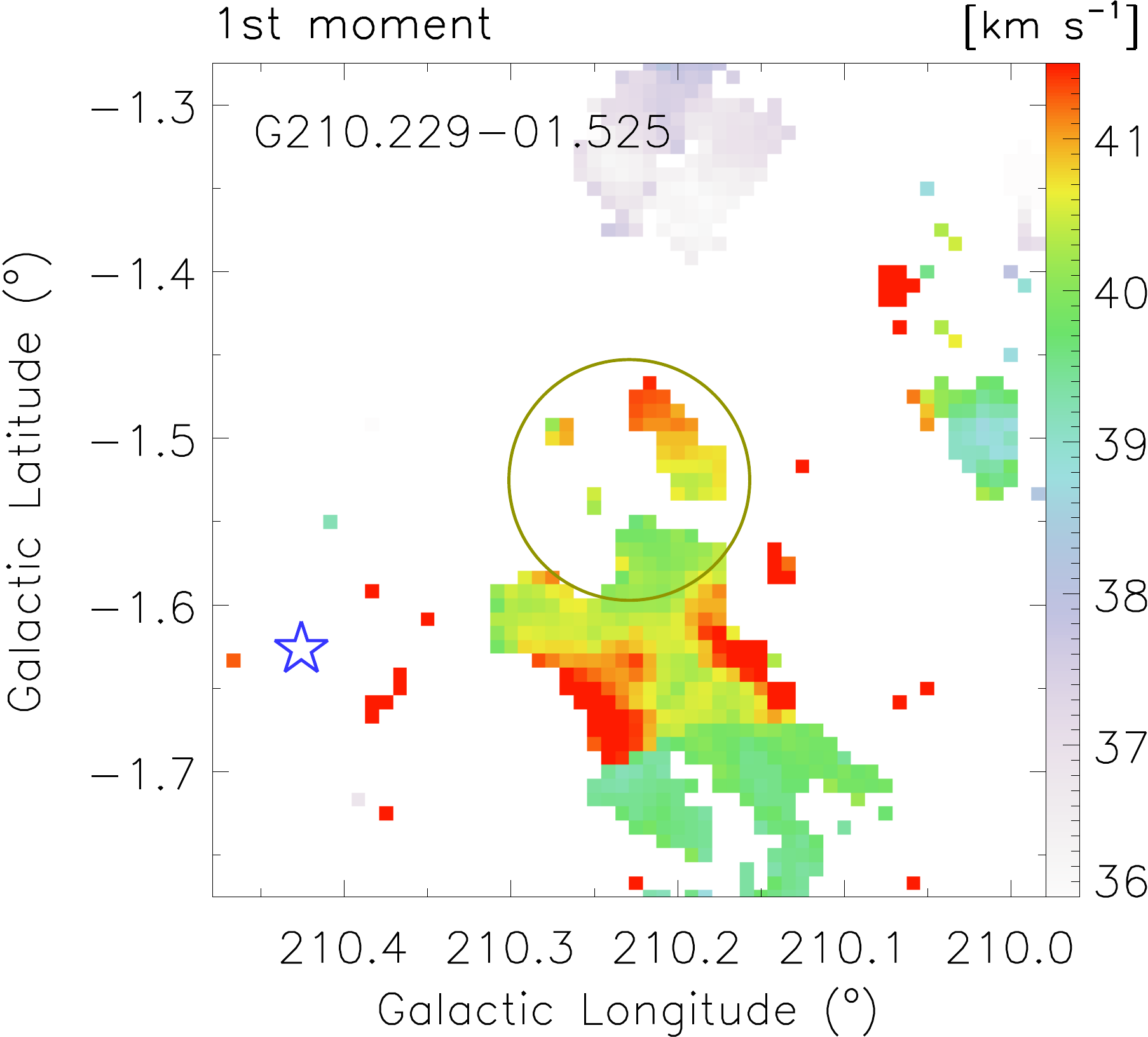}
\includegraphics[width=0.24\textwidth]{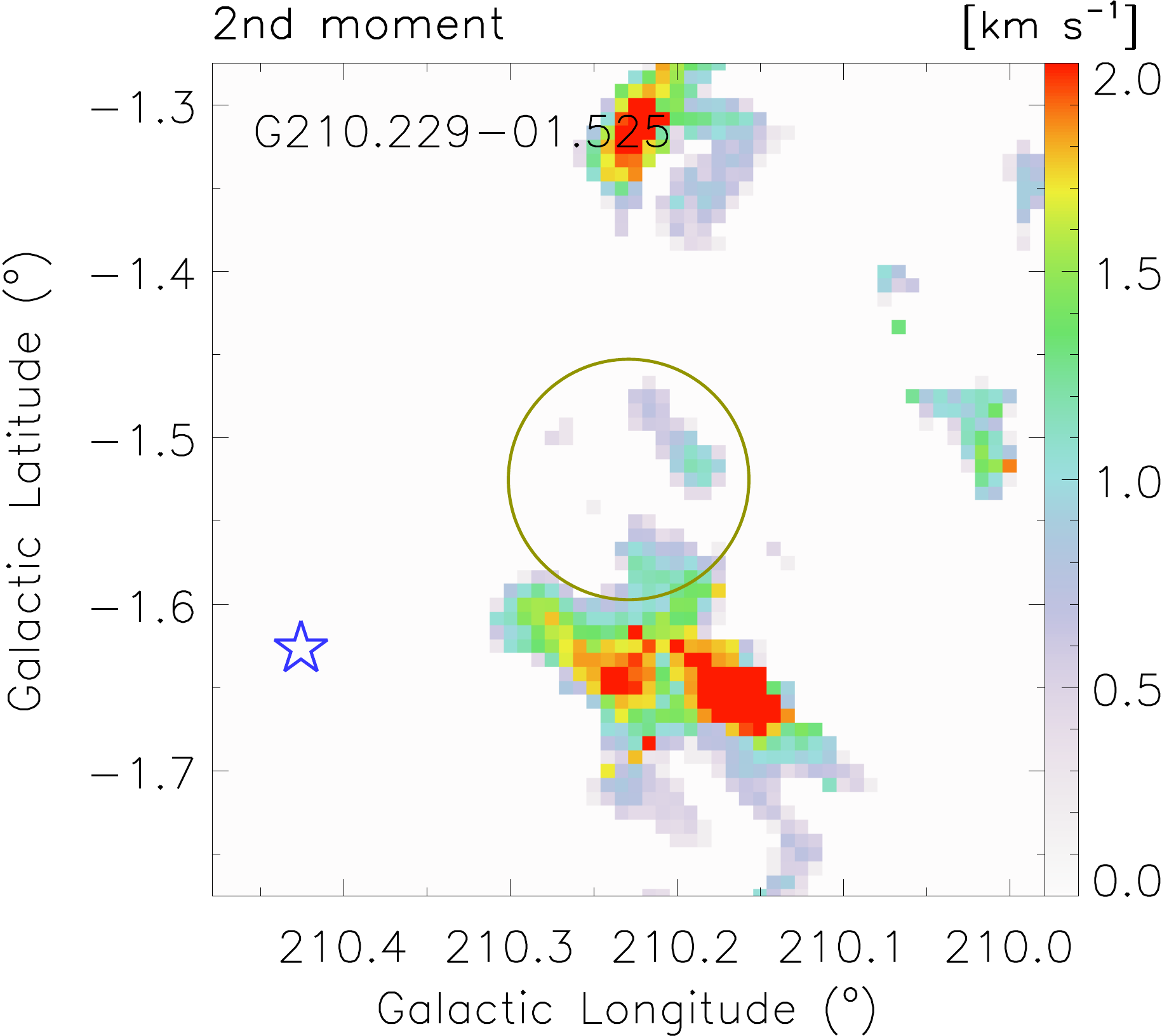}
\includegraphics[width=0.24\textwidth]{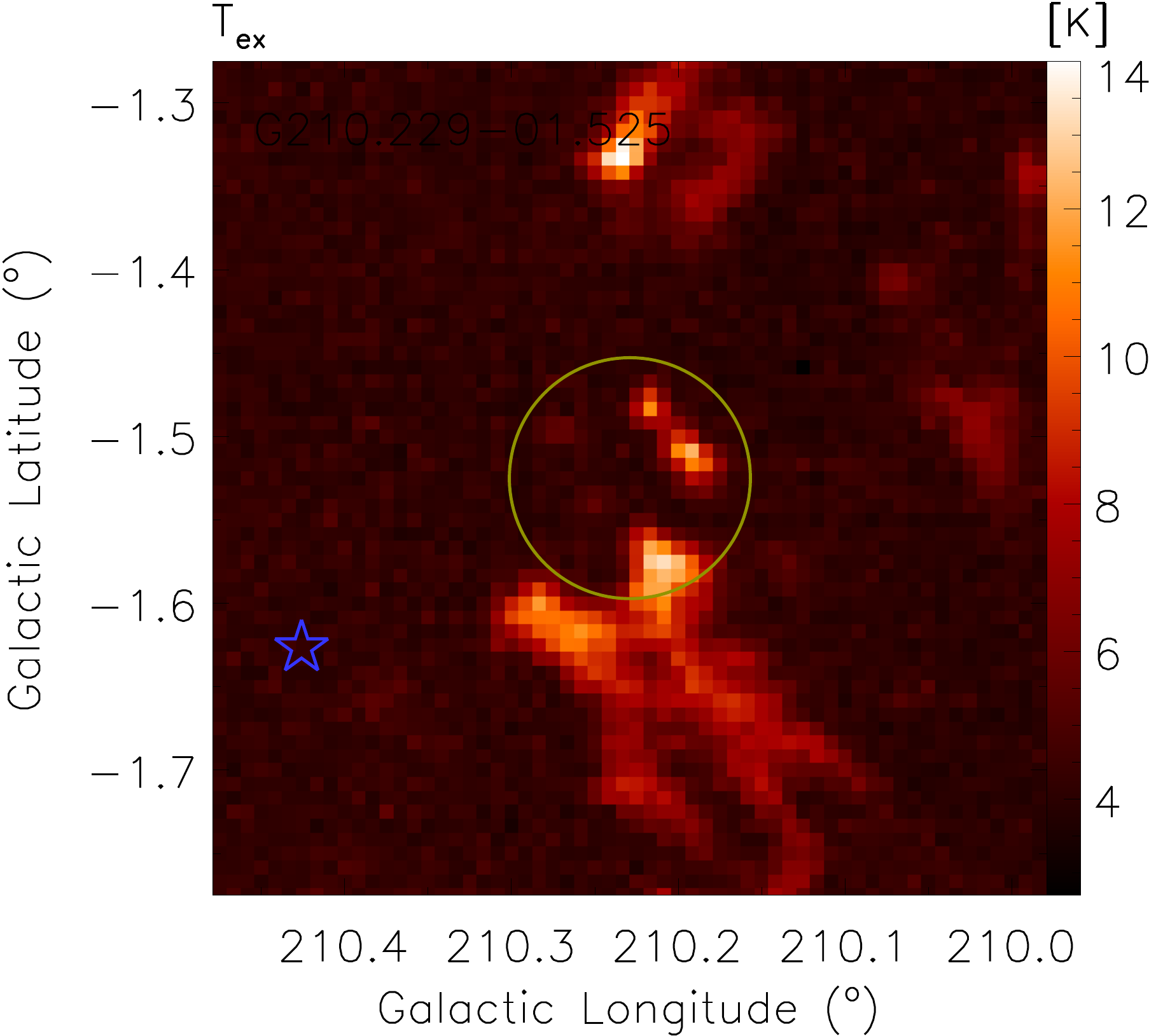}
  \caption{Morphology of G210.229-01.525 in various tracers. Middle: images  of G210.229-01.525\_near with integrated velocity range from 22 to 32 km s$^{-1}$. Bottom: images of G210.229-01.525\_far with integrated velocity range from 35 to 45 km s$^{-1}$. The blue pentagram indicates the O star in this region from the SIMBAD database. All the others are the same as in Figures \ref{fig:G2085-023}.}
  \label{fig:G2102-015}
\end{figure}

\begin{figure}[h]
  \centering
\includegraphics[width=0.21\textwidth]{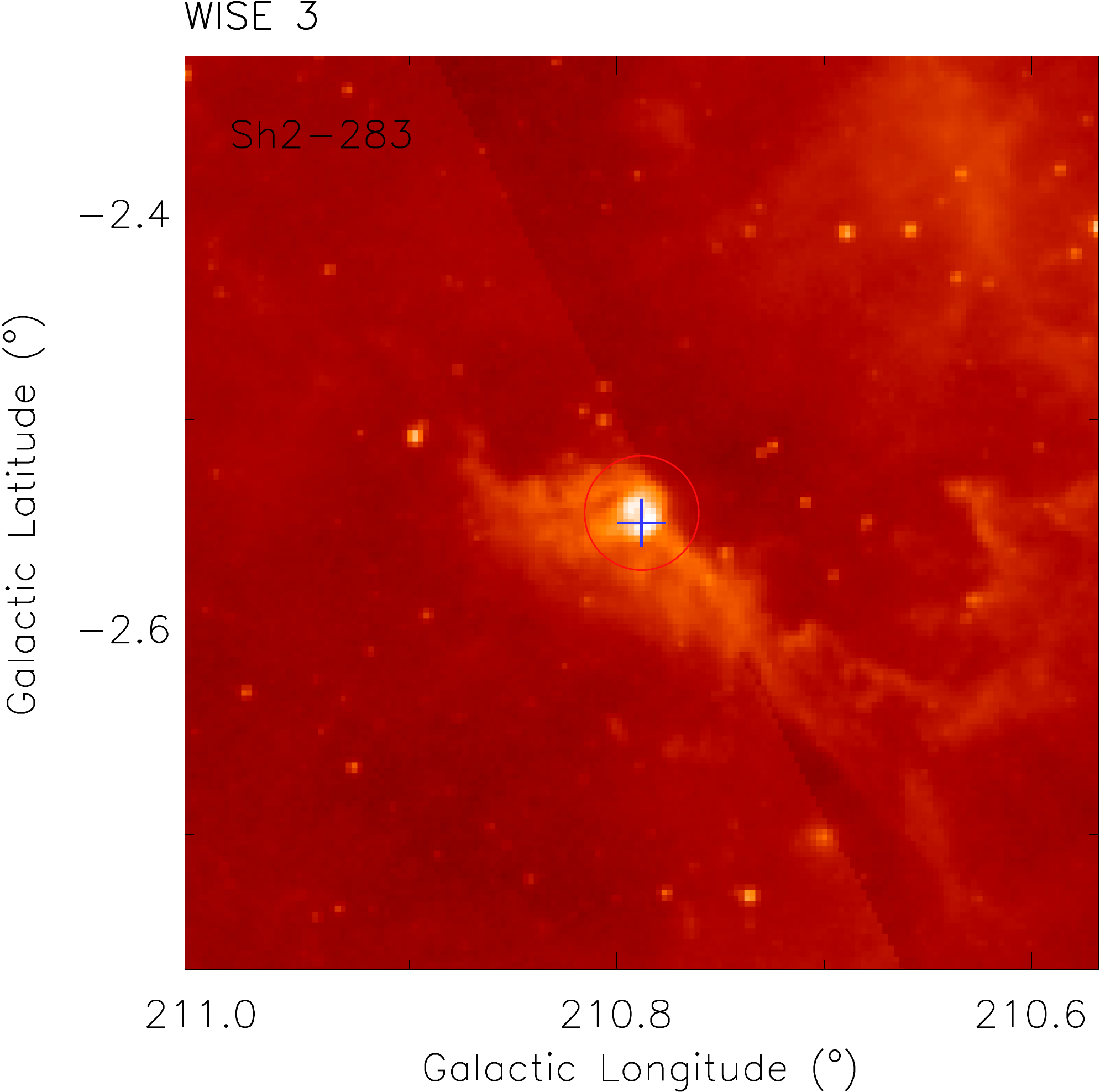}
\includegraphics[width=0.21\textwidth]{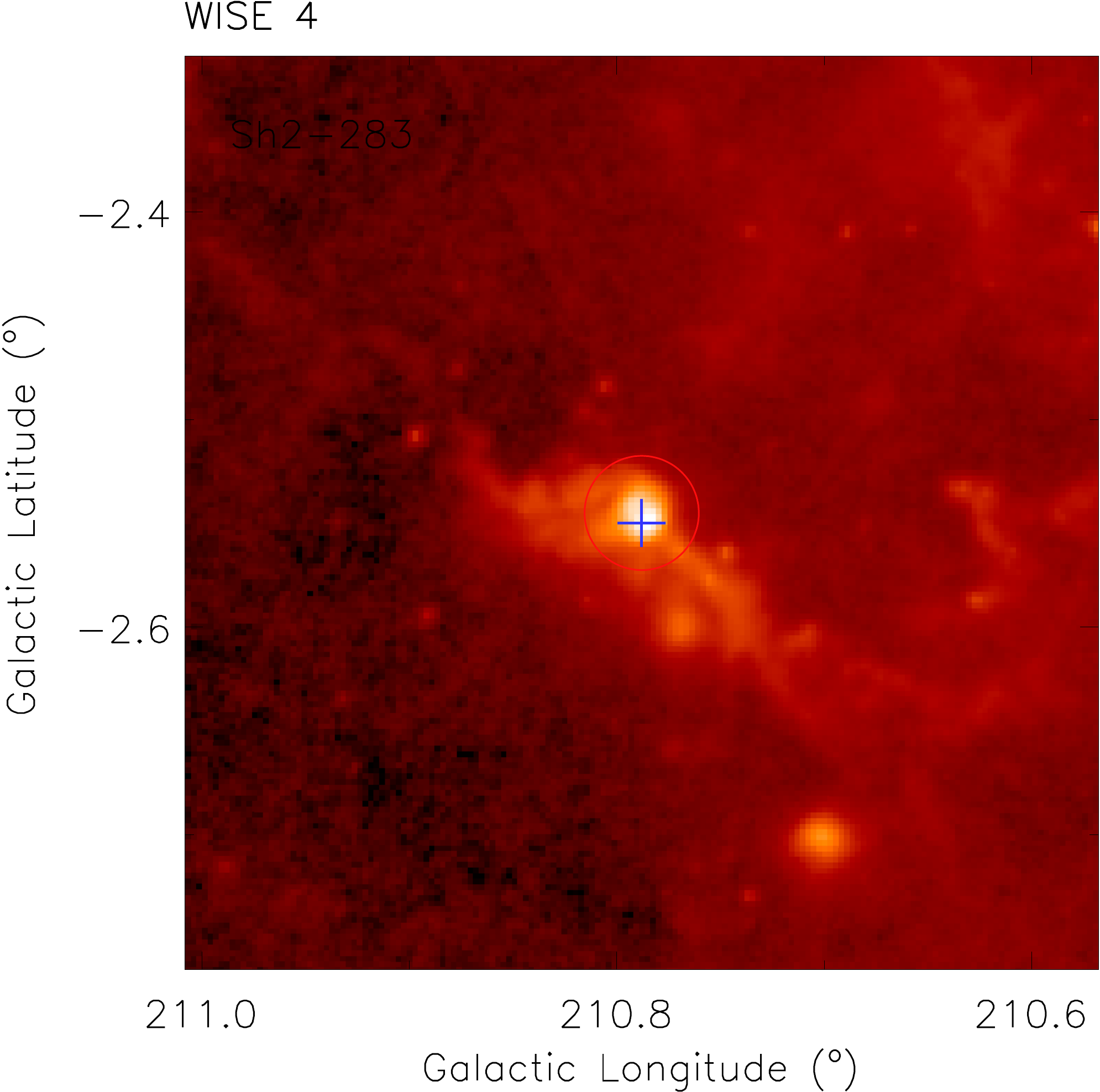}
\includegraphics[width=0.21\textwidth]{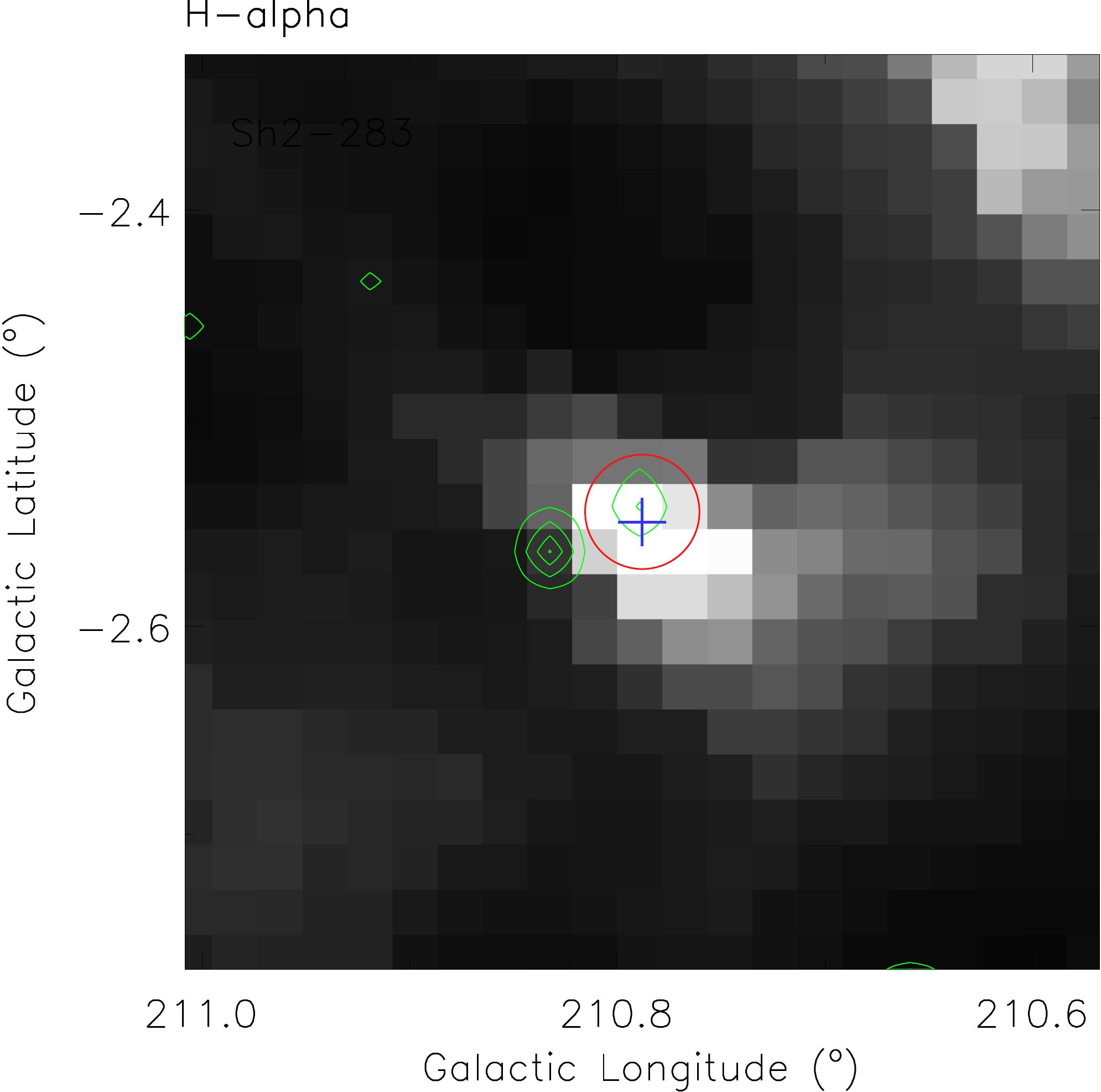}\\
\includegraphics[width=0.24\textwidth]{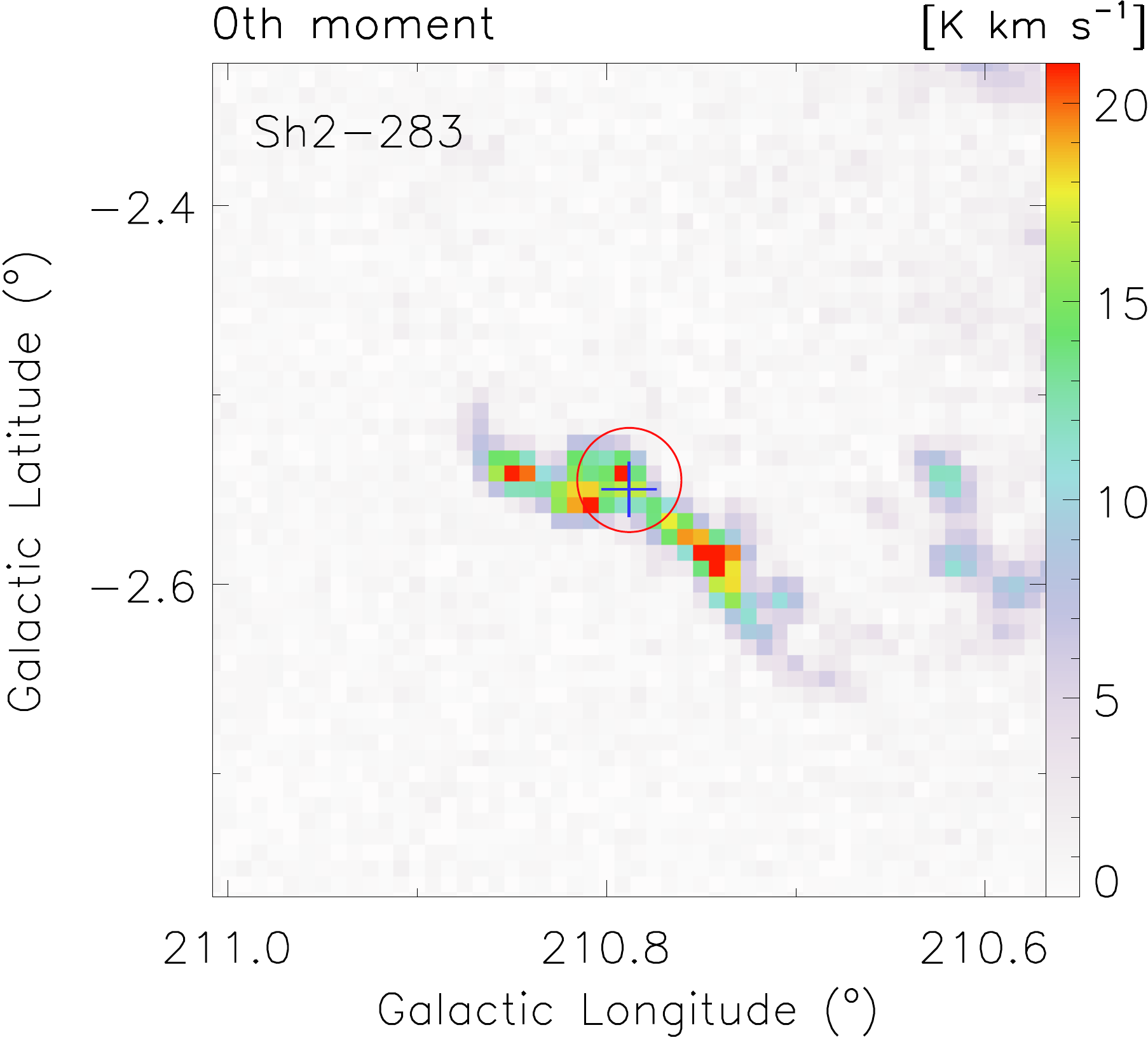}
\includegraphics[width=0.24\textwidth]{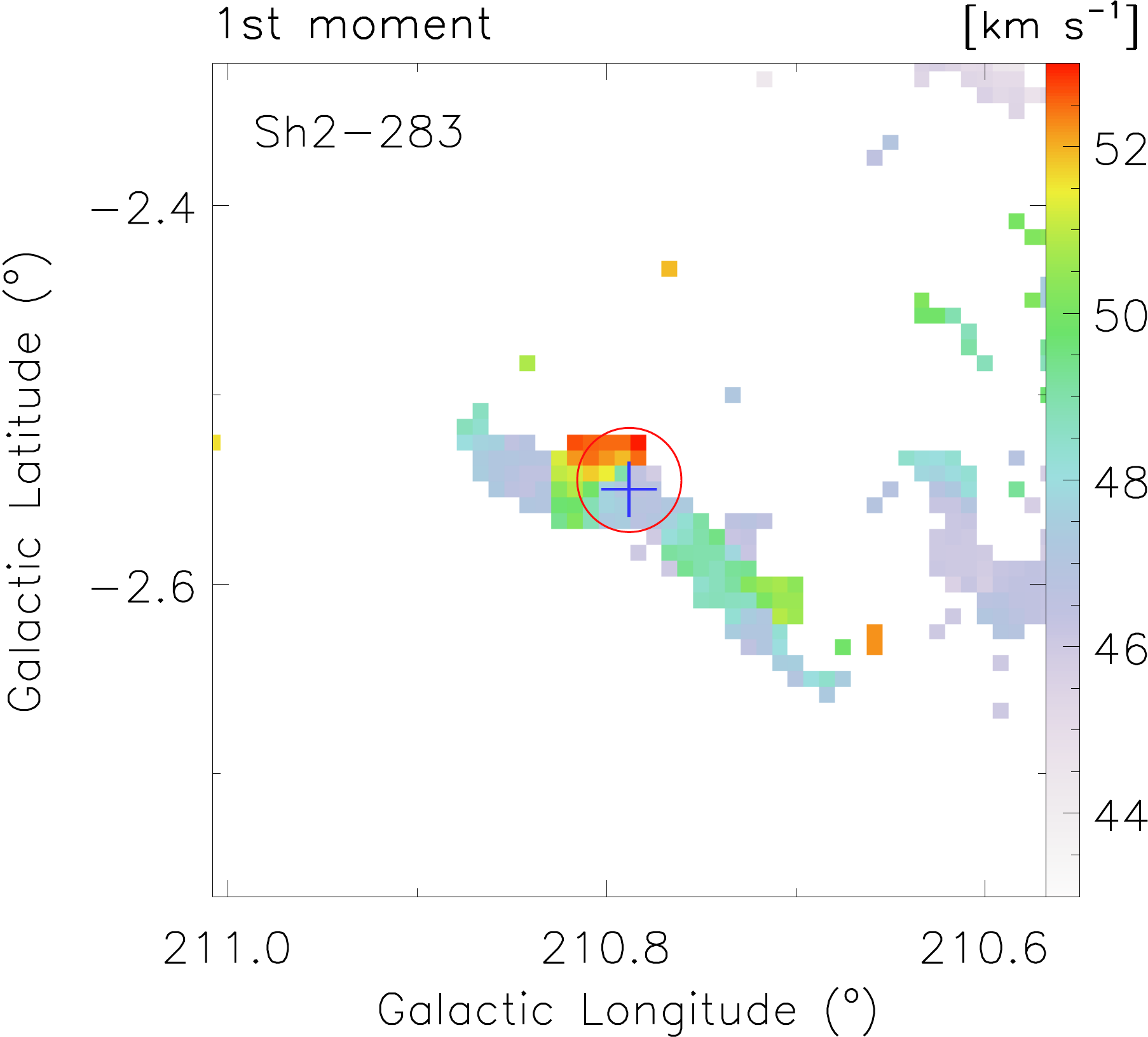}
\includegraphics[width=0.24\textwidth]{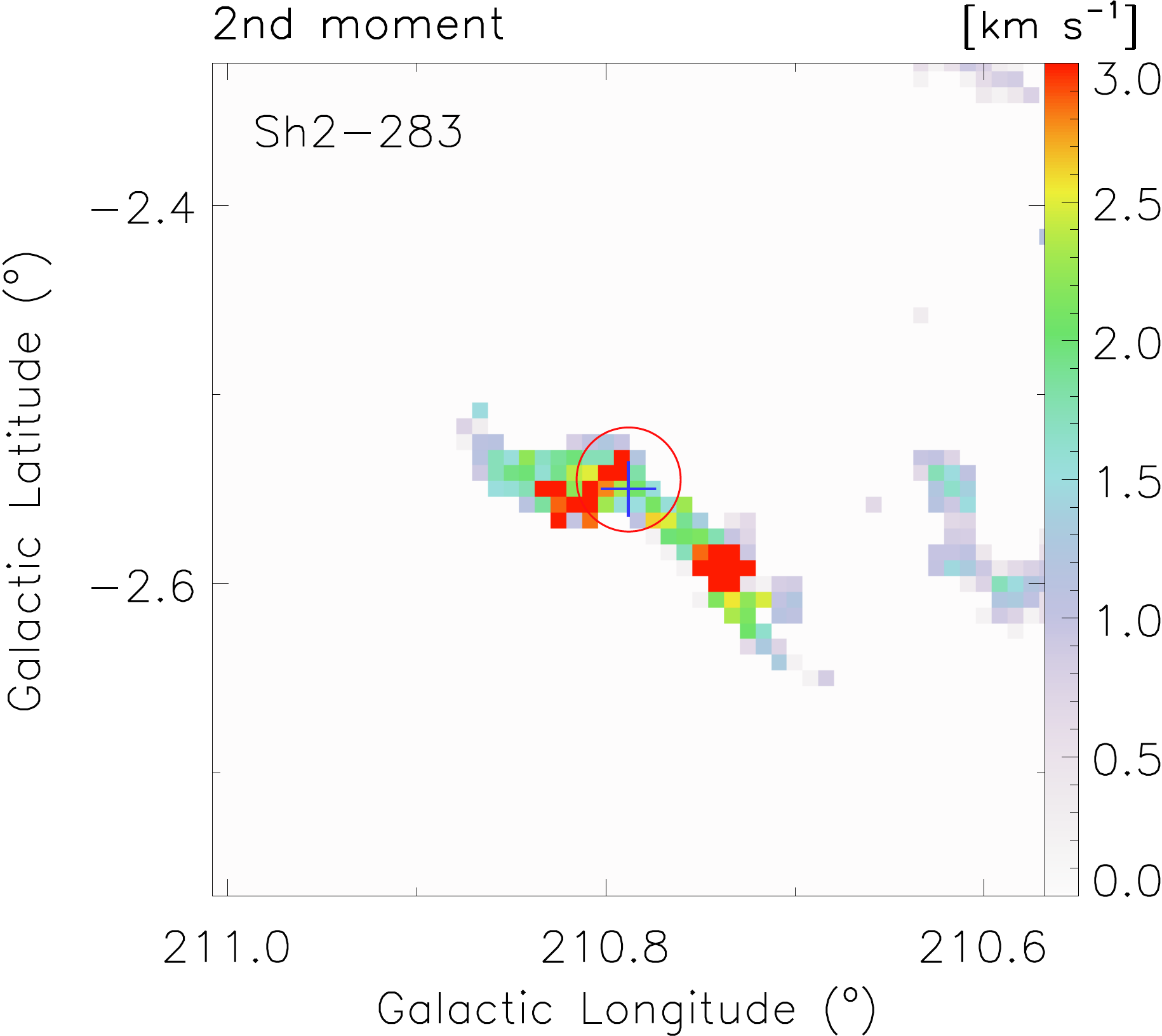}
\includegraphics[width=0.24\textwidth]{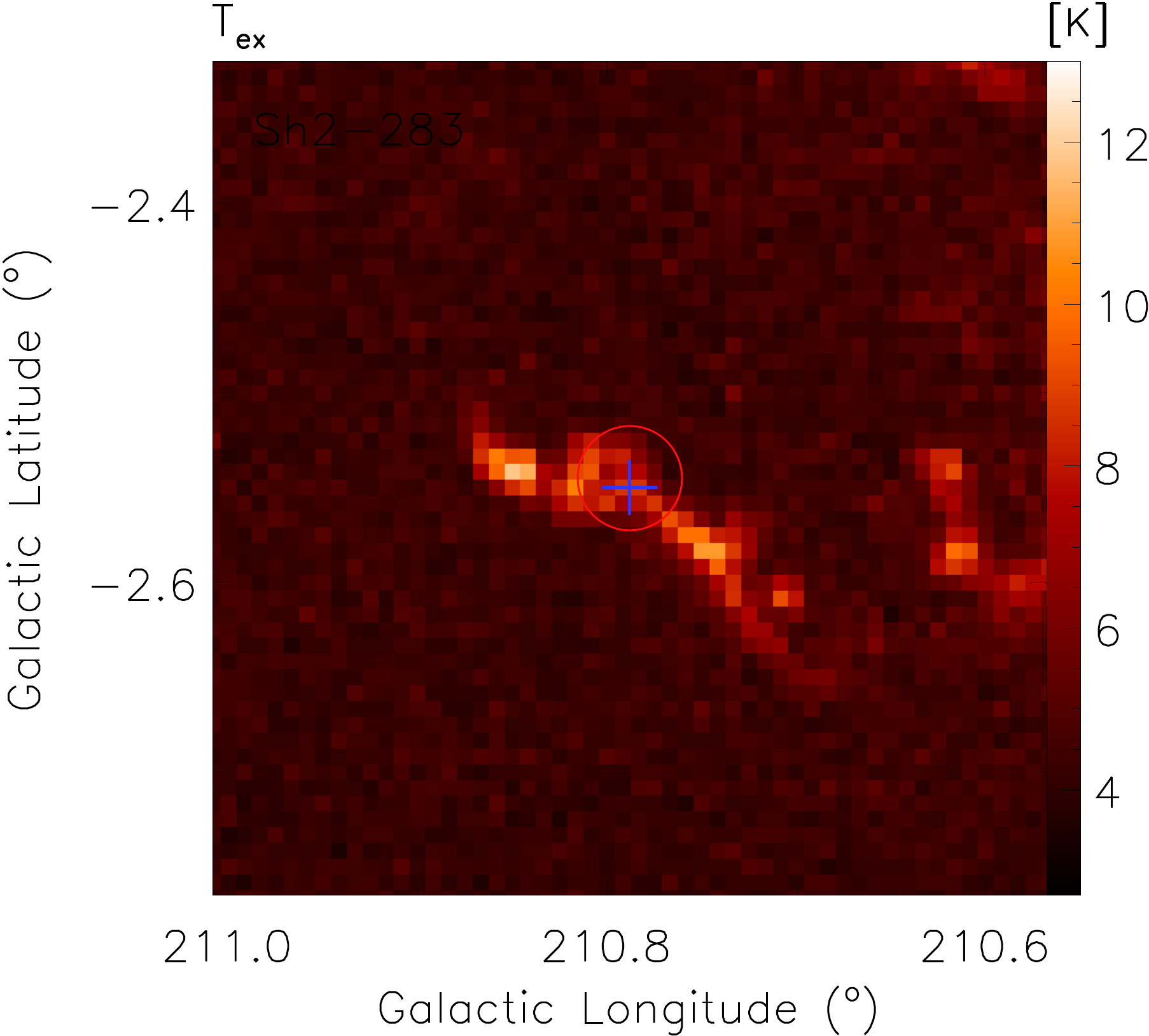}
  \caption{Morphology of Sh2-283 in various tracers. The velocity range for intensity integration is from 42 km s$^{-1}$ to 55 km s$^{-1}$. The green contours indicate the radio continuum emission, with the minimal level and the interval of the contours are 15 and 5 mJy/beam, respectively. The blue cross sign indicates the B0 star in this region from the SIMBAD database.}
  \label{fig:S283}
\end{figure}

\begin{figure}[h]
  \centering
\includegraphics[width=0.22\textwidth]{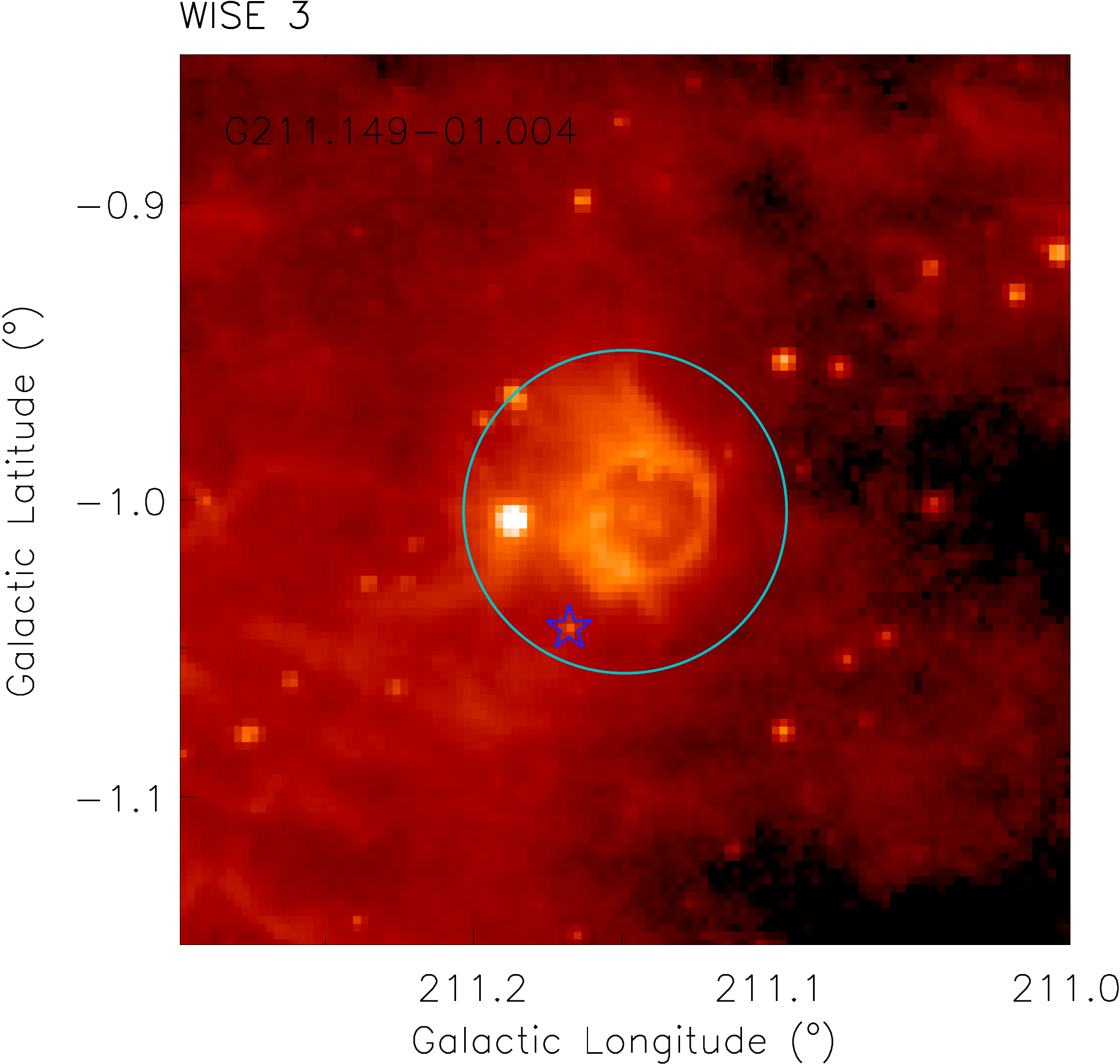}
\includegraphics[width=0.22\textwidth]{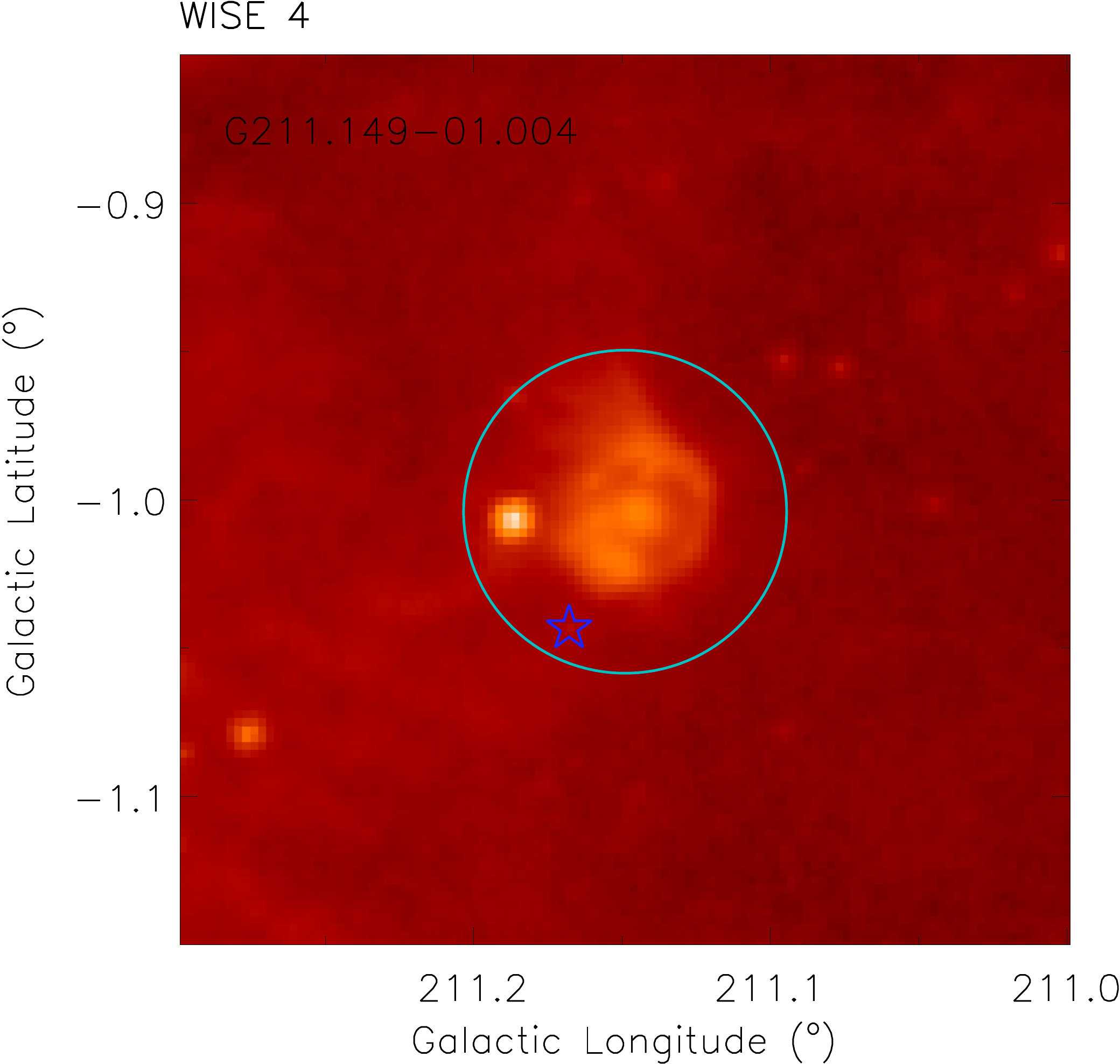}
\includegraphics[width=0.22\textwidth]{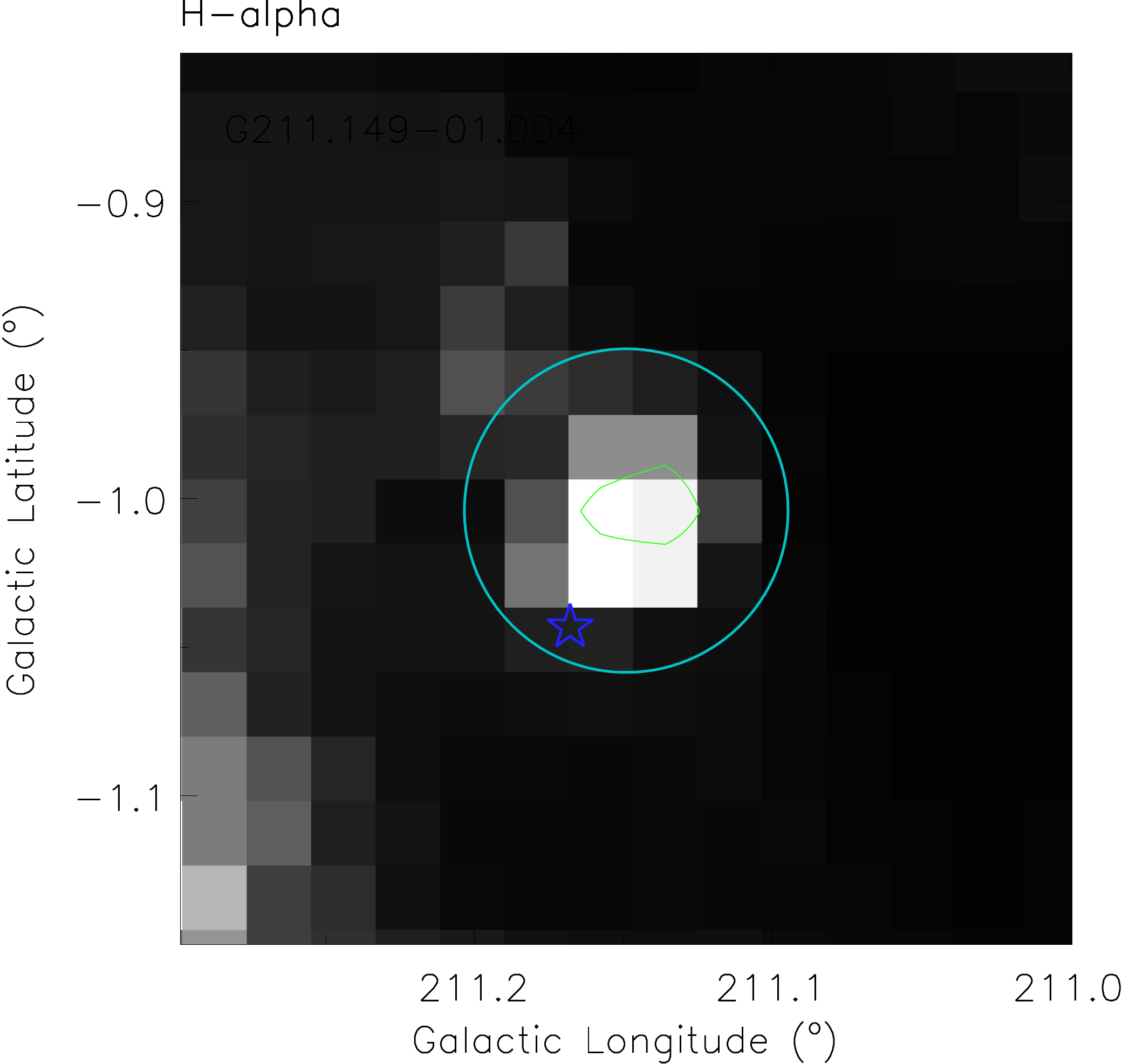}\\
\includegraphics[width=0.24\textwidth]{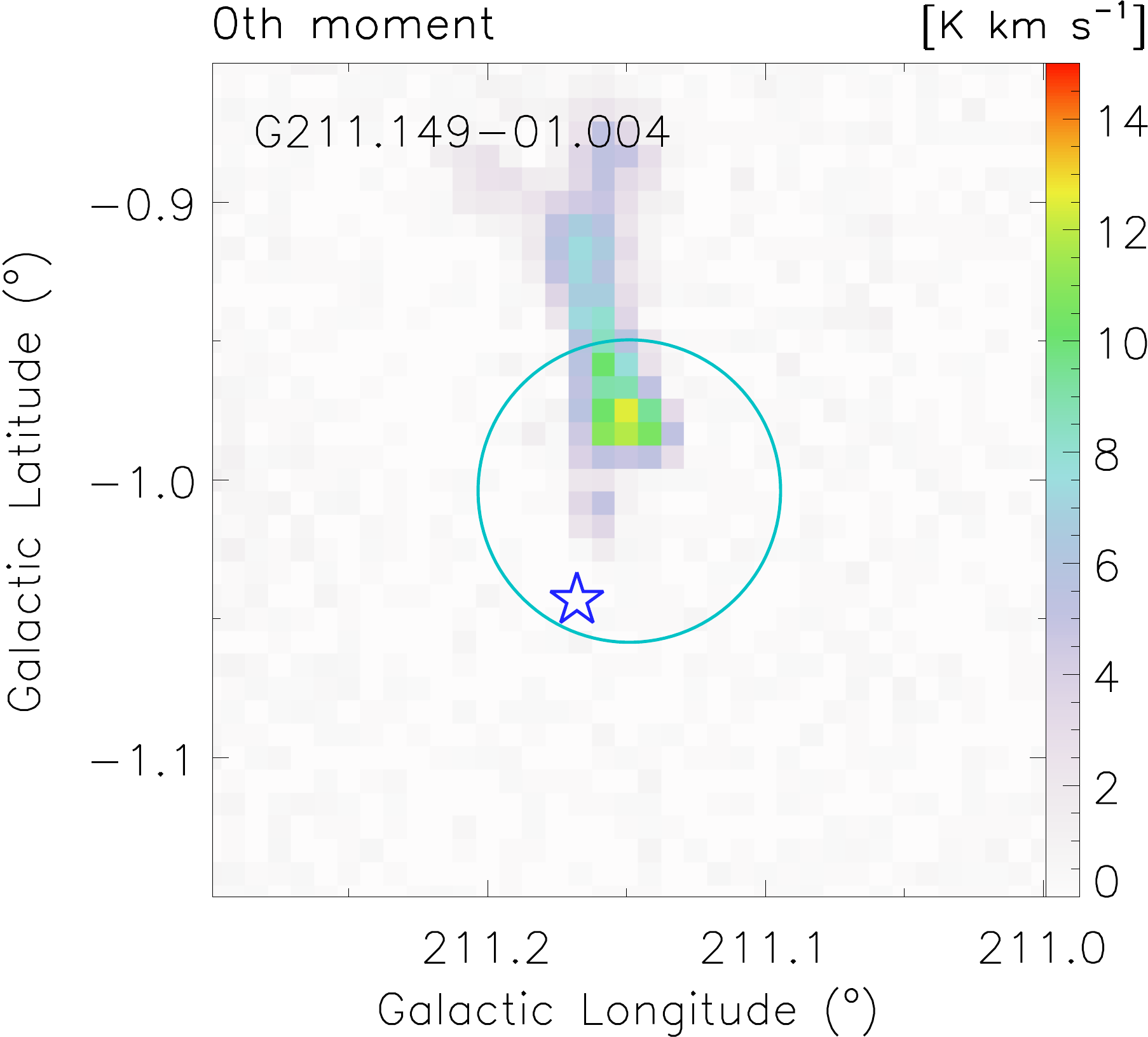}
\includegraphics[width=0.24\textwidth]{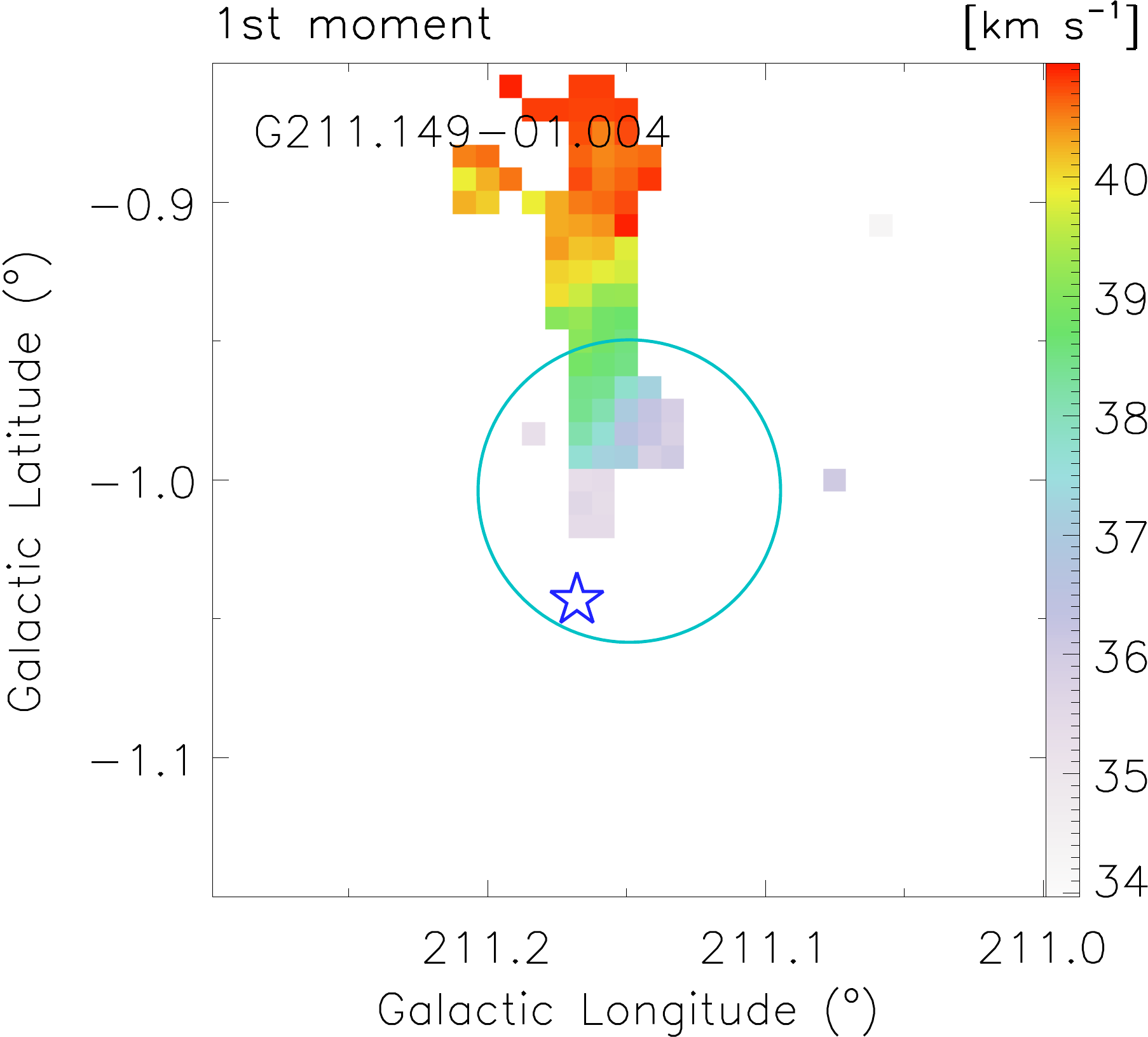}
\includegraphics[width=0.24\textwidth]{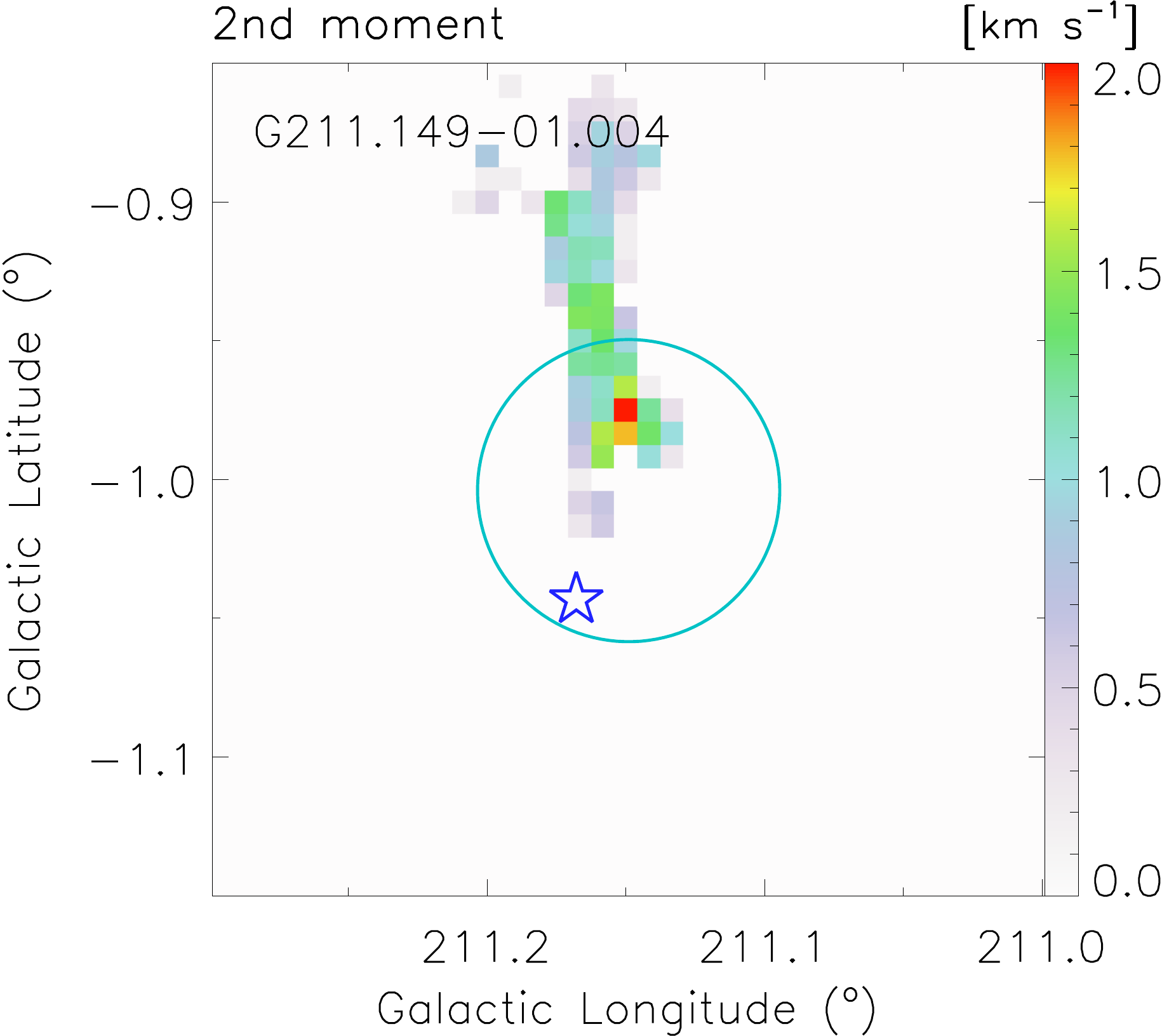}
\includegraphics[width=0.24\textwidth]{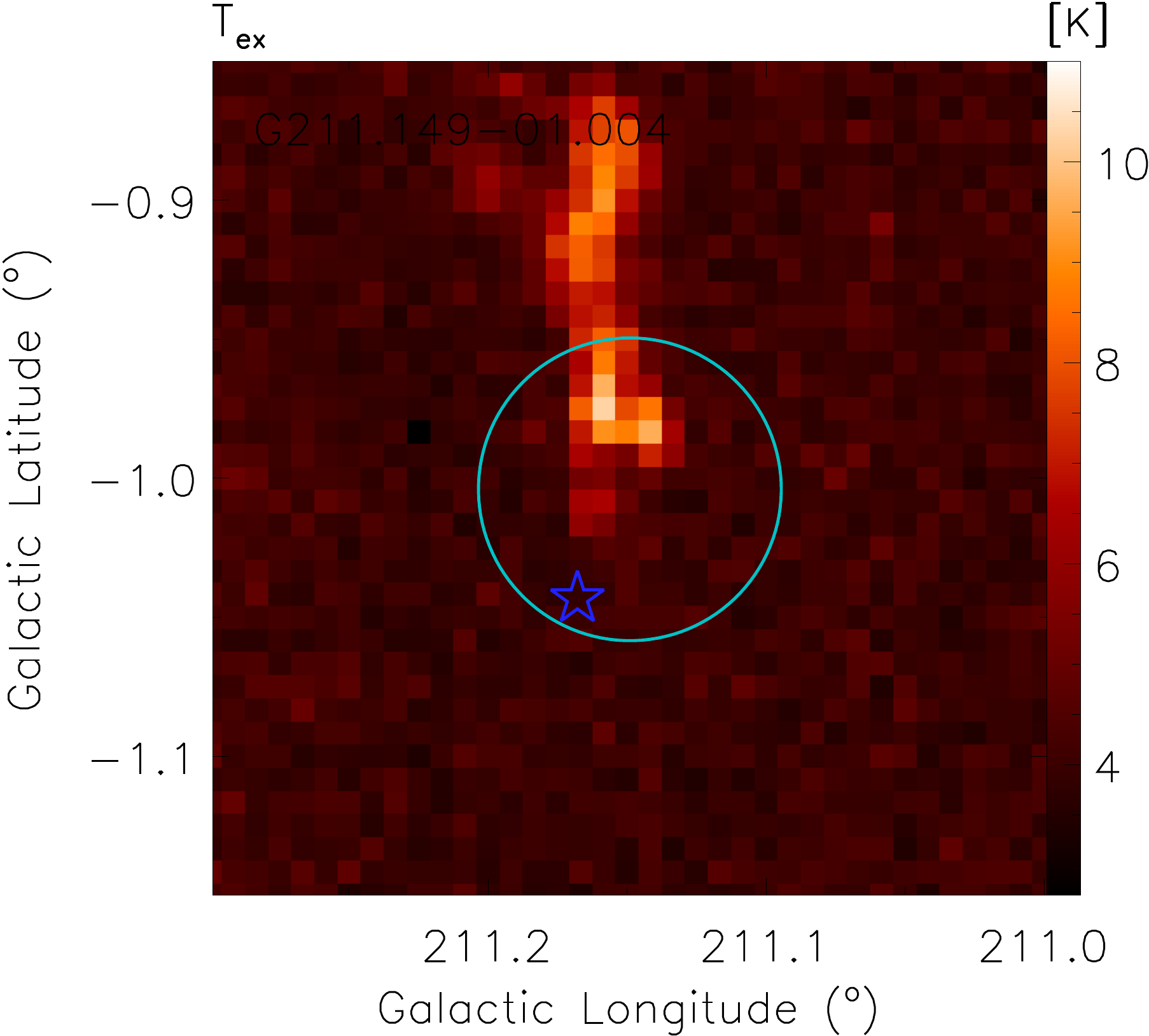}
  \caption{Morphology of G211.149-01.004 in various tracers. The velocity range for intensity integration is from 34 km s$^{-1}$ to 41 km s$^{-1}$. The green contours indicate the radio continuum emission, with the minimal level and the interval of the contours are 15 and 5 mJy/beam, respectively. The blue pentagram indicates the O star in this region from the SIMBAD database. All the others are the same as in Figures \ref{fig:G2085-023}.}
  \label{fig:G2111-010}
\end{figure}

\begin{figure}[h]
  \centering
\includegraphics[width=0.22\textwidth]{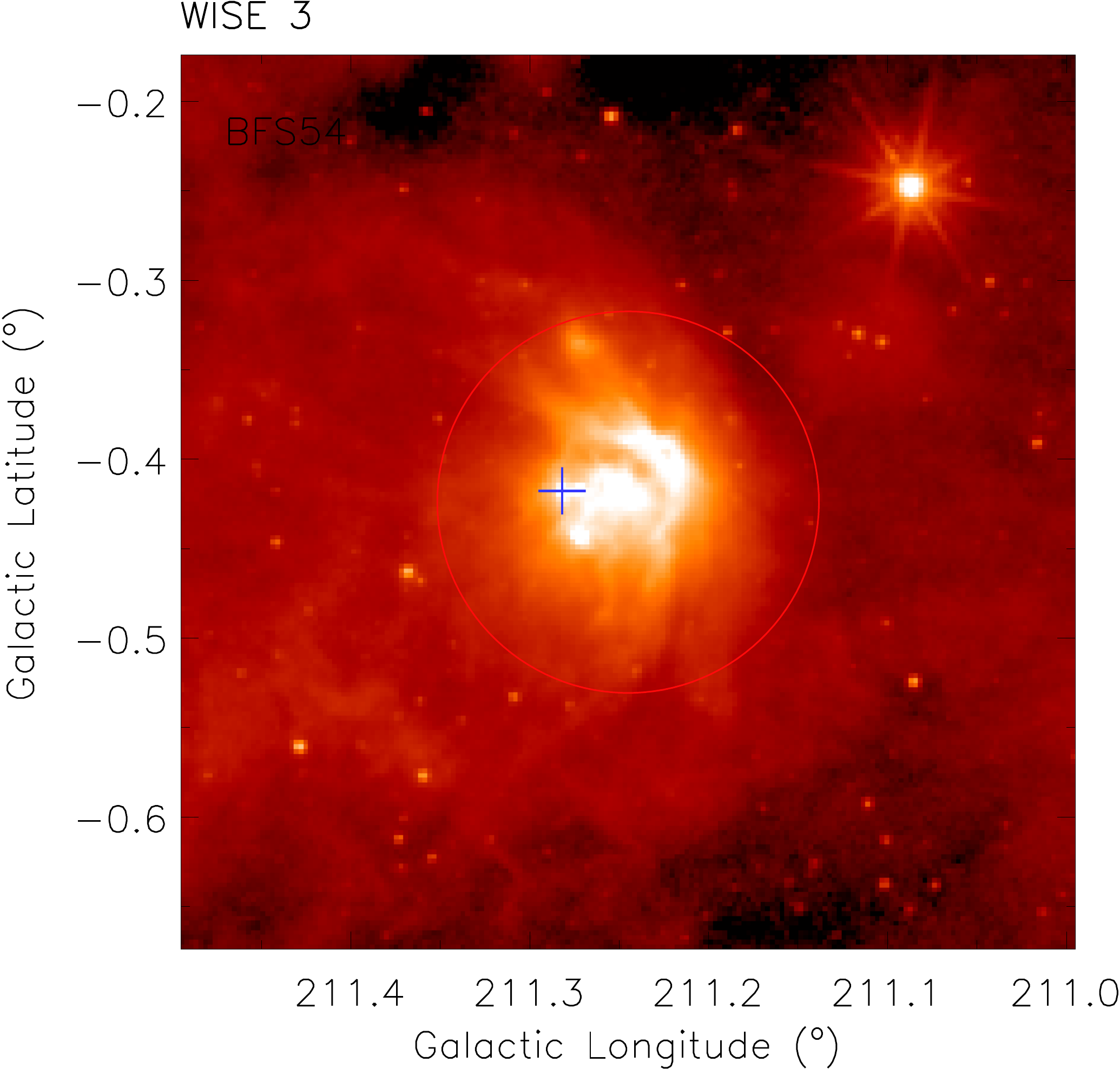}
\includegraphics[width=0.22\textwidth]{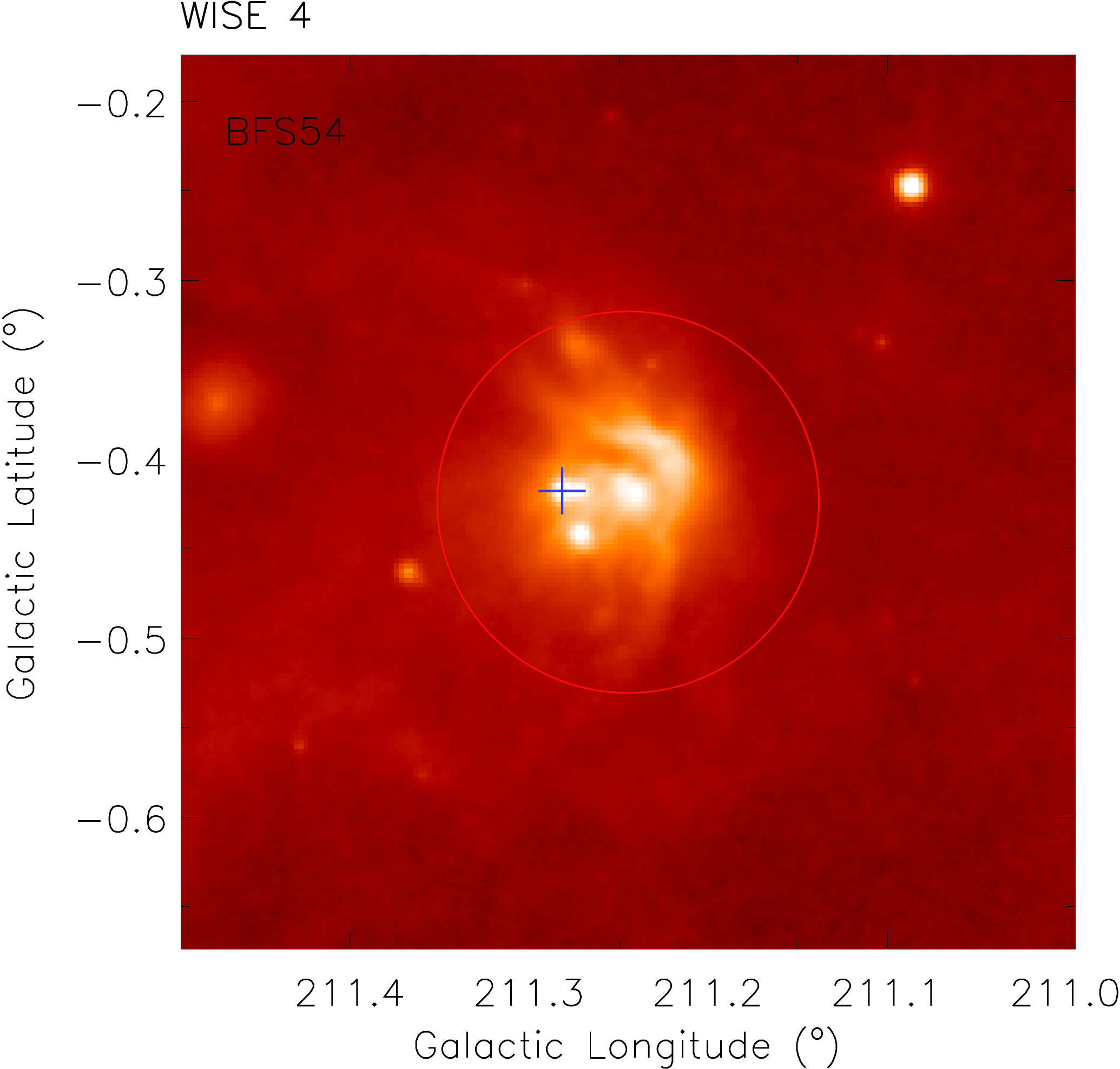}
\includegraphics[width=0.22\textwidth]{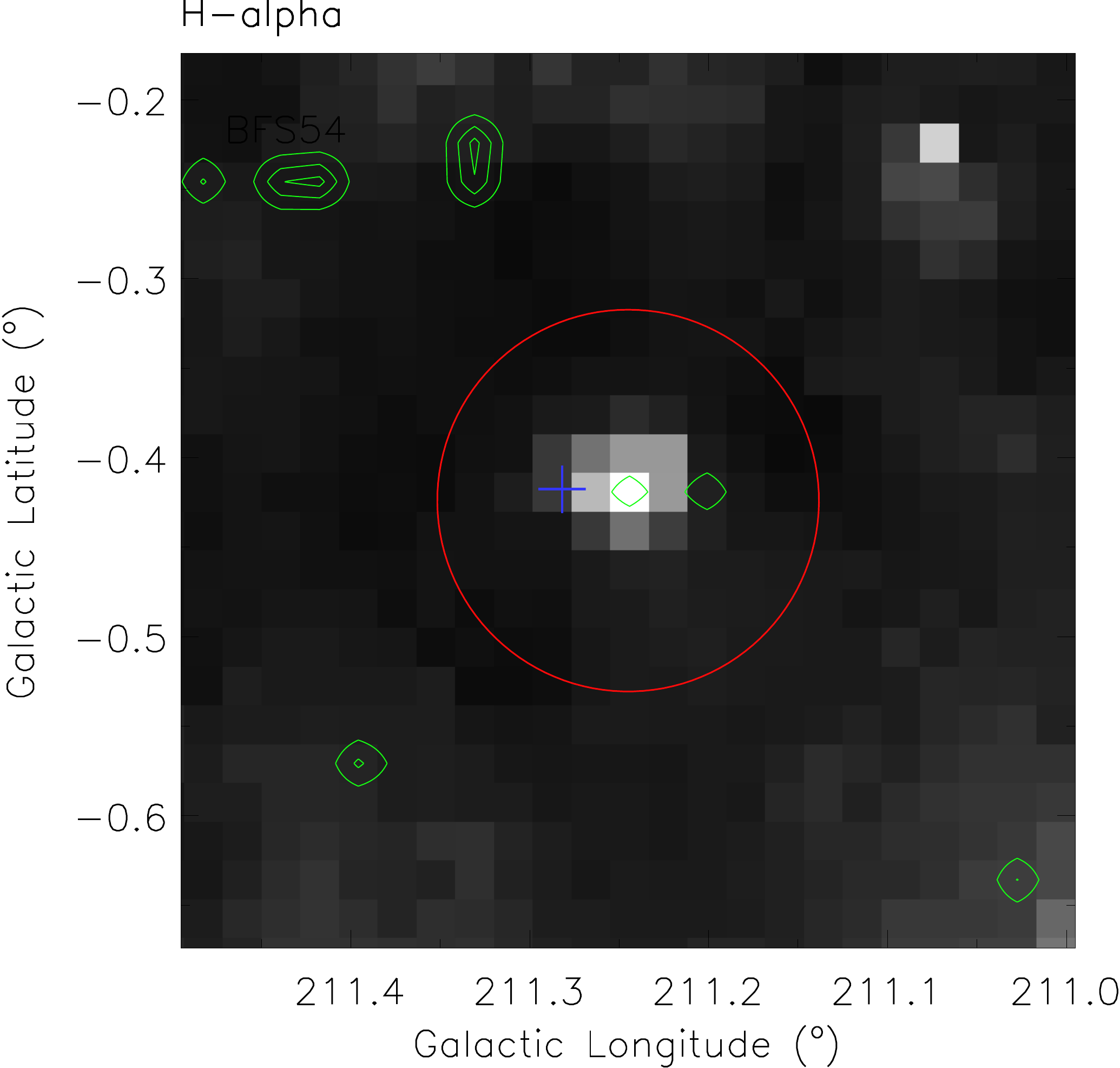}\\
\includegraphics[width=0.24\textwidth]{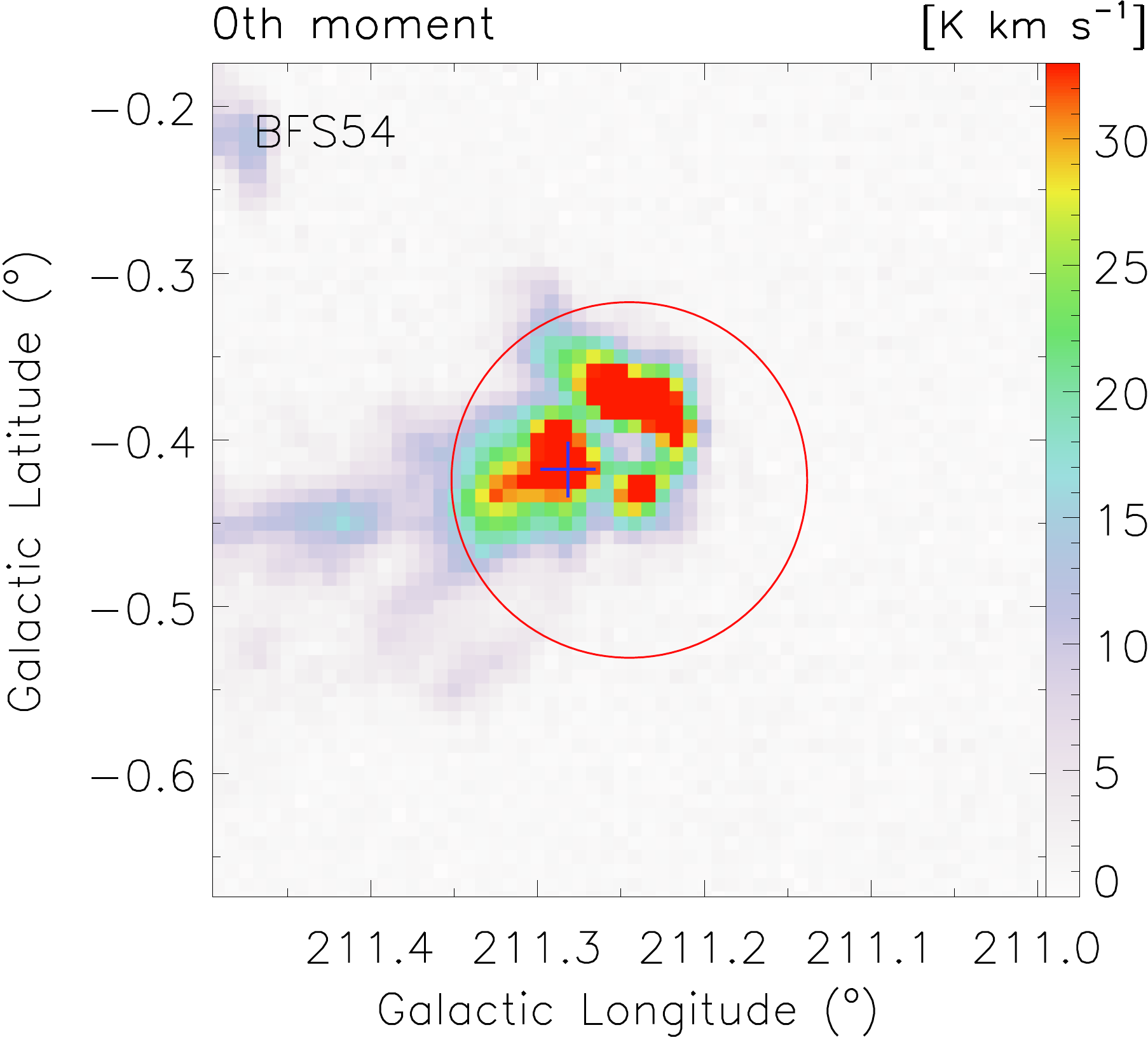}
\includegraphics[width=0.247\textwidth]{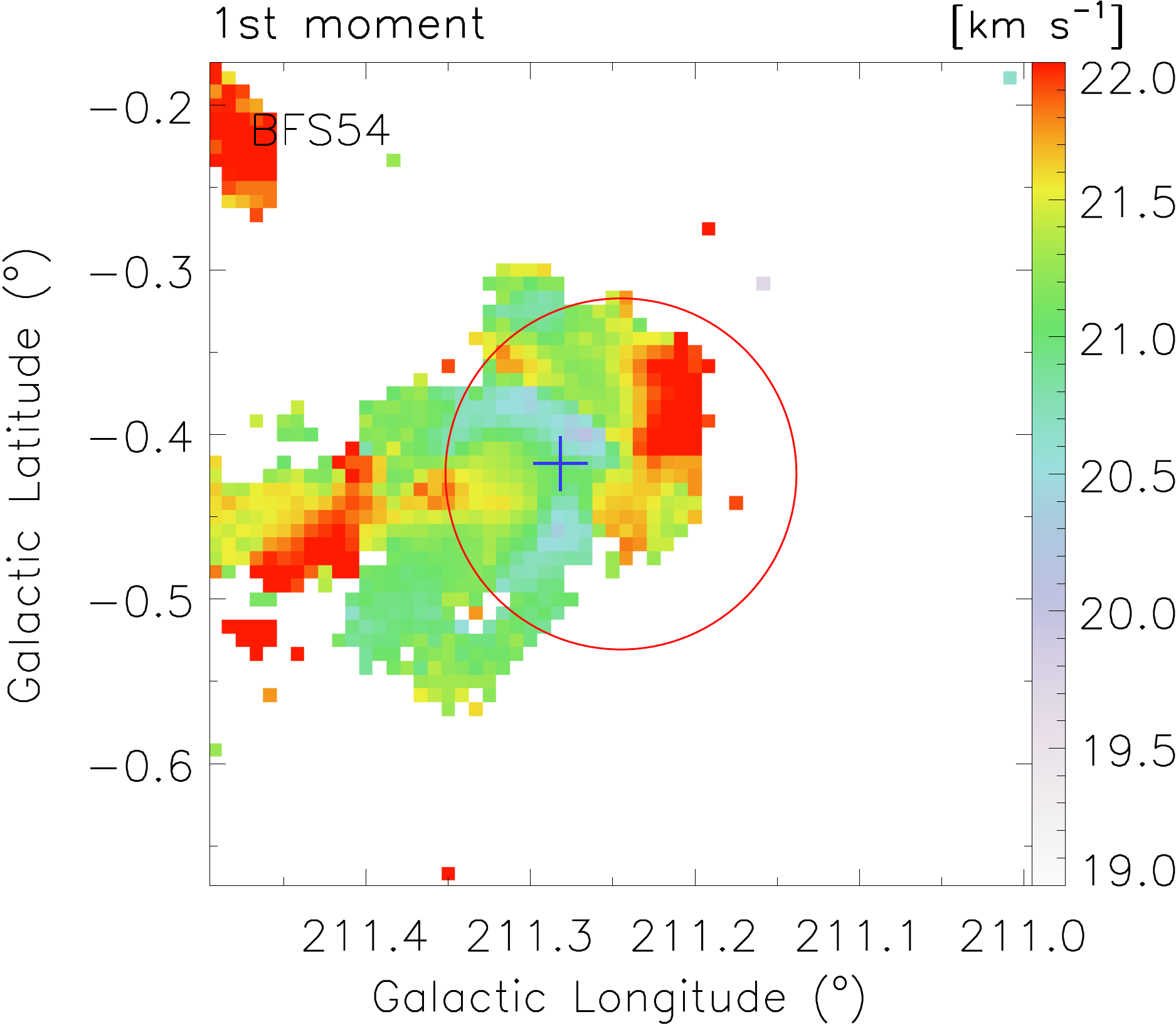}
\includegraphics[width=0.24\textwidth]{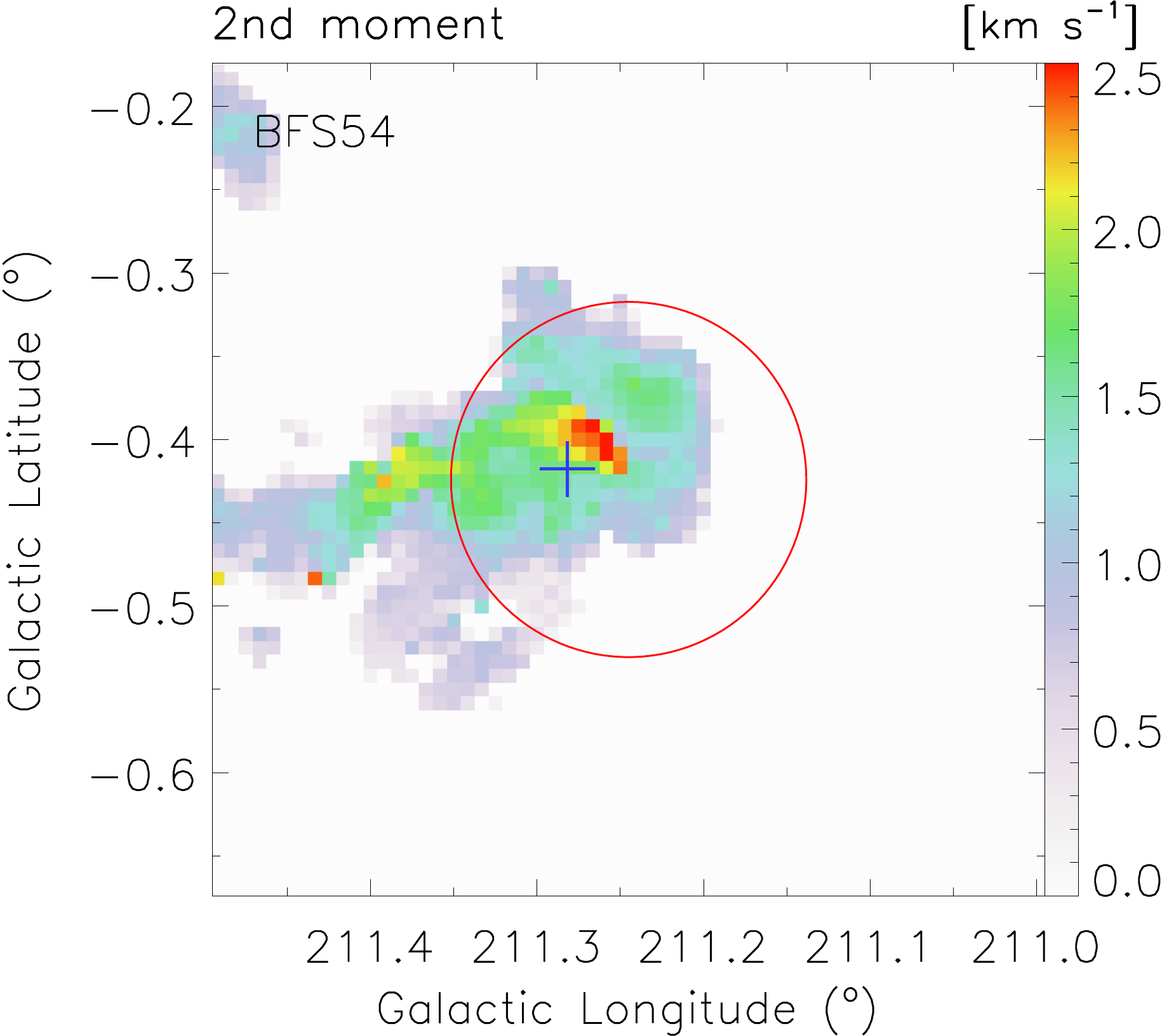}
\includegraphics[width=0.24\textwidth]{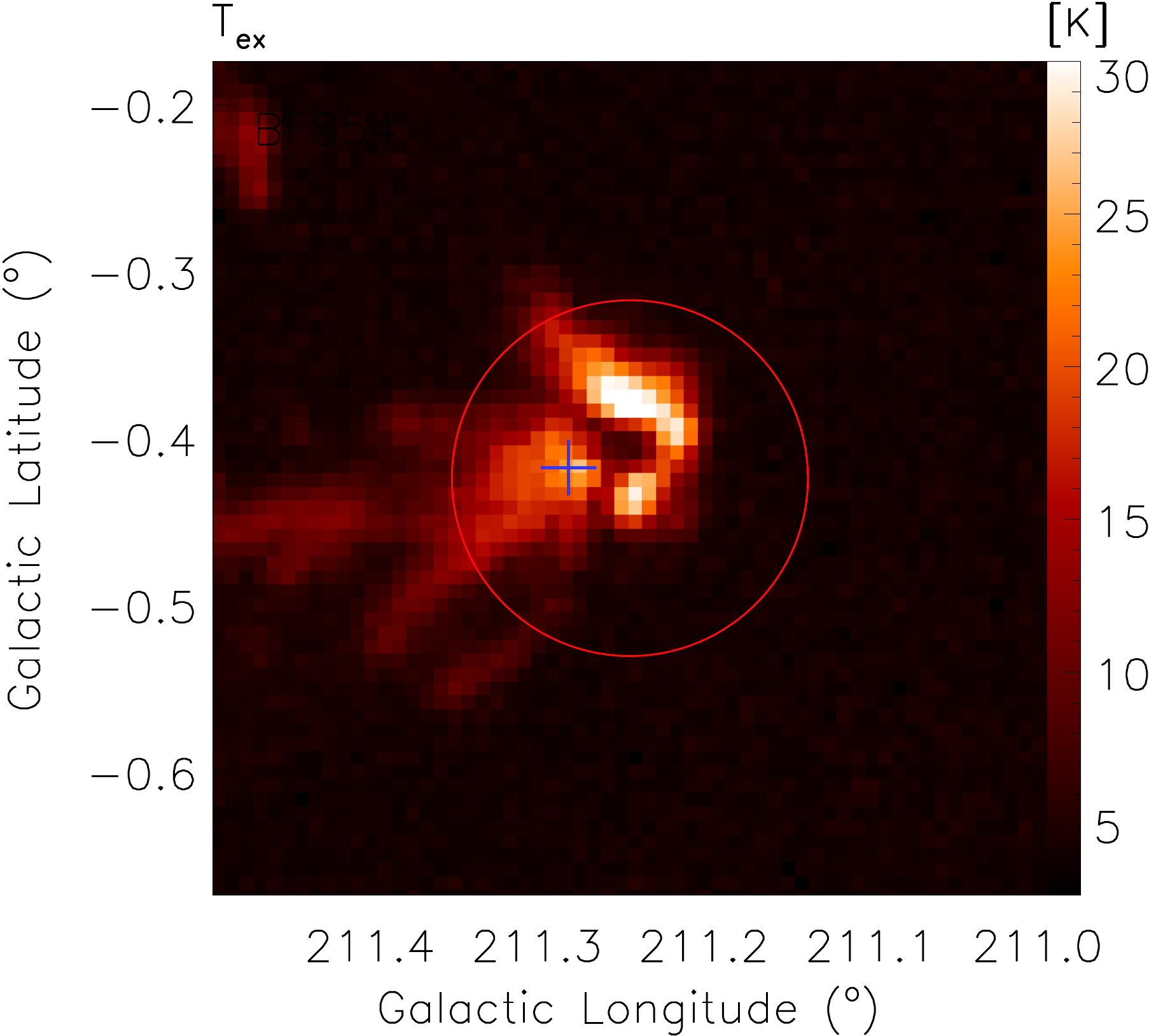}
  \caption{Morphology of BFS54 in various tracers. The velocity range for intensity integration is from 18 km s$^{-1}$ to 24 km s$^{-1}$. The green contours indicate the radio continuum emission, with the minimal level and the interval of the contours are 15 and 5 mJy/beam, respectively. The blue cross indicates the B0 star in this region from the SIMBAD database. All the others are the same as in Figures \ref{fig:G2085-023}.}
  \label{fig:BFS54}
\end{figure}

\begin{figure}[h]
  \centering
\includegraphics[width=0.31\textwidth]{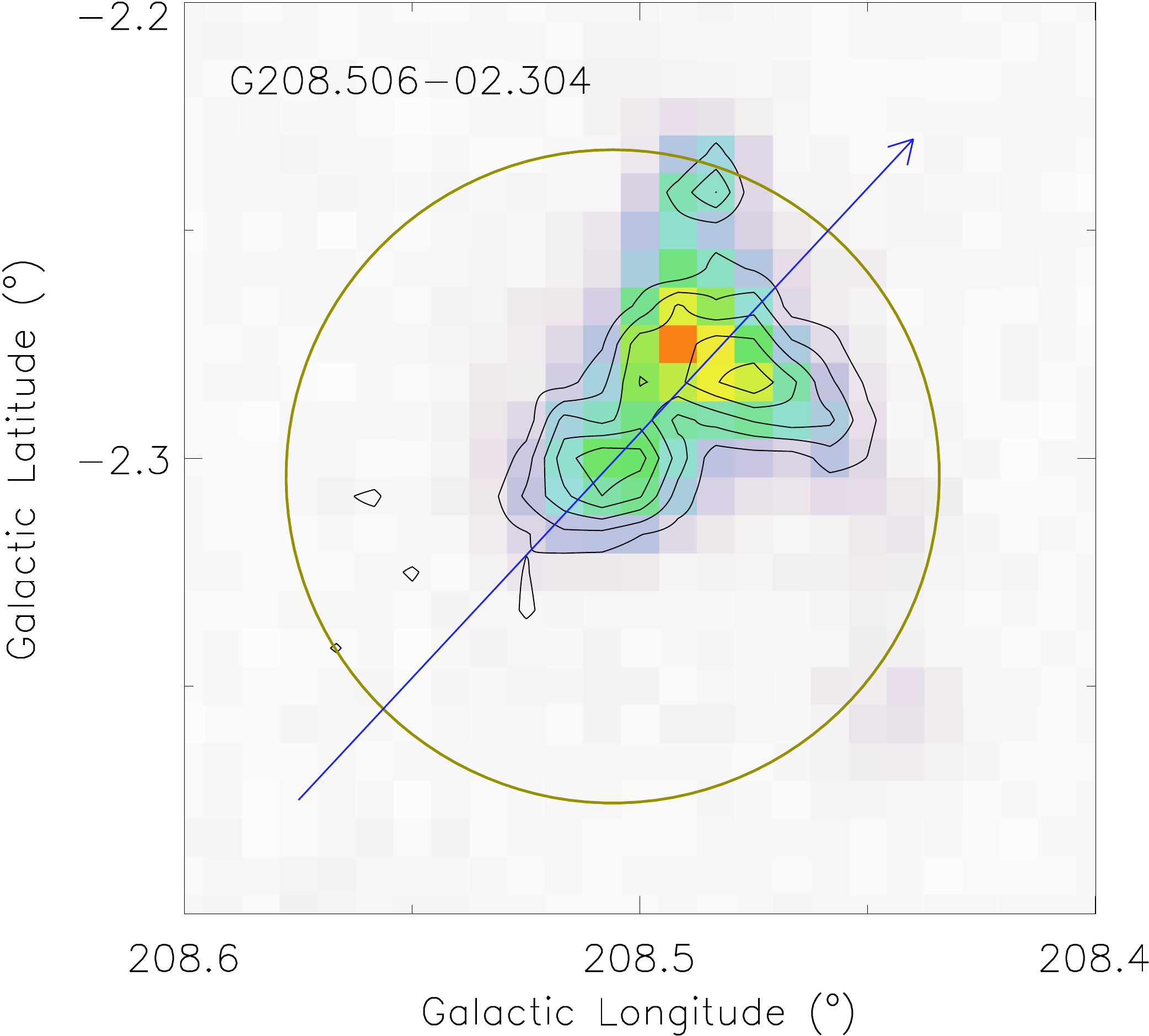}
\includegraphics[width=0.3\textwidth]{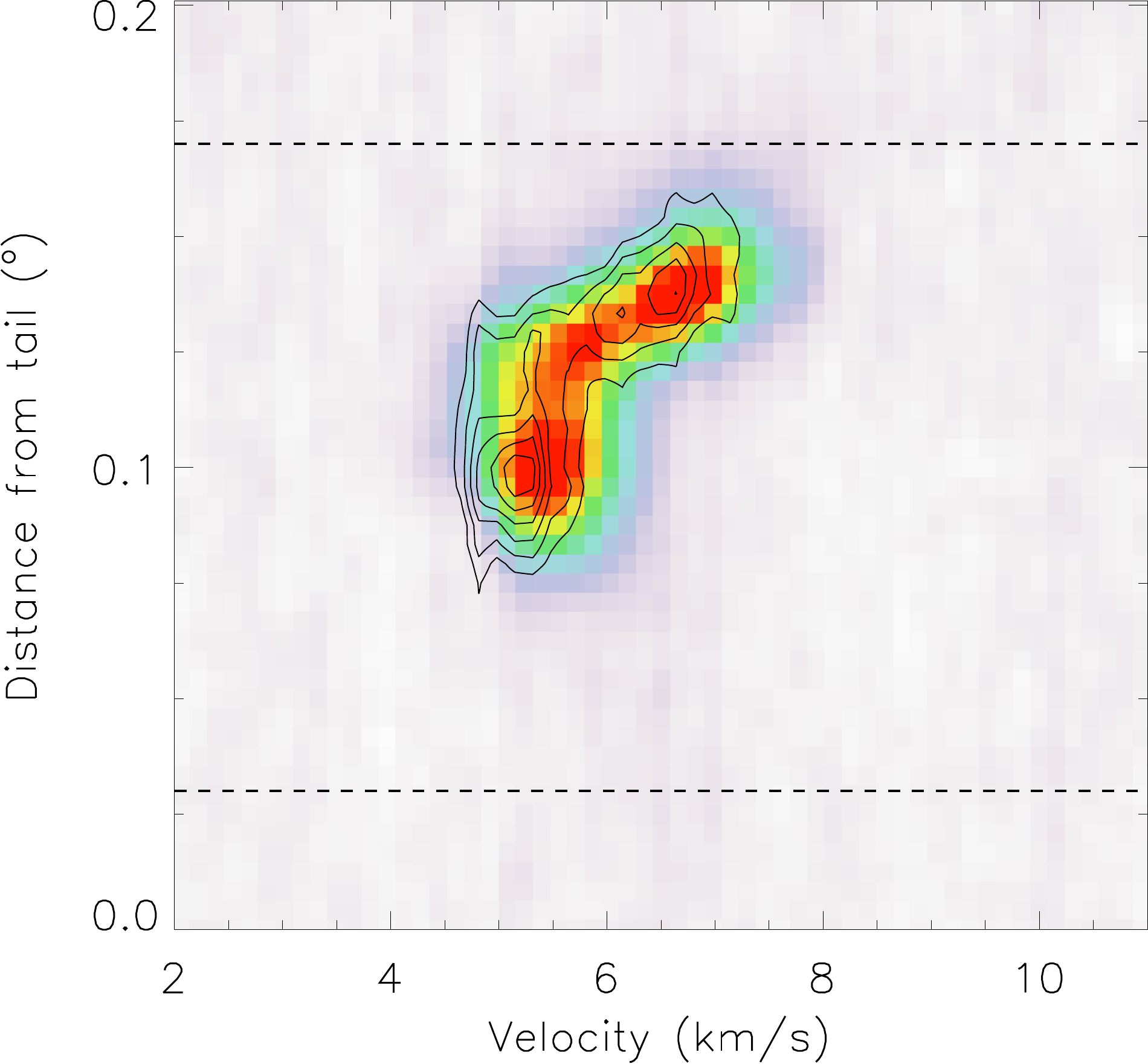}\\
  \caption{Left: integrated intensity map of $^{12}$CO of G208.506-02.304. Right: position-velocity map of $^{12}$CO emission along the arrow marked in the left panel. The overlaid contours are $^{13}$CO emission with the minimal level and the interval being 0.4 and 0.1 of peak, respectively. The dash horizontal  lines indicate the position range of the \ion{H}{2} region.}
  \label{fig:Kinematics_G2085-023}
\end{figure}

\begin{figure}[h]
  \centering
\includegraphics[width=0.3\textwidth]{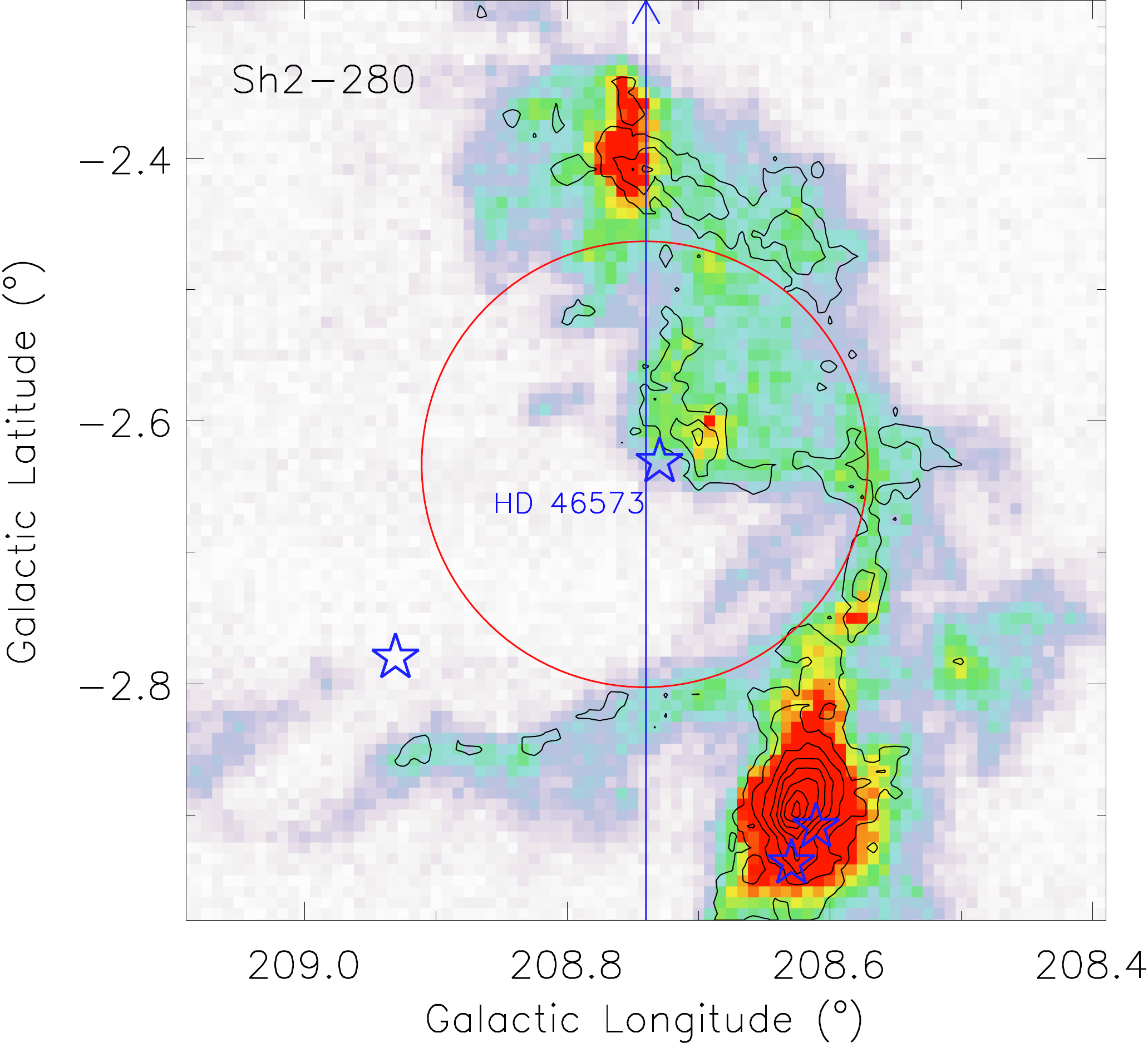}
\includegraphics[width=0.295\textwidth]{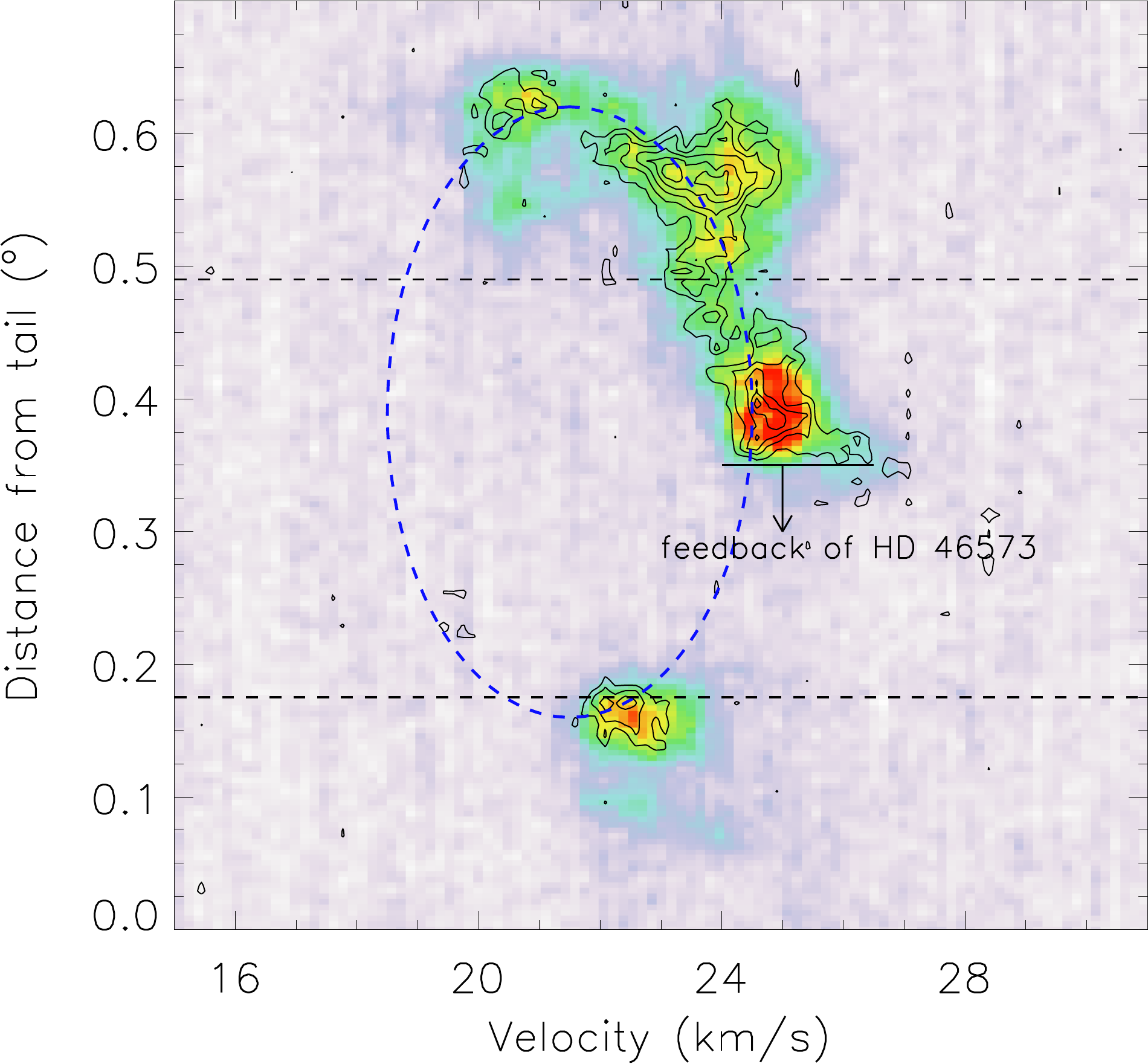}\\
  \caption{Left: integrated intensity map of $^{12}$CO of Sh2-280. Right: position-velocity map of $^{12}$CO emission along the arrow marked in the left panel. The overlaid contours are $^{13}$CO emission with the minimal level and the interval being 0.5 and 0.1 of the peak, respectively. The dash horizontal lines indicate the position range of the \ion{H}{2} region. The blue pentagrams signs indicates the O stars in this region from the SIMBAD database.}
  \label{fig:Kinematics_S280}
\end{figure}

\begin{figure}[h]
  \centering
\includegraphics[width=0.3\textwidth]{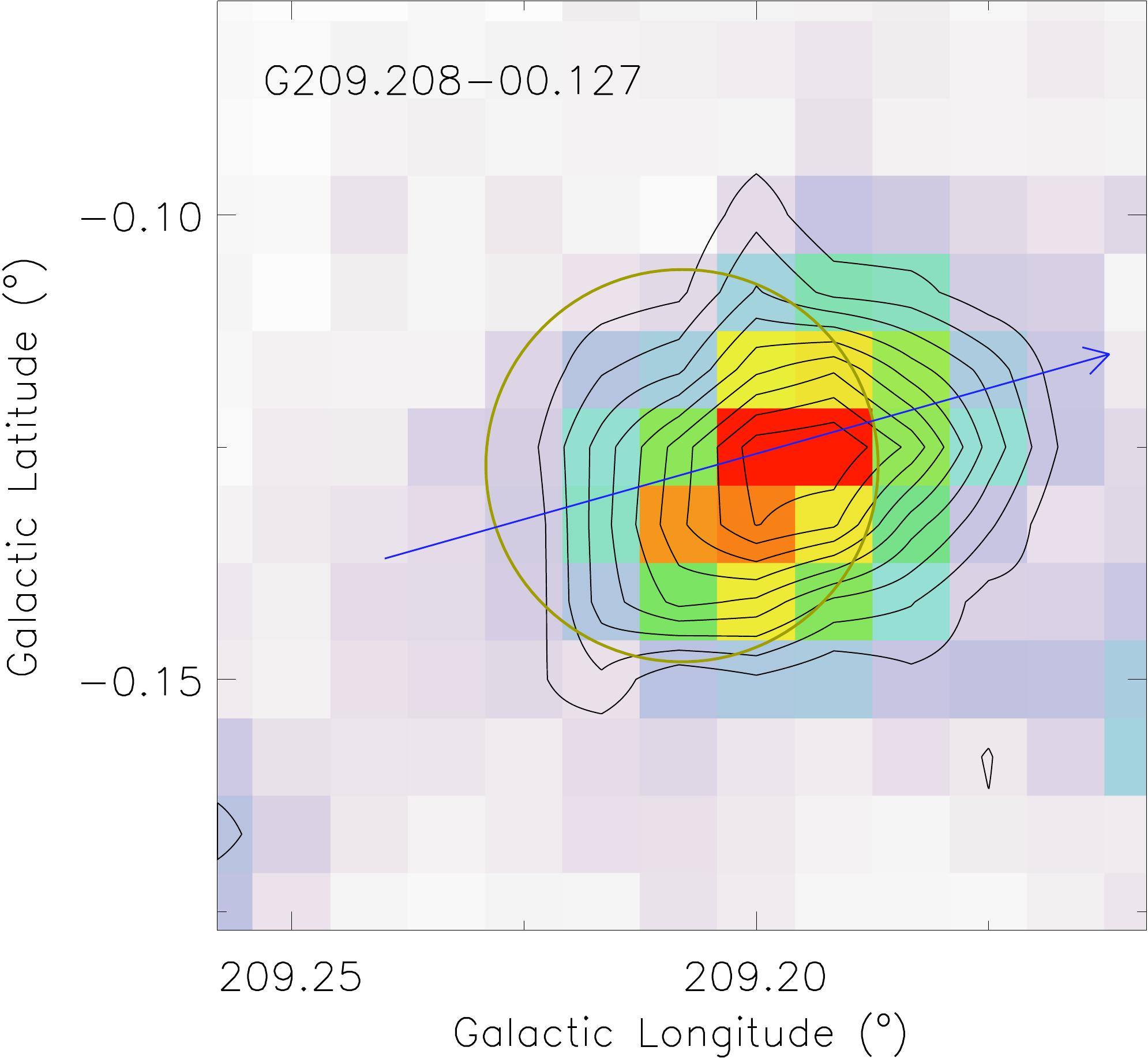}
\includegraphics[width=0.3\textwidth]{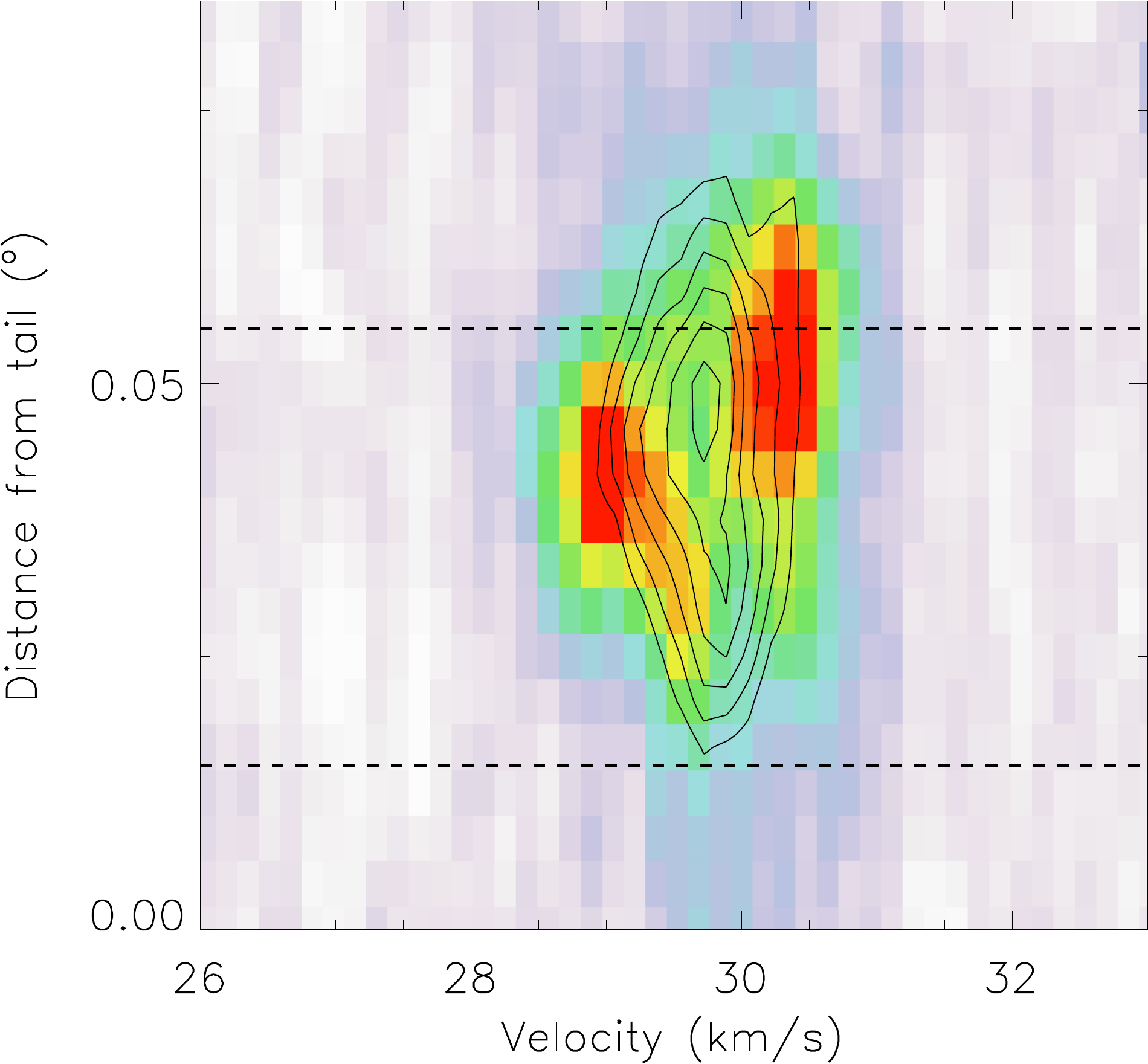}\\
  \caption{Left: integrated intensity map of $^{12}$CO of G209.208-00.127\_far. Right: position-velocity map of $^{12}$CO emission along the arrow marked in the left panel. The overlaid contours are $^{13}$CO emission with the minimal level and the interval being 0.4 and 0.1 of peak, respectively. The dash horizontal  lines indicate the position range of the \ion{H}{2} region.}
  \label{fig:Kinematics_G2092-001}
\end{figure}

\begin{figure}[h]
  \centering
\includegraphics[width=0.3\textwidth]{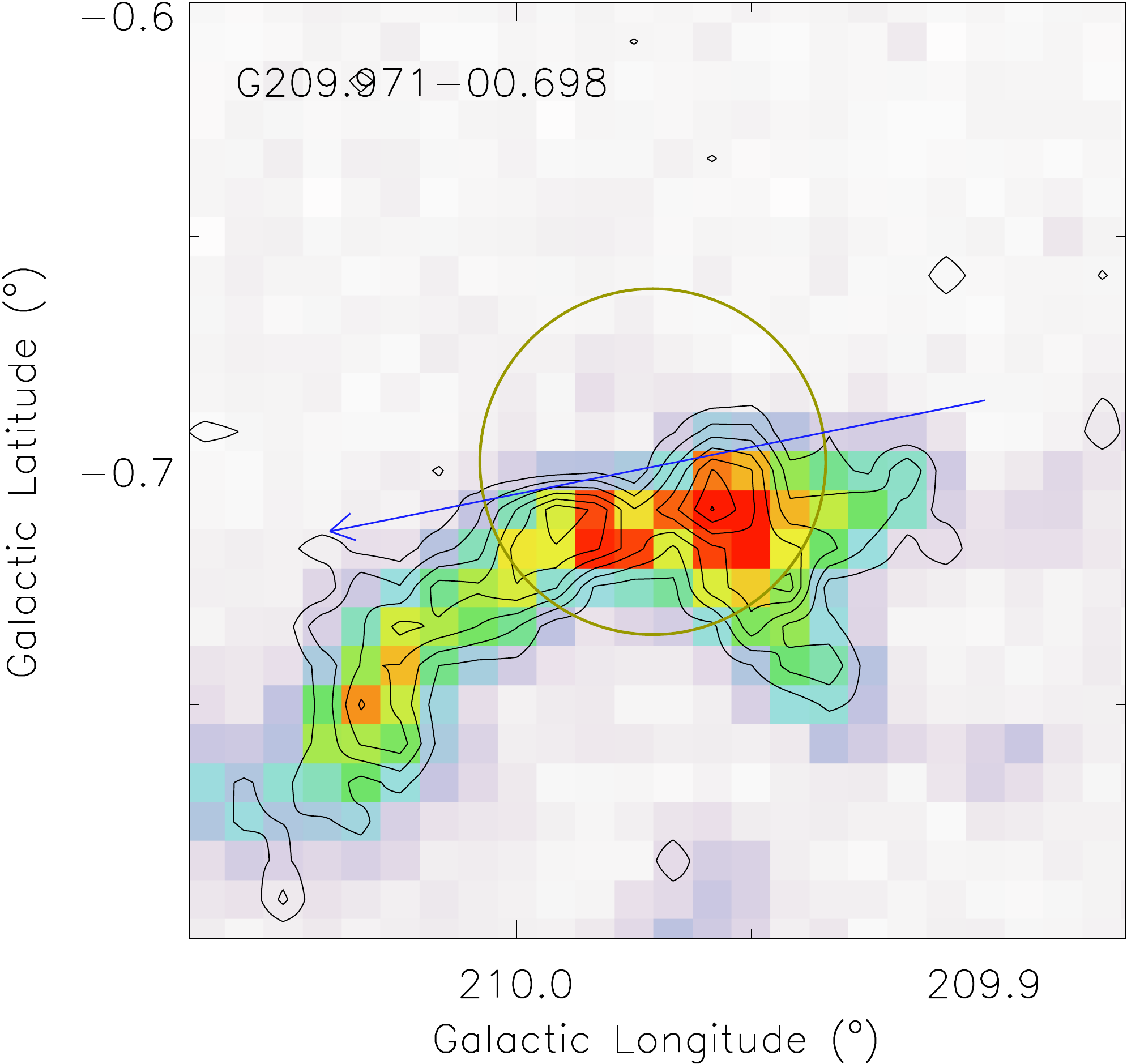}
\includegraphics[width=0.31\textwidth]{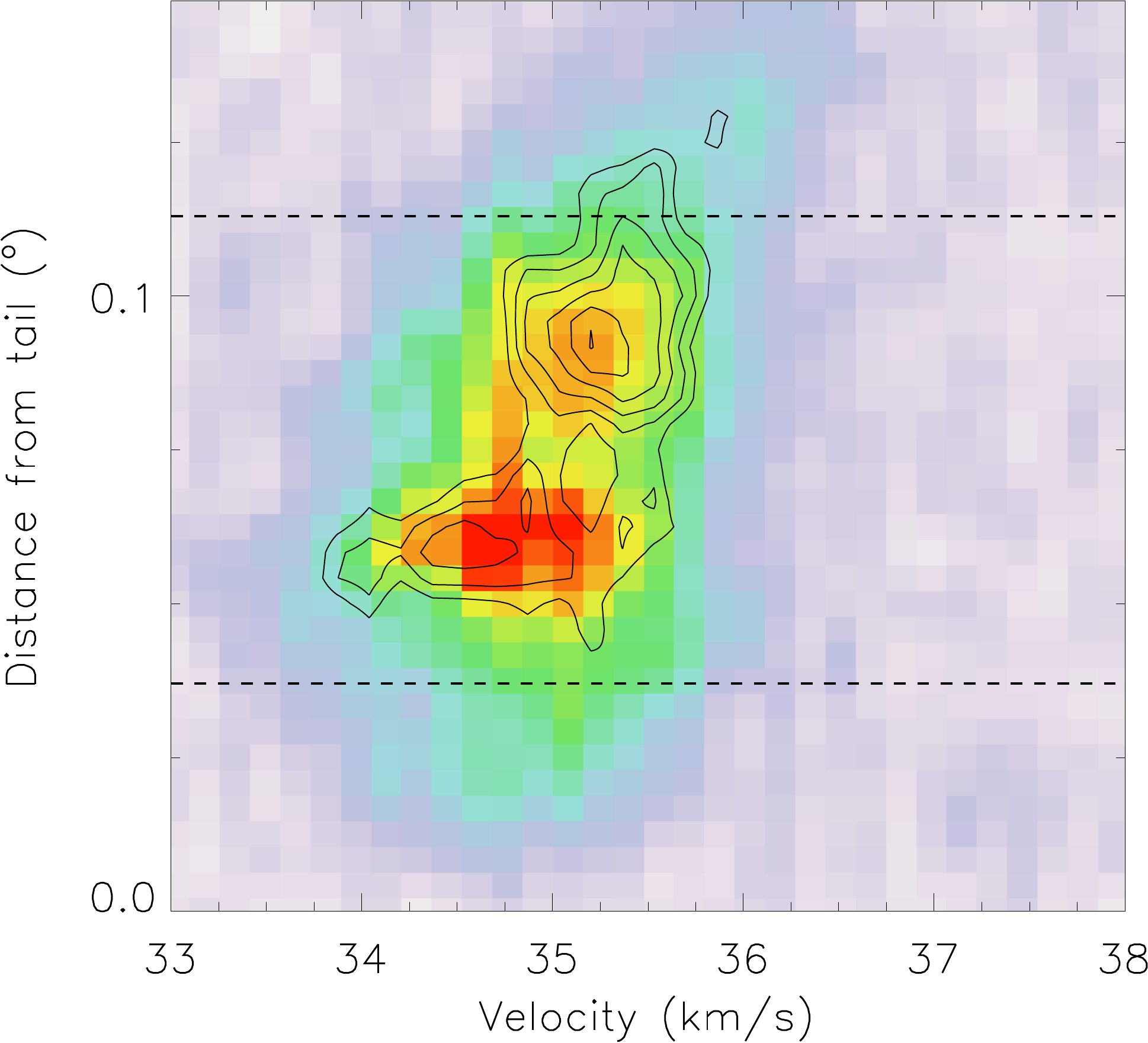}\\
  \caption{Left: integrated intensity map of $^{12}$CO of G209.971-00.698\_far. Right: position-velocity map of $^{12}$CO emission along the arrow marked in the left panel. The overlaid contours are $^{13}$CO emission with the minimal level and the interval being 0.6 and 0.07 of peak, respectively. The dash horizontal  lines indicate the position range of the \ion{H}{2} region.}
  \label{fig:Kinematics_G2099-006}
\end{figure}

\begin{figure}[h]
  \centering
\includegraphics[width=0.3\textwidth]{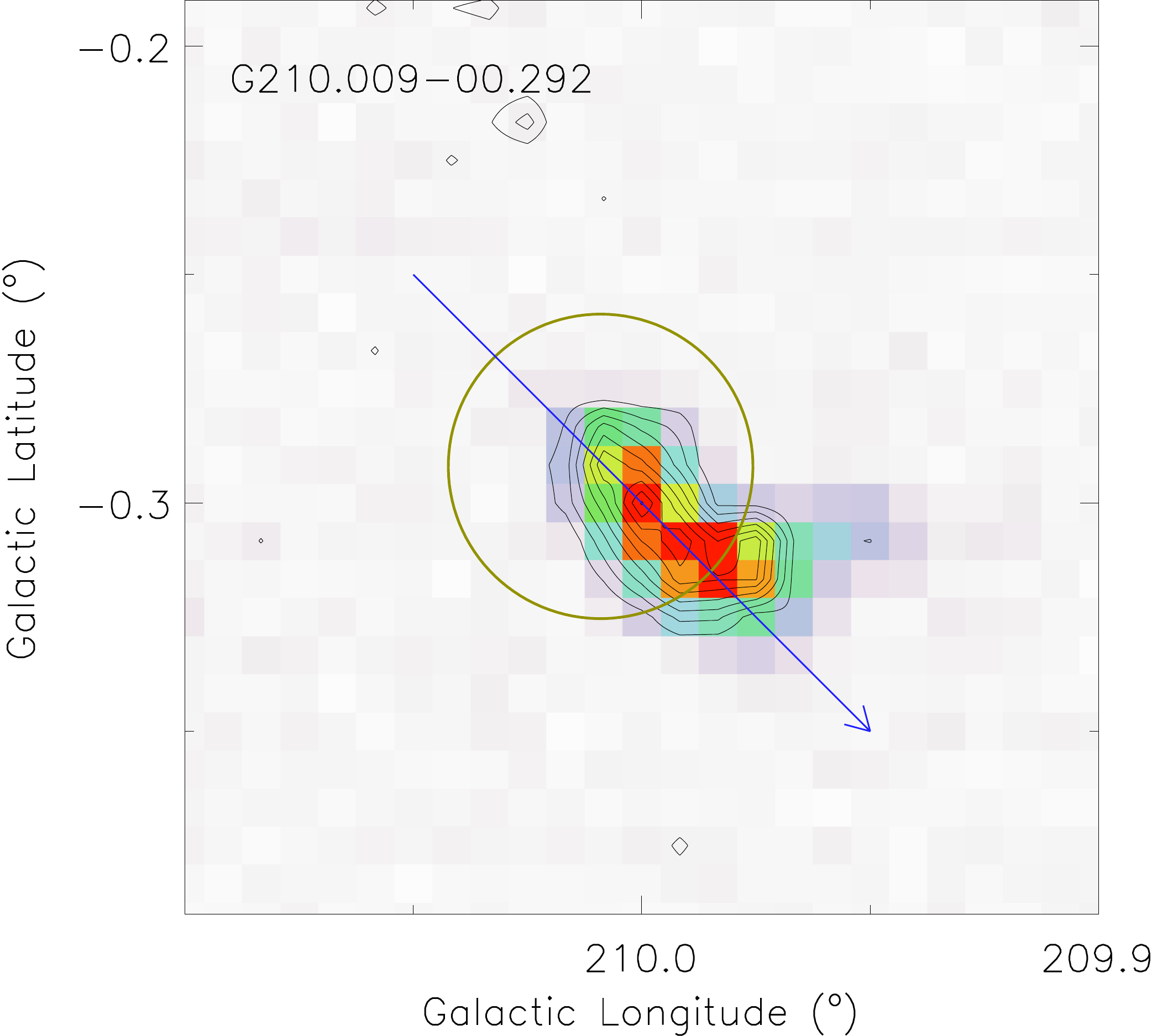}
\includegraphics[width=0.3\textwidth]{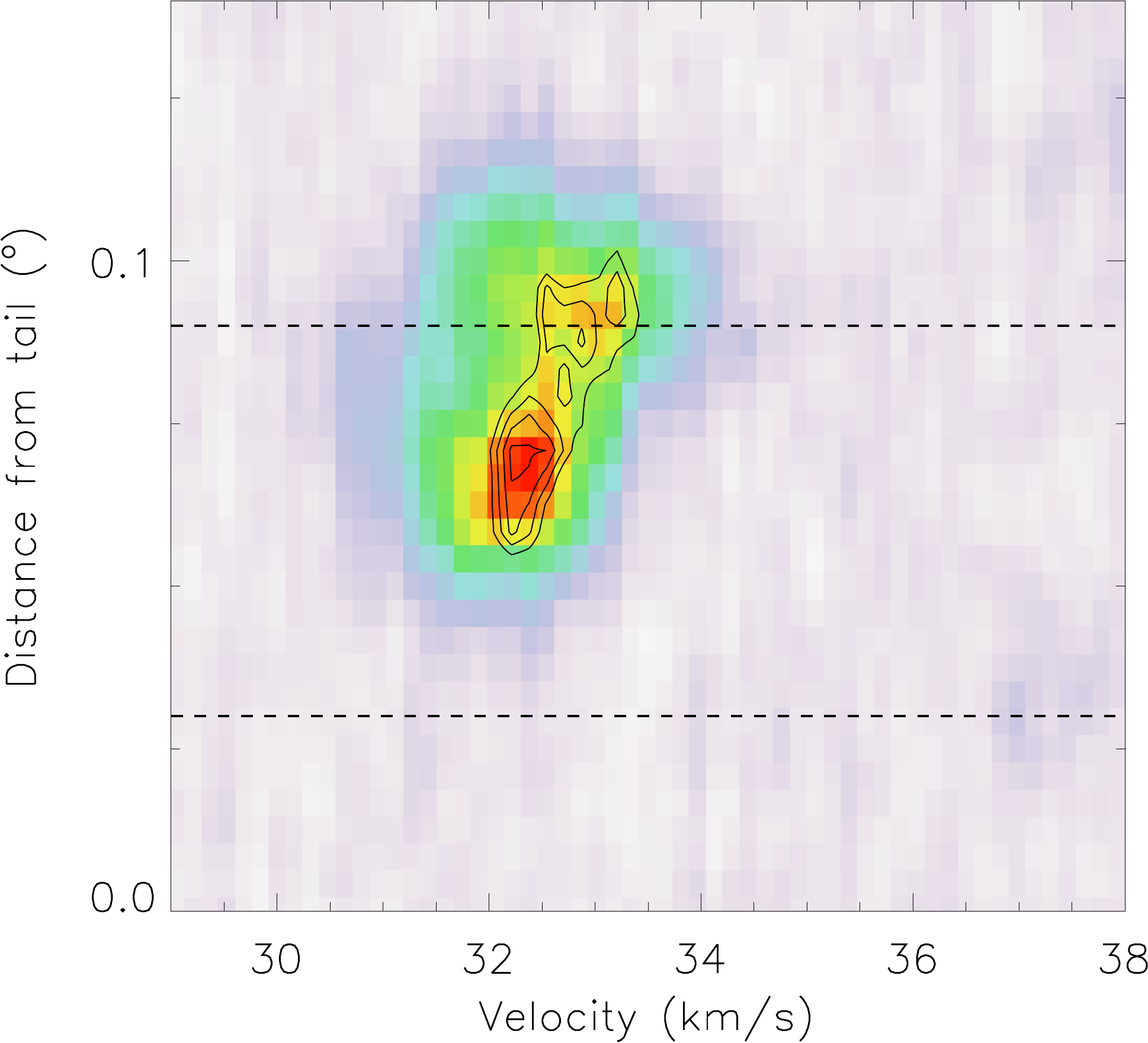}\\
  \caption{Left: integrated intensity map of $^{12}$CO of G210.009-00.292. Right: position-velocity map of $^{12}$CO emission along the arrow marked in the left panel. The overlaid contours are $^{13}$CO emission with the minimal level and the interval being 0.8 and 0.07 of peak, respectively. The dash horizontal  lines indicate the position range of the \ion{H}{2} region.}
  \label{fig:Kinematics_G2100-002}
\end{figure}

\begin{figure}[h]
  \centering
\includegraphics[width=0.3\textwidth]{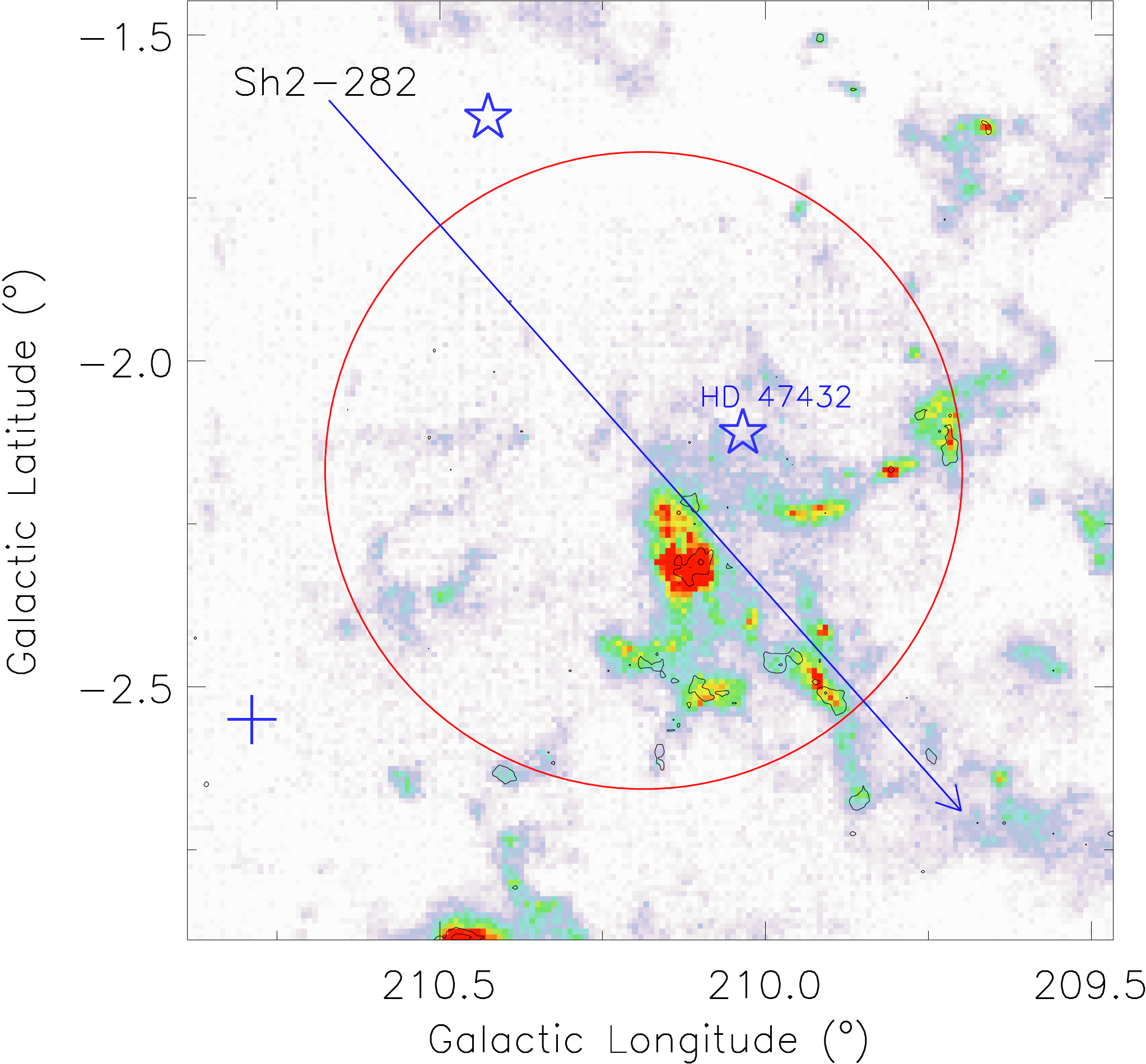}
\includegraphics[width=0.3\textwidth]{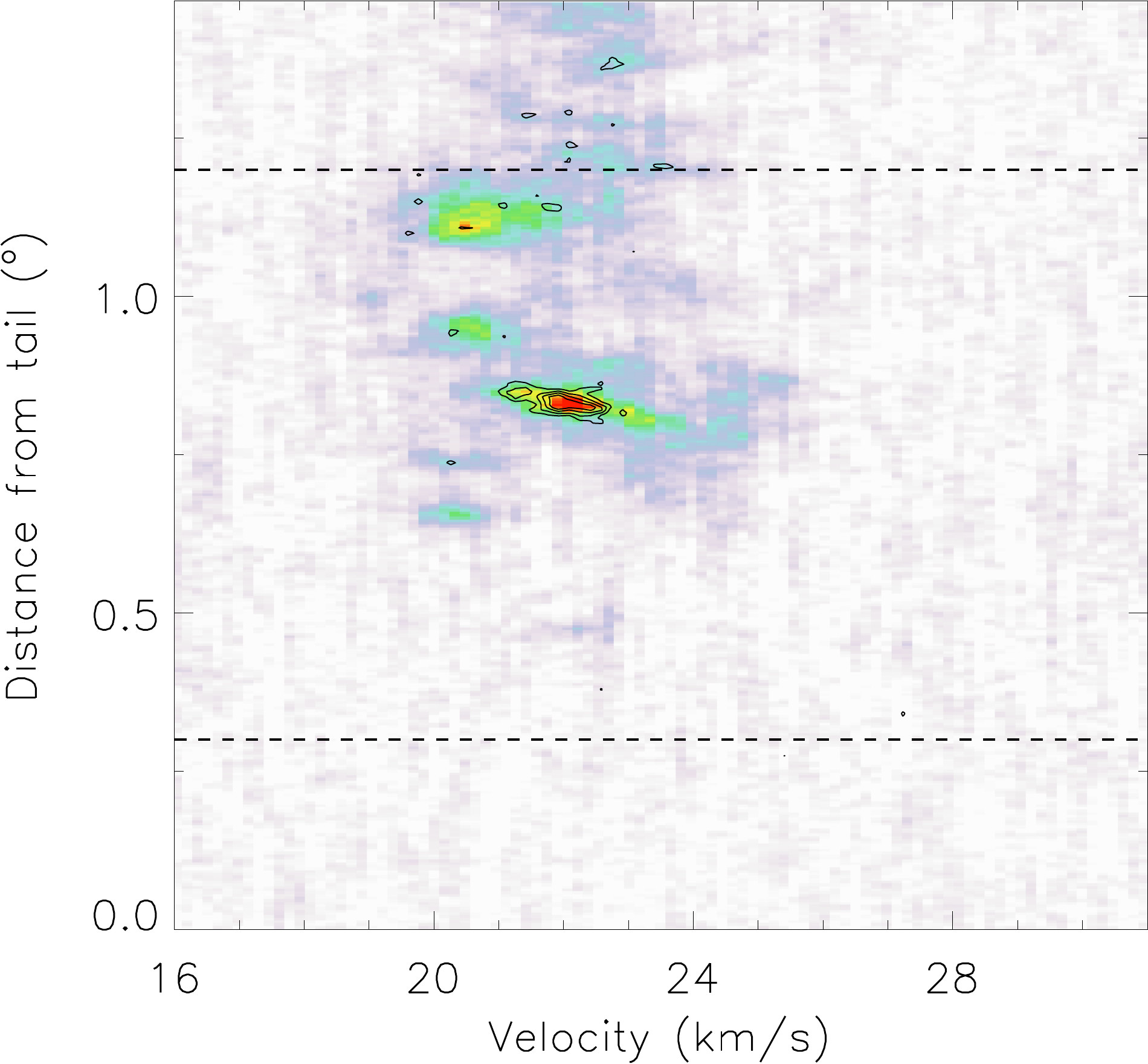}\\
  \caption{Left: integrated intensity map of $^{12}$CO of Sh2-282\_far. Right: position-velocity map of $^{12}$CO emission along the arrow marked in the left panel. The overlaid contours are $^{13}$CO emission with the minimal level and the interval being 0.65 and 0.1 of peak, respectively. The dash horizontal  lines indicate the position range of the \ion{H}{2} region. The blue pentagram and cross signs indicate the O and B0 stars in this region from the SIMBAD database, respectively.}
  \label{fig:Kinematics_S282}
\end{figure}

\begin{figure}[h]
  \centering
\includegraphics[width=0.3\textwidth]{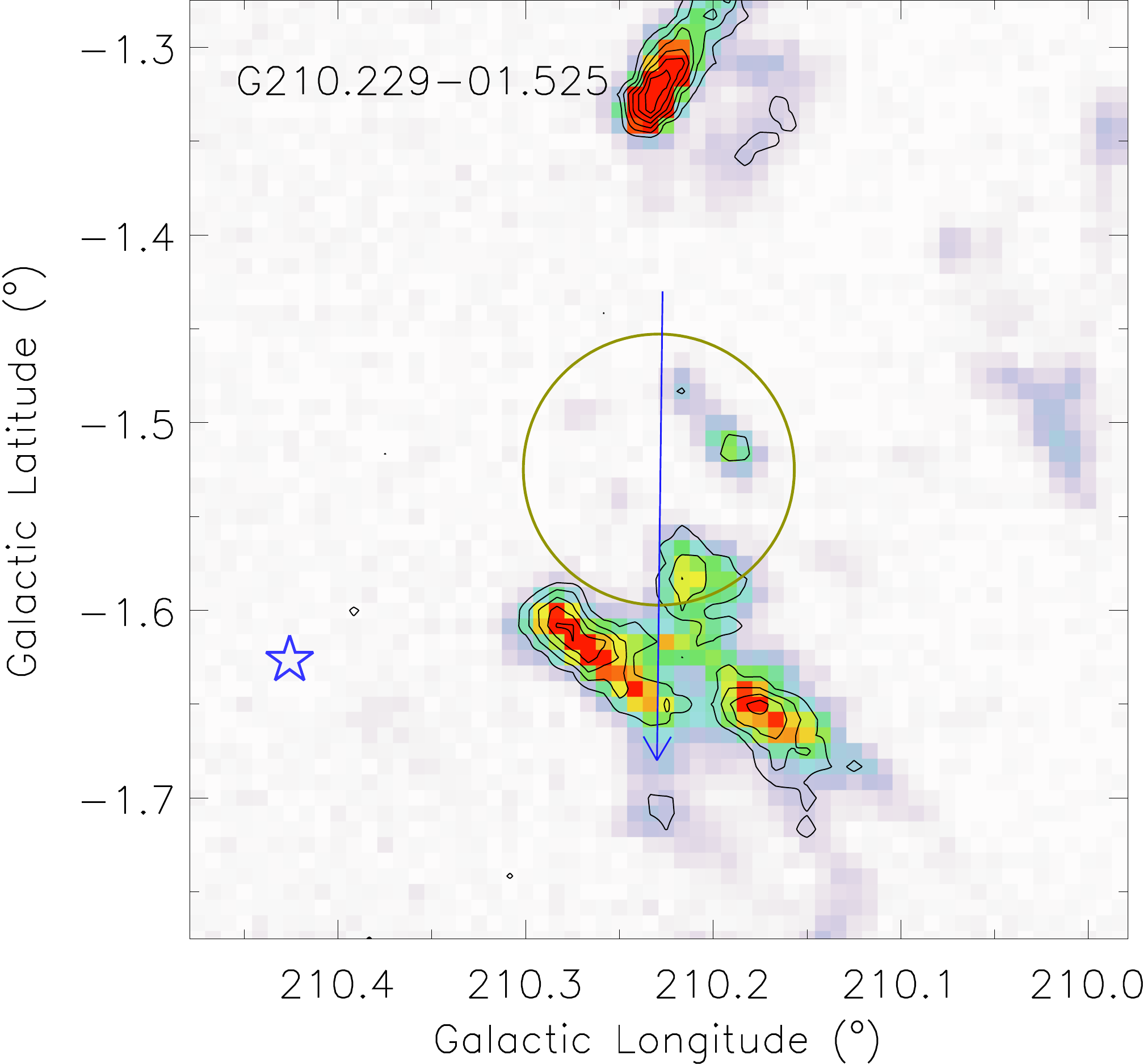}
\includegraphics[width=0.3\textwidth]{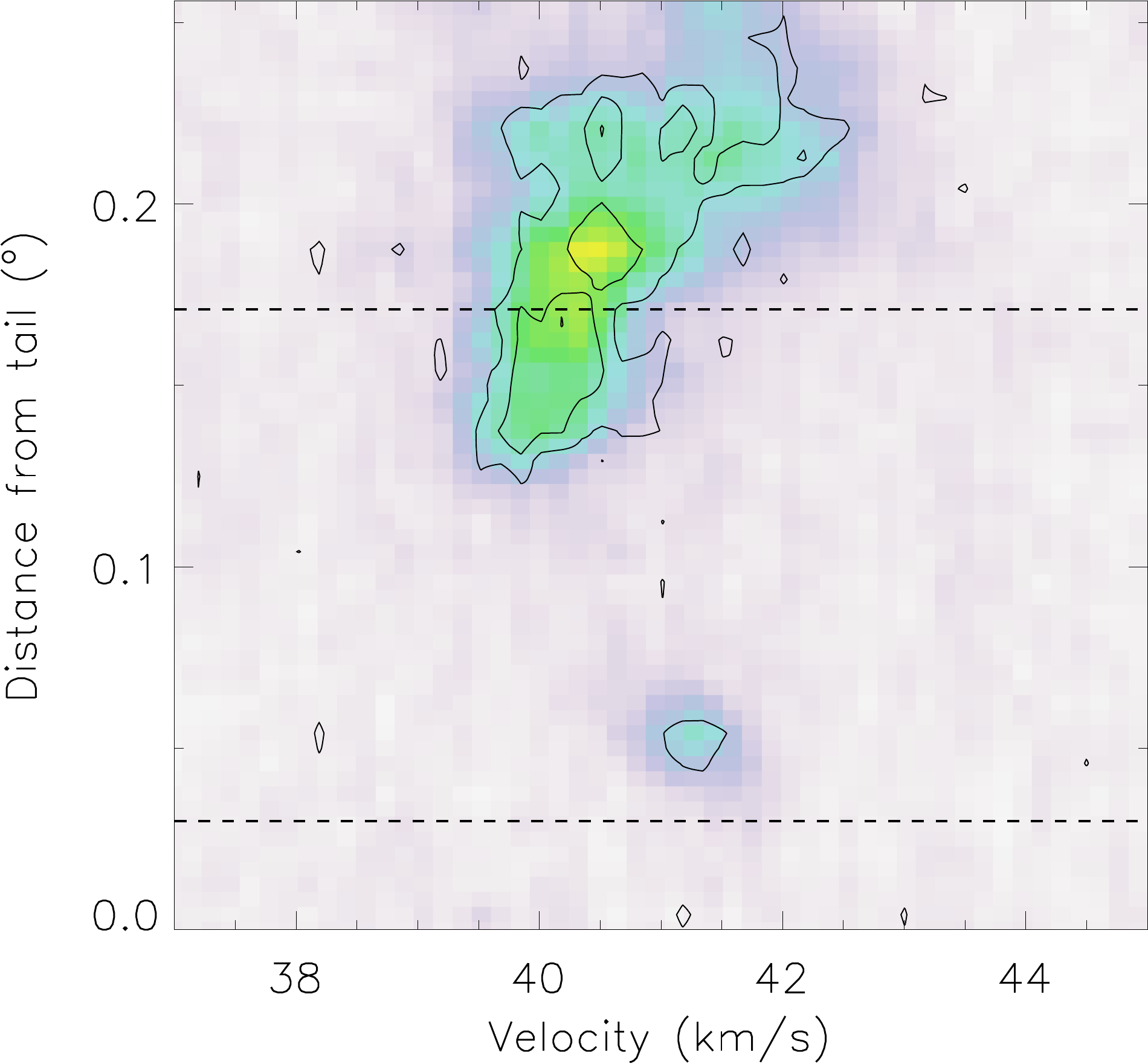}\\
  \caption{Left: integrated intensity map of $^{12}$CO of G210.229-01.525\_far. Right: position-velocity map of $^{12}$CO emission along the arrow marked in the left panel. The overlaid contours are $^{13}$CO emission with the minimal level and the interval being 0.35 and 0.1 of peak, respectively. The dash horizontal  lines indicate the position range of the \ion{H}{2} region. The blue pentagram indicates the O star in this region from the SIMBAD database.}
  \label{fig:Kinematics_G2102-015}
\end{figure}

\begin{figure}[h]
  \centering
\includegraphics[width=0.3\textwidth]{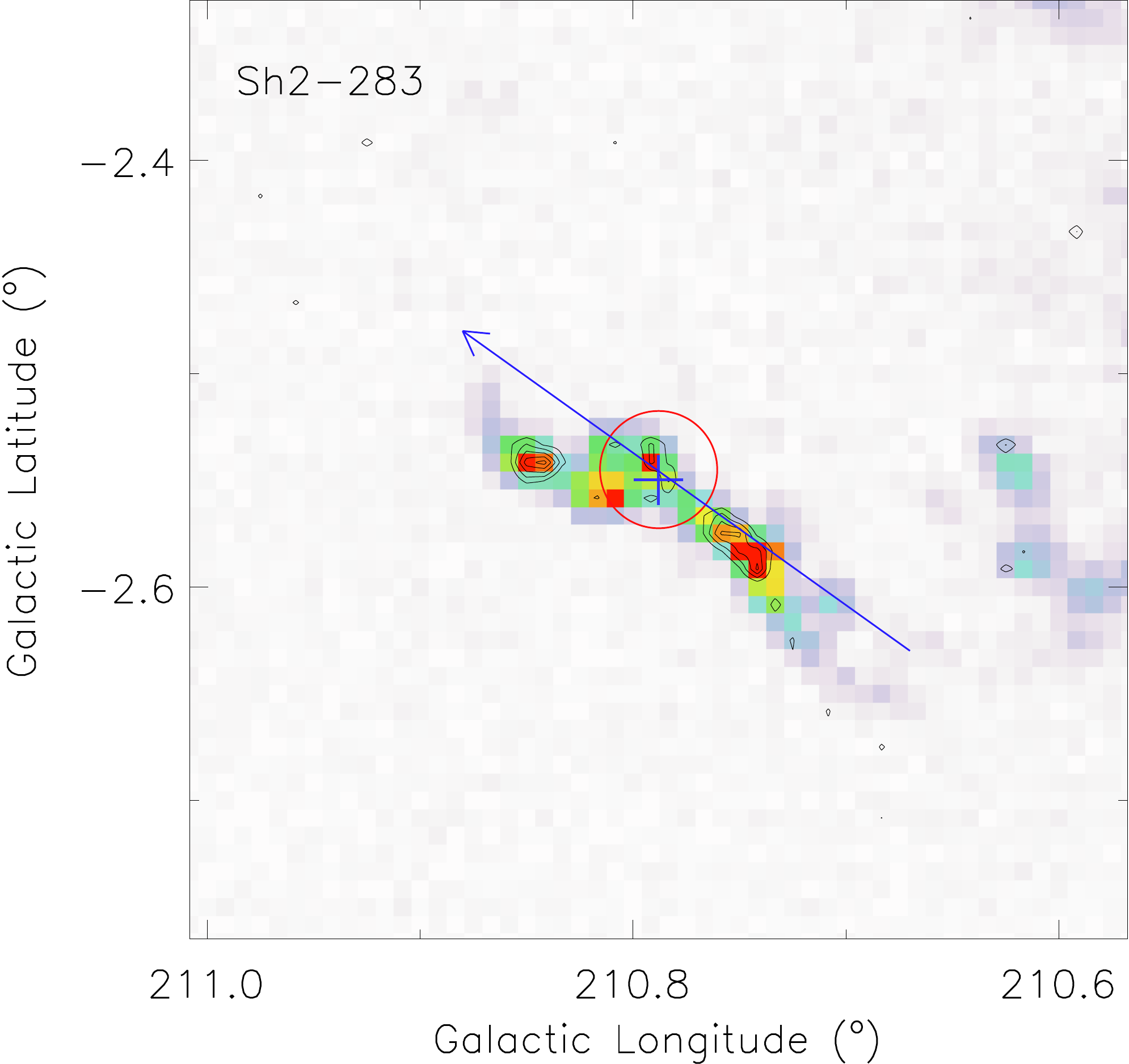}
\includegraphics[width=0.31\textwidth]{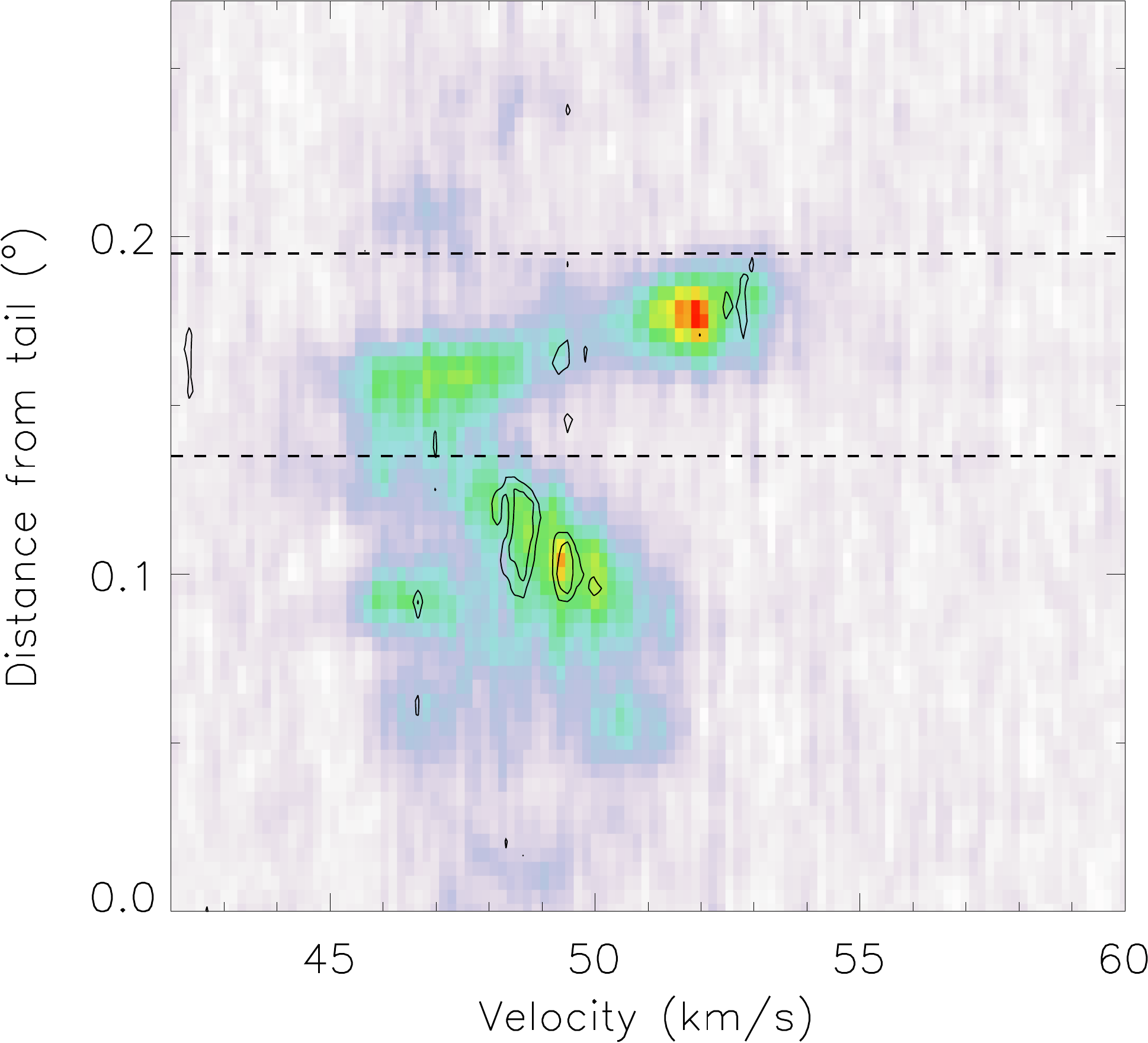}\\
  \caption{Left: integrated intensity map of $^{12}$CO of Sh2-283. Right: position-velocity map of $^{12}$CO emission along the arrow marked in the left panel. The overlaid contours are $^{13}$CO emission with the minimal level and the interval being 0.8 and 0.1 of peak, respectively. The dash horizontal  lines indicate the position range of the \ion{H}{2} region. The blue pentagram indicates the B0 star in this region from the SIMBAD database.}
  \label{fig:Kinematics_S283}
\end{figure}

\begin{figure}[h]
  \centering
\includegraphics[width=0.3\textwidth]{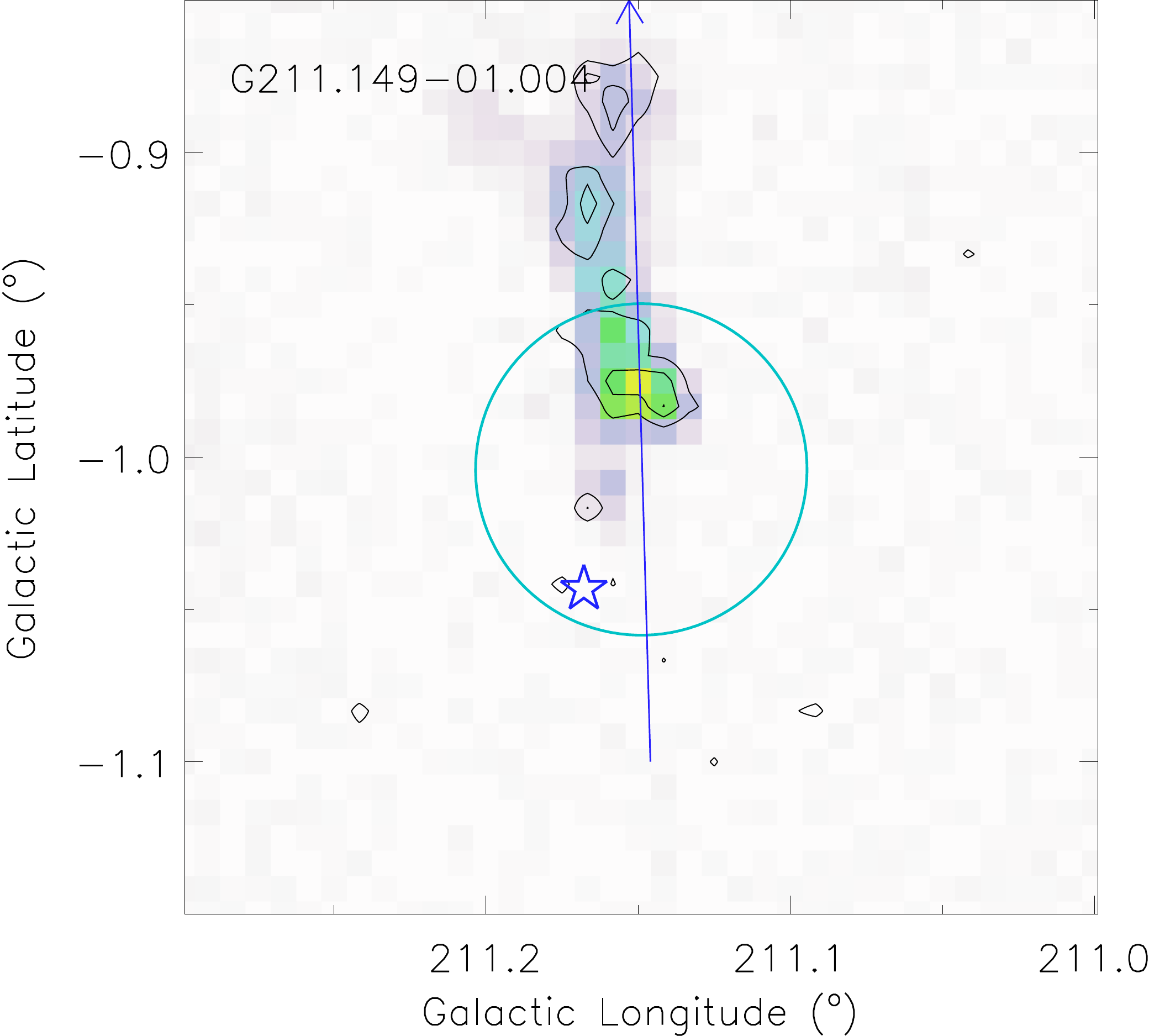}
\includegraphics[width=0.3\textwidth]{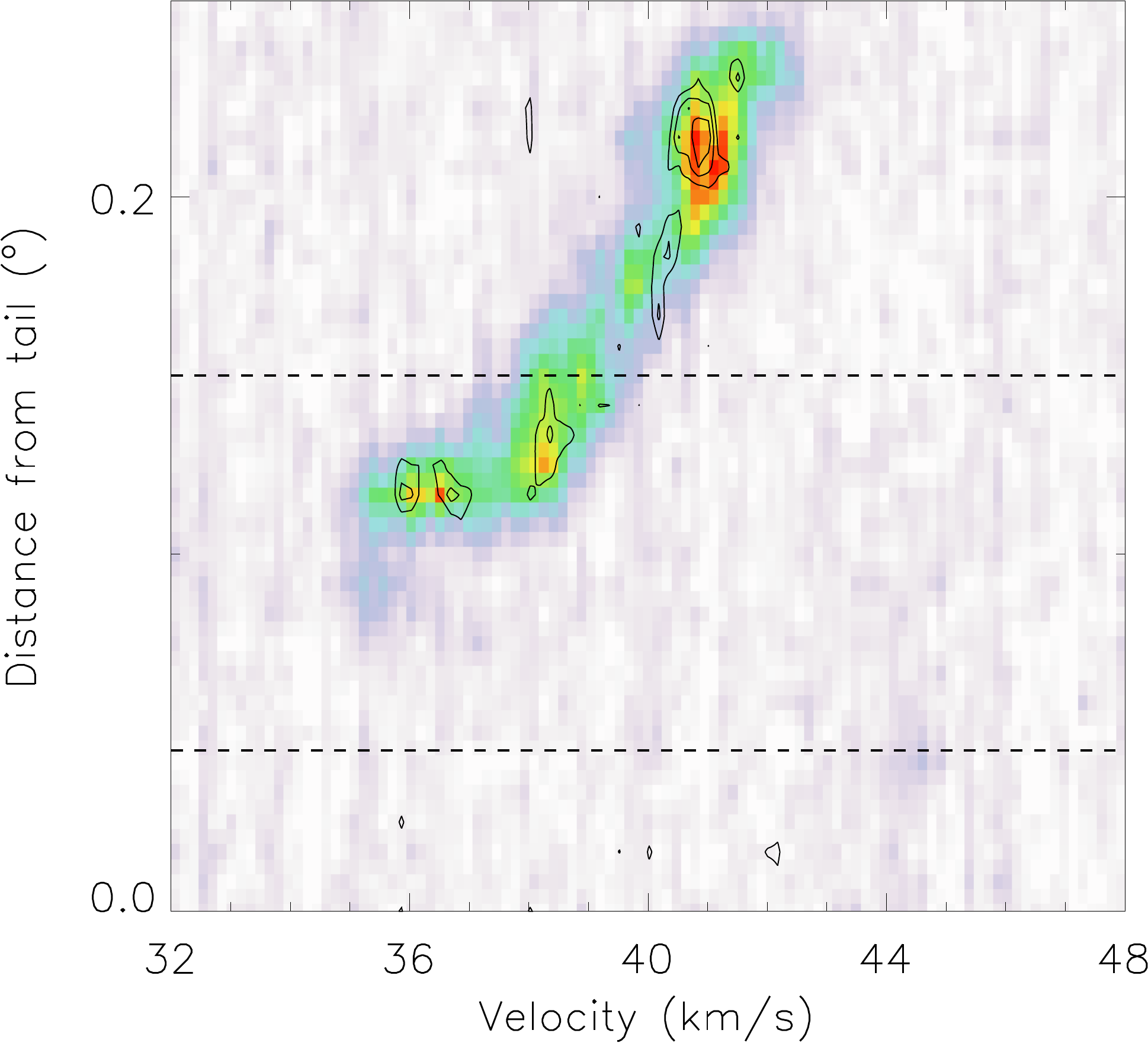}\\
  \caption{Left: integrated intensity map of $^{12}$CO of G211.149-01.004. Right: position-velocity map of $^{12}$CO emission along the arrow marked in the left panel. The overlaid contours are $^{13}$CO emission with the minimal level and the interval being 0.7 and 0.1 of peak, respectively. The dash horizontal  lines indicate the position range of the \ion{H}{2} region. The blue pentagram indicates the O star in this region from the SIMBAD database.}
  \label{fig:Kinematics_G2111-010}
\end{figure}

\begin{figure}[h]
  \centering
\includegraphics[width=0.3\textwidth]{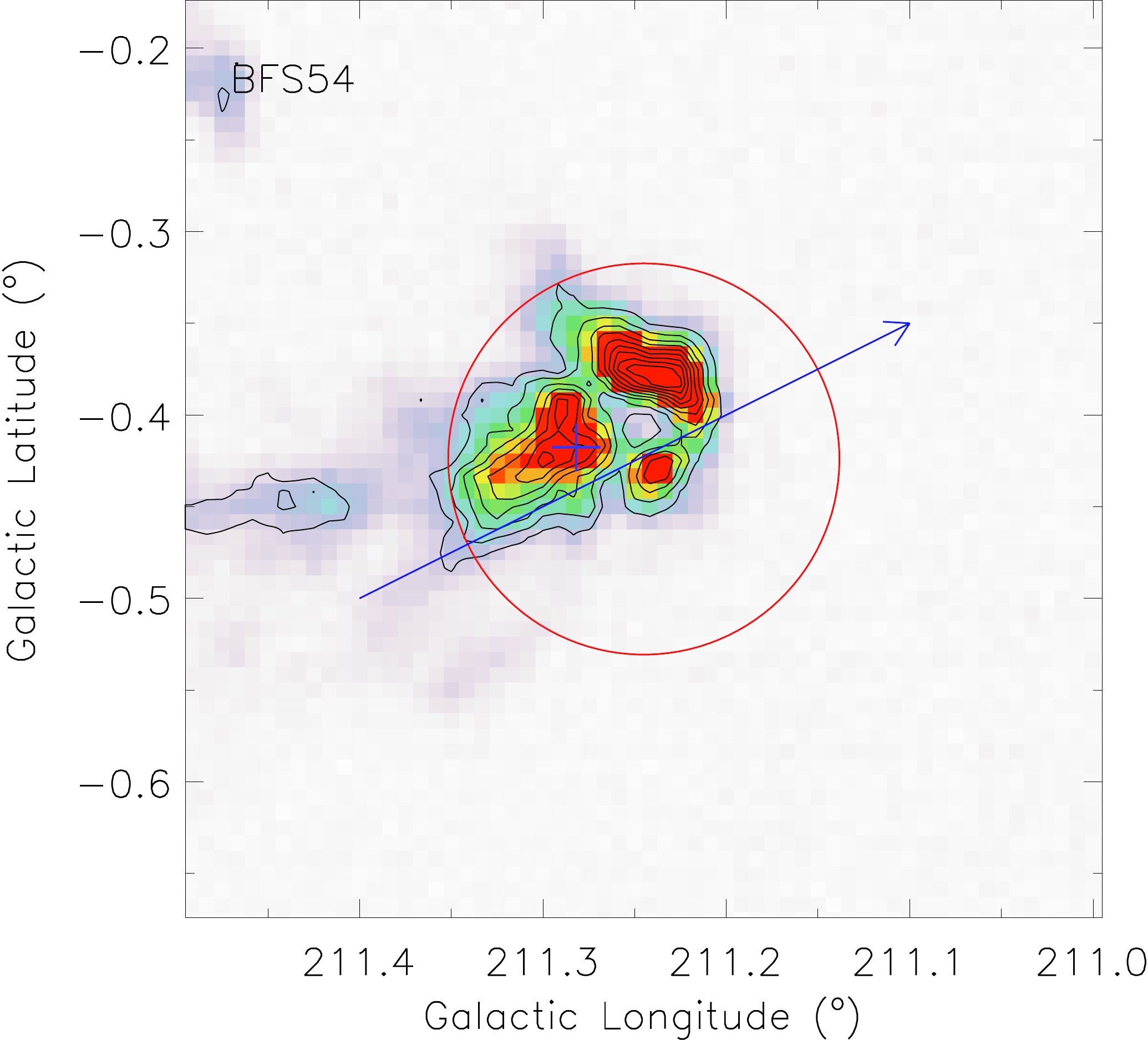}
\includegraphics[width=0.3\textwidth]{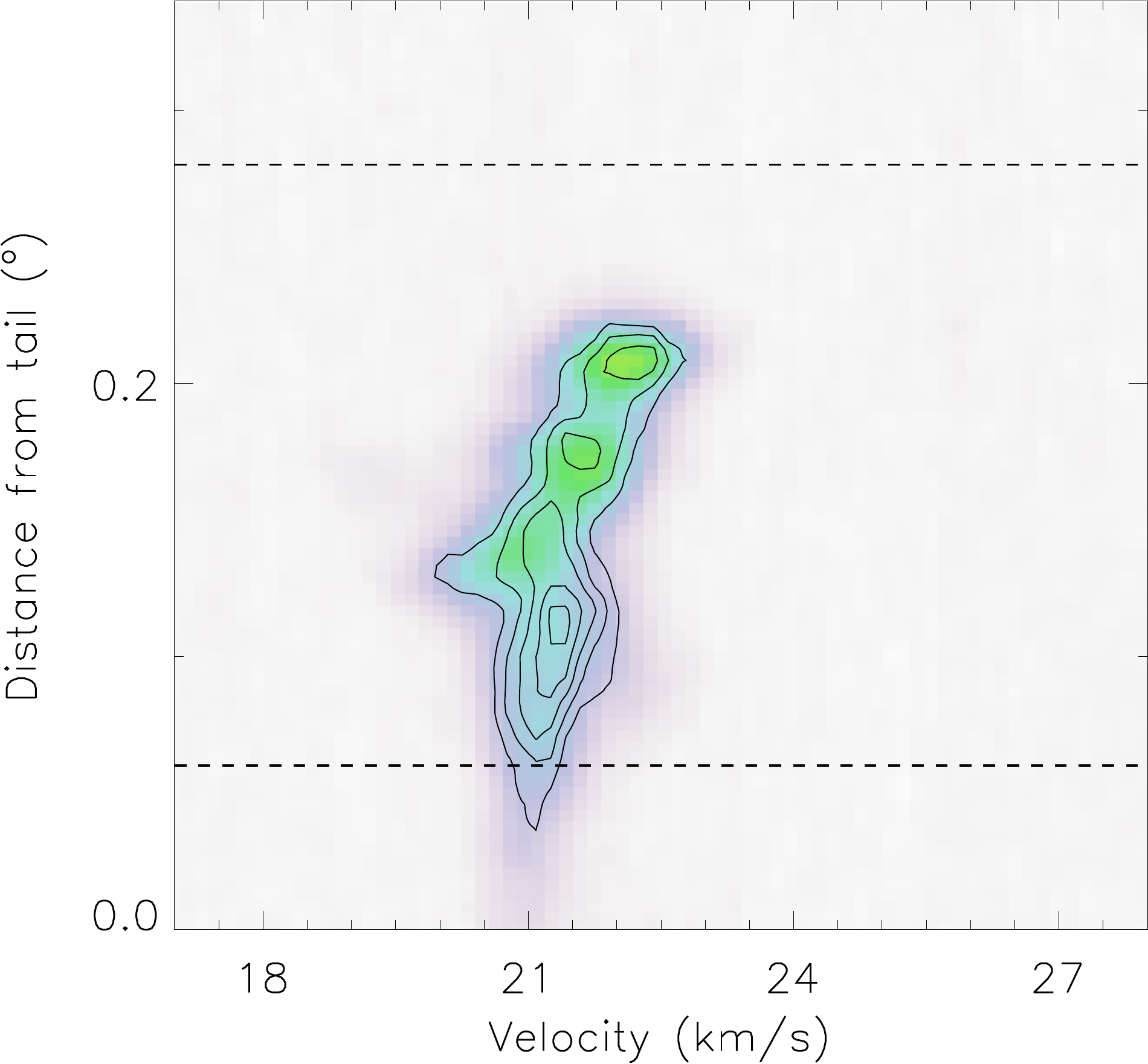}\\
  \caption{Left: integrated intensity map of $^{12}$CO of BFS54. Right: position-velocity map of $^{12}$CO emission along the arrow marked in the left panel. The overlaid contours are $^{13}$CO emission with the minimal level and the interval being 0.3 and 0.1 of peak, respectively. The dash horizontal  lines indicate the position range of the \ion{H}{2} region. The blue cross indicates the B0 star in this region from the SIMBAD database.}
  \label{fig:Kinematics_BFS54}
\end{figure}

\clearpage
\bibliographystyle{apj}
\bibliography{my_ref_all}

\begin{thebibliography}{72}
\expandafter\ifx\csname natexlab\endcsname\relax\def\natexlab#1{#1}\fi

\bibitem[{{Anderson} {et~al.}(2014){Anderson}, {Bania}, {Balser}, {Cunningham},
  {Wenger}, {Johnstone}, \& {Armentrout}}]{2014ApJS..212....1A}
{Anderson}, L.~D., {Bania}, T.~M., {Balser}, D.~S., {et~al.} 2014, \apjs, 212,
  1

\bibitem[{{Anderson} {et~al.}(2011){Anderson}, {Bania}, {Balser}, \&
  {Rood}}]{2011ApJS..194...32A}
{Anderson}, L.~D., {Bania}, T.~M., {Balser}, D.~S., \& {Rood}, R.~T. 2011,
  \apjs, 194, 32

\bibitem[{{Anderson} {et~al.}(2009){Anderson}, {Bania}, {Jackson}, {Clemens},
  {Heyer}, {Simon}, {Shah}, \& {Rathborne}}]{2009ApJS..181..255A}
{Anderson}, L.~D., {Bania}, T.~M., {Jackson}, J.~M., {et~al.} 2009, \apjs, 181,
  255

\bibitem[{{Arce} {et~al.}(2011){Arce}, {Borkin}, {Goodman}, {Pineda}, \&
  {Beaumont}}]{2011ApJ...742..105A}
{Arce}, H.~G., {Borkin}, M.~A., {Goodman}, A.~A., {Pineda}, J.~E., \&
  {Beaumont}, C.~N. 2011, \apj, 742, 105

\bibitem[{{Avedisova} \& {Kondratenko}(1984)}]{1984NInfo..56...59A}
{Avedisova}, V.~S., \& {Kondratenko}, G.~I. 1984, Nauchnye Informatsii, 56, 59

\bibitem[{{Bailer-Jones} {et~al.}(2018){Bailer-Jones}, {Rybizki}, {Fouesneau},
  {Mantelet}, \& {Andrae}}]{2018AJ....156...58B}
{Bailer-Jones}, C.~A.~L., {Rybizki}, J., {Fouesneau}, M., {Mantelet}, G., \&
  {Andrae}, R. 2018, \aj, 156, 58

\bibitem[{{Beaumont} \& {Williams}(2010)}]{2010ApJ...709..791B}
{Beaumont}, C.~N., \& {Williams}, J.~P. 2010, \apj, 709, 791

\bibitem[{{Bertoldi}(1989)}]{1989ApJ...346..735B}
{Bertoldi}, F. 1989, \apj, 346, 735

\bibitem[{{Bisbas} {et~al.}(2011){Bisbas}, {W{\"u}nsch}, {Whitworth}, {Hubber},
  \& {Walch}}]{2011ApJ...736..142B}
{Bisbas}, T.~G., {W{\"u}nsch}, R., {Whitworth}, A.~P., {Hubber}, D.~A., \&
  {Walch}, S. 2011, \apj, 736, 142

\bibitem[{{Blitz} {et~al.}(1982){Blitz}, {Fich}, \&
  {Stark}}]{1982ApJS...49..183B}
{Blitz}, L., {Fich}, M., \& {Stark}, A.~A. 1982, \apjs, 49, 183

\bibitem[{{Blitz} \& {Williams}(1999)}]{1999ASIC..540....3B}
{Blitz}, L., \& {Williams}, J.~P. 1999, in NATO Advanced Science Institutes
  (ASI) Series C, Vol. 540, NATO Advanced Science Institutes (ASI) Series C,
  ed. C.~J. {Lada} \& N.~D. {Kylafis}, 3

\bibitem[{Bolatto {et~al.}(2013)Bolatto, Wolfire, \&
  Leroy}]{doi:10.1146/annurev-astro-082812-140944}
Bolatto, A.~D., Wolfire, M., \& Leroy, A.~K. 2013, Annual Review of Astronomy
  and Astrophysics, 51, 207

\bibitem[{{Burton} {et~al.}(2013){Burton}, {Braiding}, {Glueck}, {Goldsmith},
  {Hawkes}, {Hollenbach}, {Kulesa}, {Martin}, {Pineda}, {Rowell}, {Simon},
  {Stark}, {Stutzki}, {Tothill}, {Urquhart}, {Walker}, {Walsh}, \&
  {Wolfire}}]{2013PASA...30...44B}
{Burton}, M.~G., {Braiding}, C., {Glueck}, C., {et~al.} 2013, \pasa, 30, e044

\bibitem[{{Carrasco-Gonz{\'a}lez} {et~al.}(2006){Carrasco-Gonz{\'a}lez},
  {L{\'o}pez}, {Gyulbudaghian}, {Anglada}, \& {Lee}}]{2006A&A...445L..43C}
{Carrasco-Gonz{\'a}lez}, C., {L{\'o}pez}, R., {Gyulbudaghian}, A., {Anglada},
  G., \& {Lee}, C.~W. 2006, \aap, 445, L43

\bibitem[{{Chen} {et~al.}(2017){Chen}, {Jiang}, {Tamura}, {Kwon}, \&
  {Roman-Lopes}}]{2017ApJ...838...80C}
{Chen}, Z., {Jiang}, Z., {Tamura}, M., {Kwon}, J., \& {Roman-Lopes}, A. 2017,
  \apj, 838, 80

\bibitem[{{Chini} {et~al.}(1984){Chini}, {Mezger}, {Kreysa}, \&
  {Gemuend}}]{1984A&A...135L..14C}
{Chini}, R., {Mezger}, P.~G., {Kreysa}, E., \& {Gemuend}, H.-P. 1984, \aap,
  135, L14

\bibitem[{{Churchwell} {et~al.}(2006){Churchwell}, {Povich}, {Allen}, {Taylor},
  {Meade}, {Babler}, {Indebetouw}, {Watson}, {Whitney}, {Wolfire}, {Bania},
  {Benjamin}, {Clemens}, {Cohen}, {Cyganowski}, {Jackson}, {Kobulnicky},
  {Mathis}, {Mercer}, {Stolovy}, {Uzpen}, {Watson}, \&
  {Wolff}}]{2006ApJ...649..759C}
{Churchwell}, E., {Povich}, M.~S., {Allen}, D., {et~al.} 2006, \apj, 649, 759

\bibitem[{{Clemens}(1985)}]{1985ApJ...295..422C}
{Clemens}, D.~P. 1985, \apj, 295, 422

\bibitem[{{Condon} {et~al.}(1998){Condon}, {Cotton}, {Greisen}, {Yin},
  {Perley}, {Taylor}, \& {Broderick}}]{1998AJ....115.1693C}
{Condon}, J.~J., {Cotton}, W.~D., {Greisen}, E.~W., {et~al.} 1998, \aj, 115,
  1693

\bibitem[{{Dame} {et~al.}(2001){Dame}, {Hartmann}, \&
  {Thaddeus}}]{2001ApJ...547..792D}
{Dame}, T.~M., {Hartmann}, D., \& {Thaddeus}, P. 2001, \apj, 547, 792

\bibitem[{{Dame} {et~al.}(1987){Dame}, {Ungerechts}, {Cohen}, {de Geus},
  {Grenier}, {May}, {Murphy}, {Nyman}, \& {Thaddeus}}]{1987ApJ...322..706D}
{Dame}, T.~M., {Ungerechts}, H., {Cohen}, R.~S., {et~al.} 1987, \apj, 322, 706

\bibitem[{{Digel}(1991)}]{Digel_1991}
{Digel}, S.~W. 1991

\bibitem[{{Dutta} {et~al.}(2015){Dutta}, {Mondal}, {Jose}, {Das}, {Samal}, \&
  {Ghosh}}]{2015MNRAS.454.3597D}
{Dutta}, S., {Mondal}, S., {Jose}, J., {et~al.} 2015, \mnras, 454, 3597

\bibitem[{{Dutta} {et~al.}(2018){Dutta}, {Mondal}, {Joshi}, {Jose}, {Das}, \&
  {Ghosh}}]{2018MNRAS.476.2813D}
{Dutta}, S., {Mondal}, S., {Joshi}, S., {et~al.} 2018, \mnras, 476, 2813

\bibitem[{{Felli} \& {Harten}(1981)}]{1981A&A...100...28F}
{Felli}, M., \& {Harten}, R.~H. 1981, \aap, 100, 28

\bibitem[{{Fich}(1993)}]{1993ApJS...86..475F}
{Fich}, M. 1993, \apjs, 86, 475

\bibitem[{{Gahm} {et~al.}(2006){Gahm}, {Carlqvist}, {Johansson}, \&
  {Nikoli{\'c}}}]{2006A&A...454..201G}
{Gahm}, G.~F., {Carlqvist}, P., {Johansson}, L.~E.~B., \& {Nikoli{\'c}}, S.
  2006, \aap, 454, 201

\bibitem[{{Gaia Collaboration}(2018)}]{2018yCat.1345....0G}
{Gaia Collaboration}. 2018, VizieR Online Data Catalog, I/345

\bibitem[{{Gaia Collaboration} {et~al.}(2018){Gaia Collaboration}, {Brown},
  {Vallenari}, {Prusti}, {de Bruijne}, {Babusiaux}, {Bailer-Jones}, {Biermann},
  {Evans}, {Eyer}, \& et~al.}]{2018A&A...616A...1G}
{Gaia Collaboration}, {Brown}, A.~G.~A., {Vallenari}, A., {et~al.} 2018, \aap,
  616, A1

\bibitem[{{Gaustad} {et~al.}(2001){Gaustad}, {McCullough}, {Rosing}, \& {Van
  Buren}}]{2001PASP..113.1326G}
{Gaustad}, J.~E., {McCullough}, P.~R., {Rosing}, W., \& {Van Buren}, D. 2001,
  \pasp, 113, 1326

\bibitem[{{Gong} {et~al.}(2016){Gong}, {Mao}, {Fang}, {Zhang}, {Su}, {Yang},
  {Jiang}, {Xu}, {Wang}, {Wang}, {Lu}, \& {Sun}}]{2016A&A...588A.104G}
{Gong}, Y., {Mao}, R.~Q., {Fang}, M., {et~al.} 2016, \aap, 588, A104

\bibitem[{{Gontcharov}(2006)}]{2006AstL...32..759G}
{Gontcharov}, G.~A. 2006, Astronomy Letters, 32, 759

\bibitem[{{Hensberge} {et~al.}(2000){Hensberge}, {Pavlovski}, \&
  {Verschueren}}]{2000A&A...358..553H}
{Hensberge}, H., {Pavlovski}, K., \& {Verschueren}, W. 2000, \aap, 358, 553

\bibitem[{{Heyer} \& {Dame}(2015)}]{2015ARA&A..53..583H}
{Heyer}, M., \& {Dame}, T.~M. 2015, \araa, 53, 583

\bibitem[{{Heyer} {et~al.}(1998){Heyer}, {Brunt}, {Snell}, {Howe}, {Schloerb},
  \& {Carpenter}}]{1998ApJS..115..241H}
{Heyer}, M.~H., {Brunt}, C., {Snell}, R.~L., {et~al.} 1998, \apjs, 115, 241

\bibitem[{{Heyer} {et~al.}(2001){Heyer}, {Carpenter}, \&
  {Snell}}]{2001ApJ...551..852H}
{Heyer}, M.~H., {Carpenter}, J.~M., \& {Snell}, R.~L. 2001, \apj, 551, 852

\bibitem[{{Horner} {et~al.}(1997){Horner}, {Lada}, \&
  {Lada}}]{1997AJ....113.1788H}
{Horner}, D.~J., {Lada}, E.~A., \& {Lada}, C.~J. 1997, \aj, 113, 1788

\bibitem[{{Hosokawa} \& {Inutsuka}(2006)}]{2006ApJ...646..240H}
{Hosokawa}, T., \& {Inutsuka}, S.-i. 2006, \apj, 646, 240

\bibitem[{{Jackson} {et~al.}(2006){Jackson}, {Rathborne}, {Shah}, {Simon},
  {Bania}, {Clemens}, {Chambers}, {Johnson}, {Dormody}, {Lavoie}, \&
  {Heyer}}]{2006ApJS..163..145J}
{Jackson}, J.~M., {Rathborne}, J.~M., {Shah}, R.~Y., {et~al.} 2006, \apjs, 163,
  145

\bibitem[{{Kendrew} {et~al.}(2012){Kendrew}, {Simpson}, {Bressert}, {Povich},
  {Sherman}, {Lintott}, {Robitaille}, {Schawinski}, \&
  {Wolf-Chase}}]{2012ApJ...755...71K}
{Kendrew}, S., {Simpson}, R., {Bressert}, E., {et~al.} 2012, \apj, 755, 71

\bibitem[{{Kharchenko}(2001)}]{2001KFNT...17..409K}
{Kharchenko}, N.~V. 2001, Kinematika i Fizika Nebesnykh Tel, 17, 409

\bibitem[{{Kim} {et~al.}(2004){Kim}, {Kawamura}, {Yonekura}, \&
  {Fukui}}]{2004PASJ...56..313K}
{Kim}, B.~G., {Kawamura}, A., {Yonekura}, Y., \& {Fukui}, Y. 2004, \pasj, 56,
  313

\bibitem[{{Kislyakov} \& {Turner}(1995)}]{1995AZh....72..168K}
{Kislyakov}, A.~G., \& {Turner}, B.~E. 1995, \azh, 72, 168

\bibitem[{{Kutner} {et~al.}(1980){Kutner}, {Machnik}, {Tucker}, \&
  {Dickman}}]{1980ApJ...237..734K}
{Kutner}, M.~L., {Machnik}, D.~E., {Tucker}, K.~D., \& {Dickman}, R.~L. 1980,
  \apj, 237, 734

\bibitem[{{Larson}(1981)}]{1981MNRAS.194..809L}
{Larson}, R.~B. 1981, \mnras, 194, 809

\bibitem[{{Li} {et~al.}(2018){Li}, {Wang}, {Zhang}, {Ma}, {Fang}, \&
  {Yang}}]{2018ApJS..238...10L}
{Li}, C., {Wang}, H., {Zhang}, M., {et~al.} 2018, \apjs, 238, 10

\bibitem[{{Maddalena} {et~al.}(1986){Maddalena}, {Morris}, {Moscowitz}, \&
  {Thaddeus}}]{1986ApJ...303..375M}
{Maddalena}, R.~J., {Morris}, M., {Moscowitz}, J., \& {Thaddeus}, P. 1986,
  \apj, 303, 375

\bibitem[{{M{\"a}kel{\"a}} {et~al.}(2017){M{\"a}kel{\"a}}, {Haikala}, \&
  {Gahm}}]{2017A&A...605A..82M}
{M{\"a}kel{\"a}}, M.~M., {Haikala}, L.~K., \& {Gahm}, G.~F. 2017, \aap, 605,
  A82

\bibitem[{{Milisavljevic} \& {Fesen}(2015)}]{2015Sci...347..526M}
{Milisavljevic}, D., \& {Fesen}, R.~A. 2015, Science, 347, 526

\bibitem[{{Nagahama} {et~al.}(1998){Nagahama}, {Mizuno}, {Ogawa}, \&
  {Fukui}}]{1998AJ....116..336N}
{Nagahama}, T., {Mizuno}, A., {Ogawa}, H., \& {Fukui}, Y. 1998, \aj, 116, 336

\bibitem[{{Ogura} \& {Ishida}(1981)}]{1981PASJ...33..149O}
{Ogura}, K., \& {Ishida}, K. 1981, \pasj, 33, 149

\bibitem[{{Oliver} {et~al.}(1996){Oliver}, {Masheder}, \&
  {Thaddeus}}]{1996A&A...315..578O}
{Oliver}, R.~J., {Masheder}, M.~R.~W., \& {Thaddeus}, P. 1996, \aap, 315, 578

\bibitem[{{Park} \& {Sung}(2002)}]{2002AJ....123..892P}
{Park}, B.-G., \& {Sung}, H. 2002, \aj, 123, 892

\bibitem[{{Racine}(1968)}]{1968AJ.....73..233R}
{Racine}, R. 1968, \aj, 73, 233

\bibitem[{{Reich} {et~al.}(1997){Reich}, {Reich}, \&
  {Furst}}]{1997A&AS..126..413R}
{Reich}, P., {Reich}, W., \& {Furst}, E. 1997, \aaps, 126

\bibitem[{{Reid} {et~al.}(2014){Reid}, {Menten}, {Brunthaler}, {Zheng}, {Dame},
  {Xu}, {Wu}, {Zhang}, {Sanna}, {Sato}, {Hachisuka}, {Choi}, {Immer},
  {Moscadelli}, {Rygl}, \& {Bartkiewicz}}]{2014ApJ...783..130R}
{Reid}, M.~J., {Menten}, K.~M., {Brunthaler}, A., {et~al.} 2014, \apj, 783, 130

\bibitem[{{Rosolowsky}(2005)}]{2005PASP..117.1403R}
{Rosolowsky}, E. 2005, \pasp, 117, 1403

\bibitem[{{Shan} {et~al.}(2012){Shan}, {Yang}, {Shi}, {Yao}, {Zuo}, {Lin},
  {Chen}, {Zhang}, {Duan}, {Cao}, {Li}, {Li}, {Liu}, \&
  {Zhong}}]{2012ITTST...2..593S}
{Shan}, W., {Yang}, J., {Shi}, S., {et~al.} 2012, IEEE Transactions on
  Terahertz Science and Technology, 2, 593

\bibitem[{{Sharpless}(1953)}]{1953ApJ...118..362S}
{Sharpless}, S. 1953, \apjs, 118, 362

\bibitem[{{Sharpless}(1959)}]{1959ApJS}
---. 1959, \apjs, 4, 257

\bibitem[{{Solomon} {et~al.}(1987){Solomon}, {Rivolo}, {Barrett}, \&
  {Yahil}}]{1987ApJ...319..730S}
{Solomon}, P.~M., {Rivolo}, A.~R., {Barrett}, J., \& {Yahil}, A. 1987, \apj,
  319, 730

\bibitem[{{Sota} {et~al.}(2011){Sota}, {Ma{\'\i}z Apell{\'a}niz}, {Walborn},
  {Alfaro}, {Barb{\'a}}, {Morrell}, {Gamen}, \& {Arias}}]{2011ApJS..193...24S}
{Sota}, A., {Ma{\'\i}z Apell{\'a}niz}, J., {Walborn}, N.~R., {et~al.} 2011,
  \apjs, 193, 24

\bibitem[{{Storchi-Bergmann} {et~al.}(2012){Storchi-Bergmann}, {Riffel},
  {Riffel}, {Diniz}, {Borges Vale}, \& {McGregor}}]{2012ApJ...755...87S}
{Storchi-Bergmann}, T., {Riffel}, R.~A., {Riffel}, R., {et~al.} 2012, \apj,
  755, 87

\bibitem[{Su {et~al.}(2016)Su, Sun, Li, Zhang, Zhou, Fang, Yang, \&
  Chen}]{Su_2016}
Su, Y., Sun, Y., Li, C., {et~al.} 2016, The Astrophysical Journal, 828, 59

\bibitem[{{van den Bergh}(1966)}]{1966AJ.....71..990V}
{van den Bergh}, S. 1966, \aj, 71, 990

\bibitem[{{Watson} {et~al.}(2010){Watson}, {Hanspal}, \&
  {Mengistu}}]{2010ApJ...716.1478W}
{Watson}, C., {Hanspal}, U., \& {Mengistu}, A. 2010, \apj, 716, 1478

\bibitem[{{Watson} {et~al.}(2008){Watson}, {Povich}, {Churchwell}, {Babler},
  {Chunev}, {Hoare}, {Indebetouw}, {Meade}, {Robitaille}, \&
  {Whitney}}]{2008ApJ...681.1341W}
{Watson}, C., {Povich}, M.~S., {Churchwell}, E.~B., {et~al.} 2008, \apj, 681,
  1341

\bibitem[{{Williams} \& {McKee}(1997)}]{1997ApJ...476..166W}
{Williams}, J.~P., \& {McKee}, C.~F. 1997, \apj, 476, 166

\bibitem[{{Wouterloot} {et~al.}(1990){Wouterloot}, {Brand}, {Burton}, \&
  {Kwee}}]{1990A&A...230...21W}
{Wouterloot}, J.~G.~A., {Brand}, J., {Burton}, W.~B., \& {Kwee}, K.~K. 1990,
  \aap, 230, 21

\bibitem[{{Wright} {et~al.}(2010){Wright}, {Eisenhardt}, {Mainzer}, {Ressler},
  {Cutri}, {Jarrett}, {Kirkpatrick}, {Padgett}, {McMillan}, {Skrutskie},
  {Stanford}, {Cohen}, {Walker}, {Mather}, {Leisawitz}, {Gautier}, {McLean},
  {Benford}, {Lonsdale}, {Blain}, {Mendez}, {Irace}, {Duval}, {Liu}, {Royer},
  {Heinrichsen}, {Howard}, {Shannon}, {Kendall}, {Walsh}, {Larsen}, {Cardon},
  {Schick}, {Schwalm}, {Abid}, {Fabinsky}, {Naes}, \&
  {Tsai}}]{2010AJ....140.1868W}
{Wright}, E.~L., {Eisenhardt}, P.~R.~M., {Mainzer}, A.~K., {et~al.} 2010, \aj,
  140, 1868

\bibitem[{{Xu} {et~al.}(2017){Xu}, {Xu}, {Yu}, {Zhang}, {Liu}, {Wang}, {Ning},
  {Ju}, \& {Zhang}}]{2017ApJ...849..140X}
{Xu}, J.-L., {Xu}, Y., {Yu}, N., {et~al.} 2017, \apj, 849, 140

\bibitem[{{Zhang} {et~al.}(2016){Zhang}, {Li}, {Wyrowski}, {Wang}, {Yuan},
  {Xu}, {Gong}, {Yeh}, \& {Menten}}]{2016A&A...585A.117Z}
{Zhang}, C.-P., {Li}, G.-X., {Wyrowski}, F., {et~al.} 2016, \aap, 585, A117

\end{thebibliography}


\end{document}